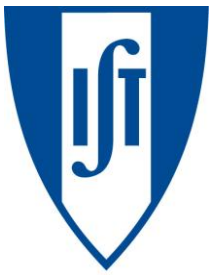

# UNIVERSIDADE TÉCNICA DE LISBOA
# INSTITUTO SUPERIOR TÉCNICO

INSTITUTO SUPERIOR TÉCNICO

# User-Sensitive Mobile Interfaces: Accounting for Individual Differences amongst the Blind

## TIAGO JOÃO VIEIRA GUERREIRO

**Supervisor:** Doctor Daniel Jorge Viegas Gonçalves
**Co-Supervisor:** Doctor Joaquim Armando Pires Jorge

Thesis approved in public session to obtain the PhD Degree in Information Systems and Computer Engineering

JURY FINAL CLASSIFICATION: **PASS WITH DISTINCTION**

**Jury**

**Chairperson:** Chairman of the IST Scientific Board
**Members of the Committee:**
> Doctor Joaquim Armando Pires Jorge
> Doctor Luís Manuel Pinto da Rocha Afonso Carriço
> Doctor Simon Harper
> Doctor João Manuel Brisson Lopes
> Doctor Daniel Jorge Viegas Gonçalves

**2012**

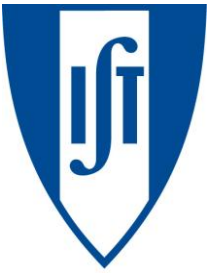

# UNIVERSIDADE TÉCNICA DE LISBOA
# INSTITUTO SUPERIOR TÉCNICO

INSTITUTO
SUPERIOR
TÉCNICO

## User-Sensitive Mobile Interfaces: Accounting for Individual Differences amongst the Blind

### TIAGO JOÃO VIEIRA GUERREIRO

**Supervisor:** Doctor Daniel Jorge Viegas Gonçalves
**Co-Supervisor:** Doctor Joaquim Armando Pires Jorge

Thesis approved in public session to obtain the PhD Degree in Information Systems and Computer Engineering

JURY FINAL CLASSIFICATION: **PASS WITH DISTINCTION**

### Jury

**Chairperson:** Chairman of the IST Scientific Board
**Members of the Committee:**

Doctor Joaquim Armando Pires Jorge, Professor Catedrático, Instituto Superior Técnico, Universidade Técnica de Lisboa

Doctor Luís Manuel Pinto da Rocha Afonso Carriço, Professor Associado, Faculdade de Ciências, Universidade de Lisboa

Doctor Simon Harper, Senior Lecturer, School of Computer Science, University of Manchester

Doctor João Manuel Brisson Lopes, Professor Auxiliar, Instituto Superior Técnico, Universidade Técnica de Lisboa

Doctor Daniel Jorge Viegas Gonçalves, Professor Auxiliar, Instituto Superior Técnico, Universidade Técnica de Lisboa

### Funding Institutions
Fundação para a Cência e a Tecnologia (SFRH / BD / 28110 / 2006)

**2012**

# Abstract

Mobile phones pervade our daily lives and play ever expanding roles in many contexts. Their ubiquitousness makes them pivotal in empowering disabled people. However, if no inclusive approaches are provided, it becomes a strong vehicle of exclusion. Even though current solutions try to compensate for the lack of sight, not all information reaches the blind user. Good spatial ability is still required to make sense of the device and its interface, as well as the need to memorize positions on screen or keys and associated actions in a keypad. Those problems are compounded by many individual attributes such as age, age of blindness onset or tactile sensitivity which often are forgotten by designers. Worse, the entire blind population is recurrently thought of as homogeneous (often stereotypically so). Thus all users face the same solutions, ignoring their specific capabilities and needs.

We usually ignore this diversity as we have the ability to adapt and become experts in interfaces that were probably maladjusted to begin with. This adaptation is not always within reach. Interaction with mobile devices is highly visually demanding which widens this gap amongst blind people. It is paramount to understand the impact of individual differences and their relationship with demands to enable the deployment of more inclusive solutions.

We explore individual differences among blind people and assess how they are related with mobile interface demands, both at low (e.g. performing an on-screen gesture) and high level (text-entry) tasks. Results confirmed that different ability levels have significant impact on the performance attained by a blind person. Particularly, otherwise ignored attributes like tactile acuity, pressure sensitivity, spatial ability or verbal IQ have shown to be matched with specific mobile demands and parametrizations. This confirms the need to account for individual characteristics and provide space for personalization and adaptation, towards inclusive design.

# Resumo

Os dispositivos móveis permeiam as nossas vidas diárias e têm papéis em expansão em variados contextos. A sua omnipresença torna-os fundamental na inclusão das pessoas com deficiência. No entanto, se abordagens inclusivas não forem fornecidas, tornam-se um veículo de exclusão. Existem soluções que procuram compensar a ausência de retorno visual mas, ainda assim, nem toda a informação chega ao utilizador cego. Boa habilidade espacial é ainda necessária para perceber a disposição do dispositivo. É ainda necessário memorizar posições no ecrã e acções associadas no teclado. Pior, a população é normalmente vista como sendo homogénea (de uma forma estereotipada). Todos os utilizadores são confrontados com as mesmas soluções ignorando a sua diversidade de capacidades e necessidades. Esta diversidade é normalmente ignorada porque o utilizador comum tem a capacidade de se adaptar e tornar-se proficiente em interfaces inicialmente desajustadas.

Esta adaptação nem sempre está ao alcance. A interacção com dispositivos móveis é altamente exigente a nível visual o que aumenta a disparidade entre pessoas que não recebem essa informação. É fulcral perceber o impacto das diferenças invidiuais e a sua relação com as exigências das interfaces móveis rumo ao desenvolvimento de soluções inclusivas. Nesta dissertação exploramos diferenças entre pessoas cegas e avaliamos como estas estão relacionadas com exigências de interfaces móveis, tando em tarefas de baixo nível (ex: realizar um gesto no ecrã) como alto nível (introdução de texto).

Os resultados confirmaram que os diferentes níveis de habilidade têm um impacto significativo no desempenho e que este impacto está relacionado com as diferentes exigências dos métodos e dispositivos. Essas variações confirmam a necessidade de contemplar características individuais e oferecer espaço para a personalização e adaptação, fomentando o desenho inclusivo. Os resultados obtidos são destilados sobre a forma de implicações para o desenho de interfaces sensíveis ao utilizador.

## Palavras Chave

Cegueira, Diferenças Individuais, Interfaces Móveis, Capacidades, Exigências, Dispositivos sensíveis ao Toque

# Acknowledgements

The last five years of my life were 'Legen........wait for it.....Dary!'. I started my PhD with such great plans which have not all come to a reality but have in several other aspects surpassed my expectations. I thought a lot, argued with others and myself, discussed ideas, collaborated, observed much, wrote and shared all I saw worth sharing. Independently from the final outcome and results achieved in this dissertation, I feel a better researcher and more fulfilled individual. It has been a hard but definitely rewarding path.

During the last half of a decade (it seems like an eternity put up like that), I married and had the greatest joy of my life, the birth of my daughter, Beatriz. This life event was sort of half-time in the big PhD game. It changed my perspective and way of facing life and my research in particular. I still do research passionately (actually I am even more passionate about what I do now that in the beginning) but I feel more balanced. Filipa and Beatriz supported me throughout the last few years and they are a beacon I can come back to at the end of the day. During the writing of this document, when strengths started to fail to write yet another paragraph, a simple look at their photos on my desk gave me the strength to carry on. My first thanks goes to them simply for being awesome.

The immense support I had from several different people can only be surpassed by the one I had from my parents and brother. For what is worth, this thesis is dedicated to them as the success I achieved and will try to continue achieving in the several vectors of my life is built upon the foundations they have helped me build. I could always count on their support and this still happens now that I am all grown up. The best way I have to repay them is to be the best I can and reach the highest. My achievements are yours.

Passing on to the work presented in this thesis, my first word of appraisal goes to my advisers. My relationship with both started some years before my PhD and the first thing to say is that I already saw them as friends before this joint travel has started. I have learned a lot with both and not only pertaining the context of the thesis but even more in what concerns more pervasive aspects and traits of life and research. I have no doubts in saying that they both were pivotal in shaping the researcher I have become. If I did not turn out the researcher I should, it is your fault :) ! Joaquim Jorge was my first role model in research. I recognize in me his motivation, ambition and continuous thirst for knowledge. I also believe that he was the one responsible for my decision to follow academia and research, particularly in the accessibility field. It is obvious for me now that I should not do anything else. Thank you. Daniel Gonçalves has been much more than an adviser. I think it is an honour to be able to work with my adviser as a colleague with whom I can share all my thoughts and ideas but also have the confidence to share my doubts and questions. A PhD can be a lonely thing: mine was not. I have improved and gained skills way beyond traditional research with Daniel. Thank you for everything. I truly hope to be able to continue collaborating with both in the projects the future will bring as I am sure that they will continue to be enriching experiences.

What one learns from doing a PhD goes way beyond what is presented in this document. I have learned so much from the people I shared space with. I would like to express my thanks to several students, colleagues and researchers of the VIMMI research group: Ricardo Gamboa, Paulo Lagoá, Pedro Santana, Jorge Sepúlveda, David Lucas, João Oliveira, Filipe Dias, Vasco Gervásio, Bruno Araújo, Ricardo Jota, Alfredo Ferreira, Manuel J. Fonseca, João Madeiras Pereira, Ricardo Dias, Gabriel Barata, among others. It has been a pleasure to have your company and to share discussions with you. I learned a lot. A special thanks to Manuela Sado for her support with bureaucratic aspects pervading the life of a PhD student.

Paulo Lagoá and João Oliveira, students that I co-advised, played a very important role in the course of my thesis. I feel that the help and collaboration I gave in their master thesis was paid back with interest. Plus, it was comforting to share the burden of my research and these two wonderful students were great at what they were purposed to do. The quality of the research done here has increased with your efforts.

Hugo Nicolau was a pervasive aid during this thesis. He collaborated in some studies but more than that I think we have been shaping and improving each other for a long time now. Hugo started as a student I co-advised but soon became my partner in research, the one I climb the different ladders of knowledge with. This collaboration has brought up the best of both and has improved drastically our abilities in different scopes. This partnership has been intense and that has lead to a great friendship, one that I will cherish forever. Hugo and I share an office with my brother João. It is motivating and fun to be able to work jointly with my brother, and share concerns and challenges on the spot in a daily basis; it could not be better and losing this is what scares me the most in the next step of my life.

This work would not have been possible without the support of institutions and professionals working with blind people. I thank Fundação Raquel and Martin Sain (FRMS), Associação Promotora do Ensino dos Cegos (APEC), Associação Promotora de Emprego de Deficientes Visuais (APEDV), Associação dos Cegos e Amblíopes de Portugal (ACAPO), Direcção Regional de Educação de Lisboa e Vale to Tejo (DRELVT), in particular, Dr. Elvis de Freitas, Dr. Carlos Bastardo, Dra. Vera Rapagão, Dr. Vitor Reino, Dr. Agostinho Costa, Dr. António Pinão, Dr. José Carlos Fernandes, Ana Ferreira, Cristina Palma, Rogério Silva, Dores Pinheiro, Vitor Graça. I also thank to Professor Graça Neves for her support in recruiting professionals working closely with blind people.

I owe a particular acknowledgement to Fundação Raquel and Martin Sain and to all its members, from the administration board to therapists and teachers, along with the students therein who have been the basis for the studies presented herein. I had total availability to perform my studies in the institution and the liberty to use their own evaluation and technological material. I felt at home and a great deal of my work was performed there. Dr. Carlos Bastardo was the most important person to me on the field. We discussed ideas and protocols, he helped with the cognitive evaluation assessments and gave me access to his own evaluation equipment. He was pivotal in the recruitment process making all possible participants feel engaged with the studies. I wouldn't be able to present studies with similar numbers and significant if it wasn't for Carlos. My deepest thanks!

I thank all participants for their availability and for teaching me some much. I will continue to work with blind people in the future as I am continuously surprised with their varying abilities and coping mechanism but also with the immense barriers they still face. Hope to be able to help and give back the attention and time you *lost* with me.

Last but definitely not least, I thank my thesis committee (Luis Carriço and João Brisson Lopes), for their valuable comments in specific occasions in my work timeline. You helped me to gain focus but still maintaining my goals, with a very positive accompaniment. I also thank Luis Carriço for his continuous motivating words for me to finish my thesis. I appreciated the support. Finally, I was delighted to have Simon Harper as my international contestant. In the last sprint, he showed tremendous flexibility and availability and that was paramount for me to cope with my expected calendar. I thank Simon Harper and Luis Carriço for the comprehensive and insightful comments to my provisional thesis and I am sure its quality has improved with the changes suggested.

I would like to thank the institutions that allowed me to perform my work. Particularly, my host institution, INESC-ID, for giving me the conditions to pursue my research goals, and FCT, for providing the 4-year scholarship that allowed me to focus on my PhD: this work was supported by FCT, grant SFRH/BD/28110/2006.

*To my Parents.*

# Contents

# List of Figures

xi

# List of Tables

*If we cannot now end our differences, at least we can help make the world safe for diversity.*

*— John Fitzgerald Kennedy*

# Statement of Collaboration

I am the primary contributor to all aspects of this research which was performed under the supervision of Doctor Daniel Gonçalves and Doctor Joaquim Jorge. Several master students assisted with the execution of this research, always under my close guidance and co-supervised with my advisers.

Paulo Lagoá collaborated in the development of the mobile system deployed for the long term evaluation, presented in Chapter 3. He developed NavTap. Pedro Santana came up with the raw idea for the vowel navigation system used in NavTap and NavTouch.

João Oliveira contributed on the study design (cognitive component) and co-ran the studies on Touch Typing (Chapter 7). Hugo Nicolau collaborated in the early stages of this research (Chapter 3), collaborated in the development of the mobile system for the long-term study, developed the NavTouch prototype, and has contributed with discussions and ideas throughout this research.

# 1

# Introduction

Mobile devices play an important role in current society. Their relevance goes far beyond the initial purpose of enabling mobile telephonic communication. They are increasingly seen as extensions of one's body [Townsend, 2000] and empower the user in several different ways, from leisure to productivity and both synchronous and asynchronous communication.

Given this relevance and the pervasive usage of mobile devices by one's social and professional networks, the inability to effectively operate such a tool is likely to be a strong vehicle of exclusion. Counteracting its initial purpose of improving communication among people anytime and anywhere, the widespread but non-inclusive usage of mobile phones is likely to widen the gap between disabled and non-disabled people. Conversely, if access to these tools is provided they can be empowering to a disabled person improving her communication, productivity, cultural and leisure opportunities. In the overall, they can improve one's independent living.

This awareness has brought and maintained mobile accessibility in the research agenda from the early keypad-based devices to the current touch screen interfaces. We can observe an effort to maintain mobile solutions accessible to disabled groups. In particular, we focus our attention on blind people. Mobile interfaces are extremely visual and this group faces several challenges in using them. Mobile user interfaces are designed to fit a common user model, shaped with a few adaptable and adaptive mechanisms, which are mostly aesthetic. However, no two persons are alike. We can usually ignore this diversity as we have the ability to adapt to the devices and, without noticing, become experts in interfaces that were probably maladjusted to begin with. However, this adaptation is not always within the user's reach.

Screen reading software was deployed in early keypad-devices as a replacement to the visual information presented on screen. However, several pieces of information are lost in this visual-audio replacement as happens with the layout of the keypad or the attribution of alphanumeric characters to keys.

With the advent of touch screen devices, the problems felt by blind people increased. While a blind person is likely to be able to interact with a keypad-based phone to place a call without the need for any assistive technology, it would be a herculean task to do so with today's touch screen devices. The magnitude of this problem increases as we load the screen with interface elements, as happens with text-entry interfaces, where all letters are placed onscreen. Assistive screen reading software, like Apple's VoiceOver[1], enables a blind person to overcome these issues by offering auditory feedback of the visual elements onscreen. Still, as aforementioned, mobile interfaces are extremely visual and a large amount of information is lost in this visual-audio replacement. Possible examples are the need of a good spatial ability to have a notion of the device and the interface components therein, or cognitive capabilities to memorize letter placement on screen. Visual feedback makes these attributes dispensable or less pertinent, while its absence makes them relevant and worthy of consideration. Furthermore, there is a lack of knowledge on which abilities play a relevant role and which should be considered.

Besides personality differences, two blind users are likely to have totally different stories to what blindness, and its implications, is concerned. The cause of the impairment, age of blindness onset, age, cognitive or tactile abilities, are some examples of the characteristics that may diverge between users. A young 'recent-blind' is different from an older one. While the former is likely to have all his other senses immaculate, the latter may have some other age-related impairments. However, he is also likely to have developed sensory compensation mechanisms [Burton, 2003]. How are they different and how will those differences affect their functional ability?

The enormous diversity found among these particular group of users makes the *"stereotypical blind"* concept inadequate. Regardless, all are presented with the same methods and opportunities ignoring their abilities and needs. Moreover, interaction with mobile devices is highly visually demanding which increases the difficulties. Even mobile assistive technologies for blind people have a narrow and stereotypical perspective over the difficulties faced by their users. A blind user is presented with screen reading software to overcome the inability to see on-screen information. However, these solutions go only half-way as the responsibility to adapt and ensure enough knowledge to operate the device and its interface is all translated to the user's end (e.g, people still have to memorize letter placement on the keypad, people still need to have a notion of the icon position on a touch application to be able to search for it). In the absence of sight other aptitudes/limitations stand up.

---

[1]http://www.apple.com/accessibility/voiceover/

To empower these users and foster inclusion, a deeper understanding of their abilities and how they relate with technology and its demands is mandatory. In this dissertation, we stressed the relationship between individual abilities and interface demands.

The thesis of the dissertation is:

> *Inclusive mobile user interfaces for blind people should be designed considering the individual differences that define the user within mobile interaction contexts.*

In the research towards the aforementioned thesis, we were able to reveal obstacles to the blind population as a whole and matchings between abilities and demands. Otherwise ignored individual attributes (in the mobile interaction setting) as were pressure sensitivity, tactile acuity, spatial ability, verbal IQ, blindness age of onset showed to be related with the proficiency level a blind person attains with mobile interfaces. Furthermore, they showed to be matched with particular mobile demands (e.g., landing on a target with spatial ability). In the overall, the studies presented in this dissertation showed that one single blind person presents measurable characteristics that determine how this person is able to overcome the demands imposed by a particular interface. This relationship between abilities and demands has shown to occur both in simple lower-level tasks (e.g., acquire a target) as in higher-level tasks (touch typing). These findings reveal that indeed further attention to variations within blind people is due to foster inclusion.

## 1.1. Scope

This dissertation explores the individual abilities' discrepancy within blind people and how those differences are situated in relation to interface demands. In particular, we explore how the demands imposed by different mobile device settings are surpassed by people with different levels of ability.

We focus our research on a particular target group: blind people. The reason for this focus is threefold: 1) mobile interfaces have a high visual component and blind people are severely damaged in their inclusion in the society; 2) mobile interfaces are mostly static (designed for the average human) and solutions for blind people are also stereotypical disrespecting the differences within the population thus leading to sub-optimal user experiences and, in some cases, exclusion; and 3) current mobile devices and operating systems detain the characteristics and possibilities to adapt and simultaneously cope with the lack of sight and a wide range of abilities.

The research presented in this dissertation has been performed in close collaboration with a handful of Portuguese organizations situated in Lisbon and several people working with blind people. In particular, in the last four years, I have been a weekly attendant to the Raquel and Martin Foundation, where the majority of the studies reported in this dissertation took place. This close experience proved immeasurably valuable as what started as the attempt to provide alternatives to the stereotypical blind user, which was, looking retrospectively, a naive approach, slowly was transformed into an enriching lesson about the idiosyncrasies and diversity within the population. The acknowledgement of such diversity shifted our focus from seeking unattainable ultimate solutions to understanding the gaps in the design space and therefore the reasons behind exclusion. Our main goal is to understand and quantify device demands and relate each with a set of required levels of ability. Ultimately, this approach enables us to quantify the inclusion of an interface and, through another perspective, as an ophthalmologist prescribes a pair of glasses, identify the most adequate interfaces for a single user.

Our work started with a focus on keypad-based devices. Results obtained from preliminary studies with the population called our attention to people with specific characteristics who were capable in so many ways but excluded due to particular insufficiencies in ability. These are reported in Chapter 3. We were witnesses to the advent of touchscreen devices and their proliferation in the market. We were also witnesses to the fear expressed by many blind people to the daunting tactile-less mobile future. On the other hand, these devices pose way more opportunities for interface adaptation to cope with individual traits. We focused the remaining of our research efforts in the exploration of mobile touch interfaces and variations within and how to design interfaces to cope with different users' abilities and needs.

Our preliminary research studies focused on text-entry both due to the recognized difficulties associated with the large number of input options and its pervasiveness (asynchronous communication, note taking, managing contacts and agenda,...).We revisit this scenario throughout the dissertation to evaluate the high-level validity of our conclusions.

### 1.1.1. Terminology

Blindness can be informally defined as the inability to see. However, the heterogeneous character of the blind and visually impaired group as well as the difficulty to define what degree of vision corresponding to *inability to see*, calls out for the need to clarify the concepts used herein [Cavender et al., 2008]. In this dissertation, I use the term *blind people* to refer to people that have at most light perception. These have a visual impairment that reduces their ability to access a computer or a mobile device. As a consequence, they use a screen reader to access a computer. People with a specific type of visual impairment, other impairments or with specific abilities/needs will be described when necessary.

An early blind or congenitally blind has been blind from birth (or lost sight in the first years of life), while those who have lost sight later in life are called adventitious blind or late blind [Hollins and Leung, 1989]. The attitude towards blindness as well as space representation may be affected by the age of onset of blindness [Levesque, 2005].

In the next section, I detail why looking at individual abilities within the blind population is required and how this approach is likely to provide the long required opportunities for the blind population as a whole (even if with an individual user-sensitive approach).

## 1.2. Motivation for our Approach

In this dissertation, we provide an in-depth analysis of the usage of different mobile device settings by blind people with different profiles, tactile, cognitive and functional abilities. Our focus is thus divided by dissecting the demands imposed by different devices, methods, layouts, primitives, and to explore the range of individual abilities found within the population and assert the relationships with the former.

One motivating factor for our research seems unquestionable: increasing the accessibility of such devices is a major benefit for the social and professional inclusion of blind people. In Portugal, the number of mobile devices (estimated in 12,1 millions in the first trimester of 2012) exceeds the population [2]. All around the world, even in underdeveloped countries where computers or wired telephony are hard to find, mobile devices (we will use this broader term from now on to include mobile phones, personal digital assistants and smartphones) are quite common [Wobbrock, 2006]. Mobile devices work as portable computers maintaining a connection between the user and his professional and social network at all times. This presents opportunities at the broader social, professional, leisure and cultural levels but can also be seen as a tool for healthcare, security and emotional well-being. Missing out on such opportunities is to be avoided.

Questions may arise whether mobile devices and their interfaces are not already accessible to blind people and why we place such a strong focus on the individual, an approach that *supposedly*[3] goes in the opposite direction as Universal Accessibility [Stephanidis, 2001] methodologies. Further, one might question why are individual differences among blind people dissimilar from the individual differences found between every sighted human.

Based on that, our research is then motivated by the following factors which answer the aforementioned questions: first, that mobile user interfaces are very strict in general and even assistive solutions are stereotypical which by turn translates into frustration and exclusion for a portion of the population; second, that individual differences among blind people are particularly relevant due to the absence of such an integrating sense along with interfaces designed for a two-dimensional consumption; and, third, that the mobile market is diverse and each mobile device presents the capabilities to recognize and adapt to different blind people, enabling accessibility and fostering inclusion.

---

[2]ANACOM - **http://www.anacom.pt/render.jsp?contentId=955213**, Last visited in 13/07/2012

[3]Ultimately our main goal is to provide accessibility for all. However, we consider that this goal is to be achieved with an user-sensitive approach, one that understands and deals with the idiosyncrasies lingering within the population.

## 1.2.1. Mobile Interfaces are Maladjusted to Blind People

Placing and receiving a call seem to be the tasks that a blind person, with no other severe impairment, can perform without the support of additional hardware or/and software. Prior to the work presented in this dissertation, we have encountered blind people which used their mobile keypad phones without any assistive technology or adaptation. They would place calls to memorized numbers and receive calls without knowing beforehand who was the caller. Some people had fast dial keys to access contacts where they would have a memorized list of less than a handful of contacts. In those cases, they would navigate between contacts by using the joypad and resorting to their memory to find the desired contact in the list [Lagoá et al., 2007].

The number of remaining options available on a mobile phone led to an endless set of screens which in turn are increasingly spatial and visual. The fast evolution we have been witnessing concerning mobile devices has been accompanied by an effort to develop solutions that fit the blind user and give him/her options beyond simple voice communication. The first approaches to accessible mobile computing featured special hardware Braille-based solutions (e.g., Braillino [4], Alva Mobile Phone Organizer [5], among others [MacKenzie and Tanaka-Ishii, 2007]). These assistive technologies have a strong stereotypical connotation as they replace commodity devices with Braille-based alternatives which imply a two-hand and hardly mobile usage, besides the additional costs (Figure 1.1). Indeed, these solutions have proved to be beneficial but only for specific professional settings (e.g., note-taking). Softer hardware adaptations relied on the decrease of possible task options and simplification of the device to lower the financial load (Figure 1.1). These approaches failed due to two main factors: 1) one does not want to be more comfortable in doing the same limited set of tasks that one can already perform with mainstream devices. Blind people also want the full package and inclusion in the current society depends on it; 2) a blind person wants to use the same devices as her sighted peers [Shinohara and Wobbrock, 2011, Kane et al., 2009].

These factors have been determinant in the overwhelming acceptance of screen reading software on commodity devices. Mainstream keypad-based devices were equipped with such assistive software layers and soon they became pervasive. One must say that accessibility to mobile devices by blind people seems to have peaked with the usage of 12-key keypad devices along with screen reading software. Still, several limitations were still imposed to blind people as a whole and to particular individuals. In our studies, reported in Chapter 3 and then again in Chapter 5 difficulties and inabilities in operating mobile devices and particular tasks were visible. Examples are the need to memorize letter placement on the keypad and dealing with spatial references to keys and actions (e.g., fast access keys near the bottom of the screen to activate special options).

---

[4]http://www.natiq.com/en/node/40
[5]http://www.indexbrailleaccessibility.com/products/alva/mpo.htm

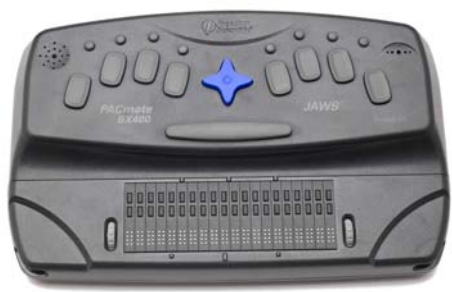  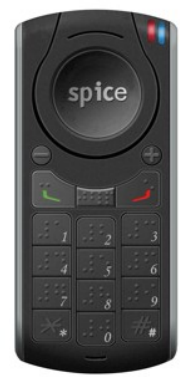

(a) PACmate BX400                          (b) Spice Braille Phone

Figure 1.1: Adapted mobile technologies for blind people: Braille note taker and Braille-based phone

The advent and fast proliferation of touch screen devices has presented new barriers to blind people. These gadgets have been considered inaccessible to a visually impaired person for several years and are still a challenge and a menace to a blind person in several settings that go beyond the mobile communication context (e.g., touch screen kiosks, flat panel microwaves among others) [Carew, 2009]. The launch of Apple's VoiceOver in the 3GS version (June, 2009) of the iPhone is a mark to what concerns mobile touch screen accessibility. From this date on, several blind people have started using an iPhone with access to a similar set of applications they would with a keypad counterpart. However, VoiceOver was also applied as a layer that is strict in layout, icons' positions and gesture directions, which is sometimes maladjusted to a non-visual input and to a single dimensional output channel (audio).

A successful example of VoiceOver's benefits is touch typing. However, as we will show in Chapter 7 the need to explore a flat panel in search for the desired key, the so-called *painless exploration*, showed to be other than painless to some participants who may lack the spatial or cognitive abilities to deal with such demand. This leads us to our next motivation factor: not only these interfaces were built with disregard for the needs of the population as a whole, as they do not give space for differences within the population.

## 1.2.2. The Paramount Role of Individual Abilities among Blind People

One might argue that the diversity we claim to exist among blind people is present among the sighted population. We acknowledge that it may be true. What we state is that the differences among blind people are likely to have a deeper effect than among sighted ones. This happens as vision dominates our current use of mobile technology, particularly, when considering mobile touch devices.

Differences may be observed in tactile, cognitive, or even in fine motor abilities, like dexterity[6]. A sighted person may experience some of the difficulties faced by blind people when trying to use a mobile device while texting on the move. In this scenario, one where the user is situationally impaired [Sears et al., 2003], the visual and cognitive resources are shared between the texting task and following the desired path without hitting other people and obstacles. Other situational contexts may happen where we see ourselves deprived from our tactile (due to cold weather [Theakston, 2007]) or visual (glare onscreen) resources.

A blind person is always deprived from receiving the information onscreen and inscribed on the device. What has been failed to acknowledge is that this calls up to other abilities, which by turn may also be limited. Figure 1.2 presents the device of a participant of our studies. This person acquired blindness due to diabetic retinopathy, a common cause, which came along with a drastic reduction in peripheral sensitivity, particularly a decrease in tactile abilities. The user had difficulties in identifying by touch the correct orientation of the device. The pendulum helps with that task. However, as long as the device is in the correct orientation, by resorting to fine spatial abilities along with training and experience, she can press the keys proficiently.

Our preliminary experiments with text-entry alternatives have shown that people with low tactile abilities could benefit from a rearrangement of the key-function attribution (presented in Chapter 3). Looking again at touch typing, our own studies presented in Chapter 7, have shown that interfaces can be designed to meet the user's abilities and requirements.

Different individual abilities are stressed in different device settings and underlining demands. There is a vast selection of devices and interface layouts and primitives along with the freedom to adapt these to the users. The challenge is to understand when and how to prescribe and adapt.

---

[6]Differences may be more severe. It is relevant to notice that in this dissertation we are focusing on blindness as a severe disability. Other differences in ability should be considered within a low range spectrum.

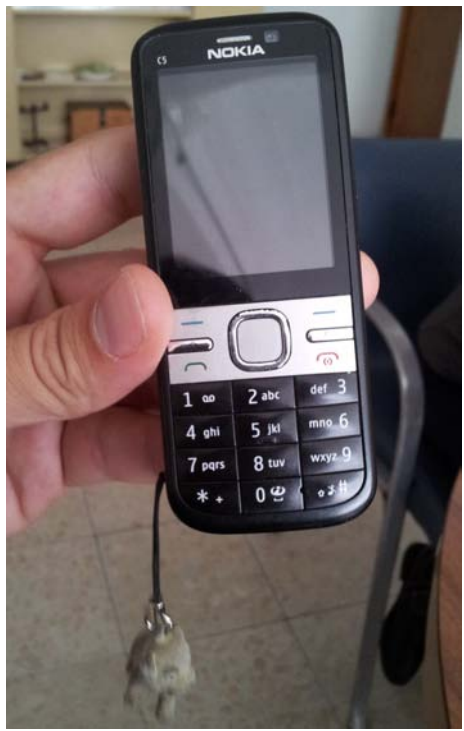

Figure 1.2: The mobile phone from a blind person with low tactile abilities equipped with a pendulum to ease identification of the correct device orientation.

### 1.2.3. Mobile Devices and Interfaces can be Made Accessible to Blind People

It is insufficient to state that mobile devices are important in our daily lives. Indeed, the aforementioned market penetration states that the functionalities provided by these devices are essential. As a matter of fact, and only with a few years of *experience*, it is hard to remember how some tasks and interactions were accomplished without mobile devices.

Mobile communication and the will of the society for its promised values have fasten the deployment of new models and even new mobile phone generations. In the last two decades, we have witnessed and continue to be spectators to an extraordinary evolution on mobile technology. New models, from different manufacturers are deployed in a regular basis and new applications and possibilities are difficult to process as new models with new capabilities appear giving no chance to explore intermediate models. However, this need for constant evolution also has its flaws. Although technological advances have been made, the begotten mobile user interfaces have not been subject of the required attention.

Figure 1.3 presents a wide set of keypad devices once available in the market. They vary in their aesthetics and technical description but also in other features like button size, relief, contrasts, colors, spacing, material, among others. However, there is an overall lack of knowledge on the variabilities and suitability of each design. Our own study on mobile keypad demands and suitability for different blind people [Guerreiro et al., 2011] asserts that there are severe differences between the demands imposed by slightly different keypad variations (e.g., key relief, spacing and size). Once again, these results point out that the main problem does not rely on the absence of suitable characteristics but yet on the lack of knowledge pertaining the characteristics that affect user performance.

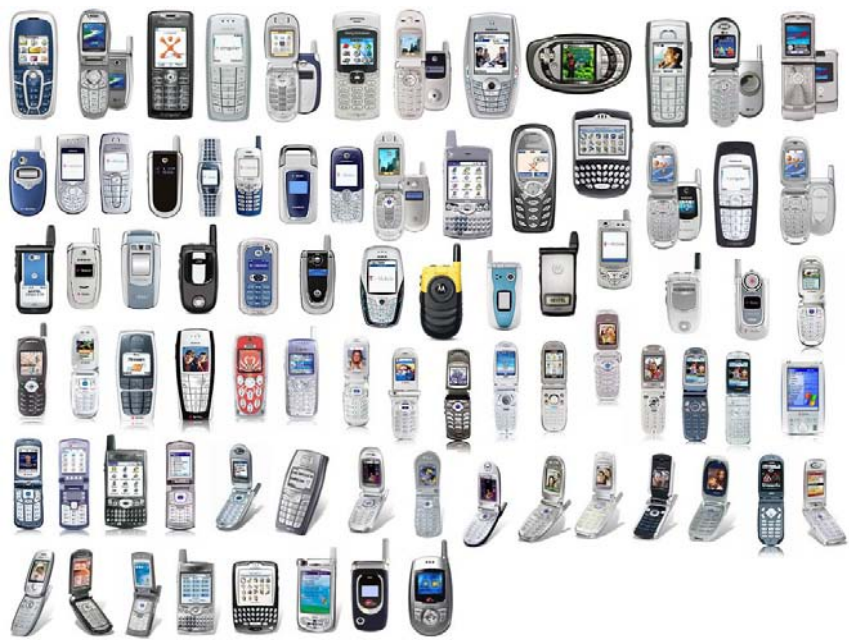

Figure 1.3: Several different mobile devices and interfaces

In the last few years, new devices have arisen. Similarly to the early approaches, also these devices have presented themselves quickly and with few studies about their effectiveness and on how to improve their usability, particularly by disabled populations. It is the case of touch screen devices.

Touch screens are increasingly replacing traditional keypads. These interfaces offer several advantages over their button-based counterparts. Particularly, they can easily display different interfaces in the same surface (e.g. 12-key keypad, QWERTY keyboard) or adapt to users' preferences and capabilities [Kane et al., 2008b]. The high customization degree of touch screens makes them amenable to custom-tailored or adaptive solutions that better fit each user's needs. However, touch screen interfaces also present challenges for mobile accessibility: they lack both the tactile feedback and physical stability guaranteed by keypads, making it harder for people to accurately select targets. Moreover, the aforementioned ability to display different interfaces may also be a strong drawback as the user is unable to predict what is presented and where.

Overall, and although mobile devices are indisputable useful, there is still the need to understand the best way to operate them depending on context, either individual or situational. Only recently, a few researchers have started to lean over the details of personalization and adaptation of mobile touch screen interfaces. This effort has been made in the context of situational disabilities (SIID [Sears et al., 2003]) to cope with interface adaptation [Kane et al., 2008b] or typing patterns [Goel et al., 2012] while walking and to cope with typing styles in general [Findlater et al., 2011, Findlater and Wobbrock, 2012].

In the particular case of blind people, [Kane et al., 2011] have presented the only study to our knowledge that has focused on assessing the differences between the gestures performed by blind and sighted people on a touch screen device. Results from this study showed that blind people perform gestures differently from their sighted peers but also that their gestures can be recognized if the devices and interfaces are instrumented towards that end.

Our own studies presented in Chapter 6 show that: 1) different devices serve different levels of ability; 2) interfaces within a device can be parametrized to include people with different profiles and levels of tactile, cognitive and functional abilities.

## 1.3. Research Goals

Our main research goal was to *acknowledge and assess the relationships between individual traits and mobile interface demands*. We expect this knowledge to be the basis for the development of a more inclusive design space pertaining blind people and mobile touch screen gadgets.

Our research started with a strong focus on text-entry, as it was perceived by us as one of the main challenges faced by blind people as well as one the more important tools for inclusion in current society (e.g, synchronous and asynchronous communication, productivity, basic contact, task and agenda management). Our first studies pertaining the usage of mobile devices and particularly the input of text with a mobile keypad phone revealed a large discrepancy within the population: some people could input text proficiently resorting to mainstream keypad devices along with screen reading software while others were unable to do so and maintained their usage restricted to placing and receiving calls. Our first research goal was to *deploy less demanding text-entry alternatives to excluded blind people and assess if by lowering the demands imposed they could use such a powerful tool and approximate their more fit blind peers in performance*. We ended up by deploying an entire application redesigned for the needs of such users and performed a long-term study to assess the benefits in their performance and social inclusion.

Working closely with blind people that were unable to use traditional methods revealed differences in ability that were impossible to ignore. From this point on, we started to look differently at these excluded people and engaged in a search for the reasons for that inabilities and differences in performance to occur. This also marks a shift in our research paradigm: we no longer looked at the population as a whole, an average, but started to look with more attention at error bars, outliers and the like.

To be able to look closely at differences that had an impact on user performance and overall ability to manoeuvre a mobile tool we had to have a starting point of attributes that might affect the blind person's ability to do so. As such, another of our research goals was to *identify the individual attributes that had an influence in a blind person's aptitudes to use mobile technology*. Only by knowing which attributes are in place we can put them to a test and assess if the variances found within the population have an impact in their use of technology.

One question that might arise is if attributes and abilities are so divergent between different people or if they can only be observed in extreme cases as the ones reported in the aforementioned long-term study. Consequently, we had to gather a large user sample and *assess the diversity within the population at different levels: profile, tactile, cognitive and functional*, taking into account the set of individual attributes previously identified.

Individual abilities showed to be highly variable between blind people. However, this did not mean that different levels of ability would translate in different performances and particularly, in predictable alterations in performance. Thus, we had to *explore the interaction design space to stress different levels of ability* and assess relationships within. This was done first at a lower level to ease the identification of ability-demand matches.

A complex task like text-entry resorts to a set of user abilities, not just one. Further, one might argue that it depends more on previous experience with mobile or other devices. One last goal was to *verify if the relationships found between low-level primitives and individual abilities were still patent at the higher level*. They were.

The knowledge gathered in the reported experienced places in position to *provide recommendations and implications for the design of user-sensitive mobile user interfaces*, the ultimate of our research endeavours.

## 1.4. Contributions

This dissertation reveals the impact of individual differences among blind people in surpassing different mobile demands. This knowledge lays the groundwork for the creation and adaptation of user-sensitive inclusive mobile interfaces. The work that led to that results also yielded the following contributions:

- *An in-depth characterization of the blind population*, in which both individual and functional attributes were studied. This analysis came to fill an existing gap in the knowledge of the abilities of blind people and how they relate with technology. Results obtained were paramount to demystify the *stereotypes* created around the blind population. This study yields results pertaining a wide set of abilities that can be of reference to the diversity found between blind people.

- *Identification of the individual attributes of a blind person relevant in a technological context*, based on a long-term study with blind people and an interview study with professionals working closely with them. This knowledge enables researchers working with blind people to focus on a reduced but meaningful set of individual attributes that define the users within mobile interaction settings.

- *A conceptual framework relating individual attributes and touch device demands*, where relationships between ability levels and interface variations are revealed. By analysing individual abilities and how levels of ability surpassed different levels of a specific demand we were able to outline existing relationships . Besides the ability-demand match, this framework reveals demands, some of which have been previously ignored in the design of inclusive interfaces.

- *A methodology for inclusive user-sensitive evaluation*, where divergent gamuts of individual abilities and device settings and their underlining demands play a key role. The procedure applied to outline match between abilities and demands in touch primitives and touch typing settings can be replicated for other low-level or applicational settings.

- *A comparative assessment of text-entry methods* and their suitability for different profiles. A variety of methods variable in demand were developed under the auspices of this research. More than the methods themselves, the comparative evaluations performed shed light about which methods should be used for each person.

- *A comprehensive set of implications for the design of user-sensitive mobile user interfaces* showing how researchers can build on top of the results presented in this dissertation towards more inclusive mobile interfaces. In sum, they include **a shift in the research methodologies employed** with greater attention on differences and outliers, learning with past experience **particularly in the design of devices that are regularly giving a step back in their accessibility**, considering **new sets of abilities and demands** and as such design for diversity, giving space for **adaptation and personalization**, as well as recognizing that **different devices should be considered differently** as they, once again, pose different demands. The ultimate implication aggregates all the remaining and **calls for informed diversity as a channel for inclusion**.

# 1.5. Publications

Since the beginning of my doctoral studies, I have co-authored 50+ national and international papers on the intersection between Accessibility and Human-Computer Interaction. All of them contributed to my formation and are somehow, directly or indirectly, connected to the outcomes presented in this document. In this section, I outline only the international ones who present contributions that are strongly connected with the context of this dissertation.

## Journal Papers

1. **Mobile Text-Entry and Visual Demands: Reusing and Optimizing Current Solutions**, *Hugo Nicolau, Tiago Guerreiro, David Lucas, Joaquim Jorge*.Universal Access in the Information Society, Springer. To appear.

2. **From Tapping to Touching: Making touch screens accessible to blind users**, *Tiago Guerreiro, Hugo Nicolau, Paulo Lagoá, Daniel Gonçalves and Joaquim Jorge*. IEEE Multimedia, vol. 15, nr. 4, pp. 48-50, Oct-Dec, 2008

## International Conference Papers with Peer-review

1. **Exploring the Accessibility of Touch Phones and Tablets for Blind People**, *Tiago Guerreiro, Joaquim Jorge, Daniel Gonçalves*. MobileHCI 2012 Workshop on Mobile Accessibility, San Francisco, USA, 09/12

2. **Mobile Text-Entry: the Unattainable Ultimate Method**, *Tiago Guerreiro, Hugo Nicolau, Joaquim Jorge, Daniel Gonçalves*. Pervasive 2012 Workshop on Frontiers in Accessibility for Pervasive Computing, Newcastle, UK, June, 2012

3. **Blind People and Mobile Touch-based Text-Entry: Acknowledging the Need for Different Flavors**, *João Oliveira, Tiago Guerreiro, Hugo Nicolau, Joaquim Jorge, Daniel Gonçalves*. Proceedings of ASSETS 2011 - 13th International ACM SIGACCESS Conference on Computers and Accessibility, Dundee, Scotland, 10/2011

4. **Blind People and Mobile Keypads: Accounting for Individual Differences**, *Tiago Guerreiro, João Oliveira, Hugo Nicolau, João Benedito, Joaquim Jorge, Daniel Gonçalves*. Proceedings of INTERACT 2011 - the 13th IFIP TC13 conference on Human-Computer Interaction. Lisbon, Portugal, 09/11

5. **BrailleType: Unleashing Braille over Touch Screen Mobile Phones**, *Tiago Guerreiro, João Oliveira, Hugo Nicolau, Joaquim Jorge, Daniel Gonçalves*. Proceedings of

INTERACT 2011 - the 13th IFIP TC13 conference on Human-ComputerInteraction, Lisbon, Portugal, 09/11

6. **Understanding Individual Differences: Towards Effective Mobile Interface Design and Adaptation for the Blind**, *Tiago Guerreiro, Hugo Nicolau, João Oliveira, Joaquim Jorge, Daniel Gonçalves*. ACM CHI 2011 Workshop on Dynamic Accessibility: Detecting and Accommodating Differences in Ability and Situation. Vancouver, Canada, 05/11.

7. **Towards Accessible Touch Interfaces**, *Tiago Guerreiro, Hugo Nicolau, Joaquim Jorge, Daniel Gonçalves*. Proceedings of ASSETS 2010 - 12th International ACM SIGACCESS Conference on Computers and Accessibility, Orlando, Florida, USA, 10/2010

8. **Assessing mobile touch screen interfaces for tetraplegics**, *Tiago Guerreiro, Hugo Nicolau, Joaquim Jorge, Daniel Gonçalves*. Proceedings of Mobile HCI 2010: 12th International Conference on Human-Computer Interaction with Mobile Devices and Services, Lisboa, Portugal, 09/2010

9. **The Key Role of Touch in Non-Visual Mobile Interaction**, *João Benedito, Tiago Guerreiro, Hugo Nicolau, Daniel Gonçalves*. Proceedings of Mobile HCI 2010: 12th International Conference on Human-Computer Interaction with Mobile Devices and Services, Lisboa, Portugal, 09/2010

10. **Assessing Mobile-wise Individual Differences in the Blind**, *Tiago Guerreiro*. Mobile HCI 2010 Doctoral Consortium: 12th International Conference on Human-Computer Interaction with Mobile Devices and Services, Lisboa, Portugal, 09/2010

11. **Proficient blind users and mobile text-entry**, *Hugo Nicolau, Tiago Guerreiro, Daniel Gonçalves, Joaquim Jorge*. Proceedings of ECCE 2010: European Conference on Cognitive Ergonomics, ACM DL, Delft, Netherlands, 08/2010

12. **Identifying the individual ingredients for a (in)successful non-visual mobile experience**, *Tiago Guerreiro, Joaquim Jorge, Daniel Gonçalves*. Proceedings of ECCE 2010: European Conference on Cognitive Ergonomics, ACM DL, Delft,Netherlands, 08/2010

13. **NavTap: a Long term study with Excluded Blind Users**, *Tiago Guerreiro, Hugo Nicolau, Joaquim Jorge and Daniel Gonçalves*. Proceedings of the ACM Conference on Accessibility and Computing (ASSETS), ACM Press, Pittsburgh, USA, October 2009.

14. **Mobile Text-Entry Models for People with Disabilities**, *Tiago Guerreiro, Paulo Lagoá, Pedro Santana, Hugo Nicolau, Joaquim Jorge* Proceedings of the European Conference on Cognitive Ergonomics (ECCE 2008), ACM Digital Library, Madeira, Portugal, September 2008.

15. **Navtap and Brailletap: Non-visual input interfaces**, *Tiago Guerreiro, Paulo Lagoá, Pedro Santana, Daniel Gonçalves, Joaquim Jorge* Proceedings of the Rehabilitation Engineering and Assistive Technology Society of North America Conference (RESNA 2008). Arlington, VA, EUA, Jun 2008

## Awards

**Best Student Paper Award**  Received at the 13th International ACM SIGACCESS Conference on Computers and Accessibility (ASSETS 2011) for the paper entitled **Blind People and Mobile Touch-based Text-Entry: Acknowledging the Need for Different Flavors**

**Best Short Paper Award**  Received at the 13th IFIP TC13 conference on Human-Computer Interaction (Peoples' Choice Award) for the paper entitled **BrailleType: Unleashing Braille over Touch Screen Mobile Phones**

**José Luis Encarnação Award 2010**  Attributed to the Best Computer Graphics & Applications international paper from a Portuguese student (1st author) to the paper **Nav-Tap: a Long term study with Excluded Blind Users** presented at the 11th ACM Conference on Accessibility and Computing (ASSETS 2009)

**INESC-ID Best PhD Student 2009**  Selected by an external international committee and based on publication records

## 1.6. Dissertation structure

This dissertation focus on the interaction of blind people with mobile devices paying particular attention to the impact of individual differences and how they affect the users abilities to cope with mobile user interfaces. Chapter 2 reviews the state of the art on mobile interfaces for blind people. Further, and laying the groundwork for the following discussions, outlines previous works that have paid attention to individual differences and their impact on the use of technology. We focus our attention on blind people. As such, Chapter 3 presents a preliminary study performed with the target population seeking to characterize their usage of current technologies and the difficulties faced. Particularly, we have leaned over text-entry as its demanding nature was likely to reveal limitations and capabilities worth exploring. This motivational chapter showed that indeed there are several blind people excluded by current technologies but also that if user-sensitive approaches are designed, inclusion is achieved.

Following this first formative research step, we then sought to have a clearer and thorough notion of which individual attributes play a role in a blind person interaction with technology. Chapter 4 presents a interview study performed with professionals working closely with visually impaired people aimed at unveiling such attributes. With this information, we were then able to explore these features and see how they affect mobile interaction. Chapter 5 presents the assessments performed with 51 blind people and explores the attributes and relations within. In this chapter, we make a thorough characterization of our participant pool (all later studies were performed with sub-sets from this group) and show the dispersion within the population in respect to the attributes explored.

Chapter 6 explores these relationships with different touch device settings and primitives therein while in Chapter 7 we revisit text-entry and explore how the relationships found at lower-level are relevant in a higher-level task.

The aforementioned studies placed us in the position to understand the impact of individual attributes and how the population and variations within stand in relation to mobile user interface demands. This enabled us to conclude by presenting implications for the design of inclusive user-sensitive mobile user interfaces for blind people in Chapter 8. In this chapter, we also revisit our contributions, present the benefits and limitations of our research and outline steps for future research.

# 2

# Related Work

Mobile interaction is still in its early stages when compared with interaction with desktop computers, that have been subject of attention for several decades. Although mobile computing is an active research theme, Mobile Human-Computer Interaction (HCI) was not an important subject until recently. In particular, only few researchers have leaned over the multitude of individuals, scenarios and situations faced by mobile devices.

In this chapter, we perform an overview on the evolution of mobile devices (with a focus on mobile telephony) for blind people, which is characterized by a lack of attention to individual traits. This leads us to surveying how individual differences have been addressed in technological settings, instead of interfaces targeted at a global *average user* model. The majority of this research is still quite narrow focusing mostly on Age and Visual Acuity within Desktop settings.

Given the paucity of research aimed at mobile contexts and targeted at blind people, we step outside the scope of individual differences and explore the area of situationally-induced impairments and disabilities (SIID) [Sears et al., 2003]. Indeed, given that the target audience of such impairments and disabilities is wider than any disabled population, the state of the art to address the limitations imposed by mobile interaction context is ahead of the one that deals with individual differences. Once may even argue that limitations imposed by particular situational impairments are comparable with the ones felt by disabled populations. As such, there are also research avenues in the exploration of technology transfer between situational and individual contexts and vice-versa [Yesilada et al., 2010a, Lucas et al., 2011], which makes worth looking at both areas concurrently. Indeed, research to tackle SIID in mobile contexts is already taking place, from which researchers leaning over individual differences can take lessons from.

Lastly, we present a discussion that outlines the main faults on mobile interaction research for blind people and motivates the work proposed in this dissertation.

## 2.1. Mobile Interfaces for Blind People

The first hand-held mobile phone was presented by Martin Cooper of Motorola in 1973. As a comment, he made the first call to a rival, Dr. Joel Engel from AT&T's Bell Labs, while walking the streets of New York (Figure 2.1).

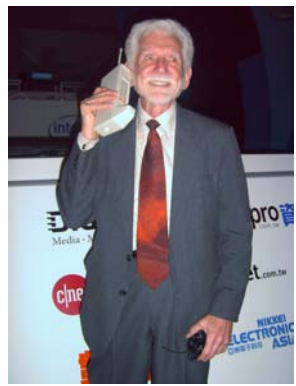

Figure 2.1: The first handheld mobile call

The first generation (1G) of mobile phones was introduced in the early 1980s. These were characterized by the usage of fully automatic cellular networks. The first 1G system was the Nordic Mobile Telephone (NMT), presented in 1981. Although cellular telephone history has started several years before with laboratory experiments and military prototypes, the first hand-held mobile phone in the US market was the Motorola Dyna 8000X, which received approval in 1983. Until the early 1990s, most mobile phones were too large to be carried in a jacket pocket, so they were usually permanently installed in vehicles as car phones. In Portugal, the first cellular telephone appeared in the late 80s by the only communications operator at that time (CTT/TLP). With the advance of miniaturization and smaller digital components, mobile phones got smaller and lighter (Figure 2.2).

These keypad-based devices were not that different from car phone rigs and the keypad itself inherited some of the characteristics of land-line phones. These devices had a basic goal: voice communication. A simple accessibility tweak (the label available in almost all phones, mobile or not, in the '5' key) was enough to assure accessibility for a blind person to place a call. With the emergence of new applications and tools, and with mobile phones

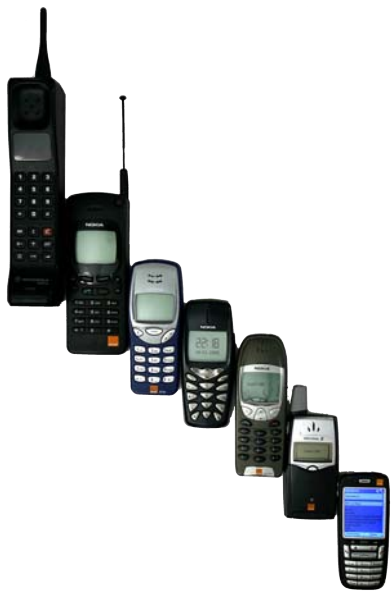

Figure 2.2: Mobile phone evolution

becoming less of a phone and more of a computer, the gap in possibilities separating
sighted and blind people started to increase. Along the years, efforts have been made
to develop alternatives for blind people. From custom-made Braille-based[1] devices to
mainstream devices with assistive screen-reading software, we offer an overview of the
approaches deployed to provide mobile accessibility to blind people.

## 2.1.1. Custom–made Hardware Technologies

A blind person, or even one with low vision, faces several limitations when interact-
ing with mobile devices. Looking at common mobile devices, whether keypad or touch
screen-based, the interaction mechanisms are convoluted to deal with the limited input
area and overall device small size. Furthermore, the mechanisms found to overcome the
lack of space (when compared to desktop computers) resort to an intensive visual-based
dialogue with the mobile device user. As an example, the majority of keypad-based text-
entry systems are based on multi-tap approaches where the user is able to see both the
relation between keys and letters (visual feedback from the physical or virtual keys), and
the evolution of the process on the display. A user with severe visual limitations is un-
able to receive this information and thus his ability to interact with these devices is highly
limited, particularly in the first attempts to do so which by turn likely leads to drop out.

---

[1]Braille was devised in 1825 by Louis Braille, a blind Frenchman, and is a method to enable visually
impaired people to write and read. Braille was the first digital form of writing [Daniels, 1996]. It consists in
the representation of a character by means of a three by two matrix where a dot may be raised at any of the
six positions to form sixty-four ($2^6$) possible subsets, including the arrangement in which no dots are raised.
This set enables the representation of simple and accentuated characters, punctuation, numbers, algebraic
signs and musical notes.

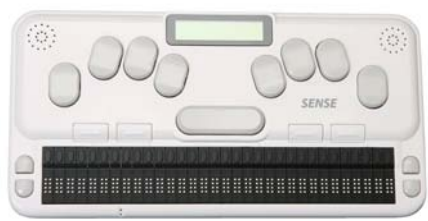

Figure 2.3: Braille Sense: a Braille-based mobile device

Special mobile devices were developed to overcome the difficulties arising from visual impairments, and particularly to offer possibilities beyond simple voice communication. What first started as simple text processors (in the early 90s) resorting to Braille keyboards and Braille screens, soon evolved to devices gathering the ability to place and receive calls, send short text (SMS) and mail messages, also incorporating speech synthesis to enrich the output possibilities (Figure 2.3). Examples are the Braillino, Braille Sense-Plus, PacMate BNS, Braille Lite, among many others very similar between each other [MacKenzie and Tanaka-Ishii, 2007]. These devices typically work as a docking station for a mobile device and enable access to functionalities like the ones provided in regular mobile phones (Figure 2.4). However, in general they can be seen as a peripheral that can be used in conjunction with a mobile device. Other approach has been to develop a full-fledged device that was itself the mobile incorporating the most basic functions in a mobile device (e.g., voice communication, text messaging). One example is the Alva Mobile Phone Organizer, launched in 2003[2].

Both approaches share the same flaws: their cost is prohibitive and they are not as portable as a mobile phone is, being too big and heavy and requiring two-hand input. Even though their cost and size have decreased, they are still not as practical as common mobile devices and have the drawback of "*looking disabled*" [Shinohara and Wobbrock, 2011].

One approach to provide a more mobile experience to the blind *nomad* has been to provide simplified mobile devices, particularly by removing the screen. Lowering the cost of these approaches was one of the goals of this simplification. Figure 2.5a presents Owasys 22C[3], a simplified screen-less mobile device, developed specifically for blind people where all feedback is offered via audio. Simplification also occurs at the functionality level: the device focuses on communication (calls and messages), phonebook, and basic service, battery and coverage status. This device has been discontinued. On the same line, the Spice Braille Phone[4], from Spice Corporation, a company in India, was presented at the Mobile World Congress[5] in 2008 (refer to Figure 1.1b in Chapter 1). It was presented as costing $20. Besides the absence of screen, it was characterized as having

---

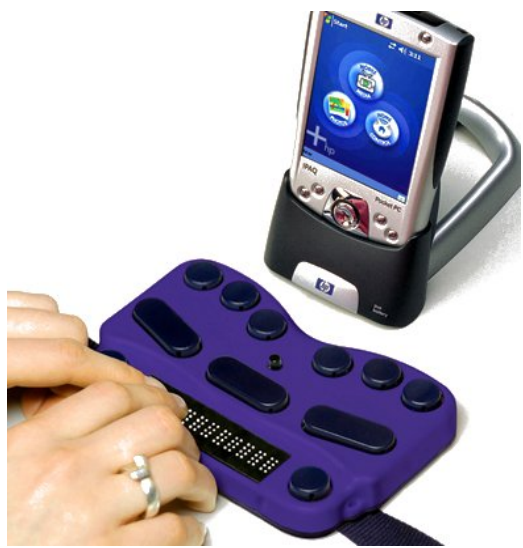

Figure 2.4: Braille keypad and screen

Braille labels in all keys. Although it has gained the media attention, to our knowledge, it has not been launched in the market.

The Touch Messenger (Figure 2.5c) was another prototype focusing on Braille which was presented by Samsung in 2006. It has been appreciated by the Design community and it was a recipient of the Industrial Design Excellence Awards (IDEA) Gold Awards, in the same year. This device enables sending and receiving text messages by incorporating both a Braille keyboard and screen. To our knowledge, it never reached the market. The Samsung Braille Phone (Figure 2.5d) builds on the same concepts and it is a design idea that features a refreshable area where Braille text is presented and another area with a Braille-labelled touch area.

These designs have created fuss in the design community and media in general. However, they seem misaligned with the needs of blind people. Looking at one almost pervasive attempt, labelling the keypad with Braille and simultaneously reducing the relief of the keys, does not likely translate in a easier recognition of the key itself. Further, selecting a key from the keypad has shown to be an easy task for blind people even without any assistive technology, which has been confirmed in our studies (Chapters 3 and 5). These ideas have also focused on reducing the available functionalities to voice communication and/or text messaging which seems rather diminishing and unrepresentative of the desires of blind people [Kane et al., 2009, Shinohara and Wobbrock, 2011].

[Plos and Buisine, 2006] describe a case study of universal design applied to mobile phone physical devices trying to integrate needs of visually-impaired, hearing impaired and the elderly. The authors pose the idea that products for disabled users should not present themselves as disabled (i.e., a mobile device with no screen). The authors studied the user's needs both on usability and design trends and from these sessions, mock-ups were created to be user-tested.

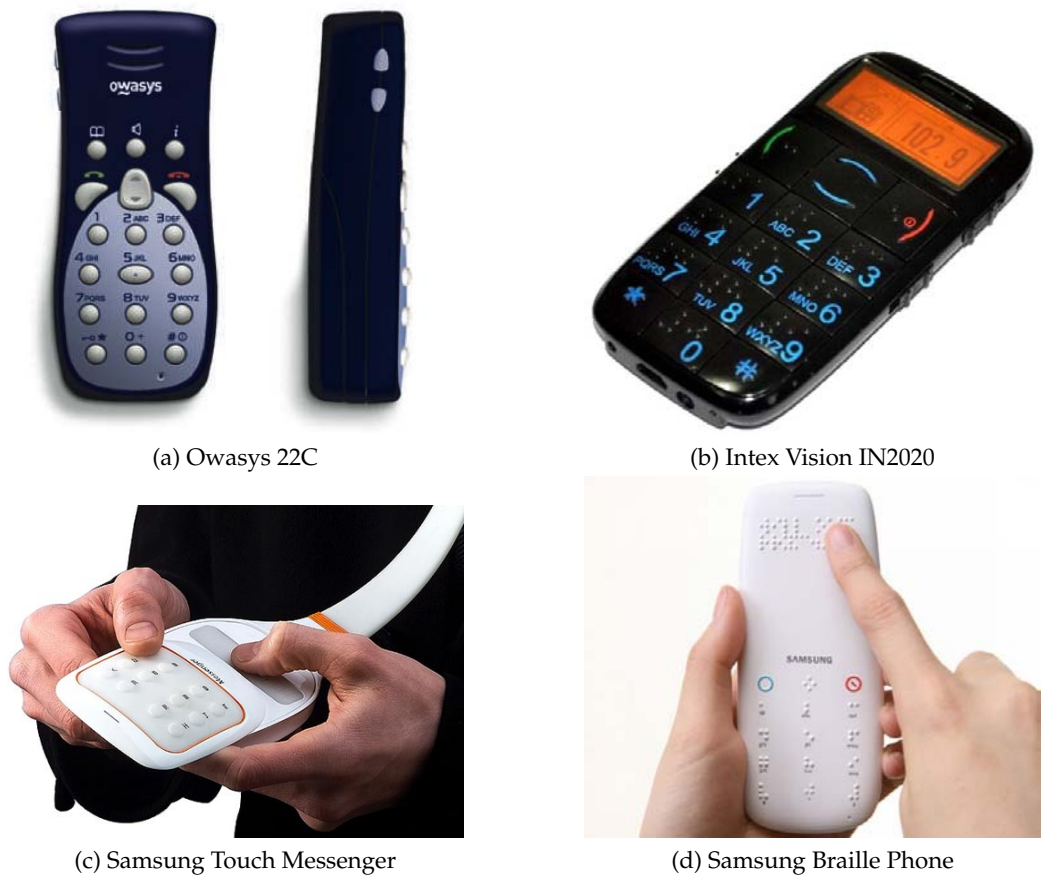

(a) Owasys 22C

(b) Intex Vision IN2020

(c) Samsung Touch Messenger

(d) Samsung Braille Phone

Figure 2.5: Custom-made *Low-cost* Mobile Devices

Concerning visually-impaired people, results showed that late-blind people encounter many more problems than early ones as the latter develop compensating capacities and strategies. Most blind users are able to operate the keypad as they can detect key '5' due to the raised dot in that key. They use both hands to operate the keypad as one holds the device and the other operates the keypad. Results stated that 18% of the users use a phone with text-to-speech receiving audio feedback on the information presented on the screen. Users stated problems concerning keypad tactile feedback and small button size.

[Amar et al., 2003] present a prototype handheld device (Mobile ADVICE) with tactile feedback and auditory display to overcome limitations on common mobile devices. The device integrates functions such as e-mail, mobile phone, personal information management as well as leisure applications. The prototype features a rotating dial for the thumb to accomplish option navigation and can be pushed to select the current option; a button for the thumb to step back in the menu; and four push buttons for the remaining fingers which are used to navigate and operate other applications' functions. To improve the system's usability, pressing buttons halfway for at least one second informs the user on the button's function. The authors used earcons ("short series of unique tones systematically mapped to menu levels and options throughout the hierarchy") and tactile feedback (button clicks and wheel stops). The system was evaluated with 3 blind and 3 sighted users both objectively and subjectively. The users performed two common tasks (playing a song and checking e-mail) and the sighted users were asked to keep their eyes closed and had no previous information on the screen or application. Sighted users showed to be more used to hierarchical navigation mechanisms but blind users stated their ability to find the location within menus higher than sighted users did. Earcons were not stated as advantageous although that is normal in a short term session. On the other hand, the halfway press was considered useful. These results suggest that there are possibilities for adaptations that may improve the blind users' experience with mobile tools. However, it is also important to notice that acceptance relies on adapting without placing a *disabled* connotation on the deployed solutions.

In the next section, we look at software solutions that resort to audio feedback but rely on mainstream keypad devices to provide accessibility to blind people.

## 2.1.2. Keypad–Based Solutions

Nowadays, a common mobile solution for blind users resorts to the usage of a screen reader, replacing the visual feedback by its auditory representation (e.g., Mobile Speak or Nuance Talks) [Burton, 2006]. The ability to use a "non-disabled" device with the same characteristics (technical, social and economical) is a great advantage of this type of solutions. The appearance of screen readers in a mobile setting has enabled the so long desired equality between blind and sighted people. With this software layer, blind people started to be able to use most applications sighted people use which is an obvious improvement towards inclusion. The ability to use the same device as a sighted person has also strong reflections pertaining cost and variety: blind people can now select from a wide variety of brands and models at lower costs than what they were able to with custom-based solutions.

This approach also presents issues. The offered feedback is restricted to the information on-screen as no information is obtained on key/function relation. Moreover, the information on the screen is prepared for visual feedback and not to be read. This happens particularly in visually rich applications: text is replaced by graphical elements that speech synthesis is not able to interpret [Huber and Simpson, 2003]. Another problem with screen readers, in general, relies in the amount of information read to the user [Pitt and Edwards, 1996]. If a large quantity is transmitted it is most likely that the person cannot process everything, sticking with the last piece of information heard (the so called *suffix effect* described in [Conrad, 1960]). A conflict between information designed to be presented bi-dimensionally and actually delivered uni-dimensionally occurs.

Considering text-entry, screen reader approaches force the user to try to find the desired letter in the keypad, committing several errors in the process, and possibly leading to situations where he simply quits trying. A person that acquires blindness in an advanced stage of life, along with the reduction of other capabilities like tactile sensitivity, is likely to face difficulties in the first contact with this approach, rejecting it before gaining the experience that enables its use [Guerreiro et al., 2008a]. In contrast to traditional interfaces, that are designed for the "average user", simple screen reading approaches are designed for the "stereotypical blind", one that has improved tactile and cognitive abilities along with motivation to learn.

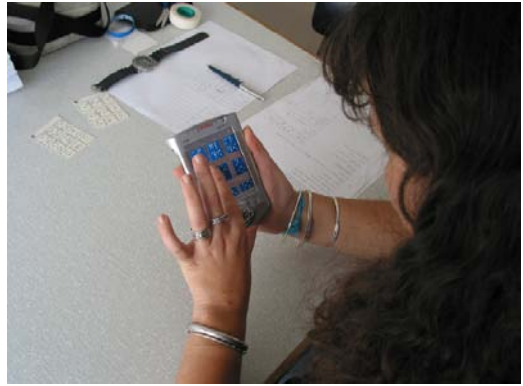

Figure 2.6: 3GM text-entry system

## 2.1.3. Touch–based Accessible Software and Adaptations

Recently, touch screen mobile devices have had a great impact in the mobile communications market and, due to the absence of tactile cues (i.e., keys), some researchers have struggled to make these devices accessible to blind people. Indeed, as touch screen devices are becoming more common among non-visually disabled people, a great deal of attention has also been dedicated to their accessibility and to answering the challenges imposed by devices that are the exponent of graphical user interfaces (visually rich). Although these devices are presented as a bigger challenge for the blind target group who are obviously unable to deal with graphical user interfaces (GUI) direct manipulation techniques on the surfaces (based on widgets' absolute position), they also present new opportunities considering alternative interaction methods (e.g, on-screen gestures).

The enlarged display surface (when compared to keypad-based devices) and the ability to easily create virtual interaction components has been explored by researchers to improve the interaction with touch screens by blind users. Some projects rely on overlays placed over the surface that are able to offer the otherwise *missing* feedback to the user. In these approaches, the interaction methods are redesigned to fit the users' capacities. Indeed, these new designs can be seen as customized keyboards instead of a new physical device; the new interaction mechanism is achieved through a low-cost overlay.

3GM [Campos, A. and Branco, P. and Jorge, J., 2004] is a text-entry system for blind users based on a quadripartite Braille keyboard overlay and a text-to-speech module (Figure 2.6). Braille-knowledgeable blind users are able to feel the surface and, by multi-tapping the required key, the letters/options are read. The great advantage underlying this approach is the offer of tactile meaningful feedback, without the need for new hardware, achieving a low-cost solution. [Gaudissart et al., 2005] used a similar approach in Sypole, their mobile assistant for blind people.

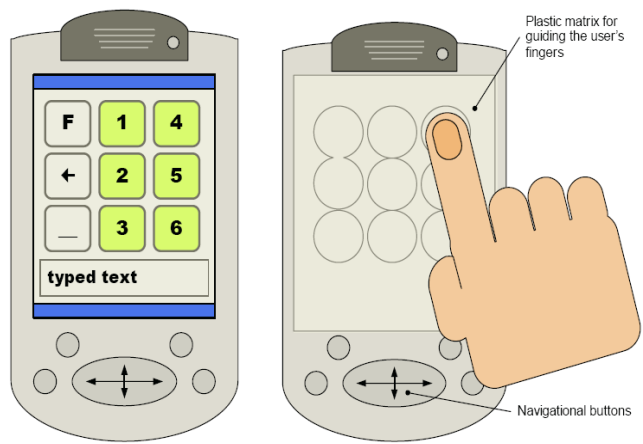

Figure 2.7: Braille Slate Talker

The Braille Slate Talker, in turn, used a fixed layout plastic guide placed over a commodity handheld PDA to allow Braille input (Figure 2.7). Inputting text works like with a traditional slate where the dots can be *punched* with the fingers.

With another approach, there have been projects to improve the navigation between and within applications, by performing gestures on the screen. These approaches also rely on audio feedback as a replacement for the visual channel. [Kane et al., 2008c] presented Slide Rule, a touch-screen multi-touch control interface for blind users which provides non-visual access to applications like the phone book, e-mail and media player applications. Slide Rule uses four different gesture types: one-finger scan, a second-finger tap to select items, a multi-directional flick gesture for additional actions and L-Select gesture to browse hierarchical items. User evaluation comparing Slide Rule with a Pocket PC with Mobile Speak Pocket (screen reader and screen 4-split) showed that users were faster with Slide Rule but Pocket PC MSP was less error-prone.

[McGookin et al., 2008] investigated the two aforementioned different approaches (overlay buttons and on-screen gestures) with a MP3 player (Figure 2.8). The overlay buttons player relied on a raised paper control panel incorporating tactile buttons superimposed on the virtual buttons on the touchscreen and the gesture driven player resorts to horizontal, vertical and tapping gestures on the screen to control the player. Evaluation studies were performed with blind-folded users and one visually-impaired person and showed overall user preference for the buttons player. However, this was probably due to a high number of false tapping positives. More than being able to compare the two approaches, these studies were useful to provide guidelines towards touch screen accessibility. The lack of blind users in the study is a great drawback. Blind folded users do not represent the blind user group. The interaction between the user and the device depends on several characteristics, and while some are damaged for some users, other are overdeveloped.

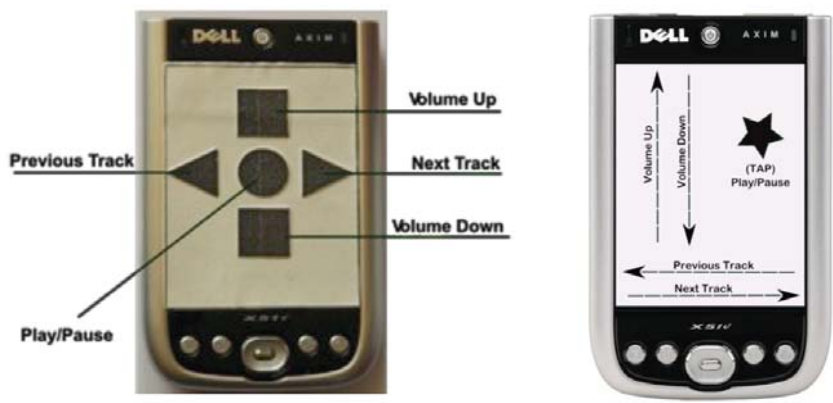

Figure 2.8: MP3 Player: Left) Overlaid Buttons Player; Right) Gesture Driven Player

Solutions with simple screen reading software and no physical adaptations are still the most accepted ones. These, like with keypad-based screen readers, aim to replace visual feedback by its auditory counterpart. Apple's VoiceOver is a successful example. Users explore the interfaces' layout by dragging their finger on the screen while receiving audio feedback. To select the item, they rest a finger on it and tap with a second finger (i.e. split-tapping [Kane et al., 2008a]) or alternatively, double-tap anywhere on the screen. Also, gestures can be performed to facilitate navigation within and between applications. This approach is application independent, allowing blind people to use traditional interfaces with minimum modifications.

Darren Burton, wrote at AccessWorld, the online journal of the American Foundation for the Blind, a review of the state of Cell Phone Accessibility[6], in June 2011. About the iPhone accesibility he mentioned:

> I highly recommend the iPhone to our readers who want a mobile device that is both powerful and fully accessible. Mine rarely leaves my side, as I use it as a Web browser, book reader, music player, and to keep up with my e-mail. With all the third party apps available, such as money identifiers, GPS tools, and bar code scanners, there seem to be unlimited possibilities for the iPhone. We're also seeing early optical character recognition (OCR) apps, and I hear that a Bookshare app is right around the corner. There is certainly a bit of learning to be done when you first try to use the iPhone's touch screen interface, but it's not as daunting as some may think. I definitely found

---

[6]http://www.afb.org/afbpress/pub.asp?DocID=aw120602 - Last visited 14/07/2012

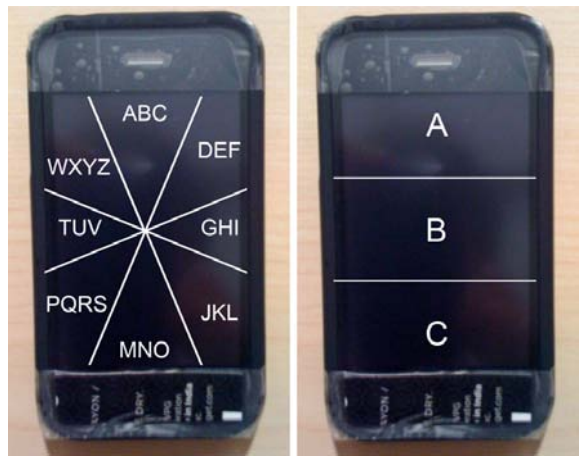

Figure 2.9: No-Look Notes Touch Typing method for Blind People

*it easier to learn to use than I did my PC screen reader. I will have to admit though, that the actual phone feature itself is probably the most difficult to use, especially when interacting with one of those annoying voicemail systems that require you to press 1 for customer service and 2 for tech support, etc. That said, it does work if you have some patience and use a headset, and all the other benefits of the iPhone make it worth it.*

Burton's feelings towards the iPhone and, in particular to VoiceOver, are shared by several blind people. It is to be recognized that a great advance has been done to offer blind people the ability to use touch screens as their sighted peers. However, the barriers for adoption are still too high for several. While we acknowledge that progresses on assistive technologies have been made, users still face problems when interacting with touch interfaces [Kane et al., 2011]. One of the major issues relates to text-entry. This is one of the most visually demanding tasks, yet common on innumerate mobile applications (e.g. contact management, text messages, email).

In the last five years several touch-based solutions were proposed. Yfantidis and Evreinov proposed a new input method that consists in a pie menu with eight alternatives and three levels. Users perform gestures in one of eight directions and select the character by lifting their fingers. The remaining levels of the interface are accessed by dwelling after the gesture until the characters is replaced by an alternative letter [Yfantidis and Evreinov, 2006].

[Bonner et al., 2010] take a slightly different approach with No-Look Notes and present a 12-key virtual keyboard with that uses an alphabetical character-grouping scheme (similar to keypad-based Multitap approaches). The layout is fixed and consists in a pie menu with eight options, which are read upon touch. Split-tapping a segment sends the user to a new screen with that segment's characters, ordered alphabetically from top to bottom (Figure 2.9). Users select the desired character in a similar way to group selection. Performing a swipe to the left or right, allows the user to erase or enter a space, respectively.

More recently, several authors have proposed Braille-based approaches, taking advan-tage of both users' knowledge and the flexibility of touch interfaces. BrailleTouch uses a multitouch approach [Frey et al., 2011] and maps typical Braille chorded typing onto the phone's screen by folding the standard 1x6 keyboard into a 3x2 keyboard. BrailleTouch's ergonomic grip faces the screen away from the user, making it a spatial mnemonic for Braille (Figure 2.10).

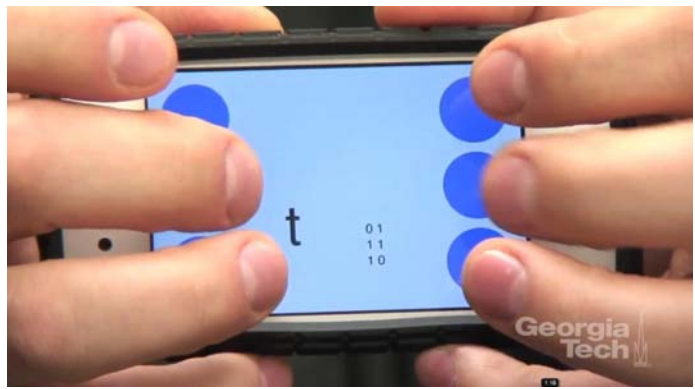

Figure 2.10: BrailleTouch Touch Typing method for Blind People

Other techniques allow the users to input a character through its Braille representation by dividing the insertion process in 3 steps, corresponding to the three Braille rows. In TypeInBraille [Mascetti et al., 2011] the touchscreen is divided into two rectangles (left and right) and four actions are available. A tap on the left part of the screen corresponds to the left dot raised and the right dot flat. Similarly, a tap on the right part corresponds to the right dot raised and the left dot flat. A tap with two fingers represents two raised dots while a tap with three fingers stands for two flat dots. Perkinput [Azenkot et al., 2012] divides the input process by column. Similarly to multitouch approaches, users enter Braille characters through chording taps on the screen.

Tablet solutions have also been proposed, but instead of being held like mobile devices, users can use all ten fingers to interact with the touch surface. Sparkins (Figure 2.11) and Touch Screen Braille Write[7] are two examples that mimic chorded input mechanisms from traditional Braille keyboards. They are also an example of the proliferation of accessible touch typing mechanisms available to be assessed and used.

Overall, much attention has been given to touch input for blind people in the last five years which illustrates the potential of low cost mainstream devices to provide accessibility. Yet, few present a formal evaluation with blind users, showing the lack of involvement of end-users in the design of these text-entry solutions. Moreover, different interaction techniques are used, from single to multi-touch primitives, directional and scanning gestures, fixed and adaptive layouts. There is little knowledge of which methods are better for each individual user. Previous works have shown this to be an important factor when considering blind people [Oliveira et al., 2011a] and touch typing tasks [Findlater and Wobbrock, 2012]. Therefore, understanding and adapting to individual needs is paramount to develop efficient and effective touch-based methods for blind people.

---

[7]http://www.wired.com/gadgetlab/2011/10/touchscreen-braille-writer/

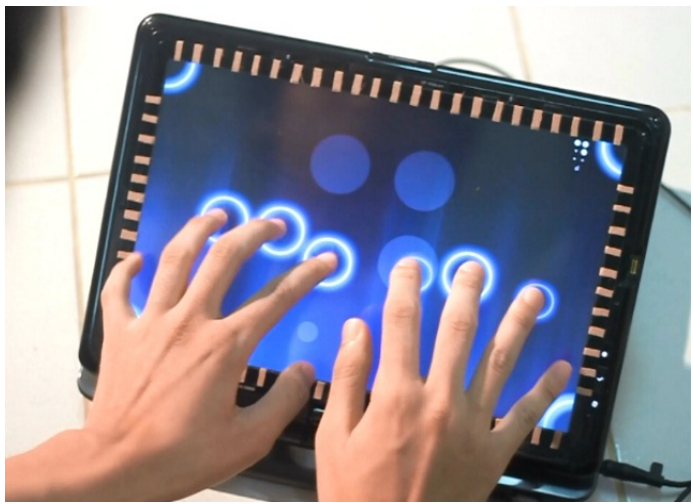

Figure 2.11: Sparkins Braille-based touch typing method for tablet devices

Besides methods for blind users, there are several approaches to tackle eyes-free interaction issues [Brewster et al., 2003, Li et al., 2008, Zhao et al., 2007, Pirhonen et al., 2002]. Although some approaches may be adequate for blind users, others may not as they may be misaligned with the specific population needs.

## 2.2. Dealing with Individual Differences

Although some differences can be found from one computer to another, its appearance, input and output layout and overall features are quite similar. Even when analysing different manufacturers and different operating systems, the interfaces are, in general, quite similar. However, every user is able to change some parameters to adjust the device to his personality, needs or preferences. [Blom, 2000, Blom and Monk, 2003] defines *personalization* as a process that changes the functionality, interface, content or distinctiveness of a system to improve its personal relevance to an individual. He defines three personalization subcategories: enabling access to information content, accommodating work goals and accommodating individual differences. These individual differences can be related to personality or, as an example, people with disabilities.

While desktop computers already offer a multitude of personalization capacities, whether considering individual differences or just taste, mobile devices are still restricted to a limited set of personalization options, that are mostly aesthetic or related to the users' personality. The mobile device is seen as an extension of the body [Townsend, 2000] but to a large part of the population it is more like a badly-prescribed prosthesis. A better understanding of the individual differences that characterize the users in the mobile interaction context is required. Only with that knowledge will we be able to "prescribe" adequate mobile devices that empower the users.

While it is important to understand that a mobile device user is different from the next one and that those differences should be considered to improve device accessibility, it is also important to understand that, even for a single user, his capacities and needs are likely to diverge across time (dynamic diversity) [Gregor and Newell, 2001]. Gregor and Newell state that most computer systems are designed for a typical younger user with static abilities over time. On the other hand, when considering disabilities, prosthesis are the alternative.

However, even when User-Centred paradigms are employed, the designers look typically at concerns such as representative user groups, without regard for the fact that the user is not a static entity. This does not take into account the wide diversity of abilities among users and it also ignores the fact that these abilities are dynamic over time. The authors propose a new paradigm, Designing for Dynamic Diversity, based on a User-Sensitive Inclusive design methodology [Gregor and Newell, 2001]. *The use of the term "inclusive" rather than "universal" reflects the view that "inclusivity" is a more achievable, and in many situations, appropriate goal than "universal design" or "design for all". "Sensitive" replaces "centred" to underline the extra levels of difficulty involved when the range of functionality and characteristics of the user groups can be so great that it is impossible in any meaningful way to produce a small representative sample of the user group nor often to design a product which truly is accessible by all potential users* [Newell and Gregor, 2000]. The authors looked at terms like "Design for All", "Inclusive Design", "Universal Accessibility" [Stephanidis, 2001], and survey some relevant projects collecting ideas in search of a new paradigm. They claim that although "Design 4 All" initiatives have been very valuable to increase the awareness to disabled people info-inclusion, it is also very difficult to achieve. Not only will the difficulties of some products increase for people without disabilities, but also people with other disabilities will hardly be able to operate the devices. The authors argue that researchers should not set impossible goals [Newell and Gregor, 2000].

Empirical investigations on the interaction between users with visual impairments and computational applications has focused mainly on desktop computers. Mobile interaction introduces new challenges as it provides access to similar applications but suboptimal interfaces: small displays and limited input/output techniques. Moreover, the interaction context is also likely to diminish the ability for interaction.

## 2.2.1. Understanding Desktop Interaction Nuances

In 1998, [Jacko and Sears, 1998] called the research community attention to partially sighted users, who were in a gap between fully sighted and blind users, as they cannot use traditional GUIs without alteration and they will not use technologies that do now allow them to make use of their residual capacities. The authors claimed that there is a lack of knowledge on how the physiology of partial vision affects computer task performance and thus, "designers are developing enabling technologies blindly". Indeed, the lack of fundamental information about how an individual's visual profile determines his strategies, behaviours and overall performance while interacting with computers, limits that ability to effective user interfaces. As an example, the authors point out the accessibility options in Windows 95 and state that it could not been constructed with an accompanying knowledge of the physiology of partial vision, disregarding facets like color and contrast adjustments. Moreover, [Jacko and Sears, 1998] state that there are no published reports on the benefits of the accessibility options provided in the aforementioned operating system.

The authors ask for a more extensive knowledge of the physiology of partial vision to understand its degree along several axes. They enumerate four facets that capture the essence of one's visual profile: visual acuity, contrast sensitivity, field of vision and color perception. They argue that, besides clinical ones, functional assessments should also be performed and coupled with performance of computer-based tasks. In the authors' opinion, this knowledge *will allow a more systematic approach to matching users with software and hardware combinations that accommodate their visual profile*.

In [Jacko et al., 1999], the authors put their intents into practice and characterize visually impaired computer users' performance by matching clinical assessments of low vision with visual icon identification. The authors evaluated the difference between full sighted users and partially sighted users when identifying visual icons in a GUI and the influence of visual acuity, contrast sensitivity, visual field and color perception on the time required to identify them Moreover, the authors evaluated the effects of icon size and background color on the time required to identify the icons. Results validated the authors' hypothesis showing a relation between visual profiles and task performance.

[Jacko et al., 2000] focus their research on a particular set of low-vision users, those with Age-related Macular Degeneration (AMD). They performed studies to characterize and assess the interaction strategies of the users and compare them according to vision loss severity. Moreover, their strategies and performance were also compared with fully-sighted users. The studies featured a cursor movement exercise. Results showed evidence for which only anecdotal evidence existed until then: fully sighted users performed better considering Velocity and Movement Time and all groups improved with Icon size. On the other hand, no significant results were obtained by changing back-ground color and set size.

The authors argue that the results achieved should motivate designers and developers to recognize significant differences in interaction styles of users with varying visual profiles.

[Edwards et al., 2005] examined the factors that affect performance on a basic menu selection task by fully-sighted users and users with Diabetic Retinopathy. The evaluation considered the presence/absence of multimodal feedback, Windows accessibility settings and menu item location as well as various visual functions and other participant characteristics (i.e., age). The studies were performed with 29 participants where 10 were fully-sighted and 19 were diabetic with evidence of retinopathy. Results indicated that Windows accessibility settings and other factors, including age, computer experience, visual acuity, contrast sensitivity, and menu item location, were significant predictors of task performance

By understanding the characteristics that may influence the effectiveness and efficiency of the interaction between the individual user and a determined device or application, researchers and designers are in a position to improve their designs and adapt devices and applications to fit the users' individual aptitudes. To this end, not only is required to understand what are the characteristics that play a significant role determining performance but also what solutions fit particular characteristics classes (i.e., age or visual acuity groups).

## 2.2.2. Understanding Mobile Interaction Nuances

[Leonard et al., 2005] explore the interaction of older adults with Age-related Macular Degeneration (AMD) with handheld computers. In their studies, both participants with AMD fully-sighted controls used a handheld computer to search, select and manipulate familiar card icons. Icon set size, inter-icon spacing and auditory feedback presence varied between trials. Severity of AMD and contrast sensitivity were found to be highly predictive of efficiency although in general the task completion rate was very high. Linear regression showed relations between task efficiency and the interface, user characteristics and ocular factors. The authors concluded that users with visual impairments are able to effectively interact with handheld computers in the "presence of low-cost, easily implemented design interventions".

[Darroch et al., 2005] studied the effect of varying font size (between 2 and 16 point) on a reading text task on a handheld computer. The experiments were conducted with 24 participants, 12 younger (18-29 years) and 12 older (61-78 years). Results showed little difference in reading performance and subjective comments showed an overall preference for sizes in the range 8-12. In the subjective analysis, the preferred sizes for older participants were slightly higher than for younger users. Based on the overall results, the authors suggest (for a small screen with 640*480 resolution) that applications should offer the choice for small (8), medium (10) and large (12) font sizes.

While the aforementioned surveyed projects and studies are focused on particular low-level tasks (e.g. reading text, visual icon identification,..), [Ziefle and Bay, 2005] examined the behaviour and performance of older and younger novice users while performing high level tasks (placing a call, send a text message, hide own number, edit entry in phone book) using handsets of different complexity. The studies were performed with 32 participants (16 between 20 and 32 years old and 16 between 50 and 64 years old) with two different mobile phone interfaces (Nokia 3210 and Siemens C35i). The mobile phones were simulated on a PC with touch screen interface. Cognitive complexity [Kieras and Polson, 1985] was defined by the number of production rules applied when processing the tasks. Results showed that the less complex phone (Nokia) was more more effective and efficient than the more complex one (Siemens). Moreover, the benefit showed with the less complex phone was much greater than theoretically predicted. This factor reinforces the doubt about the usage of production rules as the only measure for complexity as they do not account for the real difficulties faced by the users. As expected, older participants showed a lower navigation performance, although their performance matched the one achieved by younger participants when using the more complex phone. The authors concluded that both older user age and complex interfaces in mobile phones are factors which result in performance deterioration.

[Ziefle et al., 2007] have also studied how younger and older adults deal with hyperlinks in small screen devices. They explored how user characteristics like spatial ability, verbal memory, computer expertise and technical self-confidence determined user's effectiveness and efficiency of menu navigation. The participants had to solve four tasks on a simulated palm prototype. Half the users used a prototype with hyperlinks activated while the other half used a non-hyperlink condition. The two independent variables tested were users' age and hyperlink presence/absence. Considering the psychometric preliminary assessments, the most pronounced difference between age groups was revealed in the spatial abilities while verbal memory also showed significant. Moreover, in general, users with a high spatial ability also had higher computer expertise and confidence when using technical devices. Results concerning the four navigation tasks revealed hyperlink interfaces as more effective, overall. However, depending on age, efficiency showed to be different. While younger adults strongly benefit from hyperlinks, older ones seem to be more disorientated when using them.

### 2.2.3. Coping with Individual Differences

The aforementioned studies aim at being a contribution to be included in upcoming systems and applications. Like the magnifying software currently found in any operating system or any other kind of personalization, the surveyed knowledge aims to be materialized to better fit a particular individual. There are two different ways to personalize a system: customization (adaptable) or adaptation (adaptive). The difference relies in the agent that performs the personalization: the user, relying on his a priori knowledge and needs; or the system, depending on an user, task or/and context model. While in systems like [Alonso et al., 2006] the system is based on some *a priori* knowledge and configured to fit the user, others put their effort in automatically understanding the user and adapting the system to better fit his aptitudes.

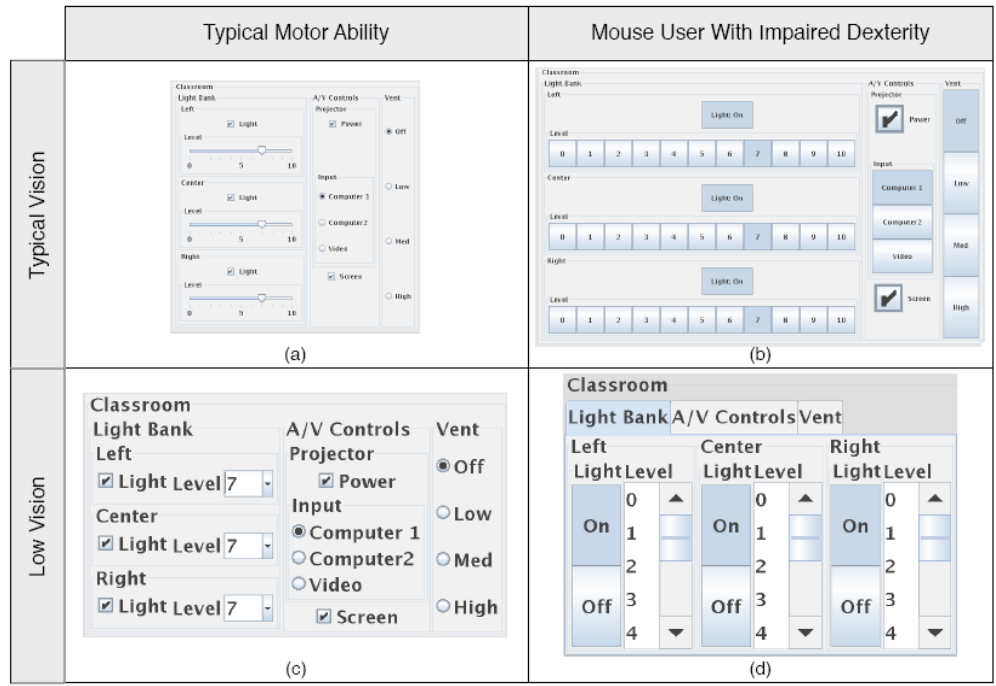

Figure 2.12: Four GUIs automatically generated under the same size constraints for four different users: (a) a typical mouse user, (b) a mouse user with impaired dexterity, (c) a low vision user and (d) a user with a combination of low vision and impaired dexterity. Extracted from [Gajos et al., 2007].

[Gajos et al., 2007] presented Supple++, a system that automatically generates interfaces tailored to an individual's motor capabilities and that can also be extended to fit varying vision capabilities (Figure 2.12). To model the user's pointing performances the authors rely on a two-step exercise ( find the best set of features and train a regression model). Models are created for pointing, dragging and multiple clicking tasks.

What is true in the aforementioned projects is that the authors try to deal with the diversity among the population. They strive to dissect large target groups in respect to particular individual differences that characterize the user in a certain interaction context. While some focus their attention in long-term characteristics, like personality [Benyon and Murray, 1993], the majority is interested in more dynamic individual differences (dynamic diversity), mostly related to age-related conditions [Jacko et al., 2000, Ziefle and Bay, 2005, Darroch et al., 2005].

However, and although it is important to assess particular user characteristics (e.g. visual profiles), it is also important to consider that conditions that make the user diverge from the so-called "average" user are likely to affect the person at different levels [Levesque, 2005]. As an example, several diseases that lead to blindness may also reduce auditory or motor capabilities (e.g., diabetes may lead to retinopathy and also reduce or the sensitivity in the person's fingertips). On the other hand, long-term blindness is also likely to sharpen senses like tact or audition.

With the goal to design products that deal with the several differences that a user may present, there are scales of user capability to capture variations in human sensory, cognitive and motor abilities (reviewed in [Persad et al., 2007]). Those can be performed at an high level (functional assessment [Price, 2006, Price, 2008], e.g., how reading capability varies with text size and contrast) or at lower level, by using low-level capabilities (e.g., visual functions) to predict high-level task performance.

[Howell et al., 2008] investigated the impact of individual differences on the usability of a speech-activated mobile city guide scenario within different contexts of use. This is one of few projects that has considered individual differences under real life scenarios. While individual differences, like age, gender and mobile device experience showed as predictors of attitude towards the application, that was also affected by the context of use (location, surroundings). Results showed that besides the individual characterization it is important to deal with other external events that may influence the user's performance and attitude towards device usage. In the next section we survey research dealing with the users' daily scenarios and momentary situations.

Assessing the device demands and user capabilities in a mobile setting is an ill-explored research area. Although we can find some efforts to classify user capabilities, they are not aligned with the requirements presented by mobile devices. Looking back to Section 2.1, where we have reviewed the state of the art in mobile accessibility for blind users, we can find clear examples where the users' individual differences have been neglected.

## 2.2.4. A word on Situationally Impairments

It is *common place* for everyone trying to read text in a mobile phone in an open space and being unable to perform the desired action due to the glare on the screen caused by sunlight. As another example, trying to input text while walking (or even in a public transport) using a touch screen can be an adventure due to tremor and lack of precision. Indeed, every once in a while, we all face situationally-induced impairments and disabilities (SIID) [Sears et al., 2003]. In a mobile setting, where every user is subject to interferences and cognitive resources are reserved for monitoring and reacting to particular contexts and events, the interaction is fragmented [Oulasvirta et al., 2005] and, in particular occasions, extreme scenarios can impair the user from operating a device (i.e., interacting with a keypad in cold environments [Theakston, 2007]).

Considering the Context definition provided by [Dey, 2001] where it is defined as anything that can influence the interaction between the user and the device, the environmental dimension plays a relevant role and has been highlighted as an important area for research [Bristow et al., 2004, Landay and Kaufmann, 1993, Dunlop and Brewster, 2002]. Recently, researchers have started to look closer at mobile scenarios and striving to improve mobile situational accessibility.

In this section, we give attention to situational impairments as they are similar to physical ones in several ways. Understanding individual differences and situational impairments as well as compatibility with devices, methods and primitives can be of great benefit for the disabled user but also to the "average" one. Overall, understanding mobile interaction contexts and its multidimensionality ultimately leads to better interfaces for all. To better understand the variables underlying situationally-induced impairments we have surveyed the major contributions in this area. Hence, the knowledge around situationally-induced impairments and disabilities can be studied to enrich the poor knowledge research base considering individual differences and their relation with mobile interaction.

One of the main differences between mobile interaction and traditional desktop interaction is the multitude of setting mobile devices can be used in. When interacting with a desktop computer the users are normally in front of the device, normally seated, with the computer on top of a steady surface. Even when considering laptop PCs, generally, a comfortable user position and a surface to put the computer are also guaranteed. On the contrary, mobile interaction supposes the ability to interact everywhere with minimum effort. However, this is hardly true. Several researchers have leaned over different mobile settings to understand how to improve the device adaptability and, ultimately, the user experience. With the advent of touch-screen mobile devices, this attention has grown as the challenges are more evident, even for a user with no disabilities.

## Acquiring Targets in Mobile Settings

[Brewster, 2002] researched the use of soft keyboards under two different mobility conditions to assess the differences between realistic scenarios and laboratory settings: seated in a lab and walking outside. Results showed that walking significantly reduced usability (less data entered and increased the perceived workload), indicating some difficulties associated with stylus-based tapping while walking. It is important to notice that the observed difficulties and damaged performance are likely to be even greater in a real mobile setting as the authors stated that the study scenario was still quite controlled ("reasonably quiet straight path").

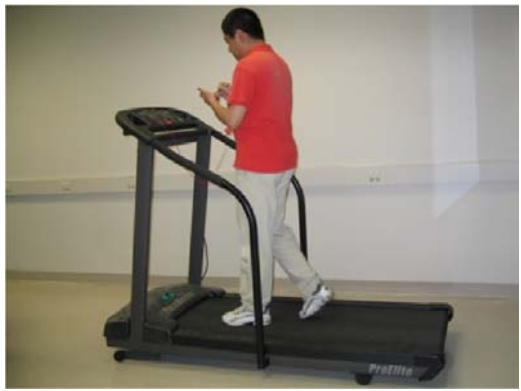

Figure 2.13: User walking on a treadmill

[Schedlbauer et al., 2006] compared stylus-based target selection while seating, standing, and walking (at a normal pace) and found Fitts' Law to be valid under all the conditions. Walking condition was the one with the highest error rate although the results also suggest that standing caused error rates for small targets to increase, without affecting selection times. However, in these studies , the participants used different devices for the different scenarios, fact that is likely to invalidate comparisons.

[Lin et al., 2005] used a treadmill to simulate walking and be able to vary and control walking speed. The classic Fitts' Law [Fitts, 1954] was demonstrated to be valid when the participants performed standard target selection tasks while walking on the treadmill at different speeds. Further studies [Lin et al., 2007] by the same authors, indicated that Fitts' Law maintains valid even under challenging conditions (walking at an obstacle course). The authors compared seated, walking on a treadmill and walking through an obstacle course condition and results showed that, although target selection times did not differ between mobility conditions, overall task completion times, error rates, and workload were significantly different. The participants achieved the worst results under the obstacle condition probably due to increased attention demands. The authors claim that the treadmill condition (Figure 2.13) is able to generate representative data for task selection times but a more realistic scenario presents other difficulties that are not considered under laboratorial walking settings.

[Bergstrom-Lehtovirta et al., 2011] investigated the relationship between walking speed and target acquisition performance, showing that to maintain selection accuracy users need to reduce speed by 26%, as compared to their preferred walking speed.

## Text–Entry in Mobile Settings

One of the most demanding mobile tasks is text-entry. On the other hand, is also one of the most common and with increasing relevance. Keypad-based devices have the keypad's haptic characteristics advantage but only a very experienced user is able to input text without the need to glimpse at the keypad and screen.

[Yesilada et al., 2010b, Chen et al., 2010] have shown that small devices, in particular QW-ERTY keypad-based devices, impose similar problems to non-impaired users as those felt by motor impaired users in a desktop setting, even in stationary seating settings. However, the magnitude of these problems showed to be smaller then the one experienced by the motor disabled population. The similarity between errors found in stationary settings was verified by the same authors in walking and standing conditions however with changes in magnitude: situationally impaired users showed comparable (and sometimes higher) magnitude of errors than motor-impaired desktop users [Chen et al., 2009]. This research has shown that the mobile device itself and the context it is used in imposes difficulties to the users that are comparable to some extent to the ones experienced by disabled people which by turn suggests that lessons, adaptations and solutions can be transferred from one setting to the other [Harper et al., 2010].

Touch-screen based devices also provide quite a challenge when in a mobile setting (Figure 2.14). With these devices, visual feedback is essential and, when in a mobile setting, the interface and the mobility scenario compete for a limited resource [Oulasvirta et al., 2005].

[Mizobuchi et al., 2005] studied the relationship between walking speed and text-entry difficulty. Four different key sizes were tested ranging from 2.0x2.5 to 5.0x6.3 mm. The participants entered text using a soft keyboard under either a standing or walking condition. Text input speed did not differ between the standing and walking conditions which might be due to the simple walking condition used. Error rates were different between the two conditions, particularly for the smallest target size under walking situation (highest error rate). The authors stated that 2.5mm is the minimum key width and 3mm is the preferred width for soft keyboards.

More recently, [Nicolau and Jorge, 2012] also study the effect of mobility and hand posture in touch typing tasks, showing that mobility decrease input quality, leading to specific error patterns. Still, the authors focused their analysis on motor demands, instead of visual demands.

To improve text-entry both on stationary and mobile settings, [Yatani and Truong, 2007] presented a two-handed text-entry software method for PDAs which takes advantage of the non-dominant hand. This chord-keyboard technique has showed to be more accurate and faster than traditional techniques like the one in Figure 2.14. The authors claim that the method is not only better in stationary scenarios but that it has also maintained good accuracy in when in a mobile circumstance. Also relevant is that people with different walking speeds preferred different input techniques.

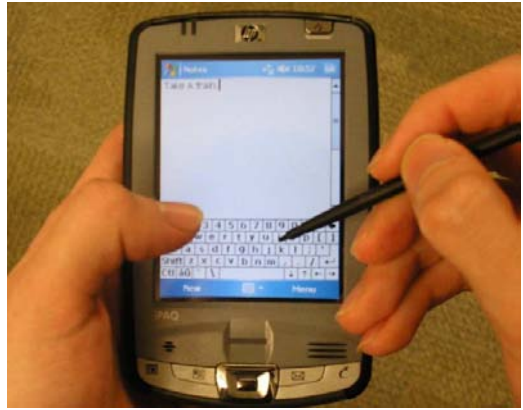

Figure 2.14: Text-entry interface

## Coping with Situational Impairments

In the last sections, we have surveyed a wide set of studies that aim at understanding mobile contexts. This knowledge will ultimately enable researchers and designers to informatively design interfaces that are able to cope with situational diversity. Some of the aforementioned studies have already featured some kind of adaptation to improve the user's experience. Several others have used this knowledge or the overall need for better interfaces to present adapted, adaptive or adaptable interfaces.

[Kane et al., 2008c] focus their contribution in quantifying the negative effects on usage due to walking and to explore interface changes that can improve performance. The authors introduce the term Walking User Interfaces (WUI) to classify interfaces that are designed to compensate the effects of walking on mobile device usability. Two user studies were performed with a music player prototype application (Figure 2.15). The first was performed with 6 users and studied the effect of different button sizes. It suggested that changes to target size may have a positive effect on performance if a device can provide the best sized interface. The second experiment was performed with 29 users and was evaluated in the field.

Besides contemplating several situational factors, this study included trials with both static simple and complex interfaces but also adaptive interfaces according to user's movement. Although some particular results can support that adaptive solutions improve user's performance, the study was not conclusive towards that goal. Indeed, the adaptive solution did not perform as well as the static-simple interface with large buttons. There is a relation between size and performance but the perfect size is likely to depend on extra variables, including individual differences.

Considering mobile settings but also the nature of particular mobile device events there have been several researchers proposing alternative interfaces, in some cases, interfaces that do not require visual attention. Indeed, some authors ask for a paradigm shift, mean-ing that if the users are typically in motion and cannot devote attention to their visually-demanding devices, other interaction channels should be explored [Angeslevä et al., 2003, Lumsden and Brewster, 2003, Pirhonen et al., 2002].

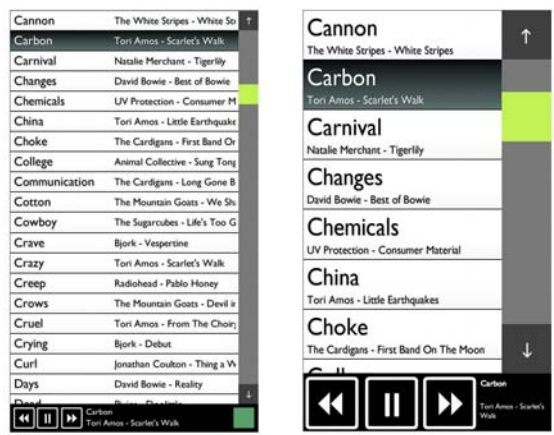

Figure 2.15: Music player user interface in two sizes. (left) The player while standing; (right) the player while walking.

Unlike individual differences, situational impairments have a wide audience and every mobile device user is likely to be unable to effectively operate a device if they are not considered. We are still witnessing efforts to understand mobile environments as well as their difficulties and uncertainties. While the environment dimension has the researchers' attention, we have not yet achieved interfaces that are able to cope with these scenarios. Meanwhile, as situational impairments may pose obstacles for a wider population than individual differences, the latter is likely to be forgotten and some people seriously damaged interaction-wise. Indeed, there are individual differences that are easily surpassed. However, other gain further relevance when paired with particular impairments like blindness.

## 2.3. Discussion on Mobile User Interfaces and Individual Differences

In the previous sections, we have surveyed research performed considering mobile user interfaces for blind people in particular or for any other person with individual characteristics that is likely to benefit from an user-centred adaptable interface. Overall, the surveyed approaches try to account, at different levels, with the diversity of devices, people and environments possibly featured in a mobile context. While a great deal of research has improved the relation between the users and the devices, mobile interaction still requires advances in different dimensions:

**Mobile User Interface Research is still in its early stages**  While desktop user interfaces have been studied for decades, mobile interaction is a relatively recent research topic. Hence, and as a resume for this section, mobile user interfaces are still not mature, and are in some cases inadequate for the users, sometimes inadequate for the situations and, in extreme cases or situations, impossible to deal with. This is not only due to the lack of research and immaturity of mobile interaction but also to its inherent complexity. As a matter of fact, mobile devices have not evolved much from their predecessors, the mobile rigs in taxi cabs or walkie-talkies available in the 40s. Considering graphical user interfaces, where a difference can be identified due to the different display capacities, the adapted metaphors are quite similar to desktop interaction but the scenarios are severely different. As stated by [Brewster, 2002], it is "clear that taking the desktop interface and implementing it on a mobile device does not work well; other methods must be investigated to make mobile interfaces more usable". Current mobile interaction disregards the users' aptitudes and situational demands.

**Individual Differences Disregarded**  Mobile interaction is physically, sensory and cognitively demanding. Due to its inherent restrictions like a small keypad (or touch screen), small display along with an astonishing set of capacities and applications, interaction designers were *obligated* to provide the user with reduced multi-tapping keypads, multi-functions keys, too deep hierarchical menus with too many options and reduced font sizes. However, "average" users are able to operate mobile devices and, with experience, they can even become proficient. As an example, younger adults who have been working with mobile devices for years and detain all required sensory, cognitive and motor capabilities, are truly connected with their mobile devices [Townsend, 2000]. However, there are also users that face several difficulties operating the devices. Indeed, even these proficient users will probably, in a near future, start to feel difficulties to operate the devices they were once a perfect match with. In our survey, we have already observed an effort from some researchers to explore individual differences but, reflecting the overall mobile interaction research maturity, it is still embryonic. [York and Pendharkar, 2004] analyzed the status and trends of Mobile HCI research. They have reviewed 68 studies and concluded that the majority (58%) focused on computer system and interface architecture issues, 23% addressed development and implementation issues, 13% focused on use and context issues, and just 6% focused on human characteristics. From those 58%, focusing on computer system and interface architecture issues, most were focused on input and output techniques. As input and output techniques make the bridge between a user and the device, additional study is required which focuses on addressing the importance of the context of use and human characteristics/capabilities. Particularly, considering blind users, one can verify that the studied individual characteristics are not fitted with the target group and their needs as few researchers have focused on understanding the difficulties faced by the users when interacting with a device. Several projects have focused on visual profiles and even on partially-sighted users which is valuable as it introduces individual characteristics as continuums instead of dichotomies [Jacko and Sears, 1998]. Others have focused their attention on age [Darroch et al., 2005] that is the exponential characteristic of dynamic diversity. However, there is more to it [Levesque, 2005]. An example already herein presented is focused on the individual's tactile capabilities and the interaction with reduced keypads like the ones present in a keypad-based mobile device: what could be an easy task for a full-sighted user is extremely hard to achieve for a blind user with low sensitivity in the body extremities. This type of characteristics should be taken into consideration when designing interfaces. Moreover, with the growing capacities considering control interfaces (touch, inertial sensing, physiological,..) and interaction methods (tap, slide, gesture,..) there is an opportunity to take advantage of the users' capacities to improve the user experience.

**Mobile Accessibility for Blind Users is Limited**  A particular population affected by the disregard for individual differences is the blind one. If in one hand, we find Braille-based devices that are aimed at a limited percentage of the population, on the other we encounter the simple substitution of the visual channel by the auditory one, a solution that is also aimed at those who were able to develop compensatory mechanisms to overcome the issues arisen by the lack of information provided. The latter only consider the replacement of the information provided by the screen, but, as an example, all the information on the keypad is missing. Considering that a great part of the population acquires blindness at an advanced stage of life, when other possible disabilities could also arise, we must conclude that current interfaces were not designed with the user in the center of the development process. At least, not with a representative sample from the target population. Solutions should be found that are able to provide an easy and subtle approach for the individual blind user, one with a distinguished set of capacities and needs, and not a stereotype.

**Capacity Assessment insufficient and out of context**  As surveyed in this document, we can distinguish two different ways to assess the users' abilities: low-level assessment, evaluating the users' motor, sensory or cognitive capabilities (i.e., visual functions); or high-level assessment, by evaluating the users functional abilities (i.e., reading a text). While there are efforts to provide general assessment frameworks, we consider that the assessment should be performed accordingly to the target group and the product demands. Despite the belief that the key to improve user efficiency relies in considering his individual differences, we also believe that the main limitations imposed to him, define the set of differences to be considered. As an example, a blind user is unable to receive information from the visual channel and thus the information is conveyed via touch or audition. These senses gain wider relevance and therefore the degree of assessment related with them is also wider. While in regular assessment scales only motor precision could be considered, in this case, tactile sensitivity is also a key feature. Moreover, it is important do note that different devices, products or applications have different demands. The required abilities are to be studied accordingly and not in a general basis.

**Mapping between demands and abilities** To our knowledge, there are not any studies that can relate blind users and their abilities with types of devices, interaction methods, interaction primitives or their parametrizations. Like an eye-doctor spectacles prescription, this type of knowledge would enable any person to improve efficiency and overall info-inclusion. This ability depends not only on understanding the users' ability levels and needs but also a characterization of the device's interfaces and primitive variations within each device.

# 2.4. Summary

In this chapter, we described background literature on mobile user interfaces for blind people. In particular, we showed how the evolution of mobile devices has been accompanied by the accessibility research community and a continuous effort has been made to deploy inclusive solutions. However, most of this research has also been characterized by a focus on a stereotype of the blind population, resulting in the development of strict one-size-fits-all solutions, an approach that is likely to leave out slices of the population. These approaches, although beneficial for the average blind person, do not take in consideration the idiosyncrasies of a particular individual, leading to the exclusion of some. Focusing on this individuality, we also presented work that has given attention to individual traits in the use of technology. Attention has been given to individual differences among older adults, particularly in desktop settings, but there is paucity of projects focusing on differences among the blind. Also, mobile technologies, due to its relative novelty, have also not been subject to much attention in these area jeopardizing the adaptation/personalization of such systems to a wide portion of possible users. In an attempt to show that mobile technologies have the potential to fit the individual and its particularities, we have also presented works in the area of situational impairments, where several projects are already addressing the vicissitudes of the context the mobile device is used in. The same has not happened still for differences in the individual ability levels. In this particular, mobile devices do not take in consideration individual differences among blind people.

# 3

# Preliminary Mobile Inclusive Attempts

In our research, we study mobile user interfaces for a particular target group: blind people. In the previous chapter (Chapter 2), we have already acknowledged that, despite its argued insufficiencies, there have been continuous efforts to provide blind people with access to mobile technology and its applications, particularly with focus on communication. Despite these efforts, it is not guaranteed that the solutions are usable and that people adopt them. In this chapter, we overview our observations of the target population and our own experience in a preliminary attempt to provide inclusive mobile interfaces. This chapter describes the first phase in our research and it is meant to provide motivation for what follows in this dissertation: a focus on individual differences.

## 3.1. Research Timeline

Mobile technology has shown a great evolution and paradigm shift in the last decade. Our research with mobile phones and blind people started in a era where keypad phones ruled the market and Personal Digital Assistants (PDA) were stereotyped as devices for a businessman.

Pertaining this dissertation, our research efforts started with the development of an alternative method for keypad-based text-entry. This was motivated by the fact that our preliminary analysis of the population and mobile inclusion, dated of 2007, showed that out of 9 people only 5 resorted to screen reading software (Nuance Talks) to go beyond basic communication functionalities. Interviews with users showed that part of these experimented such approaches but failed to achieve the desired proficiency and dropped out. One main desire was to input text which was also one of the main barriers as people reported to have difficulties with MultiTap along with audio feedback. NavTap and BrailleTap [Guerreiro et al., 2008a] intended to bridge this gap by providing methods with low demands fostering adoption. The emergence of touch-based devices called our attention for the non-existence of an accessible touch interface and thus we deployed a touch-based method similar to NavTap, called NavTouch [Guerreiro et al., 2008b] to enable touch typing by blind people. This method showed the potential of touch screens as some people tended to perform better with this version as exploring the keypad was not required.

In an attempt to understand the impact in performance and social inclusion of NavTap along with other assistive mechanisms, we performed a long-term study with a reduced set of users (5 in the study and eight in the design phase) [Guerreiro et al., 2009]. For this study, we had fourteen candidates, people who could not go beyond basic functionalities with their own devices. This was performed late 2008. In 2009, five blind people were using our solutions in a daily basis (three of them are using them still; June, 2012).

This experience enabled this set of users to become proficient with our approaches. To assess how this placed them in relation to other blind people who were able to adapt and adopt mainstream devices with screen reading software, we performed a comparative text-entry study [Nicolau et al., 2010]. Besides our 5 users, we were able to recruit 12 candidates that used the traditional methods. The experience gained in these experiments and the close following of the aforementioned group of 5 called our attention to individual differences.

Later in 2011, we revisited touch typing and developed two novel methods: BrailleType [Oliveira et al., 2011b], a touch based counterpart of BrailleTap, and MultiTap, a touch-based version of the keypad-based counterpart [Oliveira et al., 2011a].

One of the limitations of the preliminary research performed by us until mid-2010 was that we were building our knowledge on top of experiences gathered with small groups of users. All the aforementioned studies were performed with attendees of a formation centre for blind people (*Fundação Raquel e Martin Sain*). Each study was performed with those that were taking a course at the moment in the centre. However, these experiences were paramount to enable an in-depth knowledge of the population and the idiosyncrasies within. In the following sections we present the methods we developed and a studies around NavTap that gave us the motivation to look at individual traits and pursue the remaining research presented in this dissertation.

## 3.2. Text-entry: a case-study

The text-entry task is transversal to a great number of mobile applications and when no assistive technology is available, it is just not feasible for a blind user. It is one demanding task both on mobile keypad and touch-enabled phones. Screen readers are software-based adaptations that replace the visual information with its auditory synthesis (e.g. Mobile Speak[1], Nuance Talks[2]). These solutions enable blind users to operate a device as they are able to receive feedback through an available channel. However, the interaction is not adjusted to the users' needs. Indeed, they receive feedback on the screen status but, for example, no information is offered on the keypad layout, thus leading to errors and reducing, or eliminating, the chance for him to learn and improve performance.

This problem gains additional relevance when considering older blind users that are likely to face several difficulties when having to memorize the letter placement on the keypad and dealing with a trial and error approach [Luo and Craik, 2008]. Existent solutions assume a user with good spatial abilities, memorization capabilities, or even good finger sensitivity, but the reality is that *more than 82% of all people who are blind are 50 years of age and older* [3] and a great part has lost sight in an advanced stage of their lives, which by turn translates in decreasing some of the aforementioned abilities [Moschis, 1992, Burton, 2003].

To cope with these difficulties, we initially focused our research efforts on providing alternative methods that could include people otherwise excluded from resorting to text and all tasks that make use of it. In the next section, we present text-entry methods that were developed along the course of this dissertation under my guidance. Following, we focus on NavTap, a keypad-based alternative, and explore the benefits on the long run of an inclusive approach, motivating our succeeding research.

### 3.2.1. Inclusive Text-Entry Approaches

We have designed, developed and evaluated methods to enable the non-visual input of text in both keypad and touch-based phones. Herein, we outline these methods for future reference.

**NavTap** [Guerreiro et al., 2008a] is a navigational text-entry method designed to reduce the cognitive load while inputting text with no visual feedback. To this end, the alphabet was divided in five lines, each starting with a different vowel, as these are easy to recall.

---

[1]http://www.codefactory.es/
[2]http://www.nuance.com/talks/
[3]World Health Organization Visual Impairment and Blindness Factsheet - http://www.who.int/mediacentre/ factsheets/fs282/en/index.html, June 2012

This alphabet representation can be navigated with a set of keys that act like a joystick. Both navigations (vertical and horizontal) are cyclical, which means that the user can go, for instance, from the letter 'z' to the letter 'a', and from the vowel 'u' to 'a' (Figure 3.1b). The users are able to navigate the alphabet and receive audio feedback on the current letter before accepting it (in opposite to MultiTap approaches where a key press can automatically lead to an error).

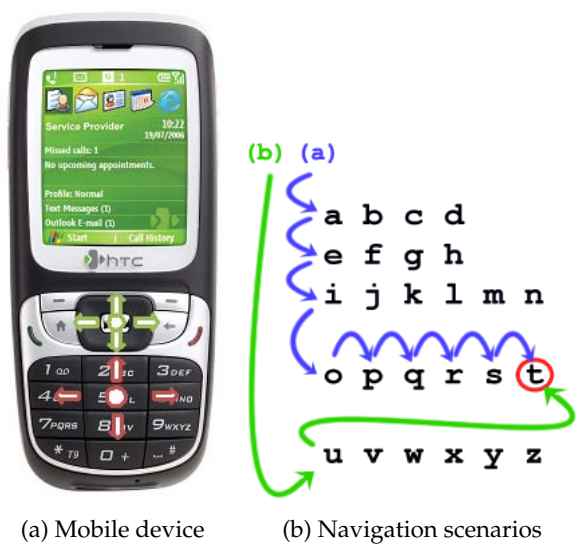

(a) Mobile device          (b) Navigation scenarios

Figure 3.1: NavTap text-entry method

As depicted in Figure 3.1b, different navigation scenarios and expertise levels can be achieved: 1) in the 1-way approach the user restricts the navigation to a single direction (straight forward), which can be classified as a naive approach; 2) in the 2-way approach the user is able to navigate through the vowels and, using them as reference points, get to the desired letter (scenario a)); and in the 4-way approach the user is able to use all 4 directions to perform the shortest paths to the desired letter (scenario b)) in Figure 3.1b. This text-entry method has been evaluated with blind users with reduced mobile device acquaintance (only placing and receiving calls) to assess the first contact with the method and the short term learning curve. The results were compared with traditional MultiTap approaches showing that NavTap, in opposite to MultiTap, enables unexperienced users to input text effectively and enables a fast performance improvement [Guerreiro et al., 2008a].

Focusing on users who know the Braille alphabet[4], **BrailleTap** [Guerreiro et al., 2008a] enables the users to input text by using some of the keys on the keypad as Braille cells.

---

[4]Being knowledgeable about the Braille alphabet and being able to read Braille is very different. One stereotype about blind people is that all know Braille and read Braille astonishingly well. With digital technologies, another stereotype is often referenced: that blind people no longer learn Braille. Neither is true. Knowing Braille and being able to read Braille are two different things. Most blind people learn the Braille alphabet during rehabilitation. In Chapter 5, we present statistics on a high level of Braille knowledgeable people in the population; on the other hand, very few read Braille proficiently as reading Braille requires other abilities like tactile sensitivity and perception

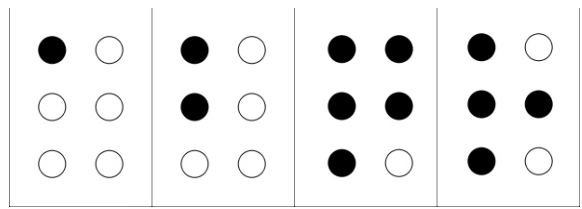

Figure 3.2: Letters 'a', 'b', 'q' and 'r' in the Braille alphabet

Thus, the user is able to select a letter by selecting the cells correspondent to the letter graphical representation, the Braille character. In the Braille alphabet, letters are formed by groups of 6 dots in a 3x2 cell (Figure 3.2). Considering the keypad of a mobile phone (Figure 3.1a) we can map that cell on keys '2', '3', '5', '6', '8' and '9'. Each press on these keys fills or blanks the respective dot. Key '4' allows the user to enter the letter or, if all dots are blank, enter a space. For example, to enter the letter 'b', the user has to press keys '2' and '5' followed by key '4'. Finally, key '7' erases the last character entered.

**NavTouch** [Guerreiro et al., 2008b] is a gesture-based approach with adaptive layout, i.e. users can perform gestures anywhere on the screen, therefore not being restricted to a fixed layout. This method is based on the same concept as NavTap: gestures to left and right navigate the alphabet horizontally; while gestures up and down navigate vertically (i.e. between vowels). Vowels are only used as shortcuts to the intended letter, thus users can choose whatever path they feel more comfortable. Speech feedback is given as users navigate the alphabet. To select the current letter users can perform a split or double tap. A comparative evaluation of NavTap and NavTouch showed that the touch-based version was faster than the keypad-based one as it implied less recognition (and likely, less tactile abilities) as gestures can be performed anywhere on-screen [Guerreiro et al., 2008a].

**BrailleType** [Oliveira et al., 2011b] also takes advantage of the capabilities of those who know the Braille alphabet and it is similar to BrailleTap but this time on a touch surface. The touch screen serves as a representation of the Braille cell, having six large targets representing each of the dots positions. These targets were made large and mapped to the corners and edges of the screen to allow an easy search. Users can perform a painless exploration, while receiving auditory feedback about each dot they are touching. To mark/clear a dot, a long press is required. After marking all the necessary dots for a Braille character, in whichever order the user desires, a double-tap in any part of the screen accepts it. A swipe to the left clears the Braille cell if one or more dots are marked or erases the last entered character if the matrix is empty. This method seeks to provide a less stressful first approach with touch screen devices by reducing the number of on-screen targets.

The **MultiTap** Touch typing method is a mixed one between its keypad-based homonym and Apple's VoiceOver. This approach uses the same exploration and selection mecha-nism as Voice Over (*painless exploration* and split or double tapping for selection). However, the layout presented is similar to 12-key keypad-based devices. We chose this method since this is a familiar letter arrangement to most users. There are twelve medium size buttons, each one featuring a set of characters, thus reducing the number of targets on screen. To enter a letter, users must split or double tap multiple times, according to the character position in that group.

During this research, we sought to deploy methods that varied in demand to cope with different levels of ability. One requirement was that all methods were deployed on mainstream devices potentiating social inclusion. In the next sections, we will take an in-depth look at how one of them (NavTap) enabled inclusion of otherwise excluded blind people. The touch typing approaches are revisited in Chapter 7.

## 3.2.2. A long term study with excluded blind users

Traditional laboratory evaluations are often performed to assess the usability of a system [Dix, 2004] in a small period of time. Systems are put to a test and a good outcome happens when the designed method outperforms a concurrent one or a pre-specified usability metric. However, several characteristics of a system are only revealed in the long run. To assess if a system serves the user as it should and enables him/her to achieve their goals, a long-term approach on the field is advised.

There has been prior evidence to suggest that NavTap can be effectively used by novices with very little training. Five users with no prior experience with NavTap were able to learn the vowel navigation method and perform text-entry tasks on a mobile device in a controlled environment [Guerreiro et al., 2008a]. However, an evaluation in a real life scenario, outside of the laboratory, was required to assess if the users could benefit from a supposedly more inclusive and less demanding approach. Further, laboratorial studies were performed with supervised pre-training sessions and validation *in the wild* was required to understand if this low demand method would indeed be adequate to the users' abilities and allow for them to learn and enjoy the mobile experience. To assess if the method was able to provide the desired social inclusion for these users we undertook a long-term evaluation of the NavTap prototype. This evaluation was performed during 19 weeks with 5 users (and 3 extra users during an iterative design phase) and, besides uncontrolled (but logged) daily usage, featured regular controlled experiments to observe the users' evolution (Figure 3.3). With these experiments we were able to collect data on mobile device performance usage, particularly text-entry, but also to observe how the improvements influenced the users' habits, interactions, and ultimately, social inclusion.

Our primary focus in this investigation was to assess the users' learning experience with NavTap in a real life scenario. This is particularly important since our system is targeted at individuals with visual impairments, who may not have many alternatives to fully control their mobile devices. Thus, an easy and autonomous learning process is crucial to the system's adoption. Therefore, our method should be both immediately effective for novice users and still offer a high degree of improvement as the users become more experienced. Moreover, this learning process should be easy and natural.

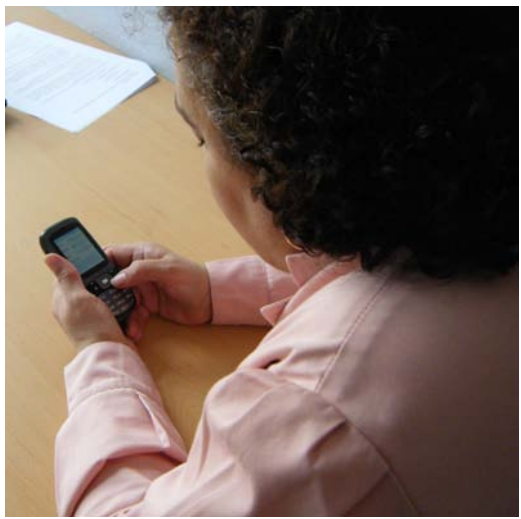

Figure 3.3: Blind person entering text with NavTap

Another focus of our study was on the daily usage of our system, particularly on the most used functionalities and communication habits. We wanted to investigate how our system influenced their habits and overall mobile and social interaction.

Although our investigation was focused in the real life scenario, we also wanted to assess the users' improvement through controlled sessions. Additionally, we were interested in understanding each participant's problems and difficulties, so we could identify the source of the issue and find the best way to address it in the future.

Therefore, with this study, we proposed to answer the following questions:

1) Can the users effectively operate a less demanding method, NavTap?

2) Do users reach an *expert* performance level on NavTap?

3) Does NavTap support the participants' social needs?

4) What issues related to NavTap and its usability are discovered in a long term analysis that otherwise were unrevealed?

In the overall, we also wanted to understand what made these users become excluded mobile-wise and if NavTap was able to lower the demands to a point they were able to surpass and become included.

## Iterative NavTap (Re-)Design

To be able to evaluate NavTap on a long term autonomous basis, we have gone beyond the text-entry task, and created a full prototype system with a simple set of applications (the ones the users revealed as essential and mostly used, in our preliminary studies). This set included contact management, messages, call management, alarm, calculator, notifications (e.g., battery), date and time. All the menu navigation and event reception (messages, calls) mechanisms were redesigned to match the absence of visual feedback and presence of auditory one (text-to-speech). Text-entry was achieved resorting to Nav-Tap with the aforementioned layout.

The prototype was developed in the Windows Mobile platform and the mobile phone used was the HTC S310 smartphone (Figure 3.1). The speech synthesis package used was provided by Loquendo. The first version of the prototype was developed accordingly to guidelines gathered in previous studies with the target population [Lagoá et al., 2007, Guerreiro et al., 2008a].

The user studies herein presented started with a preliminary (re)-design phase following a user-centred design approach. Eight blind users were selected from a group of 14 candidates at a formation center for blind people. The participants were selected accordingly to their proficiency with mobile devices: the aim of these studies was to evaluate the impact of a new text-entry method, one with lowered demands that gave space for differences in ability, and the ideal users were those unable to perform text-based tasks before, using NavTap. They had ages comprehended between 49 and 64 years old, all had a mobile device and none was able to input text. The re-design phase lasted for three weeks and was divided in modules (3 sessions per week): Navigation, Event Reception and Text-Entry. Each session with each user consisted in a 30 minute tutorial on particular aspects of the module being presented. With these sessions two goals were accomplished: 1) we were able to detect inconsistencies and adapt the prototype to better suit the users' needs and capabilities (re-design); and 2) the users were able to get some familiarity with the prototype, learning its most important concepts (training). In the iterative re-design, the prototype was modified in different aspects: missing functionalities, screen reading parametrization, input keys, sounds and earcons, among others.

In particular, and considering these studies' main scope, NavTap featured important adaptations: **1) Keypad Layout -** Prior design of NavTap linked numerical keys to directions (red arrows over the keypad in Figure 3.1) and a central key to input spaces and special characters. This design aimed at a full coverage of keypad-based devices as they all have a numerical keypad. However, the proximity and lack of distinction between the keys could be erroneous or slow if the users lacked sharp tactile capabilities (which is also a disadvantage of traditional MultiTap approaches). What is also true is that the majority of the keypad-based mobile devices now feature a navigation set of keys (joy-stick alike) (green arrows over the keypad in Figure 3.1) with closer buttons which are probably also wider and with better tactile characteristics. Thus, we have enabled their use to operate NavTap. The central key is also closer and easier to detect. This approach also enabled the remaining part of the keypad to be used as a special function repository. Once again, to ease the finding process we have placed the special functions in the corner and reference positions ('1', '3', '5', '*', '#',); **2) Letter Acceptance -** Timeouts are normally hazardous and have been criticized in the Human-computer interaction field [Raskin, 2000], although they have been commonly used in mobile text-entry interfaces due to the inherent lack of space. We identified two major problems with our previous timeout-based character acceptance mechanism. Firstly, considering a mobile context, the user is subject to interferences that can lead him to interruptions while navigating, thus leading to an error. Secondly, timeouts pressure the user, damaging confidence and the overall learning process. This was clear in the first contact with the users. An alternative was obligatory. Thus, the central key, while navigating the letter matrix, functioned as an acceptance key. If the central key is pressed after accepting a character (before entering another navigation step), a space is entered. The erase character, besides deleting the last letter, also disables an unwanted navigation and returns the system to a non-navigation state.

A third unforeseen goal was accomplished. With a closer contact with our participants we were able to identify differences that were likely to cause their exclusion. Instead of ignoring those particularities, we sought to explore them and tweak the interface to test the effectiveness of such approaches with the users. Our initial goal was to provide a method with such low demands that would be a solution for all the excluded blind people we could identify but these experience close with the population showed that different abilities are in place and an approach that takes those differences in consideration is likely to be more effective.

Particular examples are of two participants with low tactile sensitivity due to diabetes (their blindness was caused by diabetic retinopathy) that, even with a device with fine tactile cues (key size, relief and spacing), were very erroneous even when trying to place a call. By shifting the text input task from the numpad to the joypad, their problems were drastically reduced. For one of them, with severe limitations, we re-designed the application to require input just from the joypad (even to input a number, place a call, menu navigation, option selection) following a navigation approach. This changed his functional abilities drastically as he was as of that time able to communicate as never before.

One other example relies with audio feedback. What may seem obvious for most, may be a challenge for some. One participant, a late-blind person (acquired blindness one year before the study), showed very little confidence and often became confused with audio feedback. Simple changes in the order of the audio feedback along with a slower speed rate (along with the withdraw of timeouts) made this person the greatest success of our experiment.

At the end of the iterative re-design and training phase, the users were able to effectively operate the device.

## Procedure

To assess the users' learning experience with NavTap we have developed a functional prototype, which comprises the most common cell phone functionalities, previously described. After the initial learning and design period, we left the mobile devices with the users, so they could use them in their daily lives.

The evaluation was based on the analysis of the overall usage experience, which was captured through a logger (the user's privacy was totally safeguarded as no understandable personal data is collected). Apart from this, we performed weekly evaluation sessions in a controlled environment. This gave us a comparison baseline and deeper insights about NavTap. In those sessions, the participants were asked to input 3 different sentences (different across sessions). Those sentences had 3 difficulty levels based on their length and keystrokes per character for the best (KSPC) theoretical case. The chosen sentences lengths were fixed for the short (6), medium (11) and long (17) difficulty, as well as the interval of theoretical best case scenario keystrokes per character (KSPC) values, to allow evolution analysis through sessions.

The evaluation sessions took place in the formation centre for blind people over a period of sixteen weeks (thirteen sessions). Moreover, in order to compare the participant's performance before and after the daily usage experience, we performed two evaluation sessions still during the training period.

## Participants

All the initial eight volunteers for our long term study were students at the training center in which our controlled evaluation sessions took place. However, three participants had to drop out from our study because their courses at the training center ended. Table 3.1 illustrates basic characteristics about the remaining participants. The target group was composed by five participants (2 males and 3 females) with ages between 44 and 61 years old. All participants used their mobile devices on a daily basis but, typically, they could only place and receive calls. All participants used screen readers as their primary means of accessing a personal computer or mobile device. However, only three participants of our target group (P01, P03 and P04) used this kind of technology regularly.

P01 is blind since the age of 3 and has learned the Braille alphabet at the age of 8. The participant works with personal computers and speech synthesizers for sixteen years. Also, he has a degree on Psychology and good reasoning capabilities. However, he could only place and receive calls, as his cell phone did not have a screen reader.

| User | Gender | Age | Education | Time with impairment |
|------|--------|-----|-----------|----------------------|
| P01  | Male   | 49  | BSc       | 46 years             |
| P02  | Female | 44  | 4th Grade | 1 Year               |
| P03  | Female | 51  | 4th Grade | 10 years             |
| P04  | Female | 59  | 4th Grade | 12 years             |
| P05  | Male   | 61  | 9th Grade | 11 years             |

Table 3.1: Study participants' basic characterization

P02 is the youngest participant, with forty four years of age, and started to lose her sight a year ago. This progressive process of blindness has revealed to be very painful and stressful, reflecting on the participant's behaviors and moods. She has the fourth grade and, according to the formation center's psychologist, the participant had some learning and memory difficulties.

P03 had recently bought a screen reader for her cell phone but she could only hear text messages, place and receive calls. Until the time of the experiment, she was unable to learn the available text-entry method (i.e. Multitap) and perform more advanced tasks, such as contact managing.

P04 was blind for twelve years, and has never learned the Braille alphabet. She used a screen reader on her cell phone, for the past three years, but could only perform the most common tasks. Although she was able to hear SMSs, the participant was not able to reply.

P05 started to lose his sight eleven years ago, with fifty years old, due to diabetes, which is affecting both his nervous system and tactile capabilities. The loss of his tactile capabilities has already begun to affect his interaction with several devices, particularly those with less salient buttons.

Overall, our target group has a great diversity of sensory, memory and learning capabilities, mostly due to their age, diseases and impairments. Moreover, some of the participants are rapidly losing their residual vision or tactile capabilities, which is reflected in their behaviors, concentration, mood, and consequentially in the obtained results.

## Results

In the following sections, we present some of the key results regarding our weekly controlled sessions, participants' daily usage, and how the latter influences their performance on text-entry tasks. Moreover, due to the limited number of participants, our

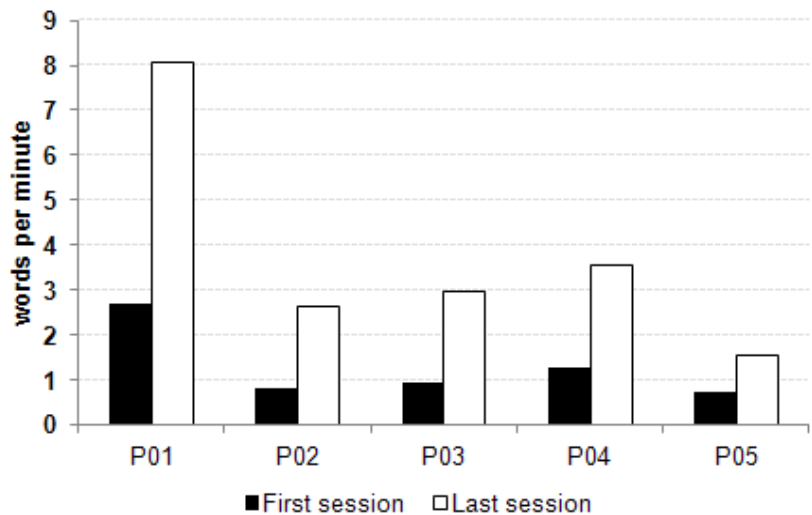

Figure 3.4: WPM on the first and last session.

goal is not to statistically analyze the data, but rather try to understand each user diffi-
culties and issues. We then highlight some key observations about each participant to
better understand specific behaviors and results. Summary tables of the results achieved
are presented in Annex A1.

**Weekly Controlled Results**

To observe the participants learning process, in a controlled environment, we weekly
assessed their performance by asking them to write 3 sentences, over a period of 13 ses-
sions.

Figure 3.4 shows the words per minute (WPM) achieved on both the first and last session
for each participant. Overall, participants demonstrated a great improvement in their
performance. Among the target group, the words per minute on the first session ranged
from 0.7 to 2.7. Over the 13-session (16 weeks) period, the participants reached, at least,
twice the initial performance with values ranging from 1.6 to 8.46 WPM. P01 had the
highest improvement from 2.7 to 8.46 WPM, indicating that the other participants still
have margin to improve.

Keystrokes per character [MacKenzie and Tanaka-Ishii, 2007] is the number of keystrokes,
on average, to generate each character of a text in a given language using a given text
entry technique . Figure 3.5 shows the KSPC on the first and last sessions for each partic-
ipant. Although some participants follow a naïve approach on the first session, as their
mental map becomes clearer they begin to follow a 2-way or 4-way approach. Com-
paring the improvement rates of KSPC and WPM, the latter is much greater, indicating
that participants begin to memorize paths and executing them faster, as they feel more
comfortable and confident using NavTap. The impact of times between key presses is
greater than the one resultant from a better navigation. However, participants do learn new paths and the ones that started with a naïve approach rapidly enrich their mental model outperforming the theoretical 2-Way scenario (excepting P02). Moreover, three participants almost reach the best case scenario, which indicates that NavTap is easy to use in a first contact and shows a good learning curve. This gave us a strong indication that indeed the burden to adapt to traditional assistive approaches is placed on the users' end and it is assumed that the users are able to cope with the demands imposed. By presenting methods with lower demands, even people with lower ability levels are likely to get over the first contact successfully and then improve with experience.

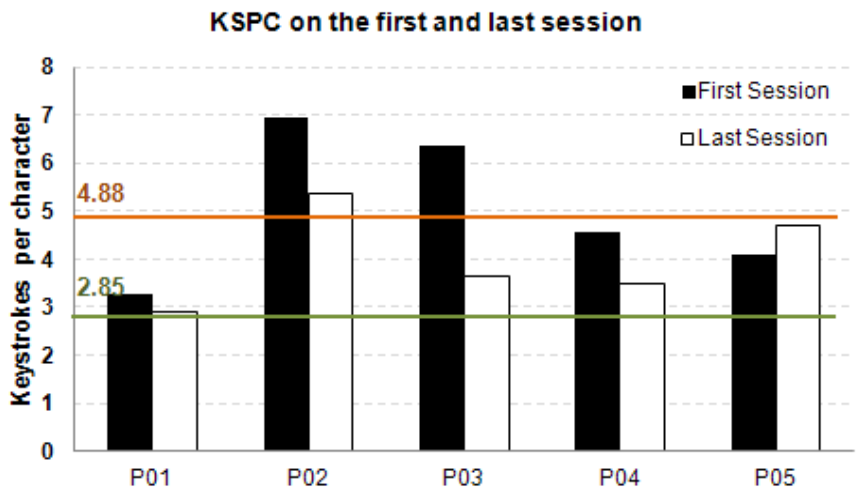

Figure 3.5: KSPC on the first and last session for each participant. The bottom (2.85) and top (4.88) lines correspond to the 4-way and 2-way theoretical approaches for the last session, respectively.

Figure 3.6 shows the improvement (%) in average preparation, navigation and acceptance times between the first and the last session for each participant. Overall, participants demonstrated a good improvement in all times, with exception to acceptance time. The acceptance time corresponds to the time between hearing a letter and accepting it, by pressing the joystick central button. On the final session the average acceptance time ranged from 0.78 and 1.69 seconds.

As we have mentioned before, as participants become more familiar with NavTap and the vowel navigation method, their navigation times between characters improved and is also reflected in the WPM chart (Figure 3.4). The navigation time between letters, when they were in the same direction had an average improvement ranging from 45% to 66% (average time on final session was 0.74 seconds). On the other hand, the average improvement in navigation time between letters on different row/column was smaller, ranging from 13% to 62%.

Moreover, an interesting fact is that preparation time had the greatest overall improvement from the first to the last session. Although participants improved their KSPC that did not affect their preparation time (i.e. spent time to begin the navigation). Indeed, one could argue that improving the paths to letters did not affect the participants' mental load, as they would discover new paths naturally.

The error rate (i.e. number of times a participant deletes a character) across sessions ranged between 1% and 4%, which indicates that participants usually did not make errors. To better understand the quality of the transcribed sentences, Figure 3.7 shows the Minimum String Distance (MSD) error rate [MacKenzie and Tanaka-Ishii, 2007]. Both P01 and P04 transcribed sentences are exactly the same as the proposed sentences for all 13 sessions. Moreover, P02 and P03 had an average MSD error rate of 3%, which is not significant, typically one error per session. P05 had the highest MSD error rate with only 8%, though. This indicates that NavTap is indeed easy to use and aids the users in their text-entry tasks, by preventing errors and consequently minimizing frustration.

Following the experiment, we performed a questionnaire to subjectively assess NavTap. The participants specified their agreement with a set of statements using a 5-point Likert scale (1 = Disagree strongly, 5 = Agree strongly). The gathered results are here present in the form of "statement (median, interquartile-range)": easy to use (5, 0), fast to use (4, 0), easy to learn (5, 0), felt in control (5, 0), improved with practice (4, 2), and makes the cell phone accessible (5, 0), increase communication (5, 0). Overall, the values are very high and consistent with exception to the statement "improved with practice." This can be explained because P01 did not feel that he had significantly improved, since he was already very good in the first session. On the other hand, P05 had a very limited usage of text messages and consequently was not able to improve.

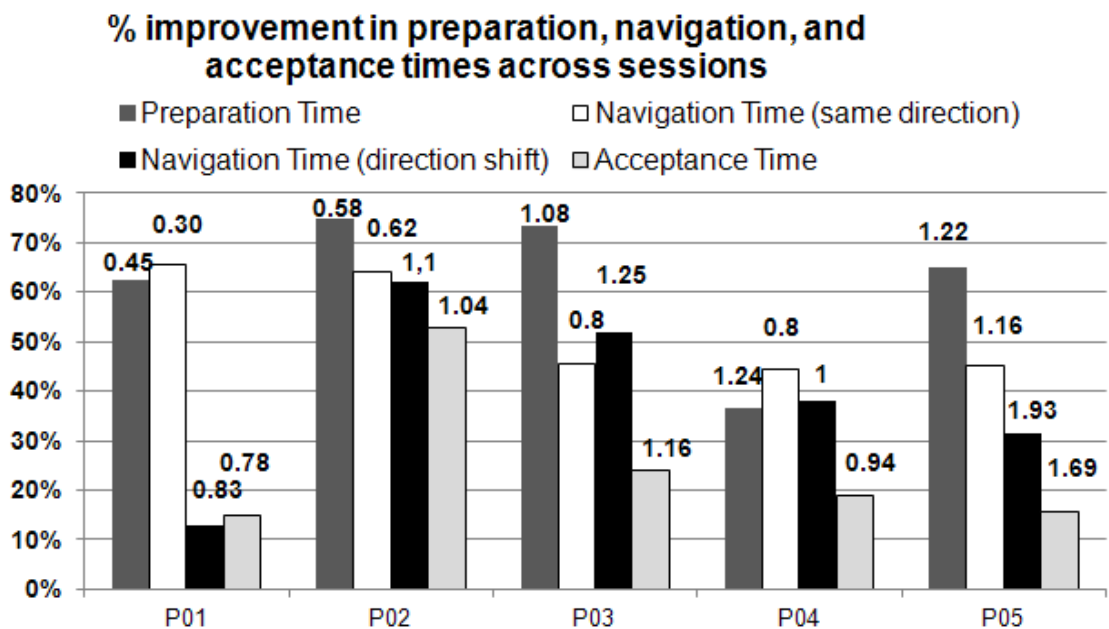

Figure 3.6: Improvement (%) in average preparation, navigation and acceptance times between the first and the last session for each participant. Numbers above the bars indicate the value in the last session.

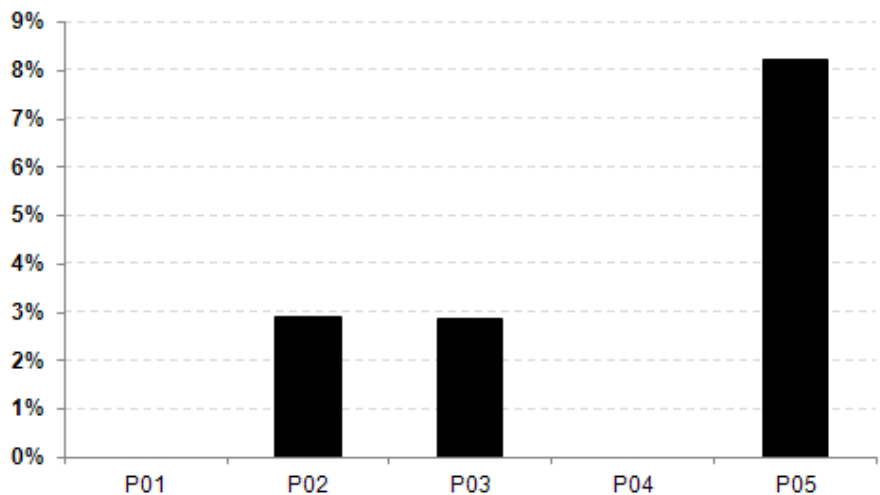

Figure 3.7: Average MSD error rate.

**Daily Results**

As aforementioned, although participants used their (old) mobile devices in a daily basis, their usage was very limited prior to the herein presented system. Most participants could only receive and place calls to a limited number of contacts, even those with a screen reader. Because they were not able to learn the traditional text-entry method (i.e. Multitap), they could not perform more advanced tasks such as contact managing (add, delete, edit or search contact) or sending text messages.

The results presented in this section and the system's usage may be influenced by a great number of factors, some of which are beyond our control, such as social and economic factors. Therefore, during this investigation we aimed at understanding the reasons for each usage pattern.

Overall, participants liked our system since day one and were very enthusiastic in using it. Our target group, with five participants, received and placed 678 and 797 calls, respectively, over a period of 16 weeks. Although this is a great result we cannot compare it to previous call usage. However, regarding text messages we know that none of the participants was able to send SMSs before they used our system. The achieved results were surprising and impressive; overall, participants received and sent a total of 1200 and 1825 text messages, respectively.

Figure 3.8 shows the percentage of communication methods used over a period of 16 weeks. Overall, SMS usage was over 20% and 3 of the participants in our target group prefered text messaging to voice calls. This indicates not only that text messages are in-deed needed by older visually impaired people to communicate with friends and family, but also that our method was able to support this need. It is noteworthy that none of the participants was able to send SMSs with their old cell phones.

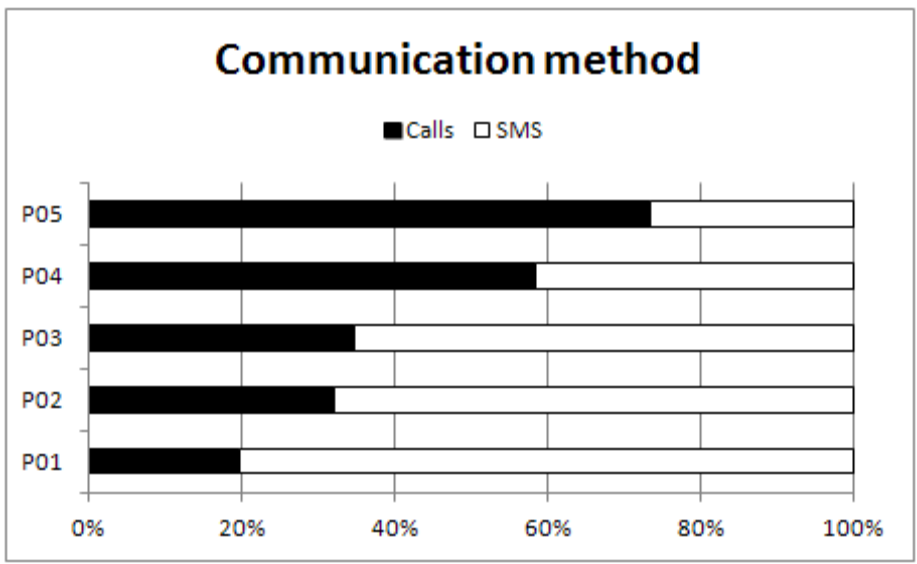

Figure 3.8: Participants' communication rate

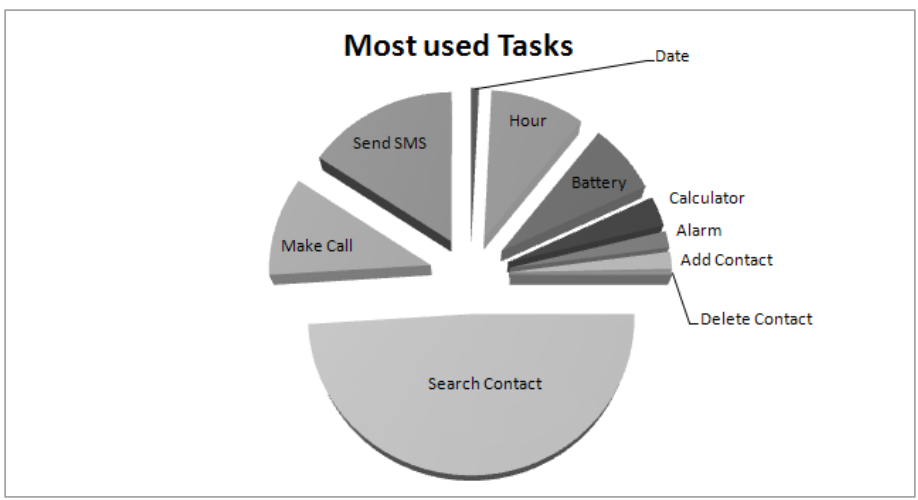

Figure 3.9: Most used tasks.

Relatively to contact managing, participants added a total of 133 contacts and deleted 26. The search contact task was the most used (Figure 3.9), which can be easily explained as this task is a sub-part of other tasks, such as placing a call, sending a text message or deleting a contact. However it also indicates that participants could easily input text and perform more advanced tasks.

Concluding, before participants began to use our system they had a very restrictive usage of their mobile devices. Indeed, they were only able to receive and place calls. NavTap allowed our target group to make a more efficient use of mobile devices, augmenting social inclusion and assisting them in their daily tasks.

**Usage Influence**

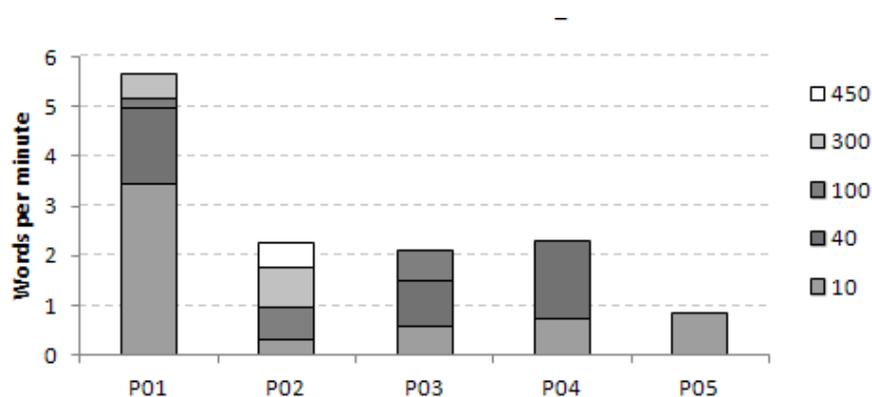

Figure 3.10: Sent text messages influence on WPM.

In our previous studies, we evaluated NavTap's learnability over a period of 3 laboratorial sessions, meaning that the participants' improvement could be somehow restricted. In this investigation we wanted to observe if their daily usage influences the method's learnability. Figure 3.10 shows the progress in words per minute (in controlled sessions) of our target group according to the number of text messages sent. Notice that the first 40 SMSs sent had the biggest influence in the participants' performance (exception has to be made to P02, which had the highest social activity but her improvement was very slow when compared to the remaining participants).

Moreover, Figure 3.10 also illustrates the diversity of our target group, both in social activity and performance improvement. P02, P03, and P04 reached the same WPM performance degree, although the number of sent text messages is very different.

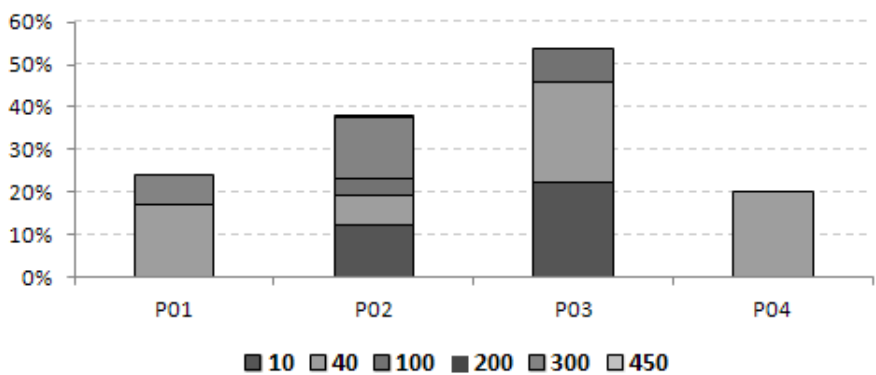

Figure 3.11: Sent text messages influence on KSPC.

Figure 3.11 shows the influence of sent text messages KSPC improvement. Overall, par-ticipants demonstrated the highest improvement in the first 40 SMSs. Again, exception is made to P02 that had a different learning curve, but a high number of sent text messages, though. P05 is not represented in Figure 3.11 because he did not improve his KSPC value(Figure 3.5). However, his performance was better than the theoretical 2-Way scenario, indicating that even with a small amount of experience (i.e. less than 15 text messages), NavTap is easily understandable and usable.

**Observing Each Visual Impaired Participant**

To better understand specific behaviors that may affect our target group, particularly their results, we highlight some key observations about specific participants.

Since day one, P01 had a good understanding of our system, particularly the text-entry method, NavTap. This specific participant had a very good mental model of the alphabet and a high literacy level. Therefore, it was easy for him to use a 4-Way approach since the training session (Figure 3.5). On the other hand, that did give him a very low margin of improvement, mostly on KSPC. His main improvements were in both navigation and preparation times, reaching 0.3 and 0.45 seconds, respectively. In our understanding, this participant is near from reaching the theoretical limit on both WPM and KSPC metrics. We do not consider that this participant lacked the abilities to use mainstream approaches. However, those still pose a challenge and his motivation to mobile phones and applications within was not strong enough for him to see an advantage in making the effort. An easier method offered him the opportunity to experiment the wins of a richer mobile experience. He used the prototype for two years. He is now a proficient user of mainstream approaches.

On the other hand, P02, accordingly to the training center psychologist, had severe learning difficulties. On the first training session this participant stated that she would never be able to learn how to input text with a mobile device. However, after a few minutes of practice she was able to navigate through the alphabet, even if using a naïve approach, and write a full sentence. Her interest in our text-entry method has only grown and this participant reached a perfect 2-way approach. Moreover, P02 was able to improve her preparation, navigation and acceptance times (Figure 3.6), indicating a comfortable usage of our system. The social inclusion of this participant was enormous, even with all her difficulties. She sent 625 text messages (half of the group's total) over a period of 16 weeks, and continually insisted in using NavTap on a daily basis after this research. Notice that before she had never been able to input text and had a very limited usage of her mobile phone. Because this was the only text-entry method she was able to learn, it became, without a doubt, a success story. She still uses our system which we have strived to maintain. However, she is already in her third device due to massive usage of the two earlier devices.

P05 did not improve his KSPC from the first to the last session (Figure 3.5), suggesting that our method is hard to learn. However, this happened because he did not practice and marginally used text-entry tasks. This was influenced by social and economic factors, which we could not control or anticipate. Indeed, this participant had a very low usage of his mobile device, as he only called his wife once or twice a day. Therefore, he maintained his navigation skills, reflected on KSPC, since the first session, which was already better than the 2-Way theoretical scenario, and also improved his performance. This suggests that NavTap is easily usable since the first contact and natural, even without practice. He still uses our prototype, more than three years after its initial deployment.

## Discussion

Mobile devices play an important role in our daily lives. However, they are still inaccessible to some blind users, due to their demanding interfaces. Current approaches, like screen readers, lack the adequacy to users' needs, especially for those with reduced levels of tactile and cognitive ability. NavTap is a solution to this problem, allowing users to easily control their mobile devices. Our main goal in this research was to assess the users' learning experience with NavTap in a real life scenario. We also wanted to see if users were able to use their mobile devices in a daily basis and what influence could it have on their performance and social inclusion.

All participants in this study were able to understand and use our text-entry method after a few minutes of practice, although with different performances. The higher improvement was seen on the first two weeks of daily usage, indicating that indeed, participants felt in control and comfortable interacting with their mobile devices. We also performed a study to assess how these users were positioned in relationship to expert MultiTap blind users. This study is presented in Annex A2 and reveals that these users achieved performances that although lower in average are acceptable and allow for their social inclusion.

In this research we also assessed participants' communication patterns and even though these were influenced by several factors (mostly economic and social), text messaging revealed to be an important communication method to blind users. Indeed, some participants adopt it has their primary communication method, due to context restrictions and economic factors.

Despite the participants' diversity of learning and memory capabilities, NavTap revealed to be accessible to all. By lowering the demands and easing the interaction we enabled access to an otherwise excluded blind population. Results suggest that those with higher tactile and cognitive levels could perform better on a first approach with the system, but still improve with experience. On the other hand, participants with more difficulties, although with less efficiency, could also control their devices and augment their social inclusion.

As a parallel take-away, one more recommendation arises: in order to fully evaluate a text-entry method or other communication solution we need to deploy it in real life scenarios with the target population. Only then will we be able to assess the users' true learning experience and impact in their social inclusion. The proximity with the users was paramount for us to understand the particularities within the population and focus our research on understanding and coping with those differences.

# 3.3. Major remarks

The laboratory studies performed for each individual text-entry method and the long-term study performed with a reduced set of blind people enriched our knowledge about the population as a whole and particularly about the differences within. These experiences called our attention to the relationship between an interface demand and a person's abilities. A portion of the users cannot use current technologies, not even the so-called *assistive*. These cases of exclusion are mostly due to the barriers faced in the first contact with such alternative interfaces which are too high and leave to drop out. Even some of the people that succeed, sometimes do so by means of extraordinary motivation and attitude towards technology or with the support of others.

Our first take at mobile accessibility was to provide methods that lowered the demands imposed by current technologies. Also, we did this without any hardware modifications something that has already been shown as relevant for social acceptance [Kane et al., 2009]. By lowering the demands, people became able to use those methods in their first attempts. Further, they proved to be able to improve as the confidence with the method increased.

This showed to drastically augment their social inclusion. Some continue to rely on our solutions while others took a leap and adopted mainstream screen reading technology. It is paramount to give people opportunities and confidence to embrace new technologies.

During this research, we sought to maintain ourselves updated with developments in the scientific, market and national contexts. Characterizing the usage of mobile devices by blind people along the course of this dissertation is hard as it evolved a lot from our preliminary studies to what we can observe nowadays. Still, the motivation for this research is still applicable and several blind people are excluded due to a mismatch between their abilities and the interface demands. Later in the course of this research, an in-depth study (presented in Chapter 5) was performed with 51 blind people to characterize and assess differences within the blind population, which revealed that several people are still facing the same problems we dealt with in the long-term study presented in this chapter.

These results motivate us to look with further attention at individual abilities and interface demands and seek for comprehensive knowledge on how to develop more inclusive mobile interfaces.

## 3.4. Summary

This chapter is a preamble to the remaining of the dissertation. In it, we start by giving an overview of the inclusive text-entry approaches developed and evaluated with the target population during the course of our research. All of them try to lower the demands imposed to the population although in different ways. NavTap is one of those methods that reduces the load on the user's end by resorting to a simple navigation method but still giving space for improvement. To assess if such an inclusive low-demand solution would enable users otherwise excluded to use a mobile device and its texting capabilities, we engaged on a long-term study with 5 blind people besides having an iterative user-centred design phase with 8 blind people. The long-term study included weekly laboratorial sessions and daily usage logging. The design phase enabled us to get a closer look at the difficulties faces by blind people when interacting with a mobile device and learning a new method. Also, it revealed that these difficulties varied widely between people. The weekly study itself showed that if methods are developed with demands adequate to the user's ability and allowing for improvement, the desired inclusion is attained.

# 4

# Individual Differences amongst the Blind Population

Previous chapters outlined a paucity of attention to the relationship between individual abilities and device demands. Without assessing abilities and evaluating their impact in surpassing device setting demands, we are failing the opportunity to deploy more personalized and adapted devices and interfaces that foster the inclusion and performance improvement of users spread along the ability spectrum on different dimensions.

Approaches to improve accessibility for blind people have a main focus: replacing the information transmitted visually with an alternative media. However, the complexity of visual information is higher than what can be transmitted via audio or touch. In the previous chapters, we called the attention to interfaces that request that the information lost in the visual/assistive replacement to be somehow complemented by the user's own abilities. This may come as an extra load on memory, spatial abilities, reasoning, tactile abilities, among others. Assuming that the users will have the abilities to bridge the gap and guarantee the accessibility of a method is erroneous. Actually, it damages the concept of accessibility and inclusion. The individual attributes that come to play in a technological setting should be taken in consideration when designing devices and interfaces for blind people. Our approach to identify relationships between abilities and demands is based on identifying relevant individual abilities and then stressing them with sets of different demands. By doing so, we put ourselves in a position to deploy interfaces that consider and take advantage of the user's abilities, with the ultimate goal of designing user-sensitive inclusive systems.

In this chapter, we present the first step at finding the most relevant individual attributes for a blind person in a technological setting. First, we provide theoretical background on the population. To enrich our knowledge about relevant individual attributes within the population, we performed an interview study with professionals from various backgrounds that work closely with blind people. This enabled us to outline the most relevant individual attributes in a technological setting. Further, we look at relationships between attributes and at how these differences are noticed and evaluated by the interviewees. Given the set of most relevant individual attributes we then provide theoretical background on each set of features (profile, tactile, cognitive, personality) along with standard procedures for their evaluation.

## 4.1. Causes

Blindness is due to a variety of causes. According to the World Health Organization[1], the leading causes of blindness are cataract (a clouding of the lens of the eye that impedes the passage of light) [47.9%], uncorrected refractive errors (near-sightedness, far-sightedness or astigmatism) [12.3%], glaucoma (a group of diseases that result in damage of the optic nerve) [12.3%], age-related macular degeneration (which involves the loss of a person's central field of vision) [8.7%]. Other major causes include corneal opacities (eye diseases that scar the cornea) [5.1%], diabetic retinopathy (associated with diabetes) [4.8%], blinding trachoma [3.6%], and eye conditions in children such as cataract, retinopathy of prematurity (an eye disorder of premature infants), and vitamin A deficiency [3.9%]. Age-related blindness is increasing throughout the world, as is blindness due to uncontrolled diabetes. On the other hand, blindness caused by infection is decreasing, as a result of public health action. It is estimated that three-quarters of all blindness can be prevented or treated.

According to the American Diabetes Association, Diabetes is *the leading cause of new cases of blindness in adults 20-74 years of age*. This is significant since diabetic retinopathy is often accompanied by peripheral neuropathy which also impairs the sense of touch [Levesque, 2005].

---

[1]World Health Organization, Fact Sheet 282, Visual Impairment and Blindness, http://www.who.int/mediacentre/factsheets/fs282/en/ (May 2009), Last Visited in June 2009

## 4.2. Worldwide statistics

There are several variations in definitions of blindness whether across states or countries, fact that difficults statistics on blindness. The American Foundation for the Blind estimates that there are 10 million blind or visually impaired people in the United States. In a survey realized in 1994-1995, 1.3 million Americans (0.5%) reported being legally blind. Of this number, only 10% were totally blind and another 10% had only light perception. The remaining 80% had some useful vision. Few statistics appear to be available about the age of onset of blindness. It is reported that 'only eight percent of visually impaired people are born with any impairment' [Harper, 1998]. Worldwide, an estimated 180 millions are visually impaired, of which 40-45 millions are blind [Leonard, 2001].

*The prevalence of blindness is much higher for the elderly* [Levesque, 2005]. It is estimated that 1.1% of the elderly (65 and over) are legally blind compared to 0.055% of the young (20 and under) [Hollins and Leung, 1989]. About 82% of all people who are visually impaired are age 50 and older (although they represent only 19% of the world's population). It is also reported that more than 50 percent of individuals with visual impairments also have one or more other impairments [Adams, 1986]. It is worth mentioning that blindness is expected to increase in the following years. It may come as a surprise, considering the advances in medicine that the number of blind people is predicted to double by 2030 [Leonard, 2001]. [Hollins and Leung, 1989] explain that the number of blind children is expected to increase because *the proportion of babies born to mothers at the extremes of the child-bearing years is increasing* and because *medical advances have made it possible for many premature infants, who in the past would have died, to survive*. The aging of the population in developed countries and the growth of the population in developing countries are also causes of concern. Increasing numbers of people are at risk of age-related visual impairment as the global population grows and demographics shift to a higher proportion of older people, even in developing countries.

## 4.3. Sensory compensation and diversity

The theory of sensory compensation states that a blind person's remaining senses are heightened to compensate for the loss of sight. This idea is somehow controversial and has long been debated. While several scientific sources on blindness take a conservative stance against the theory (e.g. [Warren, 1978]), there is mounting evidence from recent studies for limited sensory compensation in the blind [Bavelier and Neville, 2002]. As an example, [Stevens and Weaver, 2005] state that *one consequence of blindness appears to be enhancement across the broad categories of auditory perceptual and cognitive functions, particularly in cases of early-onset blindess*. [Zwiers et al., 2001], on the other hand, showed that some sound localization skills may be impaired in the early blind due to the unavailability of visual feedback for calibration [Levesque, 2005].

Also supporting the theory of compensation, a study by [Goldreich and Kanics, 2003] has shown evidence of better tactile acuity in the blind. The authors showed that *the average blind subject had the acuity of an average sighted subject of the same gender but 23 years younger*. On the other hand, [Warren, 1978] reports mixed evidence concerning pattern and form perception.

All summed up, *it is generally agreed that the blind are more proficient at attending to nonvisual stimulus and that they make better functional use of nonvisual senses* [Levesque, 2005]. For example, *the blind have, through need, learned to attend better to auditory stimuli and therefore can make more use of the available auditory information than sighted people* [Warren, 1978]. A good example is the 'obstacle sense', or 'facial vision', that allows a blind person to feel the presence or absence of obstacles. Researchers have shown that the obstacle sense is mediated by audition, from echodetection and echolocation [Warren, 1978]. However, the obstacle sense can be learned by blindfolded sighted subjects. The blind are also particularly skilled at attending to voices. Similarly, it has been shown that curvature is judged better by the blind due to better exploratory techniques [Warren, 1978]. This compensation builds on the concept of neuro-plasticity [Kaas, 1991]. The brain is no longer seen as a static entity but as one that is dynamic and reflects changes in our experience and damage-induced reorganization [Pascual-Leone and Torres, 1993].

## 4.4. Interview Study

An interview study was performed with ten (10) professionals working closely with blind people for at least 5 years. This semi-structured had the main goal to create a baseline set of individual abilities to be explored in the following studies.

### 4.4.1. Study Goals

The main goal of this study was to *identify the characteristics that have a greater impact on a blind user's functional ability*, particularly which are the individual attributes that may influence the interaction between a blind user and technology (mobile). Further, we wanted to deeply understand the *relations between them, how are they recognized by professionals* and in *which tasks are reflected the differences among users*. These attributes can be demographic, motor, cognitive or sensorial. To collect a set of meaningful features we have interviewed specialized professionals working closely with blind people. This study enabled us to assess a set of candidate attributes and include previously uncounted ones. Moreover, we have also tried to achieve the following goals:

**Assess the diversity among the blind population and acknowledge its impact** Besides
understanding the dimensions where the population varies, we also intended to
understand the scale of those differences. Further, we wanted to verify if indeed the
differences within the population had greater impact than the one felt by sighted
people when interacting with technology as this was one of the motivational as-
pects for our focus on the blind population.

**Understand the individual differences' practical implications** These professionals have
wide experience working with blind people both with technologies they are already
proficient with and with the learning and adoption of new technologies. As such,
another of our goals was to understand the impact of the individual differences in
both those aspects: the proficiency levels attained and the difficulties while learn-
ing.

**Understand how are these differences currently observed and assessed** All the people
that we interviewed require at some point, generally in the first contact, to assess
particular abilities of the blind person they are working with. This may be to de-
cide on an application to a course or to define a rehabilitation strategy. As such,
these people have wide experience in understanding and assessing the individual
abilities of blind people. Our last goal was to collect information on both standard
and non-standard procedures to assess a blind person's ability levels.

In this chapter, we describe the study and analyse the results obtained. Section 4.4.2
describes the methodology applied in this study while in section 4.4.3 we present the
results obtained outlining the most relevant attributes, relations between them and how
(if) they are evaluated.

## 4.4.2. Experiment methodology

To accomplish our goals, we selected a varied group of professionals with vast experience
working with blind people and performed a semi-structured interview study.

### Selecting the Interviewees

We tried to interview certified professionals working closely with blind users for at least
5 years. We looked for a diverse sample including psychologists, occupational therapists
and teachers. The search for interviewees was performed at public institutions (reha-
bilitation and formation centers, special teching divisions). We aimed to select a wide
coverage of interviewees to include different perspectives and experiences.

## Procedure

To recruit a diverse interviewee group we pre-established contacts with several public institutions. We looked for individuals working closely with blind people for at least five years. Also, we tried to guarantee elements from different intervention scopes thus trying to cover different points of view.

The recruited participant group (10 elements) was then composed by 3 psychologists, 2 occupational therapists, 1 IT teacher, and 4 rehabilitation technicians. Two psychologists work at Portuguese formation and support institutions for the blind while the remaining one works in a governmental department for the education of young blind people. Curiously, all these three individuals are also blind offering a different perspective and insight to their opinions. The rehabilitation technicians cover different cognitive, sensorial and motor perspectives and perform work with the target group from the most basic needs like orientation, motility, eating, dressing, posture while the rehabilitation technicians offer advanced formation in computers, telephone operator, carpentry, weave, among others. The IT teacher also works in a governmental department and works closely with children and their adaptation to technology and assistive components. Overall, the participants were recruited from five different institutions.

All of them work with the target population for at least 5 years and had contact with a minimum of 50 blind individuals. The interviewee profile has little relevance in this study; exception made for the profession and the experience with the population. Although they have all worked with several blind individuals in the latest years, some have passed through different technological epochs while others have always been working aided by computers or even teaching the blind how to use them.

Each participant partook in a 30 to 60 minutes in-person semi-structured interview. The interview was conducted at the participant's institution in a quiet room with two intervenients: the researcher and the interviewee. The interview covered four different topics: 1) the diversity among the population and the impact of those differences in relation to the differences between sighted people; 2) individual differences and their impact in interaction effectiveness, and learnability; 3) how are individual capabilities and limitations assessed/identified and how, if possible, are they overcame; 4) how are individual differences related and what type of action they have most influence on.

The detailed procedure for this interview is available, in Portuguese, in annex A3.1. The interview started with both the interviewer and the participant sat facing each other, both in a comfortable position. Upon initial greetings and accommodation, I introduced myself and gave an overview on my research theme. It is important to notice that while it was important to place the participant in context with our research, it was also mandatory that the description was brief and superficial to avoid biasing their contributions. Both in the introductory monologues as in the interview itself we fought to maintain our intervention as subtle and as brief as possible. Afterwards, I asked the participant to talk freely and in an informal tone. Also, I asked permission to record the interview for further analysis while assuring all data to be anonymous.

Before the interview, we performed a set of characterization questions to fill the participant file (Annex A3.2). The interview itself started with a set of introductory generic questions leading to an interactive dialogue. The interview guide includes a wide set of seed questions to help the interviewer when a theme finishes. This is important to maintain the dialogue and give a sense of continuity. The researcher's role is crucial and hard as he must be alert both to explore an idea brought up by the interviewee and to start new themes keeping this sense of continuity. Also, the guide features several different questions on the same theme to extract information with different angles and points of view. Thus, each user asked a subset of the presented questions as well as others that appeared during the conversation.

The major goal of this study was to identify the individual attributes that diverge the most among blind users and have greater impact in their technological functional abilities. Prior to the study, based on our limited experience and un-verified common sense, we had a set of possible characteristic candidates. With this study, we intended to verify our assumptions as well as include or reject other individual attributes. The interview form featured a table with a set of characteristics (demographics, motor, sensorial, cognitive). As the users spontaneously mentioned an attribute, we marked it as "spontaneously mentioned" (an "X" in the form). Attributes that were not considered in our table, were added in its end. Attributes featured in our table, unmentioned by the participant, were prompted in the end of the interview for an opinion. These mentions were marked as "induced". Both spontaneous and induced mentions, could be positive or negative references, meaning that the participant could identify the characteristic as relevant in the person's functional abilities or not. Also, the participants were free to skip a characteristic if they did not have an opinion about it.

The interview was finished with the interviewer (myself) thanking the participant for his/her collaboration. Afterwards, I transcribed all the interviews and performed a detailed content analysis, described in the next section.

## Analysis

We analysed transcribed interviews to identify individual attributes functionally distinguishing blind users, how they relate between each other and at what levels they affect interaction. We used open and axial coding to analyse responses [Charmaz, 2006], methods of the *grounded theory* approach [Strauss and Corbin, 1998]. *Open coding* consists identifying, naming, categorizing and describing phenomena found in the text, sentence by sentence. Axial coding consists in relating codes (categories and properties) to each other, via a combination of inductive and deductive thinking. Rather than look for any and all kind of relations, grounded theorists emphasize causal relationships, and fit things into a basic frame of generic relationships (Phenomenom, Causal Conditions, Context, Intervening Conditions, Action Strategies, Consequences).

## 4.4.3. Results

All the results presented herein are built based on data from the transcribed interviews.

### Diversity among the Blind

Blind people face an extra load when dealing with interfaces created to be visually explored. These particular users are extremely challenged when interacting with recent interfaces, even in the presence of assistive technologies. Particularly, considering mobile devices, a blind person can use screen reading software to overcome the barriers imposed by the absence of visual feedback. However, this auditory feedback, in a graphical user interface, is far from its visual counterpart. The inefficiencies are even more visible if we look to the devices physical properties and cues. A blind user has no information in respect to layout, key/action or screen area/action association.

On the extraordinary demands imposed to blind users, one of the interviewees stated:

> *The blind person is subject to higher efforts as it has less information sources or less chances of repetition*

The general idea among the interviewees is that blind users face extraordinary barriers when dealing with technology, justifying the efforts to improve their access to the interfaces and devices.

What is also true is that the blind population is highly different. This opinion was shared by all the interviewees. Most of them illustrated these differences with examples of people with contrasting levels of technological expertise. These differences were attributed to several different aspects among sensory, demographics, cognitive and motor dimensions. What was also stated was that this divergence among people has greater impact between blind users than between sighted people. Allied with the barriers imposed by devices and their interfaces, a particular individual characteristic may isolate a person technology-wise. Some participants revealed cases of extremely successful blind people as well as others with several functional limitations. As an example, a user stated the following alerting to the impact of individual differences among the blind as more significant:

*Regarding technologies, the variations between individuals have effects with extra significance than in sighted people. The intellectual variation, as an example, has a higher impact*

Examples given by the participants were in reference to memorising layouts, orders, numbers, key/action associations, among several others. These problems were associated with different individual differences.

## Mobile-wise Relevant Individual Attributes

It was consensual among the participants that the blind population is highly heterogeneous and that this heterogeneity has greater impact in functional abilities than among sighted people. The main goal of the study was to identify these differences. Overall, twenty-two characteristics were mentioned by the participants. Figure 4.1 presents a tag cloud outlining the most mentioned individual attributes. Peripheral sensitivity, motivation, spatial ability, blindness onset age, intelligence, memory and age stand out as the most relevant ones.

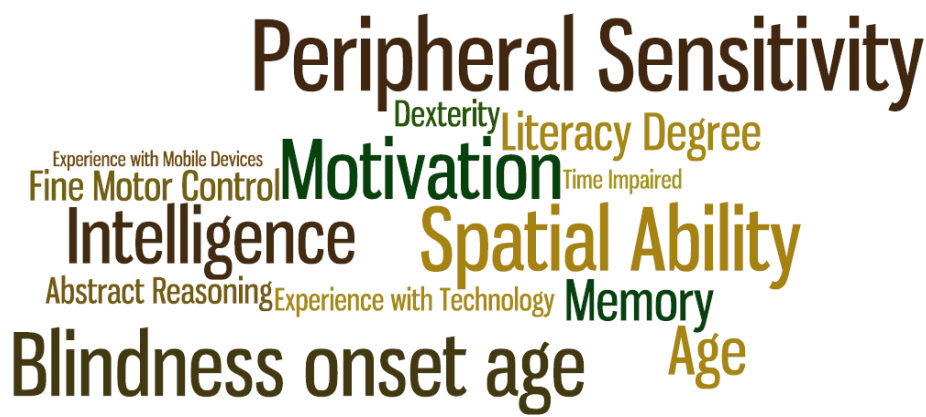

Figure 4.1: Individual attributes relevancy

Figure 4.2 presents the positive references made to the attributes, whether they were spontaneously mentioned or induced by the interviewer. This chart includes duplicates and shows the relevancy the participants attributed to each feature. Other characteristics were mentioned but we have omitted those with less than 5 mentions. Peripheral sensitivity, blindness onset age, spatial ability, age, motivation/attitude, intelligence and memory were the most referenced. Age, memory and time impaired were the ones that when induced had greater acceptance, i.e., the participants felt they were relevant although they did not mention is spontaneously. Some justified the omission of these characteristics with their obvious nature. Motivation, Intelligence and Abstract Reasoning were spontaneously mentioned several times but were not seen as relevant by those who did not mention it voluntarily.

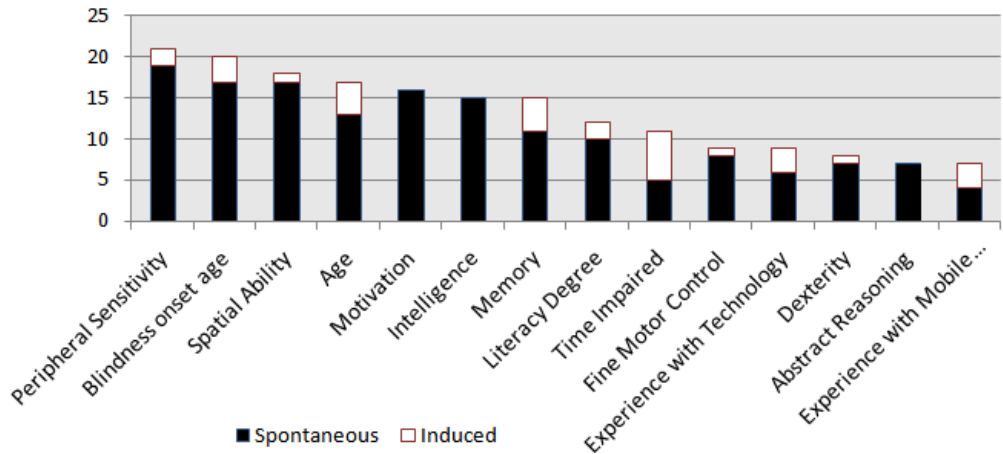

Figure 4.2: Spontaneous and induced references to relevant individual characteristics

## Relation Between Attributes

We collected references to individual attributes that are very different in type and domain. The mentioned features are not just from sensory, motor, cognitive, profile or functional contexts. They span all these areas. One interesting question lies in understanding if some of these attributes are related with each other. Such knowledge enables that work building on top of our findings may focus on particular characteristics that encompass others.

An analysis of the causal relationships between attributes revealed the following implications: blindness onset age was said to influence spatial ability (high[2]), dexterity (low), motivation (low) and technological experience (low); motivation influencing technological experience (low) and, particularly, mobile device experience (low); age having influence in peripheral sensitivity (low), motivation (low), memory (medium) and technological experience (low); literacy degree influencing spatial ability (medium) and technological experience (low) (Table 4.1). Previous experience with technology in general was only referred to influence proficiency with mobile devices and, vice versa.

In sum, greatest relevancy was given in terms of causal relationships to the cognitive component. Spatial ability, a cognitive component, seems to be highly influenced by epoch of blindness onset and literacy degree while Memory seems to be related with the person's Age. These results go in line with the background on cognitive abilities. The remaining attributes were not referred to be connected with each other, being considered independent by the interviewees.

---

[2]These classification was based on the number of people that referred this causal relationship

Table 4.1: Relations between attributes

## Levels of impact

It is relevant to notice that the pointed attributes and characteristics influence user's ability at different levels. On one hand, they are spanned along the different components of the human processing model (perception, cognition and motility), while when considering product demands, implications are also visible at different layers, for example,hardware and software, and other finer grained ones. As an example, a participant said:

> *At the basic level, previous technological and mobile experience does not have a great impact. It is relevant when one starts to explore the functionalities*

We also sought to retrieve the Context where these attributes were more relevant in a mobile interaction context. Fine motor control and Peripheral (Tactile) Sensitivity were referenced only to influence the low-level relationship with the hardware. They are thus more physical characteristics. On the other hand, literacy degree, reasoning, and previous technological (also mobile-wise) experience were only related with software components being considered independent from the device used. the remaining ones presented in Figure 4.2 were considered to be relevant in both contexts.

## Evaluating and Perceiving Levels of Ability

One other thing that we were interested to assess was if these professionals felt the need to assess the blind people's profile and abilities. If so, we were interested in understanding how they did so and in what contexts.

Given the different contexts of the interviewees' area of intervention, the knowledge shared with us was also of different nature. For example, psychologists tended to perform more standardized low-level evaluations while teachers and rehabilitation technicians resorted to a more functional approach. In this section, we outline the evaluations performed while in section 4.5 we detail each attribute and the most common assessments for blind people pertaining that attribute.

One psychologist at a formation centre for blind people performs cognitive and dexterity assessments to the candidate students there. These assessments depend on the course they are applying at. The goals were twofold:

**Acceptance**  Different courses stress different characteristics. Further, some of these courses have wider market acceptance and revenue which translates in a relevance of selecting the candidates that are most fit pertaining those abilities. As an example, the phone operating course is one that still creates job opportunities for blind people in Portugal. Assessments are made pertaining Verbal IQ, in particular Memory and Attention, and Spatial Abilities. Further, preference is given to younger people.

**Accompaniment**  Despite the course, if one gets accepted at this formation centre, a report is created and delivered to the people that will deliver the course. Particularly, in the first classes, a closer individual-sensitive learning methodology is performed, enabling the participants to evolve supported with adequate methods and tools.

The tests applied by psychologists are commonly standardized ones. The aforementioned professional resorted to a well-established cognitive assessment (from the Wechsler Adult Intelligent Scale [Wechsler, 1981]) to evaluate the verbal component of the intelligence quotient. Particularly, he focused on the evaluation of Working Memory although in specific cases he also resorted to the Verbal Comprehension Index (e.g., when he noticed a particular flaw in the person's vocabulary). This same scale also contains assessments for Reasoning and Processing Speed, among others: the problem is that these are designed for visual feedback. As such, alternatives to assess this non-verbal component are required. This same professional resorts to a standardized test specifically made for blind people to assess their spatial ability [Xydias, 1977]. This assessment is composed by two puzzle boards that the blind person is challenged to complete (Figure 4.3).

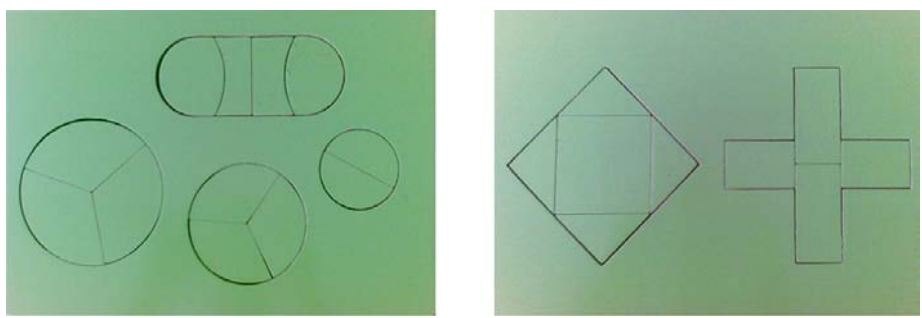

Figure 4.3: Spatial Ability Assessments for Blind People

Other than these cognitive assessments, this same psychologist often mentioned to assess people's dexterity through a test called *fil de fer* in which the blind person had to travel iron coin-shaped pieces, with a hole in the middle, through a tumultuous iron thread. This test demands that the person used multiple fingers to be able to make the coins go through the entire path and reach the end point.

One other psychologist which worked closely with young blind people also resorted to WAIS to evaluate the verbal component of intelligence but resorted to BLAT (Blind Learning Aptitude Test) [Newland, 1979], a set of tests aimed at student relying on several sheets of raised-line symbols, similar to Braille, to have a more concrete idea of the people's abilities in spatial, reasoning and processing speed components.

Contrasting with these standard assessments, one other psychologist resorted to a more functional approach. He performed open question interviews to have a general idea of the person's ability and background and then resorted to the realization of functional tasks to assess particular attributes. Detailing, he would evaluate tactile abilities by putting people reading a Braille sheet; those who did not know the alphabet he would ask them to detail the number of raised dots on each character. He also revealed that to assess dexterity he would tie knots in a thread and ask the person to untie them; he would count and register the timings until the task was completed. When asked about less subjective assessments, he showed to have knowledge of the several possibilities but stated that these evaluations were not accessible economically. He coped with that be developing his own methods. This reinforces that the evaluation is relevant as even without the proper tools, this person felt the need to find his way to assess people's ability.

Participants shared with us the need to assess individual abilities of blind people in diverse contexts. This awareness of a person's ability is valuable for vocation guidance but also to adapt teaching strategies to the idiosyncrasies of a particular individual. This feedback showed valuable as the same concepts may be applied to technology adoption. Awareness to one's abilities and how that person is able to surpass the demands imposed enable a less demanding first approach with in turn fosters adoption. With experience, the demands are likely to decrease and people tend to improve their performance. This support is paramount to assure a more inclusive use of any device or technology. In Chapter 3, we have showed that indeed by providing an inclusive method and enabling a more supported learning process otherwise excluded people are able to quickly attain an acceptable performance and improve with experience.

## 4.5. The relevant individual attributes

The interview study gave us further knowledge on the individual attributes relevant in a technological setting. This set of attributes is considered by the professionals we interviewed for selecting people to particular tasks and job opportunities as well as to enable the adaptation of learning methods and contents. Adaptation and personalization to individual traits in a technological, particularly a mobile one, is still ill-explored. However, the information gathered in this chapter suggests that attention should be given to individual differences. The set of relevant attributes spanned profile and background, cognitive, tactile and motor dimensions. We will detail each of those dimensions also revisiting related work that has addressed them and giving a first overview of how those abilities can be measured.

### 4.5.1. Basic Profile

To what we called basic profile, which is mainly related with demographics and education, the interviews revealed five attributes to be relevant: Age, Epoch of Blindness Onset, Literacy Degree (or Educational Background as an easier quantified measure), Experience with Technology in general, and Experience with Mobile Devices. Time Impaired, a sixth relevant attribute, outcomes from Age and Onset of Blindness.

**Age** is the exponent of attention in Individual Differences' related work [Czaja and Lee, 2007, Ziefle et al., 2007]. Indeed, Aging is associated with a decline in a multitude of aptitudes and several researchers have shown differences in several contexts between groups of different age groups. Concerning abilities, there is evidence that working memory declines substantially with Aging [Luo and Craik, 2008] as well as tactile abilities which also decrease with age, particularly after the age of 60, where peripheral sensitivity starts to decline drastically [Moschis, 1992, Gregor et al., 2002]. One last attribute was revealed in our study to be related with Age: Motivation or Attitude. Indeed, older people tend to resist to technology improvements when the perceived ease of use and usefulness are not obvious [Melenhorst, 2002]. They also tend to have a more conservative approach towards adoption[Renaud and van Biljon, 2008]. Related work supports the attribute relationships found (Table 4.1). Indeed, Age seems to imply and enclose a set of individual abilities that make worth looking at differences between age groups.

**Epoch of Blindness Onset** was also refereed as one characteristic prone to influence a blind person's ability to deal with technology. Indeed, an early-blind individual gets through important personal ability construction phases in a very different manner from a late-blinder who has experienced a vision-based world throughout his life. The former is also forced to pay extra attention to other senses like touch and audition, learning to perceive them better than a sighted person (and a late-blind person). On the other hand, the latter is also likely to have acquired knowledge pertaining device usage that may put her in a better position to adopt and use other technologies. Also, by being exposed to spatial information, *late-blinders* are likely to have a richer understanding of spatial information. However, this is still discussion in the research community around this effect: *Deficiency theory* states that congenitally blind individuals are unable to develop a general spatial understanding because they have never experienced the perceptual processes (as vision) necessary to comprehend two and three dimensional arrangements, scale changes and more complex concepts such as hierarchy, pattern and continuity. As a result they lack the ability to perform complex mental spatial problem. Solving involving rotations and transformations; *Inefficiency theory* states that people with visual impairments can understand and mentally manipulate spatial concepts, but because information is based upon auditory and haptic cues this knowledge and comprehension are inferior to that based upon vision. *Difference theory* states that visually impaired individuals possess the same abilities to process and understand spatial concepts, and that any differences, either in quantitative or qualitative terms, can be explained by intervening variables such as access to information, experience or stress [Kitchin et al., 1997, Monegato et al., 2007].

Despite the disagreement on the effect of epoch of blindness onset on spatial abilities, our interviewees on the field, seem to have a clear notion that it does have an influence on spatial abilities. In the relationships found (Table 4.1), Spatial Ability was highly related with Blindness Onset. Relationships were also found with Dexterity, Experience with Technology and Experience with Mobile Devices. These arose from the different patterns of technology usage between a blind and a sighted person. A sighted person is likely to have gained experience with devices and learned to cope with them; blind people are likely to have been less exposed to technologies and mobile ones in particular, as inclusion for this population is always a step behind. Dexterity was also refereed in this context as interviewees revealed that fewer experience with devices translated in difficulties in performing previously unexplored multi-finger movements. On the other hand, some also refereed that blind people which worked in hand craft could have this problem diminished.

Motivation/Attitude was also related with blindness onset age. People who were born or acquired blindness early in life tend to have a more positive attitude as they do not face the dramatic loss of a sense they were used to rely on. Attitude towards overcoming the loss of vision varies widely between people. Further, different periods are likely to be noticed from deep grief to acceptance and rehabilitation [De Leo et al., 1999].

**Educational Background** is also one aspect recurrently explored in what respects the variability between functional abilities of the general population. It is also one common aspect taken in consideration when profiling participants in the Human Factors area. The three attributes presented herein interrelate. One early-blind person receives a different education from an older blind person, which by turn goes through a phase of rehabilitation pertaining basic activities of daily living but also revives infancy has tasks like reading and writing are re-learned. This educational component was refereed to improve spatial abilities and both experience with technology and mobile devices. Early-blind people who are well accompanied in infancy (educational-wise) receive stimulus to improve their spatial abilities (e.g., puzzle solving)[Landau, 1988]. As to experience with technology, interviewees revealed that more educated people tended to have more contact with technology, particularly computers and Braille devices (also mobile ones).

**Experience with technology** was reported as improving the abilities to adapt and adopt novel methods and devices. This was also reported as to novel mobile devices and paradigms (**Experience with Mobile devices**). The main reasons presented was that many concepts underlining user interfaces were comparable between models and devices [Brewster, 2002]. Also, this relationship was also argued to exist between several technologies (e.g., desktop computers) and mobile devices (Table 4.1).

The five components pertaining Basic Profile can be retrieved via simple questionnaires. A more functional notion of the person's literacy can be evaluated resorting to functional evaluations (e.g., Braille reading speed). Experience with technology and mobile devices can be also evaluated with the assessment of functional tasks or through self-reported questionnaires [Schroeders and Wilhelm, 2011]. while cognitive and tactile abilities can be evaluated resorting to standardized and clinical assessments. Examples of those are described in the following sections.

## 4.5.2. Tactile Abilities

**Peripheral Sensitivity** was mentioned and highly reinforced during the interviews by the interviewees. The need for blind people to perform a tactile exploration of the devices they work with turn tactile abilities into a feature to take in consideration, one that is often neglected when visual feedback is available. While output from the computer or mobile device screen is often replaced with audio feedback, the input layer also needs to be perceived. Failing to do so, is likely to damage the person's ability to interact. Assistive technologies (e.g., Braille) are often more demanding concerning these abilities.

There have been efforts to identify requirements and guidelines to cope with reduced tactile abilities [Kurniawan et al., 2006, Benali-khoudja et al., 2004]. However, no work has been reported to our knowledge empirically exploring tactile abilities and the relationship with technology and attained performances.

The assessment of tactile abilities takes us to medical and clinical contexts. Measuring one's tactile sensitivity is common in two areas: measuring the loss of sensitivity due to diabetes and measuring these levels in cases of surgery, particularly, in the reattachment of extremities amputations. In the latter, the measurements are not precise as the goal is to evaluate the regain of regular sensitivity and not so much the finer grained evaluation performed in the case of diabetes. In this scenario, tactile sensitivity is measured as a way to understand the evolution of the disease, and how it is affecting neuro-peripheral sensitivity. Three key components are considered for assessing tactile abilities in a person's fingers. These components are **pressure sensitivity**, **spatial acuity** and **thickness discrimination** [Tremblay et al., 2005].

Pressure sensitivity is commonly evaluated resorting to the Semmes-Weinstein Monofil-

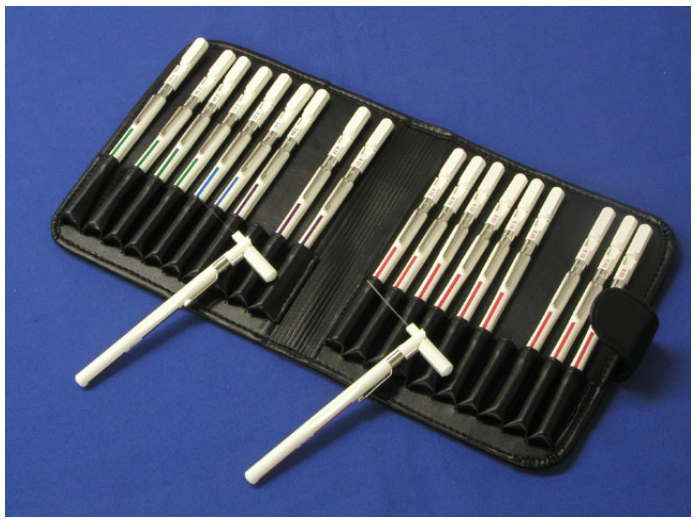

Figure 4.4: Semmes-Weinstein Monofilament Test

aments test [Kamei et al., 2005]. This assessment provides a non-invasive evaluation of cutaneous sensation levels throughout the body with results that are objective and repeatable. The basic methododogy for assessing skin sensitivity with this method is that a nylon filament will exert an increasing pressure on the skin as it is pressed harder, up to the point where it starts to bend. This allows applications of a reproducible force level, even though an imprecise hand manually applies the probes. The Semmes-Weinstein series is a standardized set of nylon monofilaments all of constant length but varying in diameter (Figure 4.4) which are then selected and labeled so as to give a linear scale of perceived intensity [Sto, 2011].

Spatial (or Tactile) Acuity relates to the ability to discriminate between two points of pressure on the skin. This acuity relates to the number of receptive fields and it is highly dependent on the body part where it is measured. As an example, spatial acuity is drastically higher in the fingertip than in the elbow. However, even in the fingertip, this acuity varies from a person to another and also varies dynamically for a single person with aging [Stevens, 1992]. Tactile acuity is often assess by means of a two-point touch discrimination test which measures the individual's ability to perceive two points of stimuli presented simultaneously. The person's acuity is smallest distance between the points that can still be perceived as two points. An example of such an assessment is the Disk-Criminator [Mackinnon and Dellon, 1985] which is an orthogonal plastic device with pair of metal filaments. These pairs present different relative distances, varying between 1 and 25 millimetres.

Thickness discrimination is regularly assessed by means of plaques of different thickness. Each plaque is held between the thumb and the index finger and the person is asked to assess differences between the plaque and another one of standard thickness. This tactile sensitivity dimension seems to have no relationship with the usage of devices as they do not resort to the ability to distinguish between different thickness. We will not consider this dimension in our research.

## 4.5.3. Cognitive Abilities

Intelligence, in general, was mentioned as a relevant attribute for interacting with main-stream technologies and applications. In detail, Spatial Ability, Memory, and Abstract Reasoning were mentioned. What is also worth outlining is that some interviewees, particularly the ones dealing with the cognitive assessment of abilities of blind people, were clear about the extra-load imposed to blind people. Examples were given considering the need to receive large quantities of information via audition and having to memorize it. Counter-examples were given for the case of sighted peers who could resort to vision to iteratively explore such contents. One example given was of a restaurant menu or a payment reference number. Examples pertaining spatial abilities were offered particularly in the usage of keypads and keyboards revealing the same discrepancies between a visual and non-visual scenario.

Cognition comprises a set of mental processes that include attention, memory, producing and understanding language, solving problems, and making decisions. It is studied mainly in the area of cognitive psychology although cognitive efforts span through all areas of human intervention.

Evaluation of cognitive abilities, or the general term *Intelligence*, focus on two main components: verbal and non-verbal. The verbal component comprises verbal comprehension (e.g., global knowledge, long-term memory, logical reasoning,vocabulary, social comprehension) and working memory (attention, short-term memory, arithmetic). The non-verbal component is mainly concerned with Spatial Ability. **Spatial Ability** is defined as the competence to manipulate or transform the image of spatial patterns into other arrangements [Pilgrim, 2007].

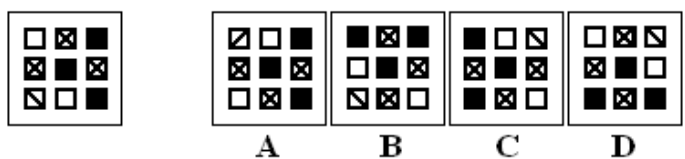

Figure 4.5: Spatial Aptitude Test traditional format: *Which figure is identical to the first?*

Test batteries for sighted people normally encompass both components. Some of these, particularly in the non-verbal component, depend on visual feedback (Figure 4.5). There have been efforts to create, adapt and validate assessments that are able to assess the intelligence of blind people. Adaptations to verbal assessments were easily deployed and corroborated [Hupp, 2003]. Examples of verbal assessments performed to blind people are the Interim Hayes-Binet, or the verbal components of the Cognitive Test for the Blind or the Wechsler Adult Intelligent Scale (WAIS).

On the other hand, the lack of adequate instruments to perform a non-verbal evaluation, often results in the generalization of the verbal component as an overall measure of intelligence [Miller et al., 2007, V. Reid, 2002]. However, spatial abilities and non-verbal reasoning are pivotal constructs of all models of human abilities. This aptitude has been highly stressed for vocation guidance [V. Reid, 2002]. Thus, some assessments were created to evaluate the non-verbal abilities of blind people. One example is the puzzle test battery presented earlier in this document [Xydias, 1977]. Another example is the Stanford-Ohwaki-Kohs Block Design Intelligence Test for the Blind [Dauterman et al., 1966] where visual stimulus are transformed in tactile ones, replacing blocks with several colors by blocks of several textures [V. Reid, 2002]. The Blind Learning Aptitude Test is another example which is by turn based on Raven progressive matrices (Figure 4.6), which consists in a series of line and dot patterns (in relief) of variable complexity from which the person being assessed has to, through touch, find a relationship or find the missing item for a determined pattern [Dai and Sternberg, 2004]. Other examples exist.

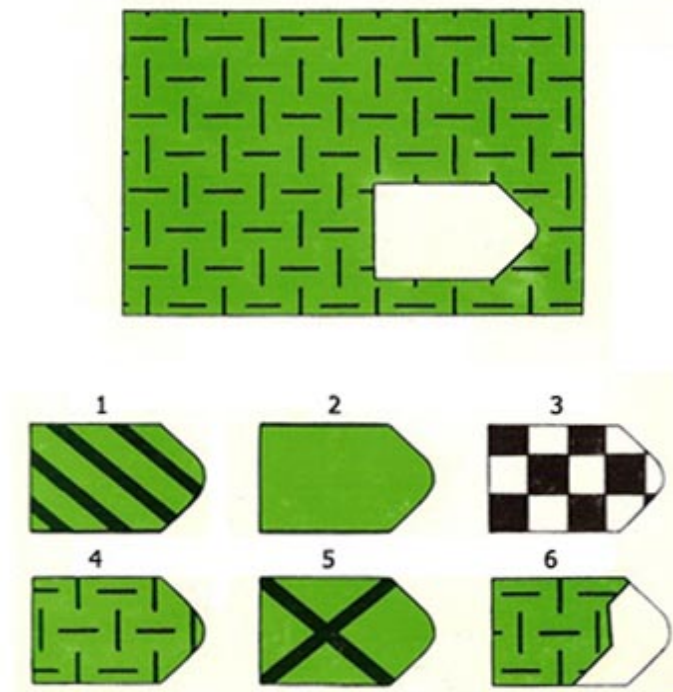

Figure 4.6: Raven Progressive matrices

Revisiting the results from the interviews, three components of Intelligence showed to be relevant in this dissertations' context: Spatial Ability, Memory and Abstract Reasoning. Other researchers have already leaned over the impact of cognitive abilities on the performance attained with technology. [Pak et al., 2006] looked at the effect of cognitive abilities, particularly spatial abilities, in web searching scenarios where became evident that spatial orientation was paramount. [Dyck and Smither, 1996] tried to understand which cognitive abilities of older people were stressed in text processing applications (on desktop computers). Deductive and logical reasoning along with spatial visualiza-tion showed to play a relevant role. Sara Czaja et al. have been focusing on outlining relevant individual abilities, particularly among older people, that have influence in different tasks like information retrieval [Czaja et al., 2001, Sharit et al., 2004], data introduction [Czaja and Sharit, 1998] and voice-based phone menu navigation[Sharit et al., 2003]. [Edwards et al., 2005] presented one of the few researches that has focused on people with visual impairments, particularly diabetic retinopathy, in menu selection tasks. Results showed that short-term memory and attention were relevant to user performance. Along with the paucity of studies focusing on the impact cognitive differences among blind people, there is also a lack of studies pertaining mobile devices which may come as a surprise as these settings are likely to be more demanding as interfaces are over-complicated to fit in a more restricted environment. [Ziefle et al., 2007] present one of the few exceptions showing that spatial ability and verbal memory are relevant in the usage of hyperlinks in a PDA.

## 4.5.4. Motor Abilities

Pertaining motor skills, two features were mentioned by several participants of the study: **Fine motor skills** and **Dexterity**. In fact, although the former is broader, in the context of using hand-based technologies (computers and mobile devices), they can be argued to represent the same, that is the coordination of small muscle movements in body parts such as the fingers. In this sense, we will just resort to the term **Dexterity**. This attribute has been proved to suffer alterations with age and tool usage [Tremblay et al., 2002] and has been subject of attention in research pertaining the effect of individual differences among people with visual impairments and handheld computer interaction [Leonard et al., 2005].

Several assessments can be found to evaluate Dexterity [Yancosek and Howell, 2009]. Examples that are commonly used in research studies are the Box and Block Test, Minnesota Rate of Manipulation Test or the Purdue Pegboard. They vary in cost, whether or not the test assesses bilateral hand use, tool use, manual versus finger dexterity (or both), and the time taken to administer the test. One thing that is common to all, and, in fact, to the definition of Dexterity itself, is that they measure the coordination between muscles and the eye. This is, they all resort to vision. There are not many alternatives to evaluate the dexterity of blind people. The Purdue Pegboard Test [Tiffin, 1948] is one successful example of adaptation of a dexterity assessment to blind people [Curtis, 1950] and it has been used to assess the abilities of blind people, for example, for vocational purposes[Kathryn and James, 1960, Tobin and Greenhalgh, 1987]. This assessment measures dexterity in two types of activity: one involving gross movements of hands, fingers, and arms, and the other involving what might be called *fingertip* dexterity. The pegboard is equipped with pins, collars, and washers, which are located in four cups at the top of the board.

### 4.5.5. Attitude and Motivation

Attitude towards Blindness and Motivation for adoption of technologies was a perva-
sive component in our interviews. Although the remaining individual attributes have
their place in the performance attained when dealing with technology, two main ground
rules arose: 1) in a presence of low-motivation and negative attitude towards the ben-
efits of technology, a barrier is imposed for adoption and improvement; 2) people with
extraordinary attitude and motivation to use technology tend to overcome their indi-
vidual limitations with experience. Pertaining the first point, one of the common re-
ported cases relates to older people, who tend to resist to the adoption of new tech-
nologies [De Leo et al., 1999] mostly due to low perceived ease-of use and usefulness
[Davis, 1989]. One other aspect relates to the resistance in the process of acquired blind-
ness [De Leo et al., 1999, Murray et al., 2010] as well as the resistance to adopt assistive
technologies, particularly by younger people [Shinohara and Tenenberg, 2007]. The sec-
ond point (overcoming lack of abilities) reveals that the limitations imposed by individ-
ual attributes can be overcome. This happens traditionally at the cost of compensatory
mechanisms and habits.

Our interviews revealed that these **drive** influences the experience with technology and
with mobile devices, which goes in line with the aforementioned. More motivated people
tend to experiment more and gain more proficiency with technologies and alternative
methods.

Given the unstable character of these attributes, particularly, motivation, and the diffi-
culty in measuring them consistently, we will not give attention to them in the remaining
of this dissertation. We acknowledge that motivation and attitude can interfere with the
correlations and implications we will take. However, they do not do it consistently. In
the remaining of this research, we focus on attributes that although dynamic, guarantee
stability along a large period of time.

## 4.6. Summary

In this chapter, we presented the findings of an interview study with people with a vast
experience in dealing with blind people under technological settings. The interviewees
clearly stated that the individual disparities between two blind people are likely to have
a larger effect than among sighted peers in dealing with interfaces designed for visual
feedback. Tactile sensitivity, blindness onset age, spatial ability, age, motivation/attitude,
IQ and memory were the most referenced. Either with standard or non-standard proce-
dures, these professionals recurrently feed the need to assess the ability levels of blind
people to select candidates for a particular position or to adopt devices and formation strategies. Further studies are required to assess how the population differs within the aforementioned dimensions as the literature presents very few details in this domain.

# 5

# Assessing Individual Differences among Blind People

The *state of the art* (Chapter 2) review shows that there is a sense of the relevance of individual attributes when dealing with different devices and their underlying demands. Words like *individual differences*, *dynamic diversity*, and *ability* are strongly connected with *inclusion* and *inclusive design*. We consider that, when dealing with technology, individual differences have larger impact among blind people than among sighted ones. This is mainly due to the paradigm followed by current interfaces which are, in general, mostly visual. This supposition has been corroborated by blind professionals working closely with other blind people (Chapter 4) who have revealed that capabilities like memory, spatial ability or tactile sensitivity are stressed when a blind person deals with user interfaces, even those, and sometimes particularly so, which are meant to be accessible.

To cope with these differences and foster inclusion, a better knowledge of which capabilities and related demands is required. It is then relevant to understand where people vary and the magnitude of those variations. Further, devices present demands and those require a certain level of ability. Does the user present the required level of ability?

In this chapter, we explore the differences within the blind population. Particularly, we seek to reveal dissimilarities in the target group that are often ignored and show that the *blind stereotype* is inadequate. To this end, we have designed a study to characterize blind people within different aspects: background profile, tactile abilities, cognitive abilities and functional abilities. These were selected based on the knowledge gathered in Chapters 4 and 3. The study was performed with 51 blind people. Empirical studies within the accessibility field are often characterized by a reduced number of participants. This was counteracted in this dissertation through an extended and careful recruitment and accompaniment phase leaded by the author with close collaboration with the staff of a local formation centre (*Fundação Raquel and Martin Sain*). This set of participants composes our participant pool. In the studies presented in the following chapters, participants are sub-sets of those characterized in this chapter.

As aforementioned, we have collected functional measures besides profile and ability-based ones. This enables us to present a first take on the impact of individual differences on user performance even though they are subject to contamination with years of experience and compensation mechanisms.

In what follows we describe our study and participants along with the characterization of the population. We present relationships between user profile, abilities and functional performance in usual technology-wise tasks. We end by discussing the diversity of the population and the impact of individual differences.

## 5.1. Recruiting Participants

Performing in-depth evaluations with disabled people is known to be a hard task. In the human-computer interaction community, there is an informal acknowledgement that studies within the accessibility context are likely to be performed with a smaller user sample than it would be expected in an evaluation with a non-disabled group. One can argue that this is acceptable for preliminary system validations or to compare design solutions.

The preliminary studies presented in the early chapters of this document were carried through with such smaller samples and the conclusions taken from them are exploratory. They were useful to identify flaws, formulate hypothesis and define our research questions. Yet, we aim to contribute with the formal recognition of the impact of individual differences in blind mobile users' performance and, as such, a less informal approach is required. Also, to explore differences and have a sample that is not sparse in levels pertaining each attribute, a larger and diverse user sample is mandatory.

Soon in the course of this research I engaged contacts with individuals and institutions working closely with blind people. Aiming at finding individual differences, this search for users was performed at several age, professional and social levels. I engaged contacts with national associations, institutions, rehabilitation and formation centres and introduced myself to the responsible person of those places. One important aspect is that most fruitful contacts were performed with the help of a pre-established relationship with a local institution, particularly, with a blind psychologist there. We have been collaborating with this institution and this person for several years. While my personal attempts to directly engage contacts with the administration of such centres were mostly unfruitful, the contacts supported by the institution psychologist were mostly directed at other psychologists and technicians of others institutions and were in general successful.

From our experience, it is clear that a first step for a successful participant recruitment starts with building the confidence with one institution and the people therein. From the start of our collaboration with this particular institution we have always discussed our prototypes with the technical staff there. Further, although monetary incentives were never offered, we always engaged attendees of the centre in the discussion over our prototypes and, when possible, seek to deploy them for their personal use and benefit. Support was always offered both for our own prototypes and their personal devices and technological difficulties. From our point of view, they have always seen us as part of the centre and not as outside people.

Succeeding contacts with other people and institutions were always performed with the support of people deeply inserted in the target community. This includes the professionals mentioned above, particularly the institution's psychologist, but also word-to-mouth propaganda from the previously recruited participants. It was with surprise that when contacting new participants for our studies they already knew who we were and were already very positive about working with us. Some came with their own ideas for research prototypes and technological innovations. The professionals recruited for the interviews presented in Chapter 4 were already recruited with this support and so were the other participants presented here.

## 5.2. Procedure

We gathered and performed evaluations with fifty one users (51), with light perception at most, within a timespan of approximately one year. We strived to guarantee such a large sample to cover a group of users from different backgrounds and with denoted individual differences (age, education level, professional activity, among others). Other individual attributes assessed in our studies were also expected to diverge but could not be verified during the recruitment phase (e.g., tactile abilities).

This evaluation was threefold: a first phase to portray the participant, gathering profile and background data; a second phase where we performed sensorial and cognitive evaluations; and, a third one where we undertook functional assessments to characterize the user's performance with mainstream devices. The evaluations with each participant were performed in two 45 minute sessions: the first one encompassing the background, sensorial and cognitive assessments and a second one dedicated to functional evaluations. The evaluations were structured as follows:

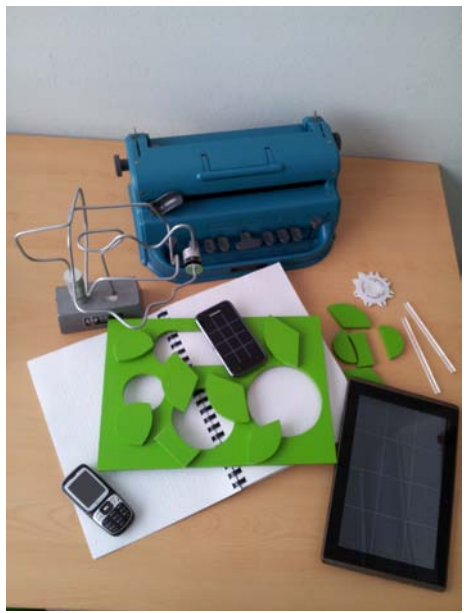

Figure 5.1: Paraphernalia of artefacts used in this study

## Greetings and Briefing

The first contact with all participants was devoted to thanking them for participating in the studies and explaining the studies' objectives, and particularly, their motivation and long-term goals. Participants were encouraged to dialogue and stepping to the following stages was never rushed by the evaluation monitor. This stage cannot be seen as a lesser one mainly because it serves a subliminal but highly relevant goal: reducing the participants' anxiety. Herein, our goal is to capture the participants' characteristics and abilities and, to do so, putting them comfortable is paramount.

## 5.2.1. User Profile and Background

The first evaluation stage with all participants was the administration of an oral questionnaire. Besides contributing to the aforementioned goal, augmenting the person's comfort, its main purpose was to collect profile information (e.g., age, educational background, age of blindness onset, type and cause of blindness), assess device experience and profi-ciency (devices and applications used), and Braille literacy level.

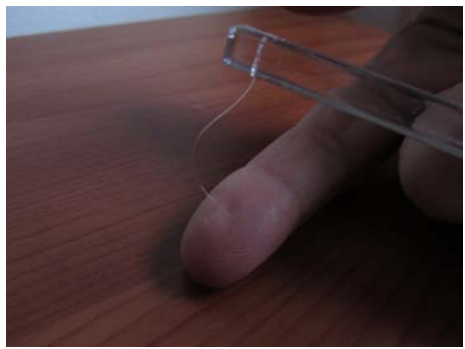

Figure 5.2: Semmes-Weinstein Monofilament (Pressure Sensitivity) Test

Related work on dynamic diversity (Chapter 2), particularly with older people, has given much attention to age-related differences. Also, studies on the blind population have shown cognitive and tactile differences depending, for example, on blindness onset age (Chapter 4). Our goal here is to capture the diversity of the population and verify the cross-relations between these background data and sensory, cognitive, and, mainly, functional abilities.

## 5.2.2. Tactile Assessments

To assess the participants' tactile capabilities, two different components of tactile sensitivity were measured: pressure sensitivity and tactile acuity. The first, pressure sensitivity, was determined using the Semmes-Weinstein monofilament test [Tremblay et al., 2005] (Figure 5.2), already presented in Chapter 4. In this test, there are several nylon filaments with different levels of resistance, bending when the maximum pressure they support is applied. This way, if a user can sense a point of pressure, his pressure sensitivity is equal to the force applied by the filament.

Five monofilaments of 2.83, 3.61, 4.31, 4.56 and 6.65 Newton were used, starting the stimuli with the one of 2.83, the least resistant one. Pressure was applied in the thumb, index and middle fingers, those generally used when interacting with devices, and in random order, so we could prevent arbitrary identification of a stimulus by the person being tested. The process is repeated with the filament with the next resistance level, until all filaments are tested or the participant correctly identifies the stimulus made. Different levels of pressure sensitivity can be found for different fingers.

The other tactile sensitivity component measured was spatial acuity, using the Disk-Criminator [Mackinnon and Dellon, 1985]. This instrument measures a person's ability to distinguish one or two points of pressure on the skin surface. The Disk-Criminator is generally an orthogonal plastic instrument that has in each side a pair of metal filaments with relative distances ranging from 1 to 25mm. When the person being tested identifies a stimulus as being two points, her spatial acuity discrimination is equal to the distance between the filaments.

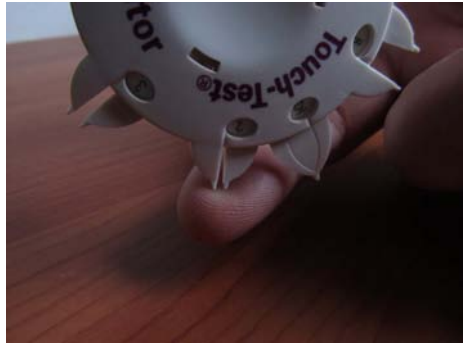

Figure 5.3: Two-Point Discriminator (Spatial Acuity) Test

The distance between the filaments of the Disk-Criminator tested ranged from 2 to 15 mm, with 1mm increments. Each of these filament pairs was, as with the pressure sensitivity, applied randomly in the same three fingers. There were made 10 stimuli per finger, randomly, alternating between a pair of filaments and a unique filament. The participants had to indicate if they felt one or two points of pressure. When they were able to correctly identify 7 out of 10 stimuli, their level of spatial acuity was registered as the distance between filaments.

One important consideration regarding tactile sensitivity is that it may change due to environmental conditions or even, in people with diabetes, vary widely during the day according to glucose level. To counteract undesired variations and capture a meaningful value for each person, two actions were performed: 1) the room temperature was warmed as one of the main problems with momentary lower sensitivity levels arises from extreme cold/hot weather. Participants were instructed to heat their hands until they felt comfortable and warm; 2) participants with diabetes were asked to measure and ascertain glucose levels. In further evaluations described along this document, the same participants were always asked to measure glucose levels. Evaluations were only performed when the participants registered their regular indicators.

### 5.2.3. Cognitive Assessments

The cognitive evaluation focused on two components of the cognitive ability: verbal and non-verbal. The verbal component was evaluated in terms of working memory, a short-term memory and main responsible for the control of attention. The non-verbal component, which consists of abilities independent of mother language or culture, was evaluated in terms of spatial ability: the ability to create and manipulate mental images, as well as maintain orientation relatively to other objects.

To evaluate working memory, the subtest Digit Span of the revised Weschler Adult Intelligence Scale (WAIS-R) was used [Wechsler, 1981]. In the first part of this test, the participant must repeat increasingly long series of digits presented orally, and on the second, repeat other sets of numbers but backwards. The number of digits of the last series properly repeated (the digit span) allows the calculation of a grade of the participant's working memory and, subsequently, to the user's verbal intelligence quotient (Verbal IQ). Figure 5.4 shows the series of numbers presented orally to the participants for both the forward and backward settings. Each series is compound of two groups: if a participant fails to recall correctly all the digits, the second group is administered. If he fails both, the Digit Span is the preceding series. From the beginning of the test, the series are presented to the participant without any further explanations or commentaries.

Spatial ability was measured using the combined grades of the tests Planche a Deux Formes and Planche du Casuiste (Figure 5.6a, 5.6b). These two tests are part of a cognitive battery for vocational guidance [Xydias, 1977]. The goal of these tests is to complete, as fast as possible, a puzzle of geometrical pieces.

## 5.2.4. Functional Assessments

To assess previous device-wise functional abilities and experience, the users were asked to input text with a mobile phone, a Perkins Braille typewriter and a desktop personal computer (PC). Once again, text-entry was selected due to its ubiquitousness and complexity. Also, it was a shared task across all the experimented devices. Participants were also asked to read Braille. Functional assessments are described below.

All users were asked to write three individual sentences in each of the devices. Each sentence comprised 5 words with an average size of 4.48 characters. These sentences were extracted from a written language corpus, and each one had a minimum correlation with the Portuguese language of 97% to be representative. We chose not to vary the sentences written between devices guaranteeing that all participants wrote the same 3 sentences with all devices. The order of the devices evaluated with each participant was randomized. To guarantee that the participants fully comprehended what they had to write, they were asked to repeat out loud the sentence and only then started writing. If the evaluation monitor noticed a miscomprehension he would correct the participant and they would repeat it afterwards until they got it right. Trials were timed and after each sentence the transcribed sentence was registered.

The Perkins typewriter and personal computer were made available by the researchers. The computer keyboard featured silicone marks on letters 'F' and 'J' to ease exploration.

**Digit Span**

| | Series | Group I | Group II |
|---|---|---|---|
| | 3 | 5-8-2 | 6-9-4 |
| | 4 | 6-4-3-9 | 7-2-8-6 |
| | 5 | 4-2-7-3-1 | 7-5-8-3-6 |
| Forward Recall | 6 | 6-1-9-4-7-3 | 3-9-2-4-8-7 |
| | 7 | 5-9-1-7-4-2-8 | 4-1-7-9-3-8-6 |
| | 8 | 5-8-1-9-2-6-4-7 | 3-8-2-9-5-1-7-4 |
| | 9 | 2-7-5-8-6-2-5-8-4 | 7-1-3-9-4-2-5-6-8 |
| | 2 | 2-4 | 5-8 |
| | 3 | 6-2-9 | 4-1-5 |
| | 4 | 3-2-7-9 | 4-9-6-8 |
| Backward Recall | 5 | 1-5-2-8-6 | 6-1-8-4-3 |
| | 6 | 5-3-9-4-1-8 | 7-2-4-8-5-6 |
| | 7 | 8-1-2-9-3-6-5 | 4-7-3-9-1-2-8 |
| | 8 | 9-4-3-7-6-2-5-8 | 7-2-8-1-9-6-5-3 |

Figure 5.4: Digit Span (Memory) forward and backward series

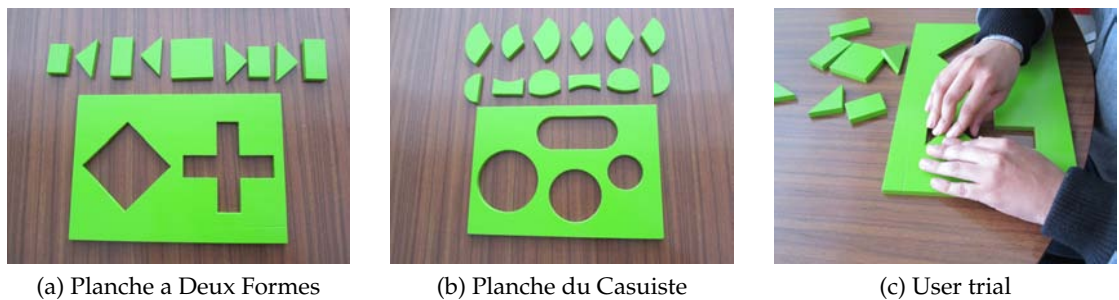

(a) Planche a Deux Formes          (b) Planche du Casuiste          (c) User trial

Figure 5.5: Spatial Ability assessments

JAWS 10 screen reader was installed. The mobile task was performed with the user's own device. All participants, except two, owned a device with a screen reader. Also in this trial, we were able to observe and register the dominant finger used. This was relevant to select single pressure sensitivity and spatial acuity levels for each user for later analysis.

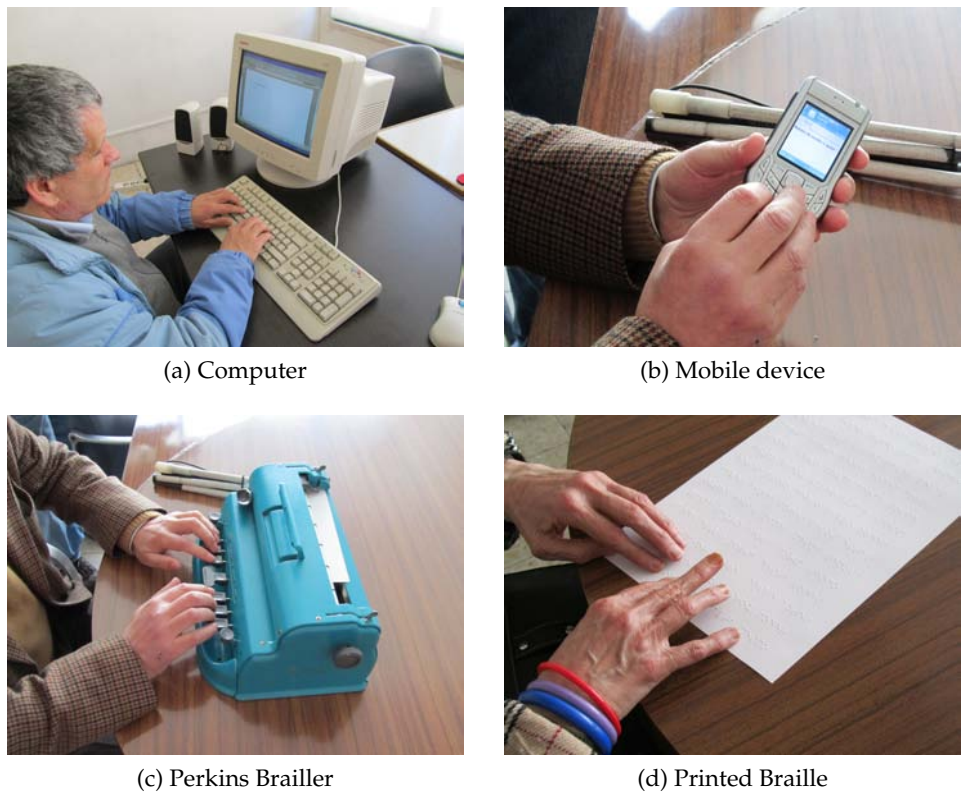

(a) Computer                                          (b) Mobile device

(c) Perkins Brailler                                 (d) Printed Braille

Figure 5.6: Functional Ability assessments

Still in the functional domain, but in what respects to reading, we also evaluated the participants' proficiency in reading Braille. To this end, we selected the 70 most common 5 letter words in the Portuguese language. Four different versions were created with the order of the words randomized. Each version was composed by 14 lines each composed of 5 words. Consecutive reading of a Braille cell decreases its quality. To guarantee that all participants performed the evaluation in the same conditions, ten A4 exemplars of each version were kindly printed by the Braille printing services at Casa Pia de Lisboa. Each exemplar was put aside after being used in two trials.

## 5.3. Location and Material

The evaluation sessions were performed in a quiet meeting room at the Raquel and Martin Sain Foundation with exception to two participants who could not attend during opening hours. These two performed both sessions at INESC-ID, in Lisbon, in a quiet pre-booked meeting room. The room at the foundation was booked in advance to guar-antee that the sessions were not subject to any disruptions and both attending students and employees were aware of the tests. At the start of day, the table in the room was set up with all the required material.

Both tactile assessments were accompanied by an administration manual and no further assembly was required. In regards to cognitive evaluation, both spatial ability and verbal IQ evaluations were provided by the foundation Psychology Department. The *planche du casuiste* and the *planche a deux formes* were also made available with a manual both for the administration and classification of the test [Xydias, 1977]. The Wechsler Adult Intelligence Scale (WAIS-R), and its subtest Digit Span, was also made available by the Psychology Department in the form of administration and classification manuals. When required we were instructed by people in the department to assure the correct administration of the tests.

## 5.4. Collected Data and Analysis

Profile and background information was collected from the administered questionnaires. These contemplated the following components: Basic Profile (Age[c], Sex[n], Professional Activity[n], Education Level[o], Tactile Impairment[n]), Blindness (Cause[n], Type of Blindness[n], Age of Onset[c], Years with Impairment[c]), Braille Literacy (Knows Braille[n], Writing and Reading Self-assessments [o], Onset Age of Braille-Learning[c]), Technology (Computer usage[n], Mobile phone usage[n], Screen Reader used[n], Mobile Phone used[n], Time using Mobile Phone[c], Used Mobile Phone before going Blind[n], SMS sent per day[c], Mobile User Level[o]). Nominal variables are tagged (between brackets) with an *n*, ordinal variables with an *o* and continuous variables with a *c*.

Two measures for tactile sensitivity were collected: Spatial Acuity [c] and Pressure Sensitivity [o]. These measures were performed in three fingers of the dominant hand. We limited the analysis to the dominant finger performing the interaction with mobile devices (this information was asked to the participant and verified during the functional assessment performed on mobile text-entry.

As to cognitive attributes, we collected and analysed both Spatial Ability [c] and Digit Span (DS) scores [c]. We chose to analyse the raw DS scores instead of Verbal IQ values as the latter are a age-weighted measure of the former and herein we intend to look at each attribute separately.

For the functional tasks, we collected both input speed and accuracy measures. To assess speed, the words per minute (WPM) text entry measure calculated as (*transcribed text* - 1) * (60 seconds / *time in seconds*) / (5 characters per word) was used. Accuracy was measured using the the MSD Error Rate (MSD ER), calculated as MSD (presentedText, transcribedText) / Max(presentedText,transcribedText) * 100, where MSD is calculated resorting to the Levenshtein distance between the target and transcribed sentences. Input speed and Accuracy in a particular device, for each participant, were calculated as the average values of the WPM and MSD ER values of the three transcribed sentences.

As to Braille reading, we simplified it to a single Braille Reading Speed measure, calculated as the number of correct words read within the task time (until the end of the sheet - 70 words - , or within a 5 minute maximum). This metric was also simplified as WPM by dividing the number of correct words read by the time spent reading them.

Normality of data was assessed resorting to Shapiro-Wilkinson normality tests. Concerning bi-variate correlations, when in presence of continuous variables and normal distributions, the Pearson correlation was used. The Spearman non-parametric alternative was used either when in presence of non-continuous variables or non-normal distributions.

As to Testing Groups, when in presence of normal distributions, we resorted to ANOVA (repeated measures or between groups depending on matching of groups). When in presence of non-normal or ordinal data, the non-parametric alternatives were used (Mann-Whitney test to compare two unpaired groups, Wilcoxon Test to compare two paired groups, Friedman test to compare more than two matched groups and Kruskal-Wallis test to compare more than two matched groups).

We report statistical significance with an $\alpha$ value set at 0.05. Nonetheless, when higher level statistical significance is achieved ($p<.01$, $p<.005$ or $p<.001$) we report results at that level, in agreement with the output tables of SPSS statistical analysis application and as it is common in social sciences [Pallant, 2007]. Besides this and given the exploratory scope of this research, we set the $\alpha$ value at .1 and report results at this level as minor significant. The statistics procedures presented in this chapter have their main tables, including p-values, presented in Annex A4.

In such a long process of data collection, there have been cases where participants had difficulties in maintaining their engagement with the study or staying for complete sessions, which translated in the existence of missing data. We decided not to exclude these participants as all of them were tested in at least half of the assessments. Further, from the whole group, only eleven have some kind of missing data. Concerning the analysis, when possible (e.g., correlations), we performed Pairwise case exclusions meaning that the subjects were dropped only on analyses involving variables that have missing values. Throughout this chapter, all the analysis are presented with the number (N) of subjects included in the analysis, i.e, those that had not missing values to what that analysis concerns.

# 5.5. Results

In this chapter, our goal is to have an overview of the differences among blind people and how they are related. In this sense, we will give an overview of the collected measures at the profile, sensorial, cognitive and functional levels. Where we see fit, we analyse and correlate measures from different attributes. Particularly, we try to understand if relationships can be found between individual attributes and functional abilities. Even though contaminated with experience and several unmeasured external factors, correlations between some determining abilities are likely to be found. Table 5.1 shows the basic profile along with tactile, cognitive and functional abilities of the participants of the study.

| # | Gender | Age | Onset | Tactile Acuity | Pressure Sensitivity | Spatial ability | Digit Span | PC WPM | PC ER | Mob WPM | Mob ER | BR WPM | BR ER | BW WPM | BW ER |
|---|--------|-----|-------|----------------|----------------------|-----------------|------------|--------|-------|---------|--------|--------|-------|--------|-------|
| 1 | f | 59 | 4 | 4 | 3.61 | 1.75 | 24 | 11.82 | 1.04 | 3.67 | 0.00 | 34.71 | 9.68 | 0.00 | |
| 2 | m | 26 | 10 | 2 | 3.61 | 1.75 | 66 | 45.78 | 0.00 | 15.82 | 0.00 | 49.41 | 26.41 | 1.28 | |
| 3 | m | 32 | 15 | 2 | 2.83 | 10.00 | 72 | 44.63 | 2.32 | 11.89 | 0.00 | 21.32 | 13.35 | 0.00 | |
| 4 | f | 31 | 23 | 2 | 2.83 | 14.25 | 96 | 49.28 | 0.00 | 10.11 | 0.00 | 9.60 | 4.29 | 16.40 | |
| 5 | f | 38 | 0 | 2 | 4.31 | 3.25 | 36 | NA | NA | NA | NA | NA | NA | | |
| 6 | f | 53 | 5 | 2 | 4.31 | 10.00 | 36 | 11.50 | 1.28 | 3.97 | 2.08 | 8.80 | 14.88 | 1.04 | |
| 7 | m | 58 | 10 | 4 | 3.61 | 1.75 | 84 | 0.00 | NA | 2.55 | 1.04 | 44.52 | 4.85 | 0.00 | |
| 8 | m | 66 | 2 | 2 | 4.31 | 1.75 | 60 | 0.00 | NA | 0.00 | NA | 56.71 | 11.88 | 0.00 | |
| 9 | m | 38 | 6 | 2 | 3.61 | 3.25 | 24 | 22.83 | 2.08 | 9.67 | 0.00 | 2.00 | 8.82 | 17.47 | |
| 10 | m | 61 | 10 | 4 | 4.31 | 2.50 | 36 | 1.62 | 25.26 | 5.08 | 13.67 | 0.00 | 0.00 | NA | |
| 11 | f | 69 | 0 | 2 | 3.61 | 1.75 | 54 | 27.49 | 0.00 | 4.40 | 1.04 | 54.47 | 13.40 | 0.00 | |
| 12 | f | 34 | 27 | 2 | 3.61 | 8.50 | 60 | 41.84 | 0.00 | 12.62 | 0.00 | 2.60 | 8.24 | 1.04 | |
| 13 | m | 24 | 2 | 2 | 2.83 | 5.50 | 24 | 45.29 | 0.00 | 14.17 | 0.00 | 63.69 | 27.34 | 1.01 | |
| 14 | m | 45 | 20 | 2 | 2.83 | 7.75 | 72 | 21.75 | 0.00 | 6.72 | 1.04 | 9.40 | 11.57 | 1.04 | |
| 15 | f | 30 | 9 | 2 | 2.83 | 7.00 | 66 | 38.05 | 5.69 | 11.48 | 1.04 | 82.35 | 35.11 | 0.00 | |
| 16 | m | 63 | 3 | 2 | 3.61 | 4.75 | 60 | 23.66 | 0.00 | 7.89 | 0.00 | 64.69 | 25.82 | 1.04 | |
| 17 | f | 35 | 32 | 2 | 2.83 | 0.25 | 60 | NA | NA | 11.89 | 0.00 | 0.00 | 7.46 | 1.23 | |
| 18 | f | 40 | 33 | 2 | 2.83 | 2.50 | 42 | NA | NA | 3.11 | 4.47 | 0.00 | 0.00 | NA | |
| 19 | f | 62 | 14 | 2 | 3.61 | 4.75 | 54 | 2.68 | 2.56 | 0.00 | NA | 2.40 | 0.48 | 42.31 | |
| 20 | f | 53 | 51 | 2 | 4.31 | 0.00 | 0 | NA | NA | NA | NA | NA | NA | NA | |
| 21 | f | 38 | 30 | 2 | 4.31 | 1.00 | 42 | 8.27 | 16.67 | 5.09 | 0.00 | 0.00 | 0.00 | NA | |
| 22 | f | 42 | 20 | 2 | 3.61 | 10.85 | 84 | 36.56 | 3.85 | 8.70 | 0.00 | 7.20 | 9.74 | 2.32 | |
| 23 | f | 45 | 25 | 2 | 2.83 | 6.25 | 42 | 20.34 | 0.00 | 10.36 | 1.23 | 26.54 | 17.82 | 1.28 | |
| 24 | m | 64 | 0 | 3 | 3.61 | 6.25 | 66 | 10.25 | 2.02 | 0.00 | NA | 33.60 | 18.17 | 1.01 | |
| 25 | m | 41 | 29 | 4 | 4.31 | 0.00 | 0 | NA | NA | NA | NA | NA | NA | NA | |
| 26 | m | 41 | 0 | 2 | 2.83 | 10.00 | 114 | 60.07 | 0.00 | 17.74 | 0.00 | 91.30 | 51.00 | 0.00 | |
| 27 | m | 61 | 0 | 2 | 4.31 | 4.00 | 90 | 24.76 | 0.00 | 9.64 | 0.00 | 80.77 | 13.41 | 0.00 | |
| 28 | m | 48 | 26 | 2 | 4.31 | 4.75 | 42 | 33.87 | 0.00 | 10.63 | 0.00 | 19.18 | 21.97 | 2.32 | |
| 29 | m | 42 | 41 | 2 | 3.61 | 4.75 | 54 | NA | NA | NA | NA | 0.20 | 4.71 | 14.74 | |
| 30 | m | 49 | 34 | 2 | 4.31 | 3.25 | 36 | NA | NA | 0.00 | NA | 0.00 | 0.00 | NA | |
| 31 | f | 47 | 44 | 2 | 4.31 | 1.75 | 24 | NA | NA | 3.43 | 3.37 | 0.00 | 0.00 | NA | |
| 32 | f | 50 | 17 | 2 | 3.61 | 5.50 | 36 | 26.73 | 0.00 | 7.09 | 0.00 | 3.80 | 7.90 | 0.00 | |
| 33 | m | 60 | 34 | 2 | 4.31 | 8.50 | 42 | 15.13 | 3.33 | 5.27 | 2.32 | 19.34 | 17.94 | 2.32 | |
| 34 | f | 59 | 9 | 2 | 3.61 | 8.50 | 54 | NA | NA | NA | NA | 14.38 | 15.93 | 1.04 | |
| 35 | f | 41 | 34 | 2 | 4.31 | 7.75 | 24 | 7.20 | 3.57 | 0.00 | NA | 6.00 | 6.79 | 3.61 | |
| 36 | f | 54 | 32 | 2 | 4.31 | 3.25 | 42 | 11.60 | 1.28 | 7.73 | 1.28 | 15.05 | 4.34 | 3.37 | |
| 37 | f | 54 | 20 | 3 | 4.31 | 1.00 | 36 | NA | NA | NA | NA | NA | NA | NA | |
| 38 | f | 57 | 15 | 2 | 3.61 | 1.75 | 54 | 11.39 | 0.00 | 1.48 | 7.53 | 11.40 | 12.93 | 12.82 | |
| 39 | f | 48 | 5 | 2 | 3.61 | 7.00 | 24 | 5.87 | 4.41 | 2.41 | 3.61 | 16.53 | 5.19 | 10.90 | |
| 40 | f | 38 | 8 | 3 | 3.61 | 0.00 | 0 | NA | NA | NA | NA | NA | NA | NA | |
| 41 | m | 27 | 19 | 2 | 2.83 | 5.50 | 24 | NA | NA | NA | NA | NA | NA | NA | |
| 42 | f | 40 | 0 | 2 | 2.83 | 2.50 | 72 | 0.00 | NA | 0.00 | NA | 87.50 | 41.26 | 0.00 | |
| 43 | m | 46 | 3 | 2 | 4.31 | 7.00 | 42 | 4.71 | 4.65 | 0.00 | NA | 9.00 | 11.70 | 0.00 | |
| 44 | m | 39 | 38 | 2 | 2.83 | 2.50 | 36 | 14.94 | 0.00 | 7.68 | 1.28 | 0.00 | 2.72 | 19.39 | |
| 45 | m | 42 | 0 | 2 | 3.61 | 1.00 | 72 | NA | NA | 0.00 | NA | 68.85 | 37.33 | 0.00 | |
| 46 | f | 31 | 8 | 2 | 2.83 | 8.50 | 60 | 26.01 | 0.00 | 11.44 | 4.17 | 41.58 | 22.49 | 2.47 | |
| 47 | m | 53 | 35 | 3 | 4.31 | 6.25 | 42 | 40.36 | 0.00 | 8.77 | 1.28 | 0.00 | 0.00 | NA | |
| 48 | m | 49 | 37 | 2 | 3.61 | 8.50 | 54 | 20.46 | 0.00 | 6.78 | 0.00 | 3.60 | 7.46 | 15.98 | |
| 49 | f | 27 | 0 | 2 | 3.61 | 1.75 | 66 | 48.16 | 0.00 | 14.27 | 0.00 | 52.50 | 15.46 | 1.28 | |
| 50 | f | 26 | 14 | 2 | 2.83 | 10.00 | 90 | NA | NA | 20.57 | 0.00 | 20.70 | 9.86 | 0.00 | |
| 51 | m | 41 | 22 | 2 | 3.61 | 11.70 | 54 | 55.71 | 1.23 | 5.33 | 1.04 | 52.50 | 19.99 | 0.00 | |

Table 5.1: Profile, tactile, cognitive and functional measures of the 51 study participants

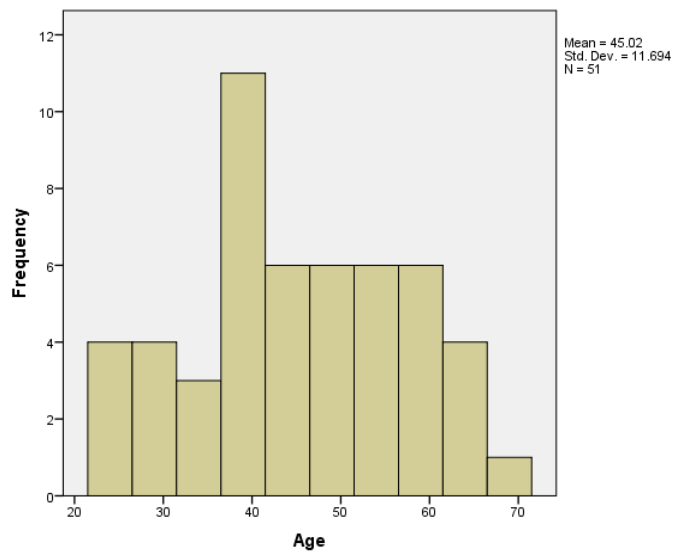

Figure 5.7: Age distribution

## 5.5.1. Profile and Background

One of the main assumptions for our goals in identifying individual differences was that the sample group should be as diverse as possible. This was to guarantee that it reflects the diversity amongst the population already acknowledged in Chapter 4.

**Demographics**

We sought to recruit participants with different profiles in regards to age, sex and educational background.

Of the fifty-one (51) participants in our sample, twenty-seven (52.9%) are female and 24 (47.1%) are male. Their ages ranged from 24 to 68 years old, averaging 45 (SD=11.7). The Age histogram is presented in Figure 5.7. Visual observation indicates a normal distribution peaking at 45 years of age. Skewness indicates a slight clustering to the left side of the distribution (lower values) while Kurtosis (positive value) indicates a peaked distribution with long thin tails.

While normality can be assessed to some extent by analysing the aforementioned distribution values, we still ran a normality assessment test, which confirmed a normally distributed sample in regards to Age (Shapiro-Wilk, W=0.97, p >.05).

As to educational/academic background, we clustered the sample in four different categories (4th grade, 9th grade, 12th grade, and additional graduation - undergraduate, graduate or post-graduate). Participants were inserted in each cluster according to highest education level completed. Twelve people (23.5%) in the sample have the 4th grade completed, nineteen (37.3%) have completed the 9th grade, ten (19.6 %) have completed the 12th grade, while the remaining 10 (19.6%) had some additional education level.

Most of the less educated participants are regular attendees of the foundation we collaborated with (with no other professional occupation) or with others of the same kind. These people *jump* between courses at different foundations normally with 3-month intervals spent at home (according to current Portuguese legislations). These courses can take from 3 months to 3 years and are as diverse as Informatics (several levels), Textiles, Tapestry, and Massage, and Phone Operation. Most of these participants already gather an impressive portfolio of courses.

On the down side, only the physiotherapy and phone operator courses are still likely to contribute to the re-insertion of the course attendees in the society as active members. There are several active protocols with companies for internships and possible subsequent full-time contracts, particularly for phone operating. The other courses are to be seen as mere hobbies. It is worth mentioning that getting accepted to participate in most courses is rather easy but proponents are subject to a profile evaluation and only the most fit cognitively are directed to the most fruitful courses (once again, phone operator).

At the time we performed the questionnaires, twenty-five participants were undertaking a course at FRMS or a similar institution (49.0%), ten were at home (19.6%) from which 8 were expecting to be called for another stay at a foundation, four were working full-time as phone operators (7.8%), five were doing an internship (5.8%) as receptionists or phone operators, 2 were studying to complete the 9th grade (3.9%) and the others were working in one of the following: one psychologist, a Braille instructor, a radio Disc-Jockey (DJ), a running athlete, and a computer technician .

**Blindness**

Blindness may be due to several causes. In our sample, Glaucoma was the major cause of blindness with fourteen (27.5%) participants, followed by *Retinitis Pigmentosa* with eleven people (21.6%) and Retinal detachment reported in 6 cases (11.8%). Other reported causes were Diabetic Retinopathy (2 people), Cataracts (2 people), Optic nerve hypoplasia (2 people), Measles (2 people), Meningitis (2 people), and Otitis (1 person). Five people reported the cause as Hereditary blindness while four others reported as caused by an Accident. Thirty-three elements (64.7%) from the participant pool reported complete loss of vision while the remaining eighteen (35.3%) stated to detain partial blindness. All of those had light perception at most, one of the study requirements.

In what regards Blindness Onset, we were also expecting to gather a diverse set of people as blindness onset detection has proven to create differences in ability [Burton et al., 2002, Voss et al., 2004]. This diversity was achieved with no particular intention (no *cherry-picking*) and can be observed in Figure 5.8a. Indeed, we were able to recruit *early-blinders* (10) as well as some others who acquired blindness in a late stage in life. The remaining cases are spread along the scale. This is relevant both to assure the wide coverage of our studies but it also denotes that attributes like Age or Blindness Onset should not be seen as binary (old vs young, early vs late-blinder). As expected, Blindness Onset does not show a normal distribution (Shapiro-Wilk, W=0.925, p <.05), peaking at birth or early years and decreasing across age ranges. However, it is important to note that, in contrary to worldwide statistics on the blind population, our sample does not reflect an high onset of blindness after the age of fifty (50). We may suppose that this happens because we were only able to reach people who are still active and try to attend (or attended in the past) formation centres for blind people. This may not happen with those that acquire blindness in a later stage in life and may not seek another craft.

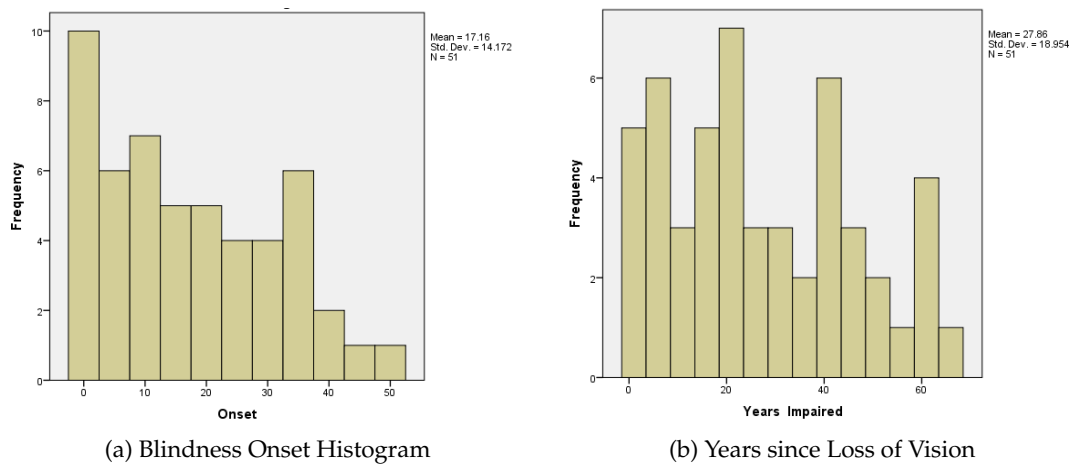

(a) Blindness Onset Histogram                          (b) Years since Loss of Vision

Figure 5.8: Blindness Onset

Resulting from different age gamuts and diverse blindness onset incidence, the sample also gathers a wide diversity pertaining the timespan that the participants have been blind. Time with impairment varies between one (1) year and sixty-eight (68) years. The adaptation to the loss of an integrating sense like vision is very hard and may take long. The time a person has been dealing with the absence of vision is likely to alter her abilities and strategies.

**Braille Literacy**

A common stereotypical assumption is that every blind person is able to read and write in Braille. Recent statistics say that this is not true [Ryles, 2000] and diminish the percentage of Braille blind literates to approximately 18%. From our sample group, forty-six participants (90.2%) stated to know the Braille alphabet while only five (9.8%) considered to be Braille illiterate. Delving into our questionnaire results, we can observe that although a great majority of people is aware of the Braille alphabet, the confidence in writing but particularly in reading skills is relatively low (Figure 5.9).

Figure 5.10 presents the histogram of the reported age the participants started learning Braille. Two patterns can be observed: the first three scales relate mostly to early-blinders who learn Braille during their first school years; the other gamuts cover the population that acquired blindness in later stages in life. Learning Braille Age presents a strong correlation with Blindness Onset (Spearman correlation, r = .662, N=46, p <.001) suggesting that the sooner a person loses sight, the sooner she learns Braille. Actually, one can observe that blind people learn Braille soon after they acquire blindness exception made for early-blinders who learn Braille when they start attending school. Low levels of proficiency in reading and writing may be due to the lack of training and by the replacement of Braille by new digital technologies (computers and mobile phones).

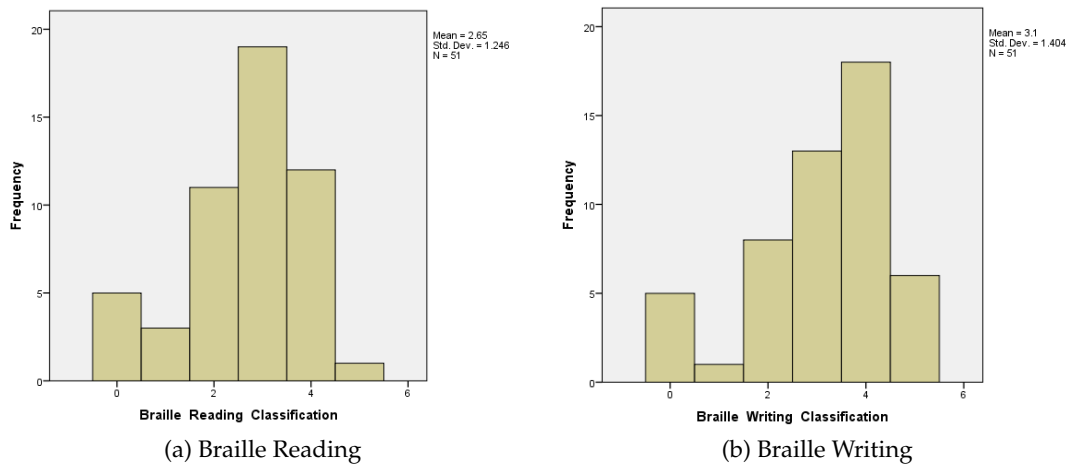

(a) Braille Reading					(b) Braille Writing

Figure 5.9: Braille proficiency self-grading

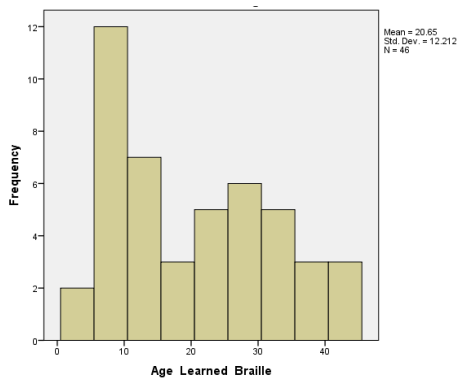

Figure 5.10: Braille learning age histograms

**Device usage**

Moving up to the usage of electronic devices, particularly, personal computers and mobile devices, they are both common and indispensable tools for a large majority of the blind population. Not only they are used for communication and productivity but also improved the life quality in what respects leisure and culture (e.g., reading e-books).

Computers are used in a daily basis by forty-three (43) of our respondents (84.3%) while

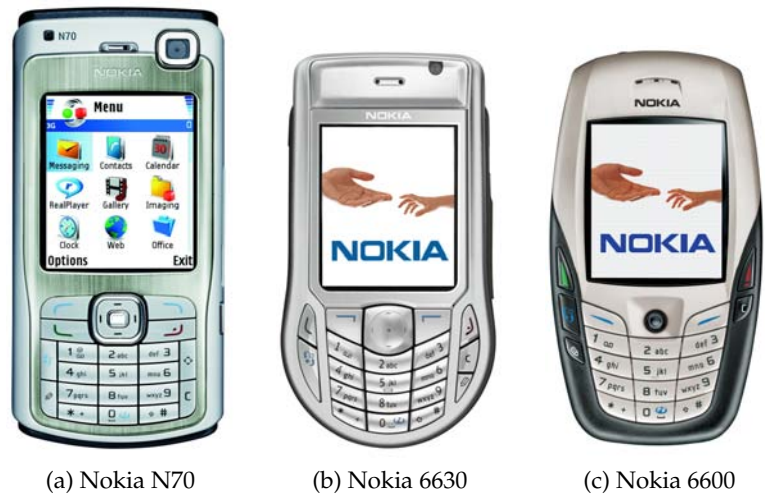

(a) Nokia N70            (b) Nokia 6630            (c) Nokia 6600

Figure 5.11: Most common mobile devices

fifty (50) out of the fifty-one own and use at least one mobile phone. Only one partici-
pant (**P42**) said to reject mobile devices as they were *addictive and changed the way people
communicate*.

Only five (5) participants use a mobile phone without any assistive technology while
the others resort to screen reading software with a major quota (42 participants) using
Nuance Talks [1], one using Apple's VoiceOver [2], and two using BloNo with NavTap
[Guerreiro et al., 2008a], our own assistive method presented in Chapter 3.

Only one respondent used a touch screen device (Apple's iPhone 3GS). This individual is
a computer technician and he is the one fixing the computers at the foundation. Thirteen
people used a Nokia N70 as their main mobile device. This was the device used by most
people. Reasons pointed were that it was compatible with the screen reading software,
it was cheap as it was covered by a national assistance program and because the keypad
was easy to perceive. Other common devices were Nokia 6630 (6), Nokia 6600 (3), Nokia
6680 (2) and Nokia C5 (2). This last one was the preferred device for those who were
thinking of buying a new one. Two reasons were pointed for that: 1) It was simple,
aesthetic and easy to perceive and, 2) it was one of the lasting keypad devices in the
aid funding program. A large set participants stated to be apprehensive with the state
of things as the devices they were comfortable with are no longer being fabricated. The
same people showed to have little or no knowledge about assistive technologies for touch
screen devices and foresee a daunting future with touch screen technology as the basis
for interaction with mobile phones.

Exception made for P42, all other participants have owned a mobile device for at least
three years. **Years using Mobile Phones** showed to have a normal distribution (Shapiro-

---

[1]http://www.nuance.com/for-individuals/by-solution/talks-zooms/index.htm (Last Visited: January
30th, 2012)

[2]http://www.apple.com/accessibility/voiceover/ (Last Visited: January 30th, 2012)

Wilk, W= 0.972, N = 50, p >.05), peaking at ten (10) years. Sixteen (16) users had experience with mobile devices prior to acquiring blindness.

As to text-entry, forty-one participants were able to input text with their mobile devices. Number of text messages written and sent daily varied from zero (0) to three hundred (300). Yes, 300 per day (an outlier, though). This person not only resorted to SMS but it was also the only active Twitter [3] user. Excluding this and another participant (150 messages per day), the average number of messages for the remaining texters was of 5.0 (SD=5.5) per day.

Depending on the type of usage given to their mobile phones, we classified participants into three categories. Level 0 users are those that use their mobile phones just to place and receive calls, as a clock and alarm. Twelve (12) participants fell in this category. These participants are not able to text. Level 1 individuals are those that are able to do all the above plus resorting to text to write messages, notes and reminders, taking photographs, listening to radio among other medium level applications / tasks. Twenty-three participants (23) belong to this group. Fifteen users (15) regularly transfer files between their computer and mobile device, either by USB or Bluetooth, and/or use Wi-fi/3G, and/or *play* with their mobile phones definitions, and/or use Facebook[4], instant messaging or Twitter clients. These are Level 2 users. Mobile User Level showed to have a positive medium correlation with the average number of SMS messages sent daily (Spearman correlation, rho = .571, N=50, p <.0001).

## 5.5.2. Sensory attributes

The role of tactile abilities has been under-looked in mobile interaction. We hypothesize, based on observations (Chapter 3) and expert interviewing (Chapter 4), that differences in pressure sensitivity and spatial acuity may lead to different mobile performance levels. Here, we explore the diversity in tactile abilities and how they relate with the user's age and blindness onset profile.

Neither pressure sensitivity nor spatial acuity presented normal distributions and no significant correlation was found between **Pressure Sensitivity** and **Spatial Acuity** suggesting that these two characteristics should both be contemplated separately. Figure 5.2 presents the number of users whose dominant interaction finger falls into each of pressure sensitivity and spatial acuity categories. It is noticeable that the variability is wider in pressure sensitivity than in spatial acuity. This may suggest that the selection of a dominant finger is related with the acuity of that finger. Also, pressure sensitivity is likely to decrease with use (i.e., callosities) while perception is likely to increase, and consequently, discrimination between two points.

---

[3]http://www.twitter.com (Last Visited: January 30th, 2012)
[4]http://www.facebook.com

**Spatial Acuity**

| | | Frequency | Percent | Valid Percent | Cumulative Percent |
|---|---|---|---|---|---|
| Valid | 2 | 38 | 74.5 | 74.5 | 74.5 |
| | 3 | 6 | 11.8 | 11.8 | 86.3 |
| | 4 | 6 | 11.8 | 11.8 | 98.0 |
| | 5 | 1 | 2.0 | 2.0 | 100.0 |
| | Total | 51 | 100.0 | 100.0 | |

**Pressure Sensitivity**

| | | Frequency | Percent | Valid Percent | Cumulative Percent |
|---|---|---|---|---|---|
| Valid | 2.83 | 12 | 23.5 | 23.5 | 23.5 |
| | 3.61 | 18 | 35.3 | 35.3 | 58.8 |
| | 4.31 | 20 | 39.2 | 39.2 | 98.0 |
| | 4.56 | 1 | 2.0 | 2.0 | 100.0 |
| | Total | 51 | 100.0 | 100.0 | |

Table 5.2: Tactile Sensitivity Assessment Frequencies

No significant correlations were found between Age or Blindness Onset Age and Spatial Acuity. Conversely, a positive large correlation was found between Age and Pressure Sensitivity (Spearman correlation, rho=.514, N=51, p <.001) which is again consistent with the aforementioned idea: pressure sensitivity decreases with use.

### 5.5.3. Cognitive attributes

We identified two main cognitive attributes possibly influencing mobile interaction abilities: spatial ability, and a more general cognitive component, Digit Span (enclosing both Memory and Attention). Neither Spatial Ability (Shapiro-Wilk, W=.941, p<.05) nor Digit Span (Shapiro-Wilk, W=.942, p<.05) assessment results presented normal distributions. No significant correlation was found between these two components revealing that they should both be considered as they assess different components of cognition.

Spatial ability presented results from 1.75 to 14.25 (M=6.40, SD=3.56) while Digit Span showed values between 24 and 114 (M=55.92, SD=23.67) revealing Verbal IQ values (this calculation considers the participant's age) amidst 64 and 151 (M=97.36, SD=22.69). No correlation was found between Age or Blindness Onset Age and Spatial Ability or Digit Span scores. A medium positive correlation was found between Education Level and Spatial Ability (Spearman correlation, rho=.457, N=48, p<.005) and a strong positive correlation was found between Education Level and Digit Span scores (Spearman correlation, rho=.518, N=48, p<.001).

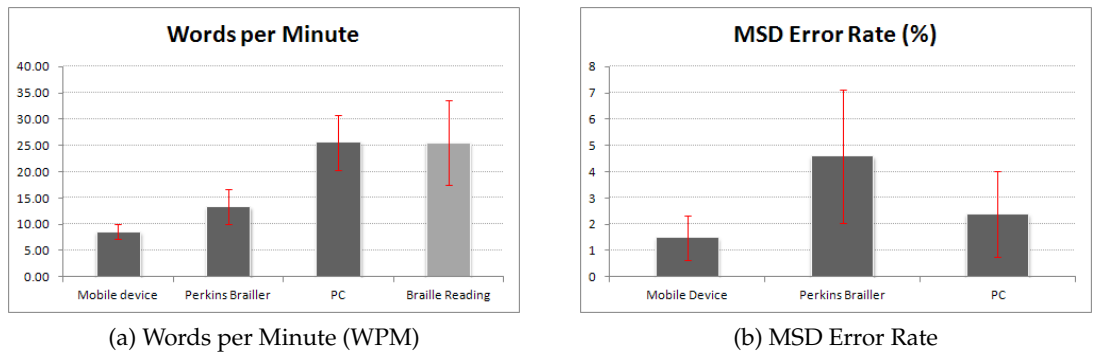

(a) Words per Minute (WPM)                    (b) MSD Error Rate

Figure 5.12: Functional Assessments Speed and Transcribed Sentence Quality. Error Bars denote 95% confidence intervals.

## 5.5.4. Functional abilities

The functional assessments performed evaluate performance in tasks the users already perform and are thus less controlled than the remaining. They are subject to the experience and compensatory mechanisms created by each participant. This may hide some relationships that might be encountered in early approaches with new technologies. Nevertheless, we still looked at how these functional assessments were influenced by other variables.

Figure 5.12 presents the overall performance results attained in each functional setting. It is not our goal to compare performances between these different devices as they have different natures. However, one thing is relevant to observe: the dispersions found. Indeed, the mobile device shows a small dispersion which means that people have performances that are more similar than in the remaining. Large dispersion can also be found in the quality of the transcribed sentences regarding the Perkins Brailler setting, meaning that people are not very slow in writing but tend to vary widely in accuracy.

**Mobile keypad**

Mobile WPM showed to be strongly correlated with Age (Pearson correlation, r=.618, N=43, p<.001) suggesting that older people tend to perform slower, which was an expected result. Also, a medium strong correlation was found between Education Level and Mobile WPM (Spearman correlation, rho=.375 , N=43, p<.05). No correlation was found with Blindness Onset but a medium negative correlation was found with the numbers of years with blindness (Spearman correlation, rho=.375 , N=43, p<.05) which is likely to be due to experience with mobile telephony prior to acquiring blindness. Also, the younger participants are likely to be blind for less time than the older ones which may contribute for this correlation.

Concerning mobile experience no correlation was found between the numbers of Years using Mobile Phones and Mobile WPM. Positive strong correlation were found between Mobile WPM and the number of SMS sent daily (Pearson correlation, r=.626, N=43, p<.05) and Mobile User Level (Spearman correlation, rho=.674, N=42, p<.001).

Some of the participants were already mobile phone users before acquiring blindness. This could increase their familiarity with this type of devices. A one-way ANOVA did not show significant differences between the participants that used mobile before going blind and those that did not (F(1,33)=1.48 p>.05).

As to tactile abilities, a significant medium correlation occurs between Mobile WPM and Pressure Sensitivity (Spearman correlation, rho=-.355, N=43, p<.05). No significant correlation was found with tactile acuity.

Spatial ability has shown to be minor significantly correlated (small) with Mobile WPM (Spearman correlation, rho=.285 , N=43, p<.1). Also, Digit Span scores showed a minor significant positive medium correlation with Mobile WPM (Spearman correlation, r=.298, N=43, p<.1). Although minor significantly, results suggest that these two cognitive components still play a role in the performance the users attain. The minor significance can be partially explained with the fact that users were tested with their own devices and as such experience is likely to play an important role. Muscular memory is likely to overcome spatial and memory limitations.

Mobile MSD Error Rate showed to be mildly correlated with Age (Spearman correlation, rho=.328, N=35, p=.054) suggesting that besides slower, older people tend to be more erroneous. Also, a significant negative medium correlation was found between Mobile MSD ER and Education Level (Spearman correlation, rho=-.342, N=35, p<.05) suggesting that those that are higher educated commit fewer errors.

A significant medium negative correlation was found between Mobile User Level and Mobile MSD ER (Spearman correlation, rho=-.465, N=34). No other correlations were found with collected experience metrics. These results indicate that those that use the mobile device more proficiently are the ones that commit less typing errors although this seems to be disconnected from experience or the attention they dedicate to mobile communication. No significant differences were found on Mobile MSD ER between those that used mobile devices before going blind and those that did not. In opposition to typing speed, the amount of errors committed has not shown to be related with experience prior to the onset of blindness.

As to tactile abilities, a minor significant medium correlation occurs between Mobile MSD ER and Spatial Acuity (Spearman correlation, rho=.322, N=35, p=.059). No significant correlation was found with Pressure Sensitivity.

Digit Span scores showed a negative medium correlation with Mobile MSD ER (Spearman correlation, rho=-.376, N=35, p<.05). Spatial ability did not show to be related with the quality of the sentences produced. These results need to be taken in consideration carefully as it is not clear if the amount of errors produced by the participants is due to lack of attention or memory abilities or to low education levels (who is, as mentioned before, correlated with the accuracy levels attained).

**Desktop QWERTY Keyboard**

Strong positive correlations were found with Desktop WPM and Age (Pearson correlation, r=-.585, N=37, p<.001), Education Level (Spearman correlation, rho=.463, N=37, p<.005) and number of Years Impaired (Spearman correlation, rho=-413, N=37, p<.05). Older blind adults are slower typists as it would be expected and those that are blind for fewer years are faster. Once again these two correlations may be related as Age is strongly correlated with the number of years impaired, in our sample (Spearman correlation, rho=613, N=51, p<.001). The most educated ones are also faster typists. A one-way ANOVA showed significant differences between daily computer users and those that stated to rarely use the PC (F(1,35)=11.763, p<.005).

Concerning tactile abilities, a minor significant negative medium correlation was found between Desktop WPM and Spatial Acuity (Spearman correlation, rho=-.307, N=37, p<.1). No significant correlations were found with Pressure Sensitivity. The ability to discriminate between two points seems relevant in the task of searching for keys in the keypad.

Both Spatial Ability (Spearman correlation, r=.399, N=37, p<.05) and Digit Span scores (Spearman correlation, r=.303, N=37, p<.1) showed to be significantly (although minor in the Digit Span score) correlated with Desktop WPM.

As to Desktop MSD Error Rate, no correlations were found.

**Perkins Brailler**

Blindness Onset Age was found to be negatively strongly correlated with Braille WPM (Spearman correlation, rho=-.633, N=45, p<.001) while Education Level showed to be medium correlated with Braille WPM (Spearman correlation, rho=.36, N=45, p<.05). Also, the number of Years with Blindness showed to be significantly correlated with Braille writing speed (Spearman correlation, rho=.382, N=45, p<.05).

Braille WPM showed to be strongly correlated with the Age of Learning Braille (Spearman correlation, rho=-.569, N=41, p<.001) and medium correlated with the numbers of Years writing Braille (Spearman correlation, rho=.442, N=41, p<.005). Braille writing self-assessment is strongly correlated with performance achieved with the Perkins Brailler (Spearman correlation, rho=.709, N=45, p<.001).

Spatial acuity showed to be significantly medium correlated with Braille writing performance (Spearman correlation, rho=-.439, N=45, p<.005). Conversely, Pressure Sensitivity showed not to be correlated with Braille writing performance. Once again, Spatial acuity seems relevant to distinguish the keys of the Braille keyboard, even such large ones.

Spatial ability presented no significant correlation with performance. The small number of keys seem to be low demanding in respect to what spatial orientation is concerned. Conversely, Digit span scores showed to be significantly correlated with skills with the Brailler (Spearman correlation, rho=.395, N=45, p<.01).

Just like with input speed, the accuracy is also strongly correlated with the Age of Blindness Onset (Spearman correlation, rho=.562, N=39, p<.0001). Conversely, no correlation was found with Education Level. The number of Years with Blindness showed to be significantly correlated with Braille MSD Error Rate (Spearman correlation, rho=-.395, N=39, p<.05). People that have been living most time without sight are likely to be faster and achieve better phrase quality.

Braille MSD Error Rate showed to significantly strongly correlated both with Age of Learning Braille (Spearman correlation, rho=.659, N=37, p<.0001) and Years Reading Braille (Spearman correlation, rho=-.547, N=37, p<.0001). Those that learn Braille sooner are likely to be faster and more accurate. This is also related with the experience (number of years) reading Braille. Braille writing self-assessment is medium correlated with the quality of the sentences achieved with the Perkins Brailler (Spearman correlation, rho=-.474, N=39, p<.005).

No correlations were found between tactile abilities and transcribed sentence quality.

A signicative medium negative correlation was found between Digit Span scores and Braille MSD Error Rate (Spearman correlation, rho=-.346, N=39, p<.05). On the other hand, Spatial Ability did not show to be relevant for Braille typing, this time to what accuracy is concerned.

**Braille reading**

Reading Braille is stated by all participants as harder than writing. This can be explained with the need, besides knowing the alphabet, to perceive the dots in the paper sheet. This ability requires a lot of training and is likely to impose demands in terms of tactile sensitivity. Concerning profile, Blindness Age of Onset showed a significant strong positive correlation with Braille Reading Speed (Spearman correlation, rho=-.740, N=45, p<.001) as did the Number of Years with Blindness (Spearman correlation, rho=.514, N=45, p<.001).

Strong correlations occur between Braille Reading Speed and both the Age of Learning Braille (Spearman correlation, rho=-.725, N=41, p<.001) and Years Reading Braille (Spearman correlation, rho=.612, N=40, p<.001). These strong correlations indicate that experience seems to have a very significant role. This also suggests that other individual differences may be overcome by experience.

Braille reading self-assessment is strongly correlated with the readig speed (Spearman correlation, rho=.813, N=45, p<.001) which indicates that the participants are aware of their performance in relation to others.

A significant medium negative correlation was found between Tactile Acuity and Braille Reading Speed (Spearman correlation, rho=.-298, N=45, p <.05) which shows that discrimination of points at the fingertips play a role in the ability to read Braille. This was expected. Opposed to our expectations, no correlation was found between Braille Reading Speed and Pressure Sensitivity.

A significant difference was found between those that mentioned to be tactile impaired (due to diabetes) and the tactile-healthy. The mean ranks of tactile-impaired and tactile-healthy were 8 and 23.95, respectively; Mann Whitney U Test, Z = -2.50, p <.05).

A significant positive correlation between Digit span scores and Braille Reading Speed (Spearman correlation, rho=.653, N=45, p <.001) was revealed going in line with the previous analysis where Digit Span scores showed to be relevant across all functional assessments. These results may indicate that Verbal IQ works as a baseline ability for surpassing interaction demands.

Interesting enough, it seems that some abilities are transversal to all functional assessments performed (e.g., Digit Span) while other seem to vary depending on the demands of device and task. More to it, the strength of the correlations seem to augment when the demands are higher. Braille reading is stated by the participants as the most demanding task at several levels and the results showed it to be demanding at the profile, tactile and cognitive levels with highly significant strong correlations.

## 5.6. Clustering participants

We intended to relate individual attributes with device demands. At this point, we were not able to predict how many individual characteristics would define the user's ability in relation to mobile user interface demands. The number of possible individual variables along with the varying demands among devices and interfaces makes it hard to provide a relationship between variables based on individual measures. Indeed, to be able to correlate in such a multidimensional space the user sample would have to be enormous [Tabachnick and Fidell, 2006]. To cope with such variability, we created groups within the sample given the values obtained in respect to a set of measures. A univariate example would be creating groups pertaining Age (younger vs older) and use these groups to understand if older adults perform better or worse than younger ones. A multi-variate approach is also feasible as groups can be made pertaining more than one variable. An example would be grouping in respect to Age and Education: younger users who are less educated; younger users who are more educated ; older users that are more educated; and more educated older users. The number of variables and the number of resulting groups needs to be carefully selected. The latter were later used to perform comparisons between groups and discriminant analysis. Taking this in consideration, cluster analysis was performed with a sub-set of 41 participants from the original 51 participant sample. These 41 blind people were the ones included in the mobile performance studies presented in Chapter 6.

**Concepts.** *Cluster analysis is a major technique for classifying a "mountain" of information into manageable meaningful piles. It is a data reduction tool that creates subgroups that are more manageable than individual datum* [Burns and Burns, 2008]. The goal of cluster analysis is to group observations on the basis of similarities or distances with no prior assumption regarding the group structure or the numbers of existing clusters [Johnson and Wichern, 2002]. This technique searches the data for a structure of natural groups. Groupings is done on the basis of similarities or dissimilarities of participant measures.

**Clustering Method.** We do not know in advance the appropriate number of groups. As such, we used a two stage sequence of analysis towards an optimum solution within the sample, one that can differentiate the most between groups of cases. We resorted to Agglomerative Hierarchical clustering as this is the major statistical method for finding relatively homogeneous clusters of cases based on measured characteristics. This method starts with each case as a separate cluster and then combines the clusters sequentially, until only one cluster is left [Burns and Burns, 2008]. Figure 5.13 shows the linkage points between cases and clusters which are linked at increasing levels of dissimilarity. The grouping is based on a distance measure. We selected the Squared Euclidean distance as it is the most straightforward and generally accepted way of computing distances between objects in a multi-dimensional space. The clustering algorithm used was Ward's method: it uses an analysis of variance approach to evaluate the distances between clusters. Thus, cluster membership is ruled by calculating the total sum of square deviations from the mean of a cluster [Tinsley and Brown, 2000]. When performing clusters from multiple variables with different scales normalization was performed by standardizing cases to Z-scores.

**Selecting the number of clusters.** Selection of the number of clusters was performed by analysing the change in the agglomeration coefficients after a first run of the clustering procedure on SPSS. The optimal number of clusters is that when the change in the agglomeration coefficient stabilizes, i.e., succeeding clustering adds very much less to distinguish between cases. This selection was supported by the creation of a scree plot and by inspecting the dendogram presented by SPSS. Looking back to 5.13 one can verify, by inspecting the length of the horizontal lines, that the major changes are present between one and three clusters and then changes start to diminish. The dendogram also helped us to verify the number of cases of each cluster maximizing discrimination between groups

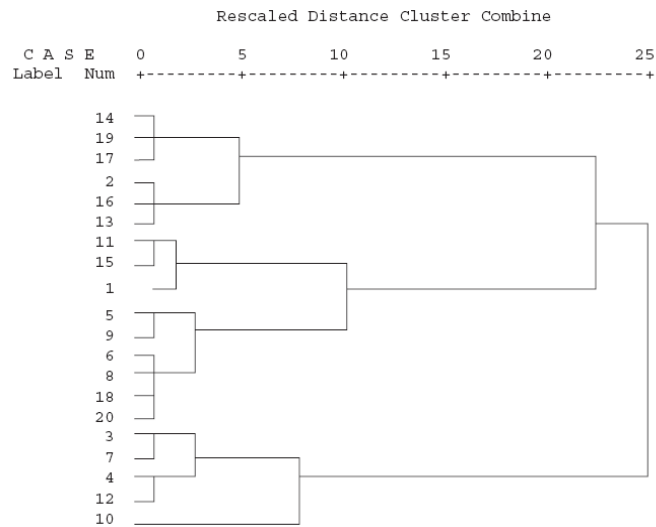

Figure 5.13: Dendogram showing Agglomerative Hierarchical Clustering of cases

but still assuring a considerable number of cases on each group.

**Selecting the variables.** Given the correlations found between individual characteristics and the knowledge gathered we were able to focus our cluster analysis on a reduced set of attributes. Concerning background, we will look at Age, Blindness Onset, and Education group. As to mobile technology experience, Mobile User Level seems to reveal the amount of experience and the daily usage one gives to a device. Individual attributes seem relevant in different perspectives: tactile acuity, pressure sensitivity, memory and spatial ability. This group of 8 attributes composes our basic set of individual features to consider. We will cluster them individually and in meaningful compositions, e.g., tactile abilities, cognitive abilities, profile abilities.

The clustering of these variables aims at enabling the comparison between groups but it serves another purpose: the clusters themselves help at understanding relationships within the data, relationships between variables and people. In what follows, we detail the clustering procedures performed and the resulting clusters:

## Clustering within Variables

Our first take at finding groups among our sample follows a univariate approach. These clusters will enable the later assessment of the impact of a single variable on the usage of mobile user interfaces. Mobile User Level (1,2 and 3) and Education Group (4th Grade, 9th Grade, 12th Grade, Bachelor and further) are already clustered and were not subject to further grouping.

The first variable to look at is **Age**. A cluster analysis was run on 41 cases. A hierarchical cluster analysis using Ward's method produced four clusters. By looking at the dendo-
gram (Figure 5.14), one can verify that selecting two or three clusters was also a possibil-
ity. We decided to select a number that still presents large in agglomeration coefficients
but presents a more fine-grained division of the sample. The first cluster comprised peo-
ple between 24 and 34 years old (M=28.89, SD=3.48, N=9). The second was composed of
people between 38 and 42 (M=40.09, SD=1.51, N =11) while the third had people between
45 and 52 years old (M=48.0, SD= 2.41, N =11). The cluster composed of the oldest in the
sample gathered ages from 56 to 65 (M=60.0, SD=2.71, N=9). The clusters created are
balanced in terms of number of cases (minimum of 9, maximum of 11). Central clusters
are more concentrated, i.e., smaller range for more participants, which is consistent with
the central tendency around 45 years old.

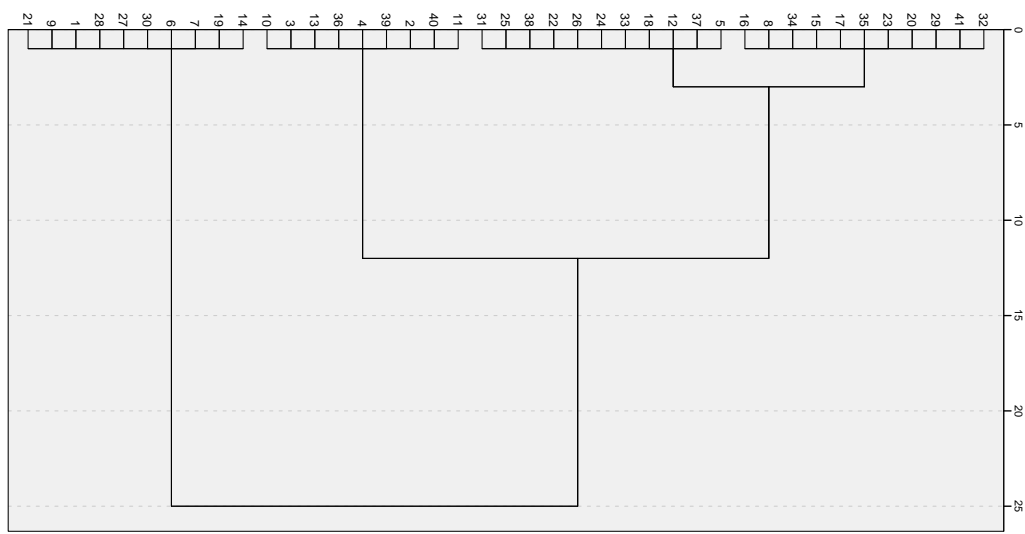

Figure 5.14: Dendogram showing the Hierarchical Clustering of Age cases

**Epoch of Blindness onset** also varies among the user sample presenting a non-normal
distribution (Shapiro-Wilk, p¡.05) with 2 peaks, one around age of birth and another one
near the age of 40 (M=16.34, SD=13.78, N=41). A cluster analysis was run on the 41 cases
in respect to Blindness onset age. A hierarchical cluster analysis using Ward's method
produced four clusters, between which the variables were significantly different. Clusters
ranged from 0 and 6 years old (early-blinders) (M=2.14, SD=2.21, N = 14), 8 to 17 years old
(M=11.7, SD=3.2, N=10), 20 to 30 years old (M=24.13, SD=3.52, N=8), and from 33 to 44
years old (M=36.67, SD=3.74, N=9). Once again, it is noticeable that our sample does not
comprise people with blindness onset ages beyond 44 years old. However, we consider
that the last two clusters are already to be seen as late-blinders and can thereby represent
those that have acquired experience based on vision during most of their lives and are
now experiencing the world differently, and in this particular topic, had to re-adapt to
dealing with technology without visual feedback.

Passing on to tactile abilities and starting with **Tactile Acuity**, the raw data is already grouped in three groups: those with a discrimination of two points separated by two millimetres (33 participants), three millimetres (4 participants) and four millimetres (4 participants), in their preferred interaction finger. Our effort here was to join the less accurate groups in a single one to improve the significance of later hypothesis testing. Two clusters were created, one with 33 people and another with 8 people. This shows that few people have low tactile acuity values. Nevertheless, it is relevant to assess if those who have not have more difficulties interacting with mobile user interfaces.

Concerning **Pressure Sensitivity**, people are more diverse. Results in the preferred finger span four values in the scale. Once again, we created two groups of which one was composed of those with a sensitivity to 2.83 Newtons [N] (10 people) and those with one of 3.61 N (15 people), while another cluster was created from those with pressure sensitivity of 4.31 N (15 people) and 4.56 N (1 people). Pressure sensitivity of 3.61 N is considered to be a regular level so these clusters differentiate between those that have regular or better sensitivity and the ones with slightly limited pressure sensitivity.

In the domain of cognition, we start by looking at spatial abilities. Cluster analysis was run on **Spatial Ability** data. A hierarchical cluster analysis using Ward's method produced three clusters. Figure 5.15 presents the raw data values for all participants along with a visual representation of the clusters created. Clusters ranged from 1.00 to 4.00 (Cluster 1, low spatial abilities, 16 people), from 4.75 to 8.50 ( Cluster 2, medium spatial abilities, 18 people), and from 10.0 to 14.25 (Cluster 3, high spatial abilities, 7 people).

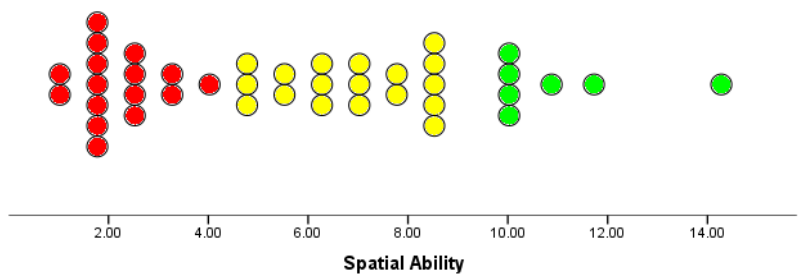

Figure 5.15: Spatial ability values. Colors denote the different clusters created.

In the other cognitive assessment, **Digit Span**, which is an indicator of Verbal IQ, enclosing both Memory and Attention, cluster analysis was also performed and resulted in the creation of four clusters. Figure 5.16 presents the raw data values for all participants along with a visual representation of the clusters created. Clusters range from those who ranked 24.0 in the digit span test (Cluster 1, low memory, 6 people), from 36.0 to 42.0 ( Cluster 2, medium-low memory, 12 people), from 54.0 to 72.0 (Cluster 3, medium-high memory, 17 people), and from 84 to 114 (Cluster 4, high memory, 6 people). In these metric, the dendogram shows large changes in the agglomeration coefficients both in the two and three-cluster solution. In Figure 5.16, the three-cluster solution is presented by the grey square which shows the agglomeration of the first and second cluster. The two-cluster solution is one that joins clusters 1 and 2 into a single one and clusters 3 and 4 into another.

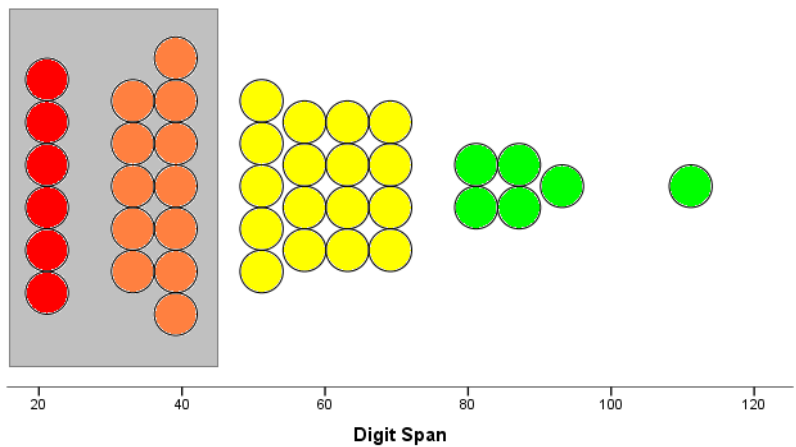

Figure 5.16: Digit Span (Memory and Attention) values. Colors denote the different clusters created. Gray area joins two clusters into a single one showing a three-cluster solution.

## Clustering groups of Variables

One of our goals in performing cluster analysis is, in some sense, to re-stereotype the population in finer grained groups. We do not wish to see the blind as a stereotype but the number of variables considered along with a sample that fails to reach the hundreds or the thousands makes it difficult to seek relationships without previously finding *types* of blind people. However, we intend to create groups that are meaningful and enable us to find relationships with deviations in mobile performance. The aforementioned clusters are univariate and deal with variations within a single dimensions hiding the differences in other variables. In this section, we seek to identify patterns of people and see if the combination of measures in different variables, the so-called *re-stereotypes*, are meaningful in what concerns mobile performance. In what follows we present the multi-variate clusters created at different levels. A single all-in-one cluster attempt was not performed as the ratio between the number of variables and the sample size dis-advises it. Clusters were created from profile and ability data.

In what concerns background and profile, Education group was considered just for univariate clustering. We then clustered the cases pertaining the two age-related variables: Age and Blindness Onset. A cluster analysis was run on 41 cases, each responding to items on demographics (age and blindness onset). A hierarchical cluster analysis using Ward's method produced three clusters, between which the variables were significantly different in the main. The first cluster was characterized by younger (24 to 52 years old, M=35.8, SD=9.12, N=14) early-blinders (Onset between 0 and 15 years old, M=5.50, SD=5.125). The middle cluster was again mainly early-blinders (Onset between 0 and 17 years old, M=7.0, SD=6.1, N=10) but older (49 to 65 years old, M=59.0, SD=4.422). The third cluster was essentially composed of late-blinders (Onset between 20 and 44 years old, M=30.76, SD=7.35, N=17) with ages comprehended between 31 and 59 years old (M=43.41, SD=6.83). Figure 5.17 presents the three clusters in the two-dimensional space. It is important to notice that our sample lacks older late-blinders. As mentioned before, this can be due to the lack of exposure to the society of this user profile. Later analysis of the impact of age-related characteristics will take this limitation in consideration and conclusions should be taken with caution.

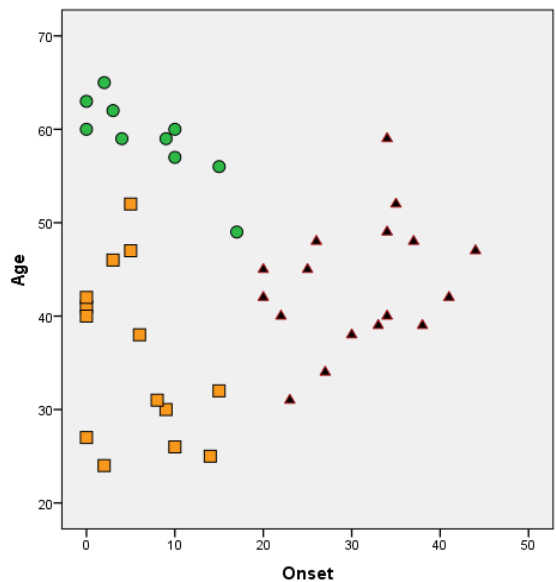

Figure 5.17: Chart presenting the combined clusters for Age and Onset

One other component worth clustering is the tactile one. Two measures were collected per user: tactile acuity and pressure sensitivity. They showed not to be correlated suggesting that they should be considered independently. Thus, we will did select one of them over the other, rather we created groups of those with low and high values on each of the combined components. Figure 5.18 shows a graphical representation of the four clusters created. The reference values for low and high classification in each dimension were the ones from the univariate clustering procedures. As such, the first cluster gathers all samples with a tactile discrimination of 2 millimetres and a sensitivity to applied forces of equal or less than 3.61 Newtons. Cluster 2 is characterized by those of similar high tactile discrimination but with sensitivity to applied forces greater than 3.61. Cluster 3, on the other hand, gathers the cases where tactile acuity is lower (discrimination of two-points separated by 3 millimetres or more) and pressure sensitivity is better (sensitivity to applied forces of equal or less than 3.61 Newtons). Clusters 4 comprises the worst tactile-wise cases: low pressure sensitivity and low tactile acuity.

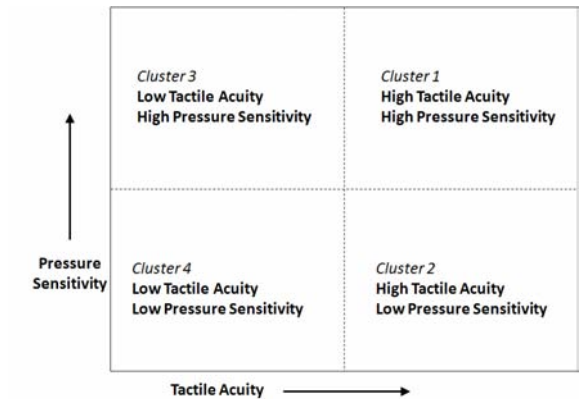

Figure 5.18: Clusters created from pressure sensitivity and tactile acuity case values.

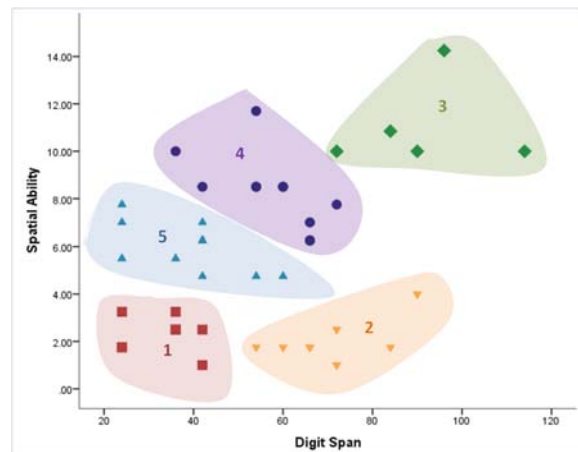

Figure 5.19: Clusters created from spatial ability and digit span case values.

Spatial ability and Memory were the two cognitive components considered in this evaluation and we undertook a clustering attempt to obtain a single cognitive grouped variable. A cluster analysis was run on all 41 cases. A hierarchical cluster analysis using Ward's method produced five clusters, between which the variables were significantly different in the main. A two-cluster solution would also be feasible but we opted for the one which created more discrimination within our sample. Figure 5.19 shows the agglomeration of cases within the clusters along. The first cluster gathers cases of people with low spatial abilities and low memory while the third cluster gathers predominantly those with greater spatial and memory abilities.

# 5.7. Discussion

**The un-stereotypical blind** In our quest for participants, no cherry-picking was made. This makes us believe that we gather a representative set of the active blind population resident in Portugal. By active we mean those that are somehow connected to the community whether by working, studying or maintaining relationships with national institutions and other blind individuals. The overall age-wise numbers herein presented are in consonance with national and worldwide statistics, although with small discrepancies. One can state a offset of the age range probably due to the inability to recruit participants no longer active and unavailable to participate. What becomes clear given the range of profile and individual attributes is that there is no such thing as a the stereotypical blind. Resorting to such a narrow view of the blind population translates in the creation of one-size-fits-all solutions that are likely to damage the first and consequent interactions by those that are unable to surpass the demands imposed. Differences between individuals were observed at the profile level but also at the sensorial and cognitive levels. The former also showed to have a significant impact in the latter, as suggested by related work in the area, presenting possible research opportunities in taking advantage of lighter and easier to collect data that, although possibly not as accurate, may fairly approximate a given capability-demand relationship.

**Layers and influence** The exploratory studies reported in this chapter were performed at different layers: the profile one, which is mostly demographic; the individual, respecting to measurable individual abilities; and the functional one, respecting to the usage of mainstream technologies where we once again focused on text-entry due to its intermediate level of difficulty. These layers are not independent from each other. Indeed, a person's background is likely to affect her abilities, either individual of functional which can be by turn related to each other. Results showed correlations between attributes of these three levels.

At this point, these relationships are not sufficient to exclude a particular level as a determinant one for the usage of mobile interfaces, particularly in the early stages of learning. However, it is important to retain that these relationships do exist and they can be considered in the creation of adaptive and predictive models of user performance. Ultimately, as an example, if profile attributes are highly correlated with individual and functional ones and these are likely to accurately model user performance with a certain interface, one would probably explore the benefit of an adaptation model based on easily measurable characteristics. In Chapter 6 we will explore different demands on several mobile devices and settings. The correlations encountered between individual traits will be revisited seeking to reduce the dimensions to explore and ease the feasibility of predicting inclusiveness of devices and interfaces.

**Divide and Conquer: fine-grained stereotyping**  One of the central hypothesis pertaining our research is that individual attributes have a determinant impact on user performance with different demand-wise interfaces. Although some of the components evaluated are correlated with each other, others are not. The overall individual ability is likely to be characterized by a few independent components. We can have a sense of the impact of each individual attribute by relating with the performance achieved in different settings with underlying divergent demands. However, to capture a more complete ability-demand model a multi-variate approach will be preferred as it would consider the interactions between the different abilities. Multi-variate approaches advise large samples [Tabachnick and Fidell, 2006]. To be able to observe the impact of individual abilities together in relation to varying different demands, a group-based approach is required. Clusters allowed us to create groups characterized with a mix of individual attributes and will enable us to observe the impact of the variations within in user performance with different demands (Chapter 6). One could argue that we are contradicting ourselves by re-stereotyping the population. It is true. A perfect solution would relate abilities and demands in a continuous space but the number of individual abilities along with the relatively reduced sample size difficult such an attempt. We will still, when possible, try to draw conclusions based on the relationship between variables.

**The relevance of individual traits**  The functional abilities evaluated in this chapter are commonplace for a majority of the population. The experience they already have with mobile phones, computers, Perkins Braillers and Braille reading is likely to reduce the impact of individual and profile differences as compensation mechanisms are created. We consider that individual differences have a great impact in the preliminary stages of using a new technology and that this by turn is a major vehicle of exclusion as some users are not able to surpass the demands imposed by the interfaces. However, other individual traits like attitude and motivation along with the accompaniment a person has is likely to support the surpass of such demands. Still, results showed that, even with the positive interference of experience, the relationship between abilities and demands is still visible. Also, these relationships showed to be stronger when the demands are higher. We conclude that it is relevant to look at individual attributes when designing for inclusion and that the impact this may have goes beyond a first approach with the system and is likely to improve the user experience in the long run.

# 5.8. Limitations

**Selection of measurable individual differences**  In Chapter 4, we were able to identify
a wide set of user attributes that were prone to affect user relationship with tech-
nology.  This set gathered characteristics from different scopes whereas we could
encounter more behavioural and abstract attributes like Motivation or Attitude as
well as others more concrete and easily measured, e.g., Tactile Sensitivity or Ver-
bal IQ. Our aims in this thesis are mainly concerned with the demands and first
barriers faced by the blind individual when starting to use the technology.  Thus,
we have restricted our studies and analysis to features that are concretely mea-
sured/assessed and do not present great oscillations along short timespans.  We
consider that Attitude and Motivation, for example, play a highly relevant role in
the long run and are determinant for improvement and adoption but have little im-
pact in the discrete abilities of a person to operate a device. However, we are aware
that failing to evaluate these characteristics may reduce the ability for our conclu-
sions to be extrapolated to long term usages and short-term variations in user's
aptitudes.  In contrast, we also believe that these abstract features are highly influ-
enced by more concrete abilities. Concerning the set of concrete abilities evaluated,
we also decided to maintain our assessments focused on the abilities stressed by
low-level primitive demands.  The reason for reducing to the most the number of
evaluated attributes was the need to take the whole evaluation timespan to an ac-
ceptable minimum. This option made us exclude Dexterity analysis from the set as
this ability is mostly stressed in composed actions where fine motility is explored.
However, this showed to be a limitation as a few primitives later explored during
this thesis are likely to be demanding regarding dexterity (e.g., double-tapping).

**Opportunistic selection of assessment tests**  The set of assessment tests for sensorial and
cognitive abilities is smaller and harder to obtain than for a sighted person. While
verbal assessments can be used by both populations, more practical, spatial or
shape-based tests, have to be custom-made for this population.  Also, while it is
commonplace to have cognitive evaluations based on shapes or visual puzzles in
simple paper sheets, the blind population requires a physical tactile-based counter-
part.  Several issues arise here as tactile abilities may interfere with the ability to
evaluate a non-sensorial attribute.The discussion and comparison between meth-
ods to sensorial and cognitively evaluate a blind person are far beyond this thesis.
Test selection was thorough but opportunistic meaning that we selected certified
tests but were also at our disposal or easily reachable.  Cost and availability of the
evaluation sets were determinant factors.

**Restricted to a minimally active population**  As shown throughout this chapter, we strived
to guarantee a wide and diverse participant sample.  This was done before the eval-uation by making contacts with several institutions and individuals from different backgrounds (e.g., schools and formation centres) but also during the evaluations as a permanent effort to enlarge the sample was issued (e.g., asking each participant for acquaintances interested in participating). I believe that a good job was performed and an uncommon user sample, in number, age and education range, was achieved. Yet, one limitation lingers: our participant sample is restricted to those that are out in the wild striving whether to work or to increase their aptitudes. For example, older late blinders may not feel the need or motivation to participate in rehabilitation, formation or any other institution-based activity. These people are out of radar and were impossible to recruit. The conclusions from this chapter and the subsequent ones are to be taken with an active blind population in mind. This might be the reason for the offset between our average age and age distribution values and those reported in national and worldwide statistics [5].

## 5.9. Summary

In this chapter, we presented an in-depth study with 51 blind people with the main goal of assessing differences in the different dimensions revealed in Chapters 4 and 3. The relative large number of participants enabled us to present representative results to attributes in the profile (e.g., age, education, blindness onset), tactile (pressure sensitivity and spatial acuity), cognitive (spatial ability and memory), and functional (pc, mobile and braille abilities) levels. Results showed that different blind people present different attributes and ability levels along the various dimensions assessed. More to it, relationships between individual attributes were also presented along with an impact of these on the participants' functional abilities which shows that, even with experience, individual traits have a large impact on how users use technology and the proficiency levels they attain. Further studies are required to assess how these individual aspects influence the usage of novel technologies, particularly the emergent touch devices, without the interference of previous experience.

---

[5]World Health Organization Visual Impairment and Blindness Factsheet - http://www.who.int/mediacentre/ factsheets/fs282/en/index.html, June 2012

# 6

# Exploring Mobile
# Touch Screen
# Demands

*No way, touch screen mobiles are not accessible to us..they have no physical keys, how could we get around?* - These words were shared with us by a blind person during preliminary questionnaires about the usage of touch screen devices. Indeed, the concerns about this technology are still common among blind people. If on one hand this can be argued as a misconception, i.e. blind people are using touch screen devices with success resorting to screen reading software like Apple's VoiceOver, on the other hand, there is still little knowledge about the difficulties these users face in their first approaches with the devices and how proficient they get in the acquisition of targets and performance of simple touch primitives. Further, touch screens come in different flavours: a tablet is larger than a touch phone and the abilities required to explore both gadgets may differ; the amount of information on-screen along with the size of the interactive elements there may also vary. There have been studies in the past to understand how to better parametrize touch interfaces for sighted people [Parhi et al., 2006, Lee and Zhai, 2009]. Blind people are faced with similar interfaces as their sighted fellows with an exploration layer above enabling them to receive feedback as they wander the screen. However, Kane et al. presented a study where differences in the interaction with touch screens between sighted and blind people became evident [Kane et al., 2011].

In this chapter, we explore how blind people interact with three different touch setting: a touch phone, a touch phone with a physical border applied, and a tablet. We explore the acquisition of targets on different sized grids and the realization of directional gestures on all three devices and aim to assess how these settings vary in terms of demand to the blind user. The study with 41 blind people is presented and the results discussed in the following sections.

# 6.1. Research Questions

Our research aims at a better understanding of the demands imposed by touch interfaces to blind people and how these are surpassed by people with different ability. Particularly, we aim to answer the following research questions.

**Pertaining Mobile Touch Demands**

In a first instance, we will look at the population as a whole and try to understand which settings are more demanding. Also, by analysing the dispersion on each setting we will be able to understand where performance varies the most. This will lay the groundwork for the subsequent set of research questions.

- How different devices affect blind people performance?

- How users cope with different sized grid-based layouts?

- How users perform different primitives on different devices?

- Which are the most prominent demands in current touch settings?

- Are blind users capable of surpassing the demands imposed by touch interfaces?

**Pertaining Abilities and Demands**

In a second phase, we want to understand if demands are surpassed similarly by the majority of the population. If not, we will try to understand how are different levels of ability related with varying demands. The ultimate goal is to show that interfaces built for the *average individual* are inadequate for a large part of the population unable to live up to the supposed stereotyped set of abilities.

- Are blind users capable of surpassing the demands imposed by touch interfaces?

- Are individual attributes related with differences in performance?

- Which individual attributes play a relevant role in a non-visual mobile touch context?

- Which relationships exist between abilities and demands?

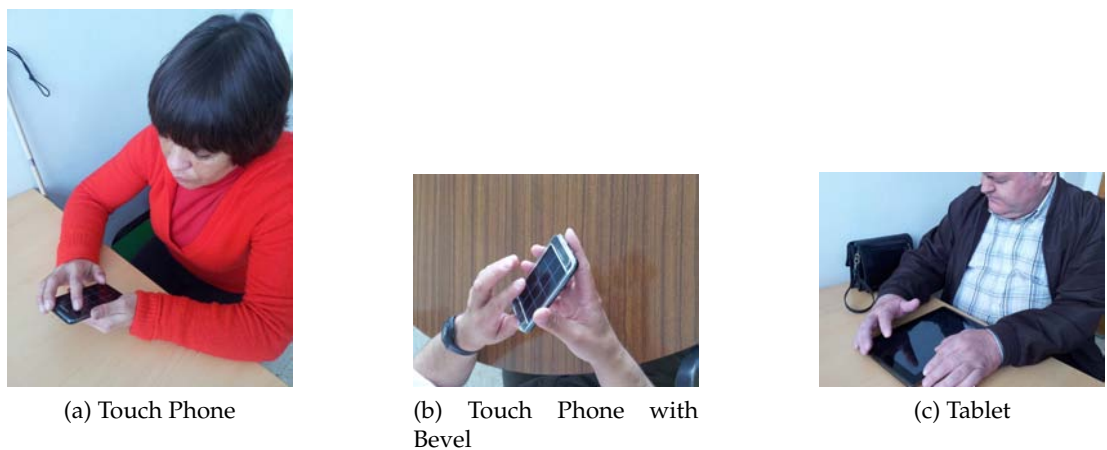

(a) Touch Phone          (b) Touch Phone with Bevel          (c) Tablet

Figure 6.1: Users in three different study device settings

# 6.2. Experimental Methodology

We conducted a controlled laboratory experiment with forty-one (41) blind people to better understand how they cope with touch screen devices and the demands imposed by these. To this end, participants in the study interacted with different devices and several primitives and target sizes therein.

## 6.2.1. Task

This study focused on the most basic tasks performed with mobile devices. The task set comprised selecting areas in a touch screen by tapping, pressing longer or double tapping, and performing simple directional flick gestures.

The tasks were performed by all participants in three touch settings (Touch Phone, Touch Phone with a physical Border, and a Tablet) and a keypad-based baseline traditional phone. For each touch setting, we evaluated two grid-based layouts and a grid-less one (directional gestures). The two grid settings selected were the ones with 6 and 12 areas (Figure 6.2). These two layouts were selected as they are familiar and thus easier to explain. In each of those settings, each participant was recalled to select all areas (randomly) three times, one for each primitive: Tap, Long Press, Double Tap. As to Gestures, all participants were asked to perform the four main directional swipe gestures in the middle of the screen (left, right, up and down) and the same gestures on the edges (in the top to the right, in the top to the left, in the bottom to the right, in the bottom to the left, in the right to the top, in the right to the bottom, in the left to the top, in the left to the bottom).

In the baseline setting, participants were asked to perform the three primitives similarly: Press, Long Press and Double Press. The 12 keys were recalled twice for each primitive.

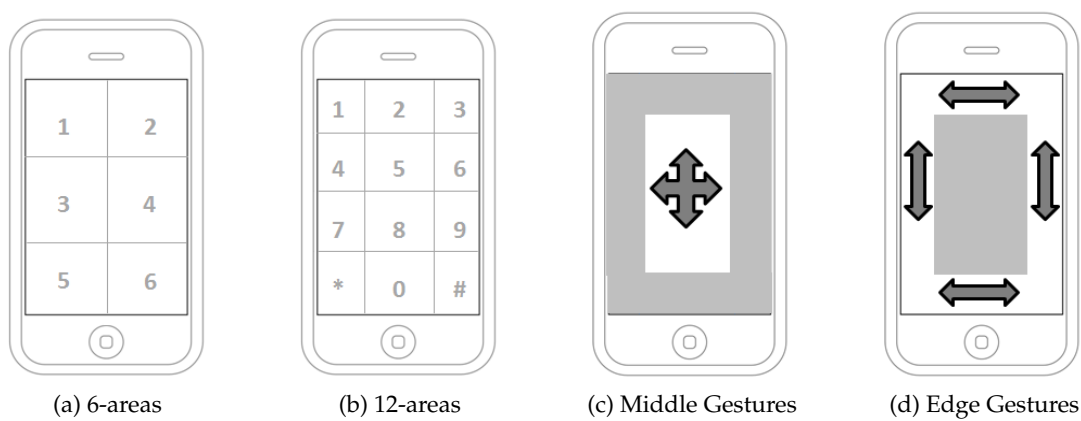

(a) 6-areas            (b) 12-areas            (c) Middle Gestures       (d) Edge Gestures

Figure 6.2: Areas recalled in the study protocol

Device (Touch Phone, Touch Phone with Physical Border, Tablet, Keypad), Method (Grid, Gestures), Area (6 or 12), Primitive (Tap, Long Press, Double Tap - Figure 6.3),and Locations and Directions were all randomized to counteract order effects. Each participant had its own Device order (algorithm run before the sessions). Automatically, for each device, each user was assigned a method order, a area order there in, primitives and so on.

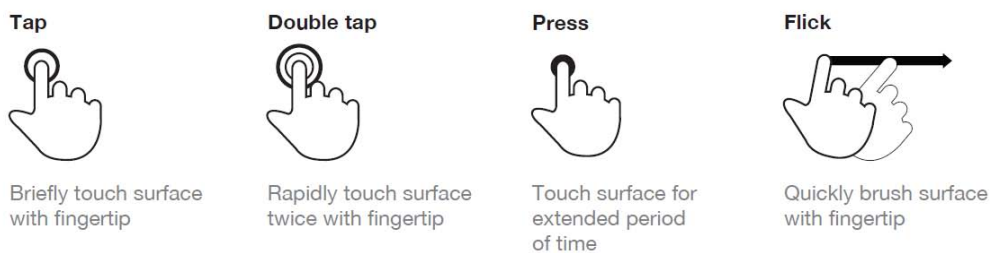

Figure 6.3: Primitives recalled in the study for all touch devices

In the overall, each user performed 18 actions on each device for each primitive pertaining area selections (162 actions) and 36 directional gestures (12 per device). This sums to a total of 8118 touch actions. Plus, each participant performed 24 keypad actions for each primitive summing to 2952 keypad actions.

## 6.2.2. Devices and Apparatus

The experiment comprised the usage of two different-sized devices, a touch phone and a tablet. As a touch phone, we selected the Samsung Galaxy S (122.4 x 64.2 x 9.9 mm) [480 x 800 pixels, 4.0 inches] while as a tablet an Asus EEE Transformer (271 x 171 x 12.98 mm) [1280 x 800 pixels, 10.1 inches] was used (Figure fig:hardware). A third setting was used but with the usage of the same touch phone with a physical border applied. The HTC S310 was used as the keypad phone.

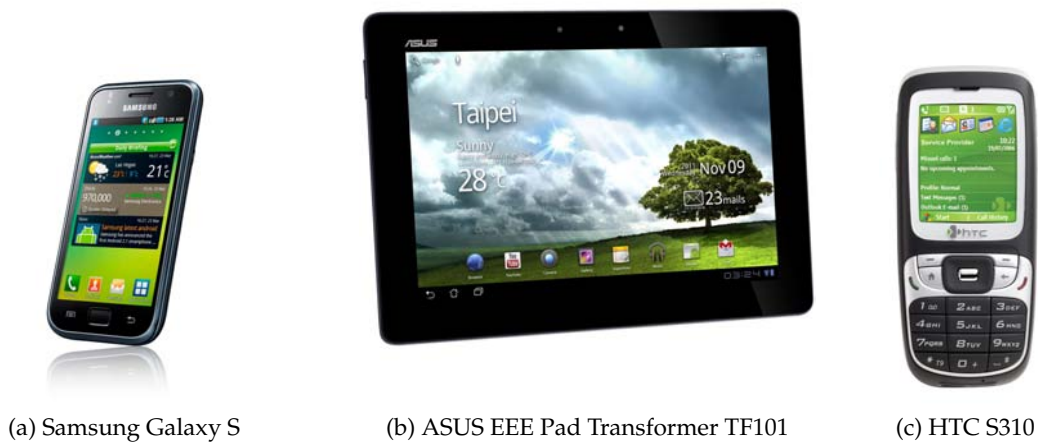

(a) Samsung Galaxy S          (b) ASUS EEE Pad Transformer TF101          (c) HTC S310

Figure 6.4: Three devices used in this study

Both touch devices run Android (2.2 and 3.0, on the phone and the tablet, respectively). The evaluation and logging software was developed in Java and was the same for both devices. Svox Classic TTS with the Portuguese language was used for audio feedback. The software was responsible for randomizing the order of the trials in respect to Method (Grid Tapping, Gesturing), Grid (6 or 12 areas) and Primitive (Tap, Double and Press) for Grid-Based Tapping, and the trial targets/directions. Also, from the start of a session with a particular device, all the sequence of steps is determined by the software. The application, for each device, can be depicted in the following steps/screens:

1. Participant Identification: the monitor introduces the name of the participant into the device

2. Method Presentation: in this first screen, the monitor and the user are presented with the method that will be tested. The application reads out loud (Text-to-speech) the name of the method and waits for a confirmation to continue. This gives time for the monitor to explain the method to be tested to the participant. When finished, the monitor unblocks the next screen. A mechanism to present a menu when this action is performed was created and is available throughout the application to enable the users to use the screen without any undesired actions and giving the monitor the ability to (safely) carry on with the study when required. The menu presented to the monitor enables to go to the next screen, repeat the current one (even if one of a trial) or terminate the experiment .

3. Screen Presentation: in this screen, the participants are able to touch the screen and hear a sound when in contact with the touch screen surface (light beeps). This exploration screen enables the user to accurately understand when the borders of the screen end. This was necessary as both the phone and the tablet (as most touch devices nowadays) fail to present tactile cues at the end of the screen. When the participant stated to be ready, the monitor would grab the device and employ the same method to move to the next screen

4. Grid Presentation (only for grid-based layouts): the participants were able to explore the grid-based layouts by touching the screen and receiving feedback about the area they were touching ("e.g., 'One', 'Asterisk',...). This enabled them to train and get in the right frame of mind

5. Primitive Explanation: the participant is presented with the name of the primitive to be tested (e.g., 'Tap', 'Double Tap', 'Flick',...). The monitor is then responsible for explaining the primitive and asking the participant to perform test trials until the primitive is understood.

6. Primitive Testing: the participant is presented with a testing screen and a message - "Please touch the screen when you are ready to start". From the moment the user touches the screen, a timer issues a request two seconds later for an action. The timer is only activated again after the user performs the on-screen action. If the user keeps pressing the screen the timer is re-started. This means that a new request is only issued if the user is inactive for two seconds. This goes on until no further requests are available. After a few seconds and a termination message, the application goes back to the Primitive Explanation screen if another primitive is still to be tested. The same happens with Grid and Method meaning that when all primitives are tested under a certain grid size, a new grid size is presented, and than when all grid sizes are finished a new method is presented to the user. Then, the following screens are, once again, presented and explained by the monitor until all combinations are tested.

The protocol for the keypad phone setting was similar. The operating system of the keypad phone was Windows Mobile 6.0 and the test application was developed in C# with the Windows Mobile SDK. Speech synthesis was provided by Loquendo text-to-speech synthesizer (*Eusebio* portuguese voice). The steps comprised in the keypad test application were the Participant Identification, Primitive Explanation and Primitive Testing. Timeouts between actions were configured to be equal to the touch settings.

```xml
<?xml version="1.0" encoding="UTF-8"?>
<sessions>
  <session method="Grid" cells="6" primitive="Tap" date="2011-10-18">
    <trial type="Test" time="01:36:02.774">
      <goal>Cinco</goal>
      <regTouch x="667" y="856" time="01:36:05.906">6</regTouch>
      <regDown x="667" y="856" time="01:36:05.908">6</regDown>
      <regTouch x="666" y="856" time="01:36:05.961">6</regTouch>
      <regTouch x="665" y="855" time="01:36:05.982">6</regTouch>
      <regTouch x="664" y="854" time="01:36:06.009">6</regTouch>
      <regTouch x="664" y="853" time="01:36:06.032">6</regTouch>
      <regTouch x="664" y="853" time="01:36:06.081">6</regTouch>
      <regTap x="664" y="853" time="01:36:06.083">6</regTap>
    </trial>
    <trial type="Test" time="01:36:07.585">
      <goal>Dois</goal>
      <regTouch x="512" y="517" time="01:36:10.653">4</regTouch>
      <regDown x="512" y="517" time="01:36:10.654">4</regDown>
      <regTouch x="512" y="516" time="01:36:10.746">4</regTouch>
      <regTouch x="511" y="516" time="01:36:10.763">4</regTouch>
      <regTouch x="510" y="515" time="01:36:10.773">4</regTouch>
      <regTouch x="509" y="514" time="01:36:10.783">4</regTouch>
      <regTouch x="509" y="513" time="01:36:10.802">4</regTouch>
      <regTouch x="509" y="513" time="01:36:10.865">4</regTouch>
```

Figure 6.5: XML log file example for touch session

Logs in both platforms were automatically created by the platform and stored in XML files (Figure 6.5). The logging was performed at two levels, particularly for the touch settings, which vary in detail. In a lower-level, we log every touch (up, touch and down events) detected by the operating system. On a higher-level, we resort to the operating system events to detect composed primitives that are already in line with some restrictions (Tap, Double Tap, Long Press, Scroll, Flick). For each detected touch or event, we store on-screen position (in pixels and detected area), velocity (in case of gestures) and time. As to the keypad, we also store all lower-level events like down and up events but also register operating system detected events like Tap, Long Press and Double Tap. Time is logged for all captured events.

## 6.2.3. Participants

Forty-one (41) blind people from our participant pool (Chapter 5) participated in this study. No selection was performed: all available blind candidates were included. Their ages were comprehended between 25 and 66 years old (M=44.6, SD=11.3). The group was composed of 22 males and 19 females. As with the complete pool, we ran a normality test which confirmed a normally distributed sample in regards to Age (Shapiro-Wilk, W=0.97, p >.05). Blindness onset failed to meet normality showing a similar shape as in the complete pool. In the overall, the sample used in this study (41 out of 51 participants) seems reflects the results achieved in the population characterization study.

## 6.2.4. Procedure

The participants of the study were already evaluated in respect to their individual attributes and abilities (Chapter 5). They were called for this study session which lasted for about one hour per person. The session encompassed a brief initial questionnaire pertaining specifically the usage and thoughts on the touch screen mobile paradigm; the experimental task for all touch and keypad settings; and a questionnaire after each setting to assess the opinions in regards to difficulty in using each primitive, device and layout.

All sessions were performed in the same training centre installations in a quiet meeting room. At the beginning of each session, the evaluation monitor explained the goals of the studies along with a description of the study procedure. Each participant performed the evaluation with all four devices. The order was randomized. For each device, the blind person had a preliminary familiarization phase. After feeling the device and its elements, the evaluation software was initialized and before each grid-based layout (also randomized), the device prompted the participant for a training session where feedback was offered depending on the touched area. When the participant felt ready, the monitor made the application step to the following screen. The test application started upon a first touch with the screen. Then, with 2-second intervals (informally fine tuned in a protocol validation phase) from the previous trial, a new request (a spoken area) was made by the device (text-to-speech). All areas were prompted only once to each participant mostly due to time restrictions. Exception is made for the keypad setting where keys were prompted twice as this device setting encloses less variations (only one Area). Upon completion of all layouts in each device, a post-questionnaire was applied to assess how the participant rated the difficulty of acquiring targets in the evaluated settings.

## 6.2.5. Protocol Validation

The protocol presented herein along with the questionnaires, logging and tools were pre-validated with three blind people that were not part of the participant pool. Problems with the software were corrected and small changes were performed to ease the deployment of the study. Further, a preliminary creation of tables, charts, descriptive and comparative statistics was performed to assess the completeness of the collected data. Analysing tools were revised to improve the automatic collection and storage of data.

## 6.2.6. Design and Analysis

A within-subject design was used. Forty-one participants performed interaction tasks with all 4 devices, varying Primitive and Area settings. The analysing tools I have developed automatically create Excel and SPSS files for all dependent variables, namely Incorrect Land Error Rate, Incorrect Lift Error Rate, Reaction Time, Tap Duration, Tap Interval, Tap Frequency, Automatic Detection Rate, and some specifically for Gestures, namely, Angle Offset, Gesture Precision, Gesture Size, Gesture Speed, Automatic Area Detection Rate, and Automatic Direction Detection Rate. Log analysers produce data ready for statistical analysis and for visual inspection. These software modules were developed in Python[1] and resorted to the COM[2] interface to automatically create excel files. Further, we resorted to matplotlib[3], a python plotting library, to automatically create point plots of all trials and overall plots of the different settings. This enabled us to recreate trials automatically and easily inspect behaviours of different participants.

Statistical analysis was performed resorting to SPSS. In the majority of the analysis presented, two-way repeated measures ANOVAs were performed to assess differences between groups (we assumed normality of data given the representative size of the sample and the power to deal with distributions that follow a non-normal distribution by ANOVA procedures [Glass et al., 1972]). As to ordinal data, we employed non-parametric alternatives (Friedman tests with Wilcoxon post-hoc multiple comparisons with Bonferroni corrections). Correlations were performed resorting to Pearson correlation when in presence of two normal continuous variables while Spearman correlation was used to assess correlations when at least one variable is ordinal or does not follow a normal distribution. Sporadically, other statistical procedures were used and are explained in the correspondent sections.

Null-hypothesis statistical testing is under severe criticism in the HCI community. Although it is used pervasively, results are often presented taking in consideration only the level of significance, normally p <.05 [Dunlop and Baillie, 2009]. However, this ignores trends in data that may approximate significance but fail to meet that hard threshold. To cope with this, we present p-values even when results fail to meet the 95% probability level. Further, we consider and state statistical significance at different levels (p¡.05, p<.01, p<.005, p<.001) besides mentioning minor significance at p<.1. In our reporting of F-statistics, where df is not an integer, we have applied a Greenhouse-Geisser adjustment for non-spherical data. All pairwise comparisons were protected against Type I error using a Bonferroni adjustment. Annex A5 gathers the ANOVA tables for further details of the analysis presented herein, particularly in the sections where p-values are not thoroughly presented.

---

[1] http://www.python.org/
[2] http://sourceforge.net/projects/pywin32/
[3] http://matplotlib.sourceforge.net/

# 6.3. Results: Mobile Interface Demands

Before trying to understand how individual attributes influence mobile performance, we present how blind people interacted with the different devices and how the different settings ranked themselves against each other. This enables us to understand which settings are more demanding and present greater variations. This knowledge will help us later by enabling focusing our attention in those demands and how they are surpassed by some and a barrier to others. The results presented in this section enable us to answer the research questions centred on the demands and differences between devices, areas and primitives.

## 6.3.1. Touch Grid–Based layouts

In this section, we will look perform an overview of the results obtained with the three touch device settings and the variations therein in performing target selections (Primitive and Grid). Exception made for primitive-related restrictions (e.g., LONG press recognition time) and an evaluation of the automatic detection performed by the OS, we assess these settings in an agnostic manner, meaning that we look at low-level features and inspect them separately. As an example, we look at land-on and lift-off positions and evaluate them separately in terms of error rate instead of making assumptions about the recognition of a Tap. Our aim is to have a deeper knowledge of the challenges imposed to blind people and such an approach enables us not only to identify the demands people are facing now but also to advice how they should be designed.

### Landing on a target

The success in operating a touch screen is highly related with the ability to acquire a target. For sighted participants, this is achieved through a high visual load. For blind people, this is likely to depend on spatial abilities, tactile sensitivity, memory, along with experience. Here we try to understand how difficult it is to acquire a target without visual feedback and how is this different between different device settings.

A three-way ANOVA revealed no significant main effect of Primitive (Tap, Long, Double) nor any interaction pertaining the type of primitive on Incorrect Land Error Rate. As such, and taking in consideration that, informally, some demands are likely to be independent from the type of primitive issued, in such cases we will present our analysis strictly for the Tap primitive. In cases where the results vary between primitives we will present the analysis for the Tap, Long and Double primitives.

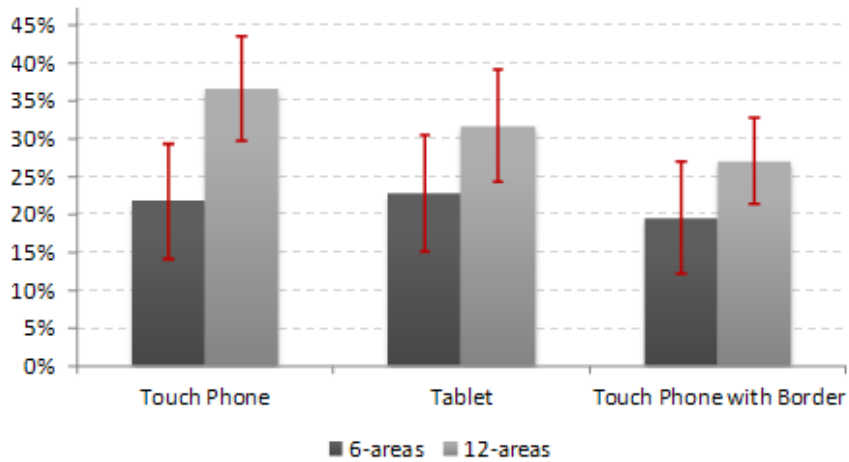

Figure 6.6: Incorrect Land Error Rate per Touch Device and Grid for the Tapping primitive. Error bars denote 95% confidence intervals.

Figure 6.6 presents the average land-on error rate for the three devices and grid settings for the tapping primitive where it is visible that in the 12-area trial differences were larger between devices. For all devices, there is an observable difference between the 6-area and 12-area settings. We will look at the differences between the touch phone and the touch phone with the border, and at the differences between the touch phone and the tablet separately, as these are the fair comparisons that can be performed. Particularly, we intend to analyse how the size of the device and consequently the targets therein affects user performance and if the presence of a physical border works as an aid for the user.

**Tablet vs Touch Phone.** A two-way repeated-measures ANOVA showed no main effect of Device **[Tablet vs Touch Phone]** on Incorrect Land Error Rate for Tapping (p=.966). Conversely, a main effect of Grid was found (F(1,35)= 11,34, p <.005). No significant interaction between Grid and Device was identified (p=.138) suggesting that the aforementioned main effect was consistent across both devices. In sum, the amount of targets on-screen seems to have an impact on land-on error rates as new demands are imposed to the user. However, a smaller or larger device, and hence smaller or larger targets, does not seem to have an impact on the users' ability to land on a target.

**Touch Phone vs Touch Phone with Border.** To assess the impact of a physical border on user performance, a two-way repeated-measures ANOVA was performed and revealed a main effect of Device **[Touch Phone vs Touch Phone with physical border]** on Incorrect Land Error Rate for Tapping (F(1,39)=5.77, p<.05). A main effect of Grid was also found (F(1,39)=9.37, p<.005). A minor significant interaction was found between Device and Grid. These results suggest that the presence of a physical border decreases the demands imposed to blind people and enable them to perform better. Also, the number of elements in a grid also has an impact on user performance. By turn, the advantage of having a border is greater when there are more targets.

Figure 6.7 shows the distribution of land-on points in all three devices. It is visible that a more sparse distribution happens on the touch phone in comparison to the touch phone with physical border. Indeed, the border works as a reference and the hit points are closer. The tablet due to the larger size of the targets also shows a larger amount of samples closer to the border as participants employ a safer approach.

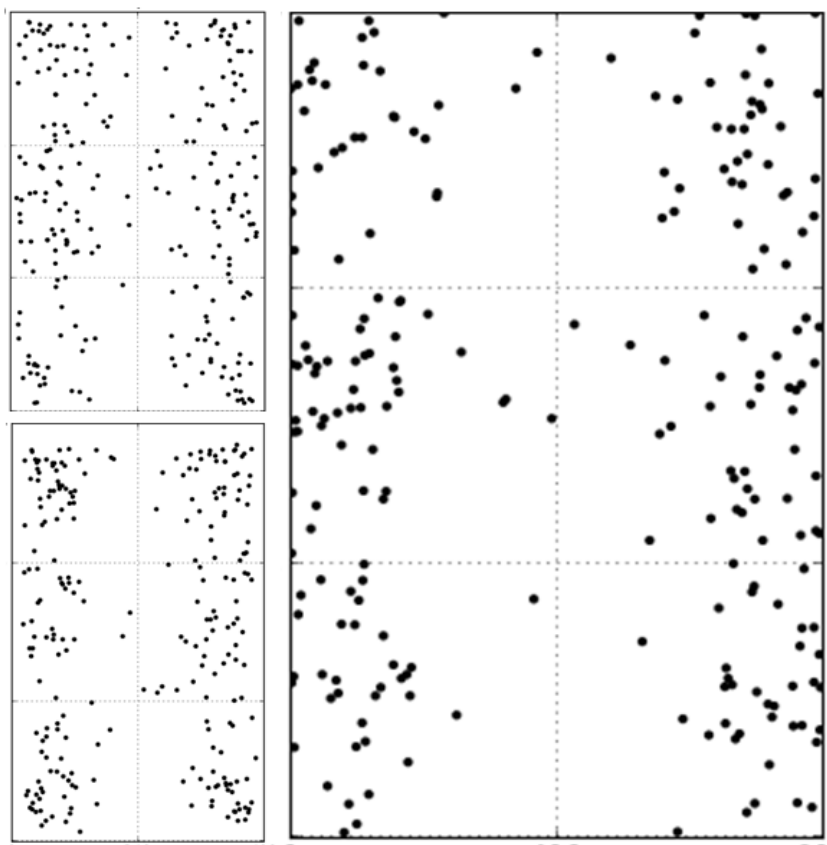

Figure 6.7: Scatter plot with all land-on events in all 6-grid tapping trials: Touch Phone (Top-left); Touch Phone with Physical border (Bottom-left); and, Tablet (right).

**Screen Areas.** To further assess the impact of size and tactile cues, we performed an analysis pertaining the position of targets on-screen for the Tapping setting. Figure 6.8 presents the error rate associated with each area on the 12-area grid setting for all devices. A two-way repeated measures ANOVA was performed to assess if it was easier to acquire a corner target (*Area*) and how did this change among devices (*Device*). Significant main effects were found for *Device* (F(2,70)=4.62, p<.05) and *Area* (F(1,35)=43,199, p <.001). No significant interaction was found (p=.882). Post-hoc multiple comparisons tests showed that tapping on a corner target is less erroneous and this is consistent across all devices. As to *Device*, differences were found between the Touch Phone and the Touch Phone with Border. The same test was applied but this time comparing target rows (Rows) and significant effect of Rows was once again found ($F(2.461, 86.12) = 26.602$, $p<.001$) between the 1st and both the 2nd and 3rd row as well as between the 2nd and the 4th row and between the 3rd and 4th row. The first and the fourth row (edge-rows) seem to be the less erroneous although an interaction was found between Device and Row suggesting that the presence of a physical border affects mainly the acquisition of the first row (worse in the border-less setting; the fourth row is consistent among all settings).

| | | | | | | | | |
|---|---|---|---|---|---|---|---|---|
| 21,62% | 24,32% | 24,32% | | 27,50% | 35,00% | 30,00% | | 12,20% | 19,51% | 12,20% |
| 32,43% | 43,24% | 35,14% | | 35,00% | 65,00% | 62,50% | | 26,83% | 63,41% | 46,34% |
| 45,95% | 48,65% | 59,46% | | 55,00% | 50,00% | 47,50% | | 31,71% | 46,34% | 37,50% |
| 13,51% | 18,92% | 13,51% | | 10,00% | 10,00% | 12,50% | | 12,20% | 9,76% | 7,32% |

Figure 6.8: Incorrect Land Error Rate per Target on the 12-area Tapping setting: Tablet (left); Touch Phone (center); and, Touch Phone with Physical Border (right).

## Lifting of a target

The selection of targets on touch screen interfaces, and graphical user interfaces in general, has been given much attention.

Various strategies pertaining target size [Sears and Shneiderman, 1991] have been explored. Also, there have been recommendations based on the strategy used to issue a selection: mainly on land or on lift. The land-on strategy only allows users to make selections where their fingers first touch the screen. Research indicates that targets 20 mm square or larger can be accurately selected using this strategy [Beringer, 1989, Hall et al., 1988, Weisner, 1988]. The lift-off strategy allows users to touch the screen, drag their finger to adjust the selection, and lift it once it is in the correct location to make the selection [Potter et al., 1989]. This strategy allows the selection of targets as small as 1.7x2.2 mm [Sears and Shneiderman, 1991]. These and other recommendations are valid for a visual-based scenario. There is a lack of knowledge of the effectiveness of such recommendations for a non-visual usage. Still, interfaces are presented seamlessly to blind people as they are to sighted people.

In this section, we will start by looking at differences in Incorrect Lift Error Rate between devices. This metric represents just the incorrect lifts performed after a correct land has been performed. In other words, it captures the slips performed by the participants. It gains further relevance in more complex primitives like Long Pressing and Double Tapping where maintaining an accurate selection without visual feedback may be difficult. It is important to notice than even state-of-the art *painless exploration* approaches rely on a preliminary exploration but do not make any improvement during the realization of a primitive.

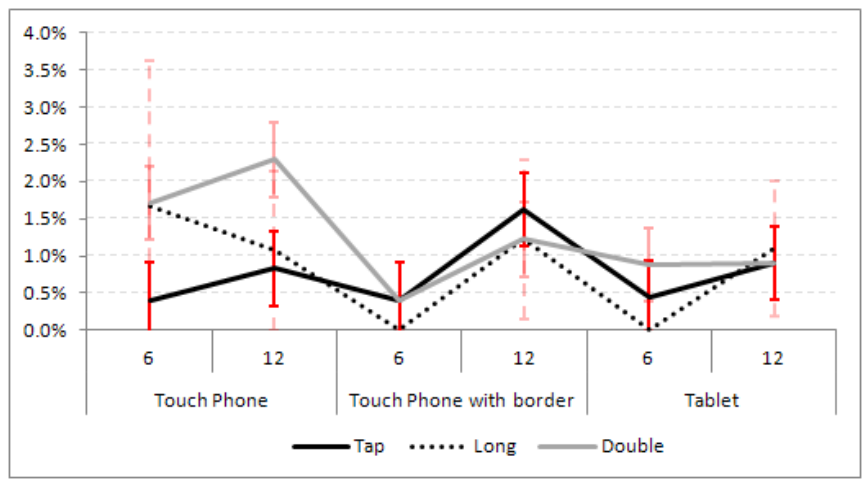

Figure 6.9: Incorrect Lift Error Rate for all touch devices and grid sizes. Error bars denote 95%confidence intervals.

**Tablet vs Touch Phone.** A two-way repeated measures ANOVA found no significant main effects of *Device* [**Tablet vs Touch Phone**] nor *Grid* on *Incorrect Lift Error Rate* for the Tapping setting. Also, no significant interaction between *Device* and *Grid* was found. Conversely, a minor significant interaction between *Device* [**Tablet vs Touch Phone**] and *Grid* was found for the Long primitive (F(1,34)=3.855, p = .058): participants performed no Lift errors on the Tablet 6-area setting and an average of 2.4% in the Touch Phone 6-area while in the 12-area setting the error rate was bigger for the Tablet (M=7.3%) than in the Touch Phone (M=4.9%). As to the Double primitive scenario, a two-way repeated measures ANOVA showed a minor significant main effect of *Device* (F(1,34)=3.804, p = .059). No main effect of *Grid* (p=.304) nor interaction with *Device* were found. This means that fewer lift errors were consistently made in the Tablet.

**Touch Phone vs Touch Phone with Border.** No significant main effects of *Device* or *Grid* nor interactions were found pertaining *Incorrect Lift Error Rate*.

Figures 6.10, 6.11 and 6.12 depict the types of target acquisition errors detected. For all devices, it is observable that most errors are related with an incorrect localization of the target. In these cases, the user landed and lifted on and from an incorrect target. The *Touch Phone* setting presents a slightly larger Incorrect Lift Error Rate than in the other two device settings meaning that the reduced size of targets along with a lack of physical stability augment the probability for errors during the action even if the land-on was performed correctly. On the other hand, the *Touch Phone with Border* presents slightly

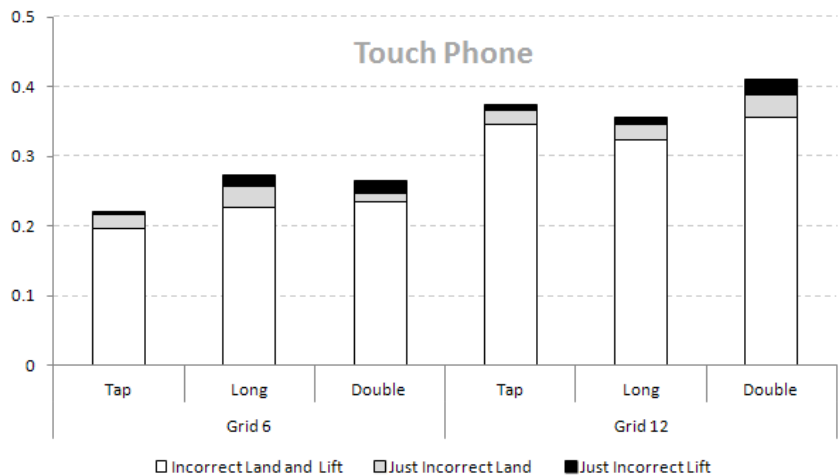

Figure 6.10: Types of errors for all grids and primitives for the Touch Phone setting.

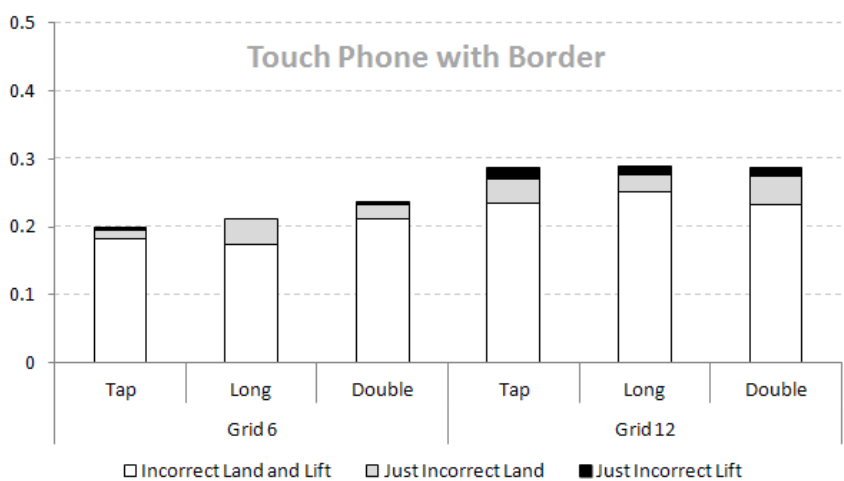

Figure 6.11: Types of errors for all grids and primitives for the Touch Phone with Border setting.

larger *Incorrect Lands* that were corrected at Lift time. This supports the idea that physical cues help the localization of the user both before (given the overall improved acquisition rates) and during the touch actions. As to the Tablet, and likely due to the size of the targets, slips and corrections are not that common.

## Reaction times

**Tablet vs Touch Phone.** A two-way repeated measures ANOVA showed a significant effect of *Device* on *Reaction Time* ($F_{(1,35)}=9.631$, $p <.005$) for the Tapping setting. A consistent effect was found for the Long ($F_{(1,39)}=6.607$, $p <.05$) and Double ($F_{(1,39)}=12.125$, $p <.005$) primitives. Grid did not show to have an Effect on *Reaction Times* and the effect of Device showed to be consistent across Grid sizes (no significant interaction). The

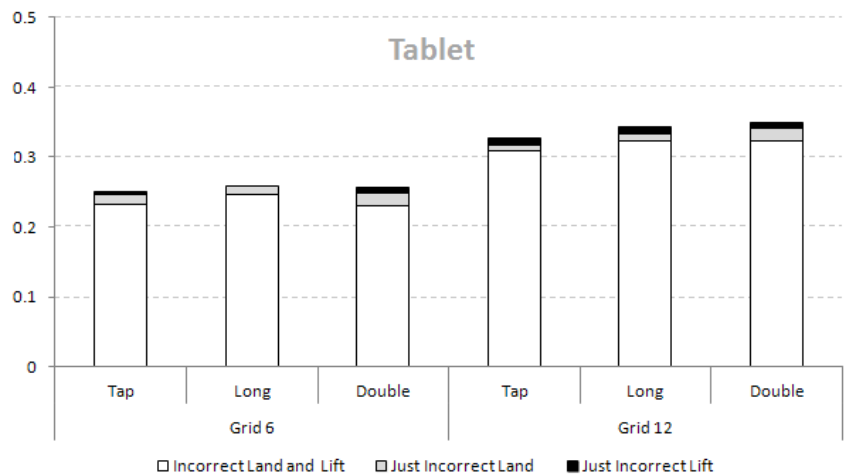

Figure 6.12: Types of errors for all grids and primitives for the Tablet setting.

Tablet presented as being cognitively more demanding as people tended to take longer to issue an attempt at the screen.

**Touch Phone vs Touch Phone with Border.** A similar analysis was performed to compare differences in *Reaction Time* between border and border-less devices and grid variations therein. For Tapping, no significant main effects nor interactions were found. As to the Long primitive, a significant main effect of *Device* was found on *Reaction Time* (F(1,38)=4.414, p <.05) showing smaller reaction times for the Touch Phone with Border. No significant main effect of Grid was found. Also, no interaction between Device and Grid was encountered suggesting that the physical border improves the reaction time of the user independently from the grid presented. This comes in line with the participants' comments where they stated that the border built up their confidence while on the other hand the large size of the tablet made them uncomfortable.

## Primitive requirements

Not all demands in touch interaction come from the location within the device. Each primitive recognition depends on a set of parameters, normally very strictly defined. Examples are the time length of a Long Press versus a Tap or the maximum interval between taps for the composed action to be considered a Double Tap. We will first analyse the automatic detection performed by the device and then delve into the parameters contributing for detection. Each one of them is likely to present a demand which we will be able to assess.

**Tablet vs Touch Phone.** Current mobile operating systems support the recognition of basic on-screen primitives [4]. Our first analysis concerning the recognition of primitives focused on the automatic detection performed by the device. As such, for each trial we collected the set of primitives detected and analysed if the one requested was performed. Notice that in this metric we do not look for errors pertaining more than one user action. If the user performed the primitive, it is marked as correct. If not, is marked as incorrect. A two-way repeated-measures ANOVA was performed to assess the impact of *Device* and *Grid* on the ability to perform detectable primitives (*Primitive Automatic Detection*). No significant impact of *Device* or *Grid* was found for the Tap or Long Press primitives. Conversely, a main effect of *Device* was encountered for the Double Tap primitive ($F_{(1,34)}=4.475$, $p < .05$) showing that it is easier to comply with primitive requirements in a larger space than a smaller one. Figure 6.13 presents the primitive automatic detection error rate. It is important to notice that these results only reflect the presence or absence of the recalled primitive; other errors may have been made and were not analysed in this particular assessment. Despite the difference between the Touch Phone and Tablet, particularly in the 12-area setting, we can also observe that Double primitive presents higher dispersion which suggests that although it can be easier for some, it can also vary in the opposite direction and be troublesome for others.

---

[4]Android SDK: http://developer.android.com/reference/android/view/ GestureDetec-

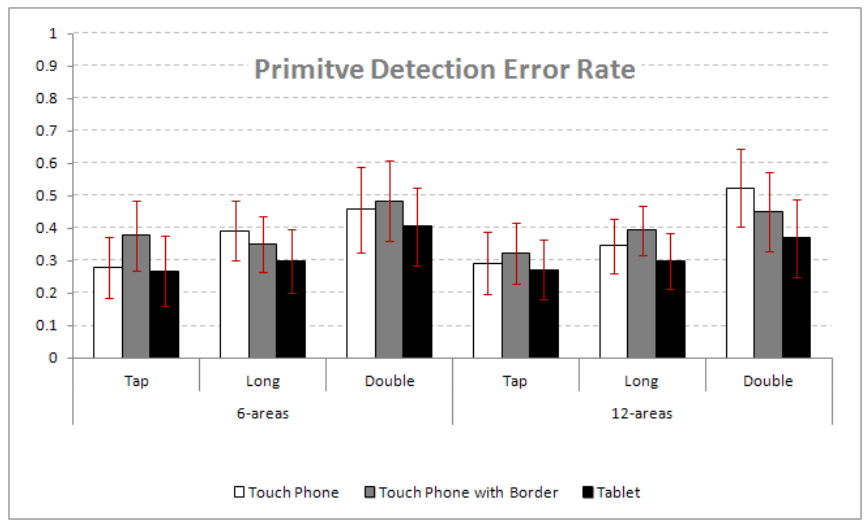

Figure 6.13: Incorrect Primitive Automatic Detection Rate per Device, Primitive and Grid. Error bars denote 95% confidence intervals.

Still concerning primitive requirements we also looked at the duration of an action (particularly to differentiate between a Tap and a Long Press) and at the interval between single actions (Double Tap). Significant differences were found between devices and grids both for the Tap (Device: $F_{(1,35)}=12.614$, $p < .05$ and Grid: $F_{(1,35)}=7.346$, $p < .05$) and Long (Device: $F_{(1,35)}=9.174$, $p < .01$ and Grid: $F_{(1,35)}=3.014$, $p < .1$ - minor effect) primitives. Participants showed to perform smaller time-wise actions in the Tablet and in the smaller Grid. This may be due to confidence or a matter of stability. Still, it suggests that the times for each action depend on the device and its demands. Also, high dispersion in

tor.SimpleOnGestureListener.html (Last visited on June 7th, 2012)

the values shows that it is also highly dependent on the user.

As to the interval between actions, no significant differences were found for Double Tapping suggesting that people perform the double tap action with similar time intervals across devices and grid sizes.

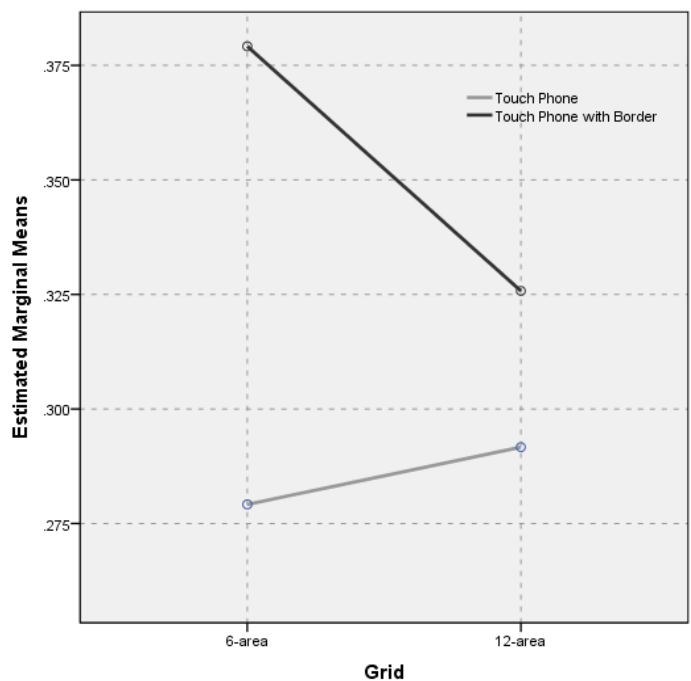

Figure 6.14: Estimated marginal means of Primitive Automatic Detection Rate for borderless and borderful Touch Phone settings. Smaller is better.

**Touch Phone vs Touch Phone with Border.** The differences in placing a physical border to aid localization were visible. People use the borders as reference and improve their spatial knowledge of the interacting area. This border also showed to have impact in the way people interacted with the device. In particular, we noticed a more relaxed posture in holding device with less worries in touching the screen undesirably. Indeed, these differences were reflected in the results of automatic detection (Figure 6.14). Particularly, *Device* showed to have a significant minor effect on *Primitive Automatic Detection* $F(1,39)=3.783$, p = .059. The Touch Phone with the border showed to have a better recognition of Tap. Nevertheless, errors also exist in this setting due to the border. Inspection on touch logs showed that the majority of errors in the Touch Phone with border were created by an attempt to *look for* the border after making contact with the screen as an assurance of having tapped in the right location. The gesture recognizer understood those actions as scrolls (small gestures). In the other primitives (Long and Double tap) no significant differences were found.

Concerning the duration of an action, no significant main effect of *Device* was found in the Tap setting. However, a significant impact of *Grid* was found showing a smaller time to perform taps in the 6-area layout ($F_{(1,39)}$=7.891, p <.01) People tended to press longer in the 12-area setting seeking to after-measure if they were in the correct location. No significant results were found for Long or Double primitives. As to the interval between taps, significance nor any other informal indications were found.

Interestingly enough results of time-based restrictions varied in some settings. Further, by inspecting the dispersion in all aforementioned results one can verify that people have different abilities and some are not even able to comply with the strict demands imposed by operating system recognizers. Considering the time restriction separating a tap from a press, a mechanism pervasive to all touch screens, operating systems and most applications, it is often set up around 500 milliseconds. We inspected how people behaved against this demand. Particularly, we averaged the tapping and long pressing values of each user for each setting and verified this value against thresholds ranging from 100 to 2000 milliseconds with 50 shifts. This enabled us the inclusion of each value as a time restriction to differentiate between Tap and Long Press. Figure 6.15 presents these results showing the range between 500 and 1000 milliseconds (topping at 800 ms) as one where we could find good general timeout values. More relevant than finding a *perfect* value to suit everyone, these results show that there are users who can perform the primitives with time restrictions in the range between 100 and 300 milliseconds while others require more than a second to make their action clear. These results present an opportunity for adaptation to each individual ability and experience.

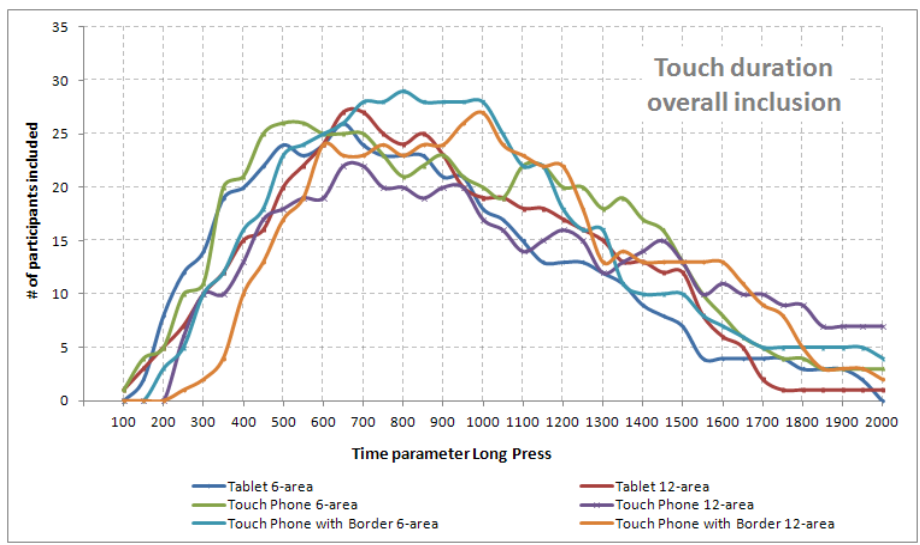

Figure 6.15: Number of participants with Tap and Long duration average measures included by different thresholds. Each line represents a different device setting.

Even more relevant than the aforementioned is the interval between taps to be able to perform a double tap. While concerning duration it may be a matter of performance rather than need, it was visible for us that some participants have little proficiency with technology and low motor dexterity in general and are thus not fit to perform fast repet-itive actions. This means that a novice user with such difficulties may never be able to perform a double tap with a touch screen which can foster his/her exclusion. Figure 6.16 presents the inclusion of each participant according to double tap intervals ranging from 50 to 1150 with 50 millisecond differences. It is important to notice that the inclusion curve comprises around 30 participants for the tablet settings and around 25 participants for the remaining at 250 milliseconds which is the most common interval for double tap recognition in current operating systems. The stabilization of the curve occurs around 400 milliseconds with inclusion of between 35 to 40 participants depending on the setting. The remaining converged until the value of 750 while a single outlier performed double taps with 1100 millisecond intervals. Once again, more than a single inclusive value, and opportunity for user-sensitive time-restrictions is apparent.

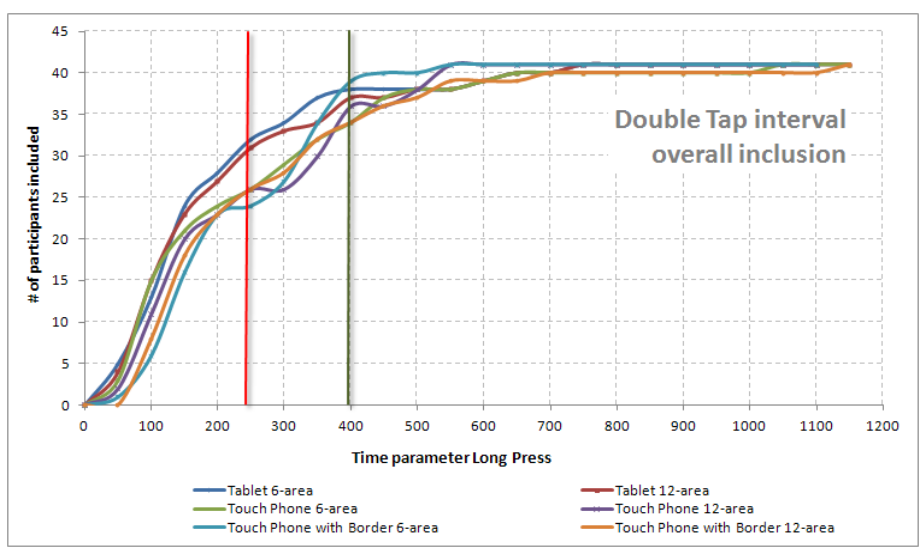

Figure 6.16: Number of participants with Double Tap interval average measure included by different thresholds. Each line represents a different device setting.

## Comparison with the baseline: a traditional Keypad

More than a third of the participants showed to be concerned with the strong prevalence of touch devices in the market and prediction of a key-less future. In general, exception made for one iPhone user, all of them stated clearly that the performances obtained with physical keypads were incomparably better. Also, most of the participants as seen before, is a keypad phone user for over a decade. Still, we also included a keypad mobile phone in our studies as a baseline setting. Participants performed the three primitives with this device. Although differences exist between devices and paradigms, these primitives can be performed similarly in touch and keypad settings. Given the differences, we will not delve into this setting but a brief comparison with the touch-based settings was performed.

We will just look at the simplest primitive and the one that is most comparable with the touch counterparts: landing on a target in the 12-key keypad. Figure 6.17 presents the average land on values for all compared settings. Results suggest that the keypad is indeed easier to use. However, a one-way repeated-measures ANOVA ($F(3,102)=6.427$, p $<.001$) showed this difference to be significant only against the borderless touch device settings. Indeed, the difference in error rate between the Keypad (M=20.9%) and the Touch Phone with border (M=27.4%) was surprisingly low and not-significant. This came as a surprise as smaller error rates were expected for the keypad setting. Conversely, it is relevant to look at the dispersion of values among participants. Deviations are visibly larger in the touch settings showing that results varied widely. On the other hand, dispersion around the central tendency is reduced in the Keypad setting showing that variations between users were smaller. This suggests that the demands imposed by the keypad are more consistently surpassed by a great amount of users while the demands imposed by touch device interfaces are surpassed at different levels by different people (and their underlining abilities). Once again, this suggests for further attention to differences between people and a better understanding of which demands are harder to surpass and which are the abilities involved in that process.

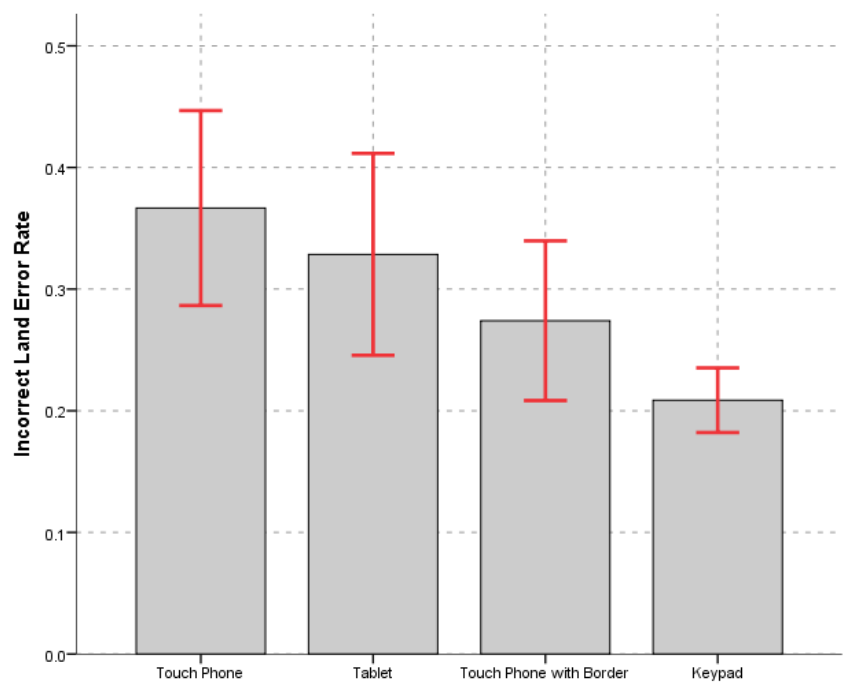

Figure 6.17: Incorrect Land Error Rate per Device for Tapping. Error bars denote 95% confidence intervals.

## 6.3.2. Touch Gesturing

One other common action performed in touch screen applications is scrolling the finger on the screen to perform a directional gesture. We focused on simple flicking gestures (fast scroll) on the four main directions. Moreover, we asked participants to also perform these gestures next to the borders (e.g., *On Top, to the right*; *To the right*; *On the left, to the top*). We will present results of the analysis comparing between devices and also results pertaining to the areas within a device. To enable a fair comparison in distance-based metrics, data was transformed to millimetres.

### Automatic detection

In this setting, the areas were not presented as strict to the participants. As such, we did not find an offset to consider as being near the edge or not. However, the Android built-in gesture recognizer identifies those areas and herein we will use that information to assess if participants performed their gestures in the correct area and how did this differ across settings.

**Tablet vs Touch Phone.** A one-way repeated measures ANOVA showed significant differences between the tablet and the touch phone on *Automatic Area Detection* ($F(1,34)=50.895$, $p <.001$) with a large advantage for the tablet. Figure 6.18 presents the results where this difference is evident. This results suggest that the larger size helps in performing a more compliant gesture. As to direction, an automatic direction recognition is also provided by the gesture recognizer in the Android SDK. No significant differences were found between both devices pertaining direction recognition accordingly to the operating system standards.

**Touch Phone vs Touch Phone with Border.** A one-way repeated measures ANOVA presented significant differences between the Touch Phone and the Touch Phone with border ($F(1,39)=15.968$, $p <.001$) with a higher recognition rate for the borderless device (Figure 6.18). Consistently, a one-way ANOVa also showed a significant advantage of the borderless setting in what concerns *Automatic direction recognition* ($F(1,34)=16.357$, $p <.001$). The borders seem to negatively affect how and where the users perform a gesture. By visually inspecting the gesture charts we were able to verify that: 1) the border also works as a barrier as it becomes impossible to perform the contact the closest to the edge. People with larger fingers have their gestures far from the real border even if they touch the physical overlay; 2) people who start the gestures far from the border tend to seek the border which translates into an incorrect direction.

**Screen Areas.** Concerning the position within the devices, a two-way repeated-measures ANOVA showed a significant impact of *Area* (Middle or Edges) in *Automatic Direction*

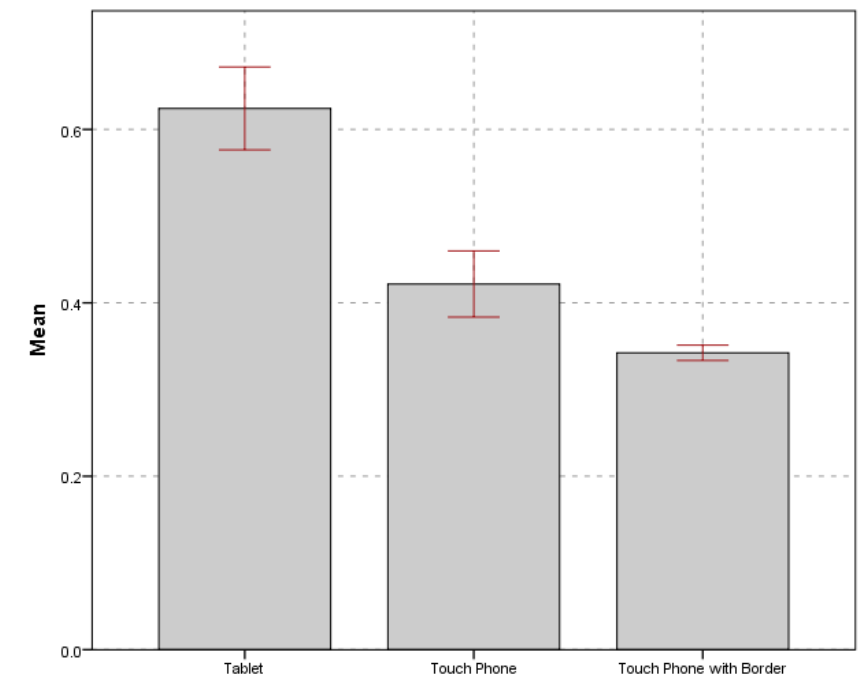

Figure 6.18: Average automatic area detection per Device. Error bars denote 95% confidence intervals.

*Detection* (F(1,34)=12.245, p <.005) revealing a consistent benefit of middle gestures over edge ones. This, along with the significant better results obtained with the borderless touch screen suggests that the borders makes it more difficult to comply with each gesture requirements. Also, a significant difference was found between Vertical and Horizontal gestures (F(1,34)=8.99, p <.01) showing that Vertical gestures are easier to comply with than Horizontal ones. In what concerns *Automatic Area Detection*, we also performed an analysis Vertical and Horizontal edge gestures. No significant differences were found.

### Reaction Times

**Tablet vs Touch Phone.**  No significant differences were found between the borderless settings in *Reaction Time*.

**Touch Phone vs Touch Phone with Border.**  A one-way repeated-measures ANOVA showed differences to be significant between the Touch Phone (M=3115.34, SD=960.02) and the Touch Phone with border (M=2740.27, SD=783.06) in *Reaction Time* (F(1,39)=8.213, p <.01). The presence of a border seems to improve the speed with which the participants start their gestures. This suggests an easier interaction.

**Screen Areas.** *Reaction Time* was significantly affected by *Area* (Middle or Edge) (F(1,34) = 81.063, p <.001) showing that Middle gestures tend to be started faster. However, a significant interaction was found between *Device* and *Area* showing that edge gestures tend to be faster in the Touch Phone with the Border than in the remaining reducing the difference to Middle gestures in this setting. No significant main effect was found between Vertical and Horizontal gestures on *Reaction Time* (F(1,34)=2.627, p=.114).

## Primitive Requirements

Gestures are recognized accordingly to a set of requirements considering the stroke performed by the user. Gesture recognizers normally use gesture speed, angle offset and travelled distance. In this section, we will explore how these metrics varied between devices and areas. The gestures performed by the participants were cleaned up before the analysis and this is particularly relevant here, i.e., the angle or the precision of a gesture would be deeply altered if an erroneous tap was performed before the intended gesture. To deal with this, our log analyser restricted the collection of values to the largest gesture performed for each trial. Further, as aforementioned, distance was normalized to millimetres to enable a fair comparison between different resolution devices.

**Tablet vs Touch Phone.** Precision was measured as the average deviation of all the points from the segment joining the first and last points of the gesture. *Precision* showed to be significantly affected by the type of device used (F(1,34)=15.251, p <.001). Gestures on a Touch Phone (M=1.32, SD=0.99) are significantly more precise than those on a Tablet (M=2.09, SD=1.39). Reasons may be the difficulty in maintaining a gesture direction with a larger gesture since people tended to perform larger gestures in the Tablet (F(1.269,43.148)=81.80, p <.001). On the other hand, the Angle Offset was not significantly difference between these scenarios although they varied (Figure 6.19). This means that although the gestures are different in their straightness they are not significantly different in their offset from the intended axis.

Gesture Velocity differed between the two devices. They are performed faster in the Tablet (M=0.16, SD=0.07) than in the Touch Phone (M=0.07, SD=0.03) (F(1,34)=70.8, p <.001).

**Touch Phone vs Touch Phone with Border.** Precision, Angle offset, Gesture Speed and Gesture Size showed no significant differences between these two settings in the overall. As such, gesture properties seem not to be affected by the presence of a tactile auxiliary cue. This contrasts with target acquisition where the borders showed a positive effect.

**Screen Areas.** Looking in detail at on-screen areas, Edge or Middle gestures showed no significant differences in what concerns *Precision*. Similarly, no differences were found for Vertical and Horizontal gestures. As to Angle offset, no significant effect of *Area* (Edges or Middle) was found although visual inspection reveals consistent differences in the mean values between areas. This statistical insignificance of the results is likely due to a large dispersion around the central tendency for the Tablet (M=8.35, SD=7.1), Touch

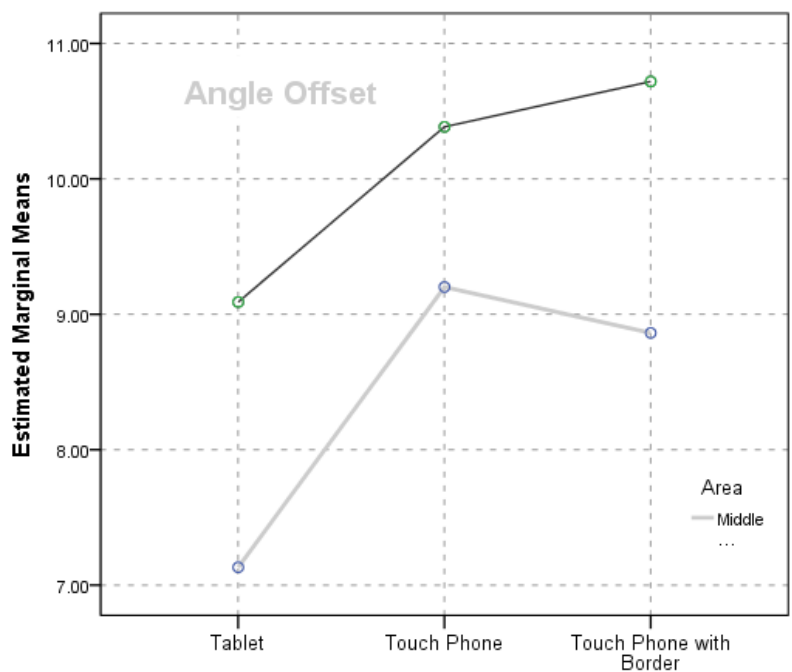

Figure 6.19: Gesture angle offset (estimated marginal means) per Device and per area (Edges and Middle).

| Gesture Angle Offset | Tablet | Touch Phone | Touch Phone with Border |
|:---:|:---:|:---:|:---:|
| Vertical | 6.08 (5.02) | 7.93 (6.23) | 7.66 (8.52) |
| Horizontal | 10.64 (11.77) | 14.51 (12.62) | 14.86 (11.92) |

Table 6.1: Vertical and Horizontal gestures angle offset per Device. Mean and standard deviation values for each setting are presented.

Phone (M=11.22, SD=8.15) and Touch Phone with Border (M=11.27, SD=8.05) settings. Conversely, a significant and consistent effect among devices was found between Vertical and Horizontal actions in the *Angle offset* (F(1,34)=16.036, p <.001). Table 6.1 presents the results for each setting.

*Gesture Size* also showed to be significantly different between Vertical and Horizontal gestures (F(1,34)=48.347, p <.001) with larger gestures being performed vertically. This is somehow expected due to the aspect ratio of the device along with the large gestures performed by some users. Tablet vertical gestures averaged 99.19 millimetres against 45.25 millimetres in the Touch Phone and 50.64 in the Touch Phone with Border. The difference between these two types of gestures is consistent between devices (no significant interaction). On the contrary, a significant main effect of *Area* (Edges or Middle) was found on *Gesture Size* (F(1,34)=4.17, p <.05) along with a significant interaction between *Device* and *Area* (F(1.464,49.764)=4.12, p <.05). Differences were significant only in the Tablet (p<.05). As to speed, *Gesture Velocity* was faster in the Middle than in the Edges of the screen (F(1,34)=20.98, p <.001). However, a significant interaction between *Device* and *Area* was also found (F(1.363,46.345)=4.97, p <.005) revealing that the aforementioned difference only existed for the borderless devices. Here, the border seems to support the user in performing the gesture more confidently as the users tended to be more careful in the other settings to maintain their trajectory within screen boundaries. *Gesture Velocity* was also faster in the Vertical gestures than in the Horizontal ones (F(1,34)=15.281, p <.001) but once again a significant interaction with *Device was found* (F(1.67,56.778)=14.1, p <.001): significant differences only occurred for the Touch Phone and the Touch Phone with Border (Figure 6.20).

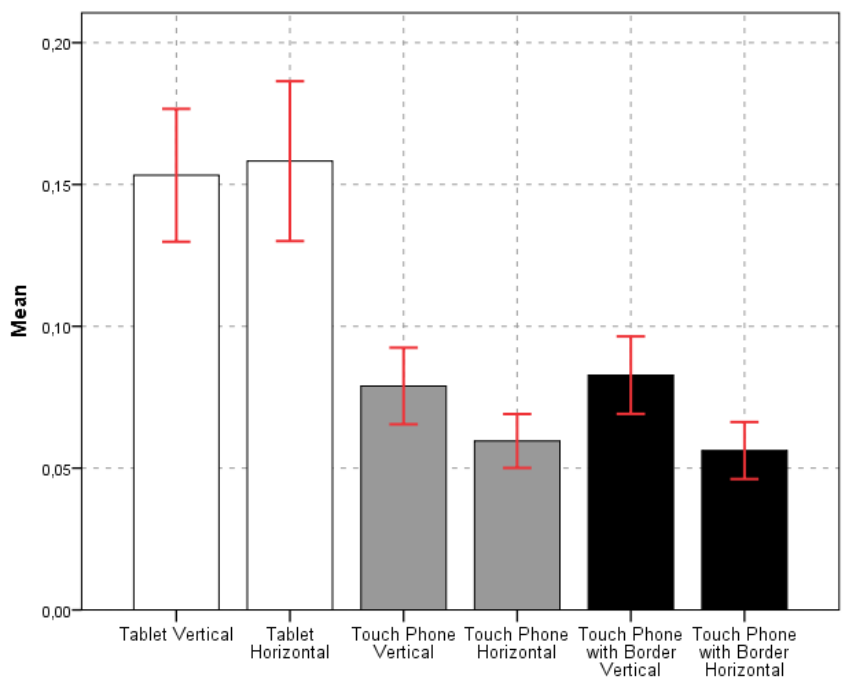

Figure 6.20: Average Gesture Velocity per Device and Type (Vertical and Horizontal).

## 6.3.3. Hand Postures, Strategies and Limitations

The protocol of the evaluation sessions comprised video and audio recording for later reference. The evaluation monitor was also responsible for, if possible, taking notes during the sessions both on observed behaviours and user comments. There were few cases where video inspection was needed as the evaluations were conducted mostly based on a *dialogue* between the participant and the devices. In this sense, the evaluation monitor was able to take notes in real time. By inspecting these notes and resolving any arising doubts with the footages we were able to summarize unpredicted and unmeasurable behaviours as follows:

**Hand Postures.** In this study, we sought to control all test variables (Device, Grid, Primitive, Target) to enable the fair comparison of setting and the later relationship with individual variables. Taking into account the latter, there was one aspect we decided not to control: the way of holding and interacting with the device. We chose to leave that choice to the users as their posture towards the device is already a result of their previous experience with technology and maybe from other individual attributes. In this line of thought, the results taken from this study reflect free usage of a mobile device, one that a person would make in an uncontrolled setting. Although this is likely to be reflected in the results we consider that its own effect is determined by other individual attributes. The prolonged training time contributed to diminish the effect of posture as the participants were able to get used to the device and maximize their performance and therefore chose their preferred posture with each setting.

Thirty (30) participants held the Touch Phone with the right hand, six (6) with the left hand, and one (1) user had to place the device on the table as one of his hands was severely impaired. From these 30, one used the middle finger to perform the actions, two used the thumb, one used both the index and middle finger, and one used the index, middle and the ring finger. This last participant employed an interesting strategy in the 12-area setting using the three fingers, one for each column, strategy that helped him with the localization on-screen. All the remaining one-hand *interactors* used solely the index finger.

Four (4) participants resorted to both hands to hold the device. Interactions with the screen were performed by using the thumb of the dominant hand. Holding the device with two hands was highly associated with a high mobile experience level (three our of four participants had maximum Mobile User Level) and was associated with a similar behaviour in their customary keypad phone usage. One participant was adverse to mobile phones and did not had experience with touch screens nor keypad phones. This participant chose to hold the device with the dominant hand and use the index finger from the non-dominant hand to interact. Both Touch Phone settings were similarly operated. Exception was made for participants who used both hands; in the Touch Phone with Border setting, they changed to using both thumbs similarly to what they do with their keypads (Figure 6.21) which was translated into better performance.

The Tablet was placed in the table in front of the test subject. All participants used the index finger. Here the difference did not come with the finger but more with the strategy used. Some participants chose to explore the position of the device during the training session and then, during the trials, employed a pure land-on strategy with few or no exploration. In several cases, this would result in a correct relative position between points but misplaced on-screen (Figure 6.22). This came as a surprise especially because this scheme was performed exclusively by early-blinders, the ones we could expect to be the most prone to tactile exploration. On the other hand, confirming this misconception were late-blinders who would use all possible tactile cues before jumping to the target. These participants employed a strategy from the outside of the device towards the inside always feeling the borders of the device first to help them with localization and then performing the mental measurements from there .

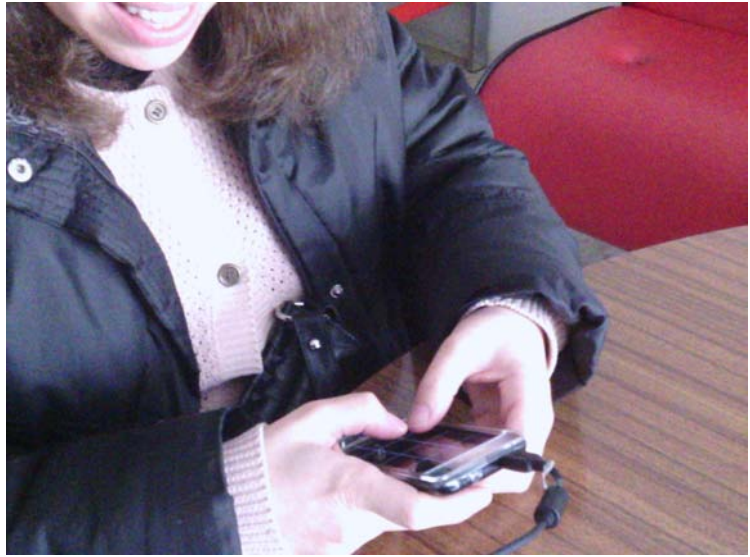

Figure 6.21: Mobile Keypad Proficient Participant operating the *Touch Phone with Border* with both thumbs.

**Flat screen.** The fast evolution of touch screen mobile phones we have witnessed in the last few years has brought us to the current panorama where touch devices are characterized by a completely flat surface. The borders of the screen have also vanished particularly since the advent of Apple touch devices (Figure 8.2). This is likely to be a problem to a non-visual interaction particularly for a non-experienced user.

In our trials, some problems were visible. Given the absence of feedback on the screen frontiers, several participants were holding the device and constantly performing touches with the hand that was holding the device. This issue vanished in the border-enriched setting as participants could safely rest their fingers and clearly identify a holding zone. The emergence of touch surfaces of different sizes and the promise of a more effective interaction give space for research to address these issues potentiating a more effective and natural interaction. As an example, [Findlater and Wobbrock, 2012] tackles undesired touches by understanding where the users rest their arms while inputting text on a touch surface and filtering the data accordingly to the user's previous interactions.

Another issue with flat screens comes with the habit to feel before making a selection. Even though participants were told that selections were performed when they first touched the screen, some were still, sometimes unconsciously so, performing exploration touches before making their attempt to select a target. This pattern of interaction supports the success of *painless exploration* approaches where the user first explores and only then confirms his intentions.

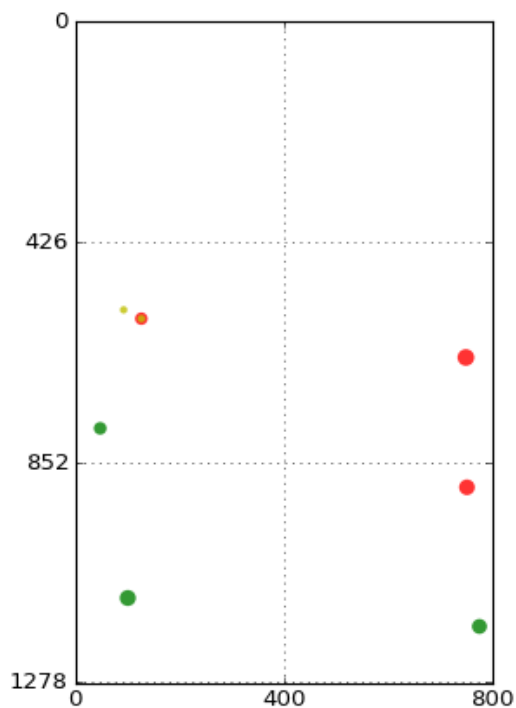

Figure 6.22: Point touch plot for the 6-area tapping setting of a single participant. Relative positioning of targets was achieved but misaligned with the targets on-screen.

**Reference Points.** As aforementioned, some participants explored the device in the training session and then tried to perform a target land-on with few or no localization before each trial. For others, each attempt was preceded by a phase of localization. Some of these subjects had previously found reference points in the device such as the charger and usb connectors, the volume buttons or even the microphone breach. They used them as reference points to ease target localization. This behaviour reinforces the benefits of tactile cues in the absence of visual feedback.

On the other hand, the physical border also presented its idiosyncrasies. In general it was beneficial but its use was sometimes exaggerated leading to errors, i.e., participants would seek the border after touching the screen and this created curved gestures in opposition to the desired straight paths and scrolls when a tap was recalled. These behaviours would be expected to decrease with experience and feedback.

**Language.** The study protocol was carefully prepared and validated with two blind people. Nonetheless, some details proved to be misaligned with the users' needs. We consider that they also reflect individual differences within the population. Particularly, we have identified issues with the language used in the automatic text-to-speech commands given to the participants. This happened with a small portion of the subjects but, given the individual focus of this research, they are worth of attention. In particular, some participants had trouble in the identification of some positioning and direction commands. This was particularly evident in the Gesturing setting where the Portuguese wording for top, bottom (*"No **Topo**", "Canto **inferior** direito", "Canto **superior** direito"*) was sometimes confused. In the training session we sought to clear all doubts but during the trials it was clear that some participants revived their doubts. Early-blinders face the challenge of living in a world where people tend to communicate with a strong basis on visual metaphors and based on the relationship between objects. Location and direction-based commands are experienced differently throughout their lives than among sighted subjects. It is worth mentioning that even some more educated participants sporadically showed limitations in the identification of directions and positions, or at least the wording used to describe them. Other studies have already shown that it is important to assure that the commands used will be comprehended seamlessly by all participants and that this is not a straightforward task when dealing with directional commands and blind people [Nicolau et al., 2009]. Once again, the validation performed before the trials did not capture any difficulty in comprehension of the commands. To circumvent this problem, we have annotated the videos of these users, identified misapprehended statements and have removed the entries from the data. The takeaway message from this issue is that these limitations should be considered when designing evaluations with blind people to assure that all data is valid. To this end, one can have an extended validation session which can be problematic due to the pervasive difficulties in finding people for the user studies or create a more verbose instruction set that presents guarantees to be perceived similarly by all.

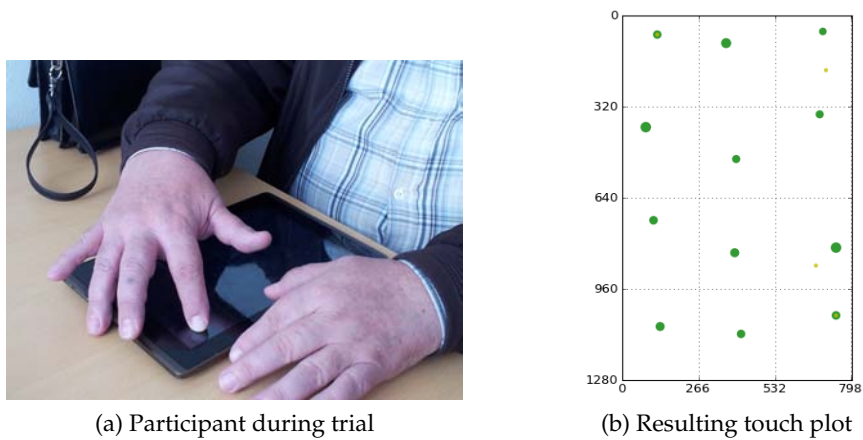

(a) Participant during trial          (b) Resulting touch plot

Figure 6.23: Participant during Tablet 12-area trials employing a device tactile exploration before touching the screen and resulting touch plot

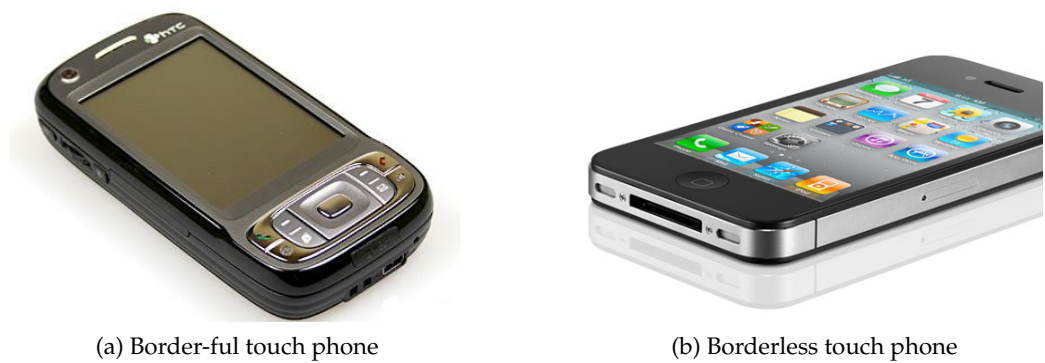

(a) Border-ful touch phone          (b) Borderless touch phone

Figure 6.24: The evolution of touch mobile phones: the disappearance of the lasting tactile cue

**Test sequence.** The 6-area layout showed to be more easily discriminated than the 12-area one, as expected. Indeed, this setting showed to present a low demand to the participants. Still, observation of the trials suggests that the 6-area results could be even more accurate as a great deal of the errors were caused by the test sequence. The training session before each setting aimed at reducing the interference between settings. However, this effect was still noticed when some participants performed the 12-area setting before the 6-area one. As it is a familiar setting, they sometimes continued to use it on the following trials forgetting they were testing another layout. This never happened the other way around which suggests that previous experience has an effect on how people behave during the studies. The localization of keys in a keypad becomes highly repetitive with the daily usage of a mobile phone as happens with inputting text with a keyboard. This process becomes highly mechanical giving place to what is called the *muscle memory*. In fact, experienced typist are likely to stumble if they are asked to think about the position of keys on the keyboard in opposition to just letting their fingers *dance* over the keyboard with a small consciously cognitive effort [Shusterman, 2011]. Comparatively, and even more than with sighted people, blind people have to memorize the keypad layout and face the pervasive challenge of knowing where a key and a letter are creating and reinforcing this mechanical process.

This interference between trial settings is also likely to be related with decreasing levels of attention during the trials due to its repetitive nature. However, no correlation of this behaviour with Attention was found.

Data was cleaned by removing trials that were clearly misunderstood (e.g., user stopped after three or four commands and stated *"ooops, I was still doing the last setting'*). Other possible similar mistakes but not clearly identified were not removed.

**Mobile keypad analogy.** In the previous item, we have identified a behaviour that is likely caused by the experience the participants have with their own mobile phones. This experience was also observed in the way people performed the primitives in the touch settings. Particularly, two participants performed double tapping by pressuring the screen as they would do with the keypad, without ever leaving contact of the surface. This issue brings one main take-away: we should take in consider-ation the previous experience people have with other devices as their behaviours and mental models are likely to be moulded accordingly.

## 6.3.4. Users' opinions

At the end of each setting, we asked the participants to rate the difficulty of selecting a target within that layout and device. We asked them to use the Tap primitive as a reference. Subjects rated difficulty using a 5-point Likert scale (1-very difficult...5-very easy). Figure 6.25 presents the median values for each setting along with the dispersion presented as the third and first quartile. Visual inspection of the chart suggests that interacting with all devices and settings was considered easy. However, some differences are observed. given the ordinal character of a Likert scale, we employed a non-parametric test to assess differences between device settings ([Pallant, 2007]). No significant differences were found in the 6-area setting. This suggests that the participants see this demand as easily surpassed independently from the device.

The results of a Friedman test revealed significant differences to occur between devices in the 12-area Tapping setting ($\chi^2$(2, n=34) = 14.265, p<.005). Post-hoc Wilcoxon signed rank tests with Bonferroni corrections (to control for Type 1 error) showed these differences to be significant between the Touch Phone and the Touch Phone with Border (z=-3.274, p<.05) and between the Touch Phone with Border and the Tablet (z=-3.061, p<.005). These results show that the participants feel that the physical border helps them perform more accurately in the 12-area setting in comparison to the borderless devices. On the other hand, device size was not seen as a determinant feature for success by the test subjects.

As to performing Gestures, a Friedman test revealed significant differences between devices ($\chi^2$(2, n=35) = 10.107, p<.01). These differences have shown to be significant between the Touch Phone and the Touch Phone with Border. Participants felt that it was easier to perform gestures with the aid of the physical border. On the other hand, device size did not have a significant effect on user opinion.

However, and contrary to the aforementioned lack of significance, regarding the size of device, some participants stated to have difficulties with the tablet. One participant said:

> **We, the blind, dominate two spaces: the fingertip and the hand palm. Thus, it is easier to use the touch phone which is smaller than the tablet despite the targets being larger**

About the tablet, another one told us **"It is such a large area. It is harder"** while another one said:

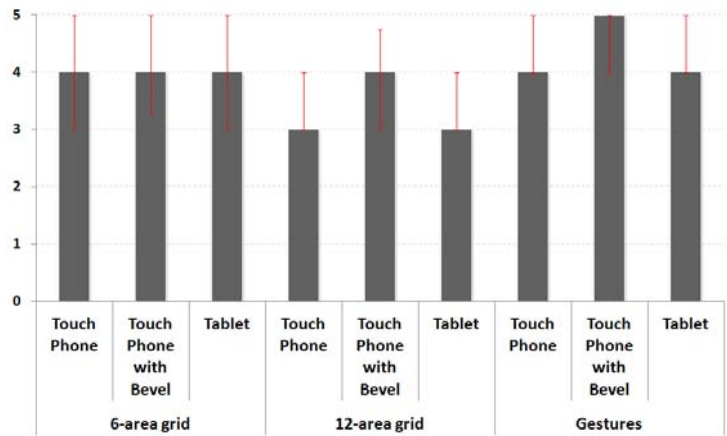

Figure 6.25: User ratings to the difficulty of performing taps in all Grid and Device settings. Bar chart presents median values while error bars present the dispersion as the first and third quartile.

> **This one [tablet] is larger but I feel lost...although the areas are very big, it gives a certain insecurity**.

On the other hand, several participants stated that the large size of the device made the targets **impossible to miss**. These contradictory opinions support the existence of different traits among the population.

As to physical borders, one participant stated that:

> **the border helps with the localization and it is also a safe place to rest the fingers**

while another said that

> **"the border makes our life easier but it also induces self-confidence and then we miss the targets"**.

However, to what concerns the border, almost all participants were positive about it and stated that it would help them in using a touch device.

In the overall, the participants did not think the trial was difficult nor do they find large differences in the settings presented. However, results show otherwise. Particular volunteers showed concerns about specific settings revealing preferences for different solutions.

## 6.3.5. Discussing Mobile Demands

Given the results presented in the previous sections, we are now in position to answer the research questions pertaining mobile demands:

**How different devices affect blind people performance?** Pertaining to devices, we varied size and the presence/absence of tactile screen limits (excluding the keypad baseline). As to size, most blind participants showed less confidence with the tablet due to the large space they had to handle with. This was reflected in reaction times and even in the timings relatively to the primitive itself. However, no differences were found in landing/lifting error rate showing that they can locate targets similarly in both smaller and larger devices if the space on screen is completely used. As to tactile cues, the effects were more evident. The physical border was positively rated by most participants and results revealed that it increased both their confidence and primitive performance. The borders are a valuable help in the localization of target. On the other hand, the borders have a negative effect when performing gestures as they affect the ability of a person to use the space on-screen.

**How users cope with different sized grid-based layouts?** The amount of targets on-screen decreases the size of each target and decreases the confidence in locating the desired one. Significant effects of grid size were found in most settings showing that this is true independently of physical cues and device size. Further studies are required to understand which grid sizes guarantee the best ratio between accuracy and number of on-screen elements. Nowadays, most mainstream mobile touch operating systems and applications present their users with grid-based icon layouts that are more dense than those tested in this study. The demands imposed to a blind user will likely be exponentiated with the proliferation of items on-screen. Further, although we have not focused on exploration beyond the first set of touches with the screen, difficulties in acquiring such targets will pervade the users' interaction, particularly before proficiency with the device is attained. This throws the question why is it called *painless exploration*.

**Are blind users capable of surpassing the demands imposed by touch interfaces?** Results showed that high error rates for all settings showing that the *raw* touch approach is not feasible for blind people. Painless exploration approaches diminish this issue as the user is able to explore the screen until the desired feedback is received but getting closer to a target is still relevant. Further, those that fail the most landing on the target are likely to be the ones that have difficulties in navigating in the touch area to find the desired item. Even more, performing a double tap or a long tap obeys to a set of restrictions that is independent from the correct localization of that target. Large standard deviations were found showing that these abilities vary among the population, fact that justifies the inclusion of some and exclusion of others. Inter-faces need to be sensitive to what a user is able or not able to do as the assumption that all can comply with the interface demands is proved erroneous.

## 6.4. Results: Capabilities and Demands

Designing interfaces for the *average human* contradicts excellence for everyone. Further, the concept itself is highly questionable. Who is the average human and how does each one of us stand in relationship to that imaginary set of abilities? In what respects people with disabilities, the problem is exponentially bigger and this is not only due to the disability in hand but mostly due to the lack of attention to wider ranges of abilities. Mobile user interfaces are designed with the *average human* in mind and follow guidelines and parameters for that model of a user. One problem is that this average is restricted to a small part of the population. Thus, assistive mobile technologies are mostly stereotypical prostheses to include a portion of the disabled population.

In sum, we have mobile user interfaces for disabled people that are no more than a stereotypical interface with a stereotypical aid over it. In section 6.3, we looked at how blind people as a whole fits in relationship to a set of selected mobile demands. Two results stand out: demands vary widely between settings and, for each demand, large dispersions are in place. In this section, we look at individual differences between the population to try to explain the variance within. This will enable us to have a better notion of which abilities should be considered when designing interfaces for blind people.

### 6.4.1. Comparing between groups

To surpass an interface demand, a minimum level of ability is required. Herein, we build on this concept and try to assess if people with different levels of ability have different performances. Given the multitude of individual attributes that may have an effect on user performance, a bivariate correlation between a single attribute and the performance in a specific setting is hard to attain. To be able to assess the impact of individual abilities, we resort to the clusters created in Chapter 5 to perform comparison between groups. Null hypotheses statistical testing is performed with mixed within-between ANOVA procedures. We are particularly interested in observing interactions between the within (Device) and between (attribute clusters) variables to understand if different demands are surpassed differently by different groups of users. Also, we will look for main effects of the between (Attribute clusters) variable to assess the overall impact of an individual attribute in user performance.

Results enable us to verify the relationships between individual characteristics and mo-bile interface settings (and underlying demands). Results are reported for Age-Related, Tactile and Cognitive groups.

| | Primitive | Grid = 6 | | Grid =12 | |
|---|---|---|---|---|---|
| | | Device * Age_Related | Age_Related | Device * Age_Related | Age_Related |
| Land-on | Tap | p=.825 | p=.754 | p=.181 | p=.877 |
| | Long | p=.294 | p=.507 | F(4,66)=2.09, p=.092 | p=.865 |
| | Double | p=.542 | p=.689 | p=.877 | p=.522 |
| Reaction Time | Tap | F(4,70)=2.069, p=.094 | p=.244 | p=.152 | p=.957 |
| | Long | F(4,74)=2.717, p<.05 | p=.103 | p=.196 | p=.824 |
| | Double | p=.163 | p=.404 | p=.895 | p=.591 |
| Duration | Tap | p=.355 | p=.910 | p=.805 | p=.325 |
| | Long | p=.829 | p=.643 | p=.257 | p=.816 |
| Interval | Double | p=.581 | p=.267 | p=.616 | p=.316 |
| | | Grid-less | | | |
| Precision | Gesture | F(4,64)=2.998, p<.05 | F(2,32)=4.647, p<.05 | | |
| Angle Offset | Gesture | p=.595 | p=.900 | | |
| Gesture Speed | Gesture | F(4,64)=2.511, p<.05 | F(2,32)=4.422, p<.05 | | |
| Gesture Size | Gesture | p=.647 | p=.540 | | |

Table 6.2: Mixed within-between analysis of variance pertaining the effect of both Age related attributes (Age and Blindness Onset epoch) on different metrics across the three different touch settings

## Age–related Differences

Previous work on individual differences puts a strong focus on Age either directly as an attribute to account for adaptation or indirectly as one causing shifts in people's abilities to perform. Concerning the blind population, a thorough discussion has been made throughout the years on the impact of blindness onset epoch. We assessed the impact of both these attributes by resorting to the compiled clusters of both variables.

Table 6.2 summarizes the results of the mixed within-between analysis of variance performed to assess the impact of Age-related (Age and Blindness Onset epoch) on different metrics across the three different touch settings. Pertaining the acquisition of targets, a significant interaction between Device and Age-related groups was found on the results to what concerns Reaction Time. This significant interaction occurred in the Long Press setting and it is depicted in Figure 6.26. Differences are encountered in the Tablet device setting while convergence between age-related groups is achieved for the remaining devices. This trend, although it has not reached statistical significance at the 0.05 level, was verified in the remaining primitives for the 6-area layout. Late-blind participants seem to perform an attempt faster than early-blind participants and older early-blinders seem to be slower than younger ones. This effect did not pervade all devices being only patent in the tablet setting. These results suggest that there is a demand in large devices and less explored layouts and that late-blinders are likely to be able to deal with such challenges. Early-blinders have a deeper understanding of the *small space*, particularly of what is on their hands than what goes outside of that space.

All other metrics seem to be unaffected by the Age and Blindness Onset of the participants. This was unexpected as most related work on individual differences places a strong focus on Age. However, we must consider that a upper limit of 65 years old may be low to collect differences related to Age. Our results suggest that, within a age range from 24 to 65 years old, the abilities to acquire a target seem to be similar (Landing on a target and Primitive Requirements).

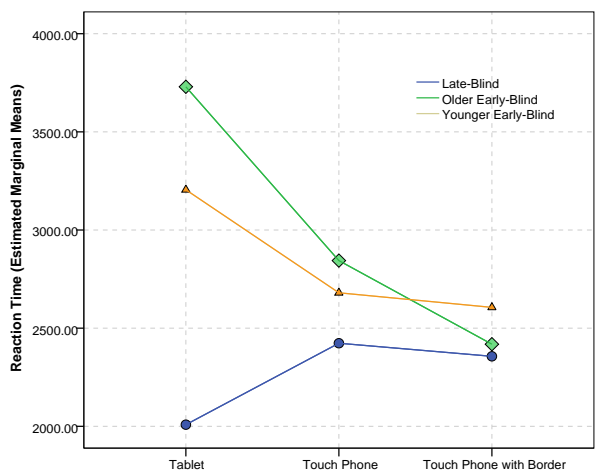

Figure 6.26: Reaction Time measures (Estimated Marginal Means) for all Devices and Age groups in the 6-area Long Press setting.

Conversely, the ability to perform a Gesture seems to vary between Age-related groups. A mixed within-between analysis of variance showed an interaction of Device and Age-related groups on Precision. Late-blinders seem, once again, to have a comparable performance on all devices. Younger early-blinders follow the same trend. Older early-blinders present large differences to the other groups: they perform less precisely in the Tablet and Touch Phone with Border settings. Further, a main effect of Age-related groups was found and post-hoc Tukey tests showed these differences to be significant only between late-blinders and Older Early-Blinders.

Delving into these results, we verified differences separately for Age and Blindness Onset clusters. A mixed within-between ANOVA revealed a significant interaction of Device and Age on Gesture Precision ($F(6,62)=3.508$, $p<.01$). Conversely to the tapping and pressing settings, gesturing seems to stress differences in the Tablet and Touch Phone with border scenarios. Post-hoc tests showed these differences to be significant between the older age group and the group of those with ages comprehended between 45 and 52 years old ($p <.05$), those with ages comprehended between 38 and 42 years old ($p <.1$, minor effect) and those in the 24 to 34 age group ($p <.1$, minor effect). As to Blindness Onset epoch, no significant interaction nor main effects of the independent variables were found on Gesture Precision.

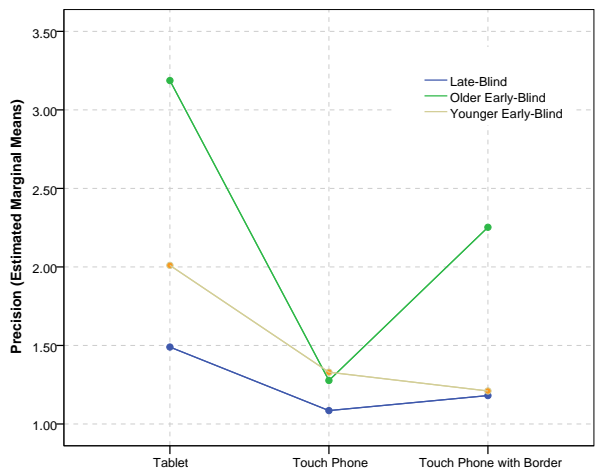

Figure 6.27: Precision (Estimated Marginal Means) for all Devices and Age groups in the Gesture setting.

A minor significant interaction was also found pertaining Gesture Speed. Differences in speed seem to be larger in the Tablet where Gesture Speed is higher for all age-related clusters. These differences decrease in the touch phone scenarios. A main effect of Age was also found and suggests that older people tend to perform faster gestures than the remaining along with Gesture Sizes not being smaller (Figure 6.28). Observation of the videos showed that older people tended to perform more abrupt gestures hence the aforementioned lack of precision and the higher velocities. Interestingly enough, no interaction or effects were found pertaining Gesture Size.

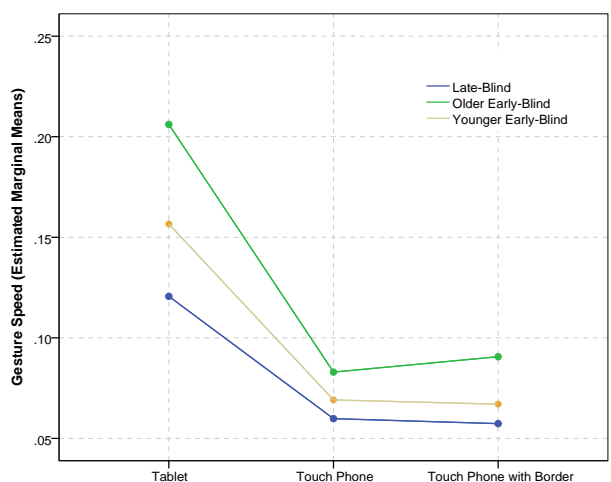

Figure 6.28: Gesture Speed (Estimated Marginal Means) for all Devices and Age groups in the Gesture setting.

*Overall, Age and Blindness Onset groups did not show differences in the different ways of acquiring targets. As to Gestures, differences were found between groups but these are only evident in the Tablet (Precision and Speed) and Border-enriched devices (Precision).*

*Gestures seem to be consistently performed in the border-less Touch Phone setting.*

## Tactile Abilities

We evaluated two components of spatial ability: tactile acuity and pressure sensitivity. In this section, we present the relationship between the demands exposed by different settings and the tactile abilities of the users all-together. When adequate, we present results pertaining the groups created for both tactile measures separately.

Table 6.3 presents the results of the mixed within-between analysis performed to assess the impact of Tactile Abilities on performance achieved on different settings. Significant interactions and main effects were found for Incorrect Land-on Error Rate and Long Press Duration of a target acquisition, and both Precision and Angle Offset of a Gesture.

| | Primitive | Grid = 6 | | Grid =12 | |
|---|---|---|---|---|---|
| | | Device * Tactile | Tactile | Device * Tactile | Tactile |
| Land-on | Tap | p=.245 | p=.519 | p=.268 | F(3,32)=3.8, p<.05 |
| | Long | p=.337 | p=.930 | F(6,64)=2.807, p<.05 | F(3,32)=3.426, p<.05 |
| | Double | F(6,64)=2.385, p<.05 | p=.369 | p=.672 | F(3,32)=4.638, p<.01 |
| Reaction Time | Tap | p=.317 | p=.522 | F(6,64)=2.047, p=.072 | p=.651 |
| | Long | p=.556 | p=.610 | p=.720 | p=.960 |
| | Double | p=.238 | p=.946 | p=.938 | p=.914 |
| Duration | Tap | p=.940 | p=.533 | p=.615 | p=.246 |
| | Long | F(6,72)=2.122, p=0.61 | p=.123 | F(6,64)=2.301, p<.05 | p=.349 |
| Interval | Double | p=.851 | p=.316 | p=.810 | p=.901 |
| | | Grid-less | | | |
| Precision | Gesture | p=.181 | F(3,31)=3.752, p<.05 | | |
| Angle Offset | Gesture | F(6,62)=4.414, p<.005 | F(3,31)=4.794, <.01 | | |
| Gesture Speed | Gesture | p=.267 | p=.333 | | |
| Gesture Size | Gesture | p=.461 | p=.293 | | |

Table 6.3: Mixed within-between analysis of variance pertaining the effect of both tactile measurements (Tactile Acuity and Pressure Sensitivity joint clusters) on different metrics across the three different touch settings

A main effect of Tactile Abilities on Incorrect Land-on Error Rate was found for all three primitives. This main effect was significant between those with *High Tactile Acuity and High Pressure Sensitivity* and the ones with *Low Tactile Acuity and Low Pressure Sensitivity*. By inspecting Figure 6.29 we can observe this main effect of Tactile Abilities. It is noticeable that people with low tactile abilities in the overall perform worse than the remaining. In the Long Press setting, an interaction between Device and Tactile Abilities was also found. In this setting it is particularly noticeable that in the Tablet scenario, people with High Pressure Sensitivity perform better that the remaining while Tactile Acuity seems to make a difference only when a low level of the former is in place. In the Touch Phone setting, the differences are clearer: the level of Tactile Acuity seems to make a large difference with Pressure Sensitivity complementing the shifts in performance.

Looking at the two attributes separately: Pressure Sensitivity (PS) seems to have a per-vasive and consistent main effect in the 12-area for the Tap ($F_{(1,34)}$=5.793, p<.05), Long ($F_{(1,34)}$=5.775, p<.05), and Double ($F_{(1,34)}$=11.323, p<.005) with low Pressure Sensitivity people always performing worse than those with High Pressure Sensitivity. No significant interaction was found revealing that there is a demand pertaining landing-on a target (independently from the device size and cues available) and that these demand is surpassed differently by people with different PS levels; significant interactions between Device and Tactile Acuity (TA) were found in the Tap and Long settings showing that tactile ability influences performance particularly in the Touch Phone setting. This suggests that the absence of screen limits (tactile cues) augments the gap between people with low and high tactile abilities. These differences are reduced if a border is applied showing improvements for people in both groups. As to the Tablet, the differences are also smaller between TA groups although we cannot guarantee that this is due to differences in size as shape and even small tactile cues (e.g., microphone) differ between both devices. Figure 6.30 presents the results of PS and TA groups for the Tap 12-area setting outlining the peak in error rate in the border-less touch phone setting, particularly pertaining the Low TA group. Participants with High TA have similar performance in all device settings.

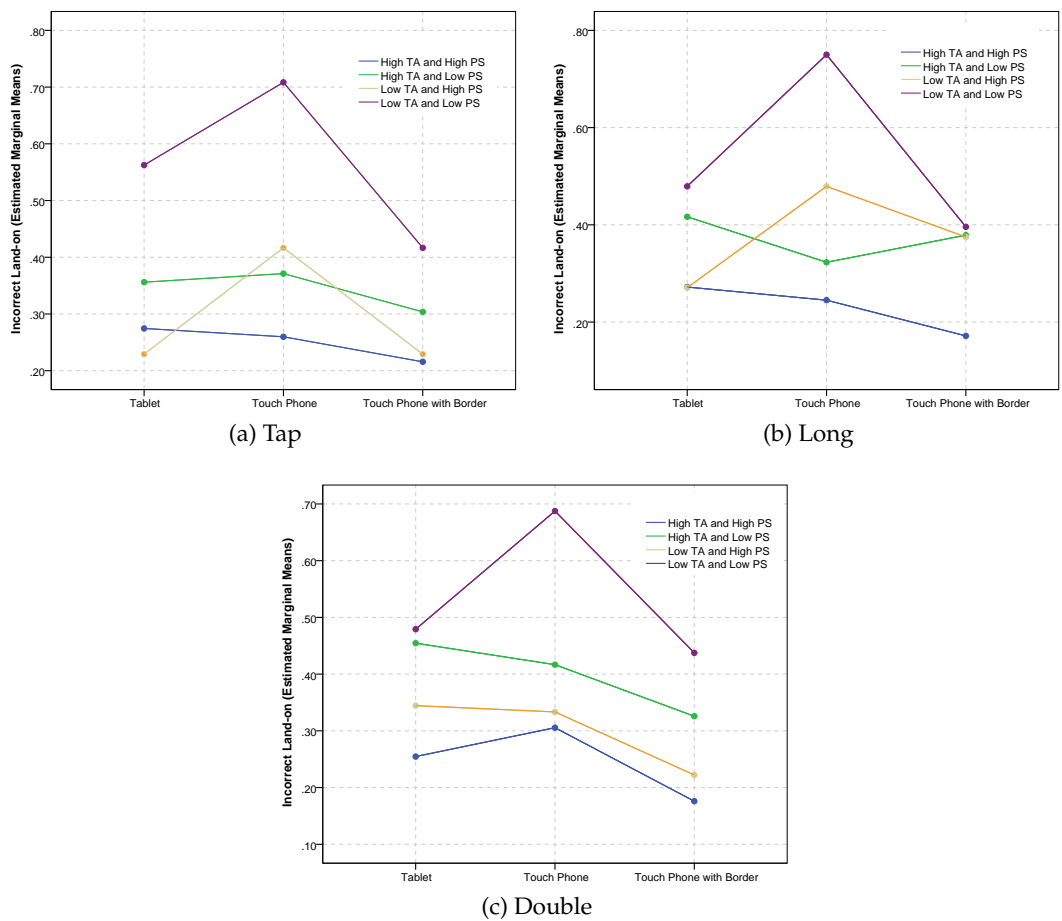

(a) Tap                                      (b) Long

(c) Double

Figure 6.29: Incorrect Land-On Error Rate (Estimated Marginal Means) for all Primitives, Devices and Tactile Ability groups (clustered Tactile Acuity and Pressure Sensitivity).

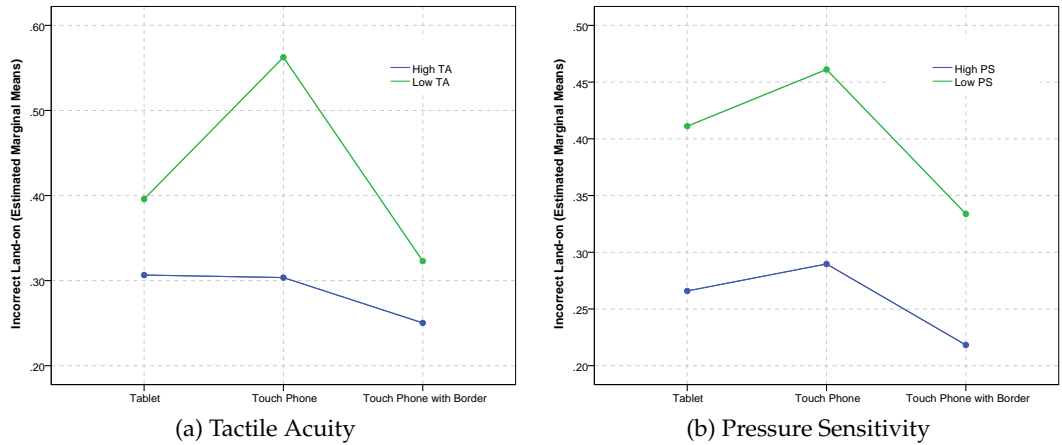

<div align="center">(a) Tactile Acuity                (b) Pressure Sensitivity</div>

Figure 6.30:  Incorrect Land-On Error Rate (Estimated Marginal Means) for Tap in all Devices in the 12-area Layout pertaining Tactile Acuity and Pressure Sensitivity clusters

Both Tactile Acuity and Pressure Sensitivity seem to affect the users' ability to surpass the demands imposed by the difference devices and primitives. We will now look in detail to both assessments separately to verify if we can have a finer relationship between each tactile attribute and the performance attained by the participants. Table 6.4 shows the correlations between both primitives and all 12-area settings (different devices and primitives). The correlation between pressure sensitivity and user performance pervades all settings: participants with higher PS measure (higher is worse) perform poorer. Stronger and significant correlations are found in the Touch Phone with Border setting (Tap and Long primitives) showing that these participants have benefits from extra tactile cues. Pressure Sensitivity also seems to play a role in distinguishing performance in the Tablet settings where, by turn, Tactile Acuity seems less relevant. We must take in consideration that the correlation between two variables is hardly damaged if an external variable impacts user performance. We already have indications that both Pressure Sensitivity and Tactile Acuity affect the ability to land-on targets and that this varies from device to device. Further analysis with multiple variables are required to assess the joint impact of both tactile measures along with others: we will come back to this later in this document.

In the 6-area layout, no significant main effects were found for the combined clusters nor the separate tactile abilities. As to interactions, a significant one was encountered just in the Double Tap setting (Figure 6.31) where it is visible that only the participants with low tactile abilities in its two components performed poorer in the border-less scenarios but their difficulties vanished with the addition of the border equalizing their results with the other tactile groups. Although not significant, the same tendency was visible in the remaining primitive settings. In comparison to the 12-area setting, we can state that the demand decreased and so did the requirement pertaining tactile abilities.

Significant differences were also found in the duration of a Long Press in the 6-area setting. People with low tactile abilities tended to perform longer selections than the re-maining (Figure 6.32). However, only a significant interaction was found revealing that this difference was only existent in the border-less touch phone setting. To understand this difference we had to look at the test videos and touch logs. One particular behaviour was noticed: people with low tactile abilities tended to perform slight scrolls on the surface probably to verify if they were touching the correct part of the device. They seemed more confident with a larger screen or the border. A minor significant interaction was also found in the 6-area setting with the same behaviours.

| | | | Touch Phone TAP 12-area | Touch Phone LONG 12-area | Touch Phone DOUBLE 12-area | Tablet TAP 12-area | Tablet LONG 12-area | Tablet DOUBLE 12-area | Touch Phone with Border TAP 12-area | Touch Phone with Border LONG 12-area | Touch Phone with Border DOUBLE 12-area |
|---|---|---|---|---|---|---|---|---|---|---|---|
| Spearman's rho | Tactile Acuity | rho | .449 | .258 | .073 | .000 | .147 | .121 | .239 | .216 | .307 |
| | | p | .004 | .108 | .666 | .999 | .385 | .452 | .132 | .175 | .054 |
| | | N | 39 | 40 | 37 | 38 | 37 | 41 | 41 | 41 | 41 |
| | Pressure Sensitivity | rho | .255 | .318 | .288 | .407 | .273 | .317 | .449 | .514 | .239 |
| | | p | .118 | .045 | .084 | .011 | .102 | .043 | .003 | .001 | .138 |
| | | N | 39 | 40 | 37 | 38 | 37 | 41 | 41 | 41 | 40 |

Table 6.4: Non-parametric correlations (Spearman's) of Tactile Acuity, Pressure Sensitivity and all 12-area settings. Table presents Spearman's rho, significance (p) and number of cases (N). Highlighted cells denote strength of the correlations accordingly to [Cohen, 1988]: small correlation - rho=.10 to .29; medium correlation rho =.30 to .49; large correlation rho=.50 to 1.0.

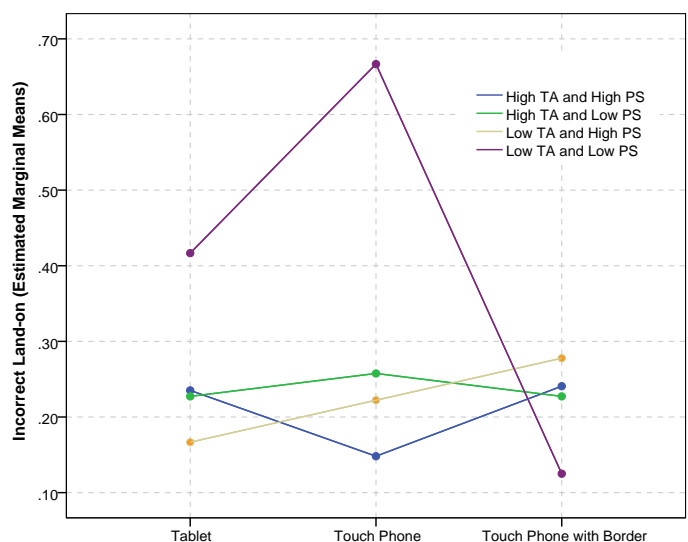

Figure 6.31: Incorrect Land-On Error Rate (Estimated Marginal Means) for Double Tapping in the 6-area layout on all Devices for all Tactile Ability groups (clustered Tactile Acuity and Pressure Sensitivity).

In what concerns Gestures, the Precision and Angle Offset also showed significant differences. Gesture Precision was statistically different between the clusters with Low and High Tactile Acuity. To verify this effect, we looked at each tactile component separately. Pressure Sensitivity showed no significant main effects or interactions in what regards

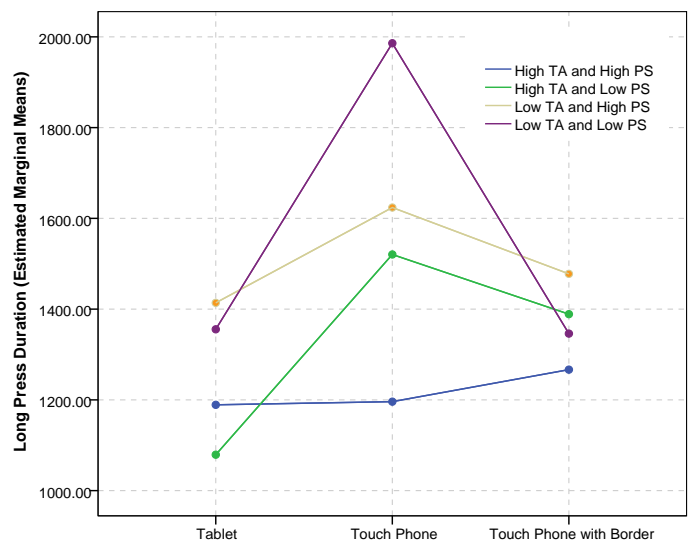

Figure 6.32: Long Press Duration (Estimated Marginal Means) for Devices for all Tactile Ability groups (clustered Tactile Acuity and Pressure Sensitivity).

Gesture Precision. On the other hand, a minor significant interaction of Device and TA was found on Gesture Precision ($F_{(2,66)}=2.499$, p=.09) along with a significant main effect of TA ($F_{(1,33)}=9.462$, p<.005). By inspecting Figure 6.33a we can verify that people with higher Tactile Acuity perform more precise gestures. We can also observe a slight difference in this effect between devices: people are more precise in the smaller device and the differences between clusters are drastically reduced in the Touch Phone setting.

Contrarily to target acquisition primitives, Gestures seem to be easier to perform in the smaller space and with no tactile aids as they seem to restrict the users' movement. This comes in line with the results obtained in Section 6.3.2 where middle gestures (away from the borders) were performed more effectively than those near the border. Gesture Angle Offset follows the same trend: both size of the device and the inclusion of a border highlight the relevance of tactile abilities, in particular, Tactile Acuity (Figure 6.33b) for which both a significant main effect ($F_{(1,33)}=6.269$, p<.05) and a significant interaction with Device ($F_{(2,66)}=7.212$, p<.005) were found. As to Pressure Sensitivity, a significant interaction was found ($F_{(2,66)}=3.283$, p<.05) once again showing comparable performances between groups in the Touch Phone setting and differences in the remaining settings.

*Tactile abilities showed to play a relevant role in a blind user interaction with different devices, primitives and their demands. Pressure sensitivity seems to have a pervasive relevance, particularly in the acquisition of targets and the complying with time-based primitive restrictions while Tactile Acuity has a stronger influence particularly in the presence of rich tactile cues. Tactile borders reduce the demands associated with target acquisition and approximate the performance of people with different tactile abilities. On the other hand, tactile borders make it harder to comply with gesture demands and they widen the gap between people with different tactile abilities.*

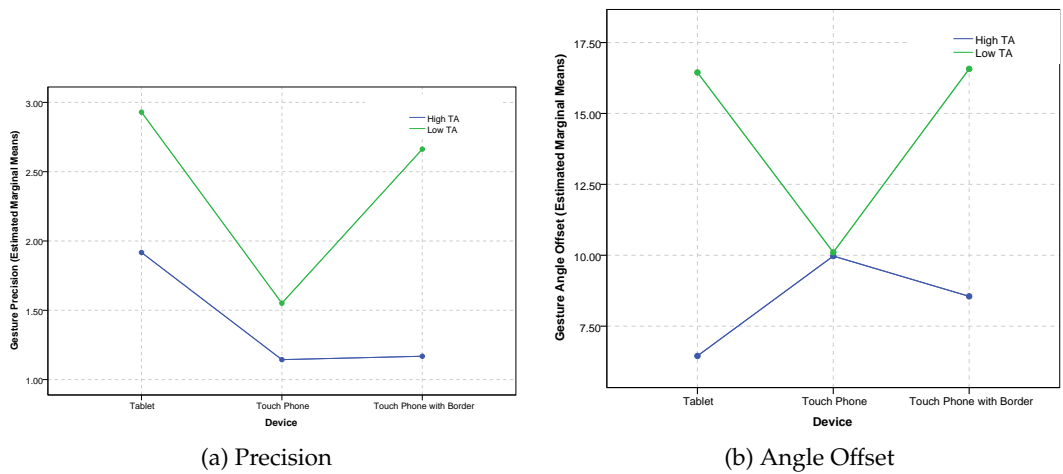

(a) Precision                                      (b) Angle Offset

Figure 6.33: Gesture quality metrics (Estimated Marginal Means) for all Devices and Tactile Acuity groups

## Cognitive Abilities

Digit span assesses a person's Attention and Memory while Spatial Ability concerns the mental representation one can create of physical space and relationships between objects and locations. We hypothesize that interface and device layouts along with many other time-based primitive requirements stress these abilities.

Table 6.5 presents the results of the mixed ANOVA procedures employed to verify the impact of the cognitive component in how a blind person is able to perform common mobile primitives in different devices and layouts.

| | | *Grid = 6* | | *Grid =12* | |
|---|---|---|---|---|---|
| | Primitive | Device * Cognitive | Cognitive | Device * Cognitive | Cognitive |
| Land-on | Tap | p=.128 | p=.129 | p=.433 | F(4,31)=4.304, p<.01 |
| | Long | p=.417 | p=.146 | p=.495 | F(4,31)=7.442, p<.001 |
| | Double | p=.531 | p=.146 | p=.988 | F(4,31)=3.65, p<.05 |
| Reaction Time | Tap | p=.363 | p=.154 | F(8,62)=1.791, p=.096 | p=.286 |
| | Long | p=.470 | p=.181 | p=.111 | p=.240 |
| | Double | F(8,62)=2.443, p<.05 | F(4,31)=2.793, p<.05 | p=.477 | p=.267 |
| Duration | Tap | p=.263 | F(4,33)=2.425, p=.068 | p=.248 | F(4,31)=3.45, p<.05 |
| | Long | F(8,70)=1.853, p=.082 | p=.193 | F(8,62)=2.232, p<.05 | F(4,31)=3.868, p<.05 |
| Interval | Double | p=.591 | p=.241 | p=.220 | p=.646 |
| | | *Grid-less* | | | |
| Precision | Gesture | p=.648 | p=.550 | | |
| Angle Offset | Gesture | p=.220 | F(4,30)=2.207, p=.092 | | |
| Gesture Speed | Gesture | p=.948 | p=.132 | | |
| Gesture Size | Gesture | p=.138 | p=.272 | | |

Table 6.5: Mixed within-between analysis of variance pertaining the effect of both cognitive measurements (Spatial Ability and Digit Span combined clusters) on different metrics across the three different touch settings

The challenge to acquire a target by landing on it showed to be cognitively demanding. A significant main effect of Cognition was verified on all 12-area Tap settings. Figure 6.34 presents how people from different cognitive clusters performed in the 12-area Tap setting for all devices. Post-hoc (Tukey) tests revealed differences between clusters 1 and both 3 and 4 to be significant. The same happens with Double Tapping. In the Long Press setting, differences are significant between cluster 1 and all the others. By looking separately at both cognitive abilities, we can observe that Digit Span presents significant differences in the 12-area setting for Tap (F(3,32)=5.443, p <.005), Long (F(3,32)=5.5, p<.005 and Double (F(3,32)=7.557, p<.005). Participants with higher levels of Attention and Memory perform better in all settings, particularly in the Touch Phone setting where they are able to maintain their performance; users with low Digit Span scores have overall lower performance and increase their error rate in this scenario. It is relevant to notice that the 6-area layout seems to be less discriminant to what Memory and Attention are concerned (no significant differences were found).

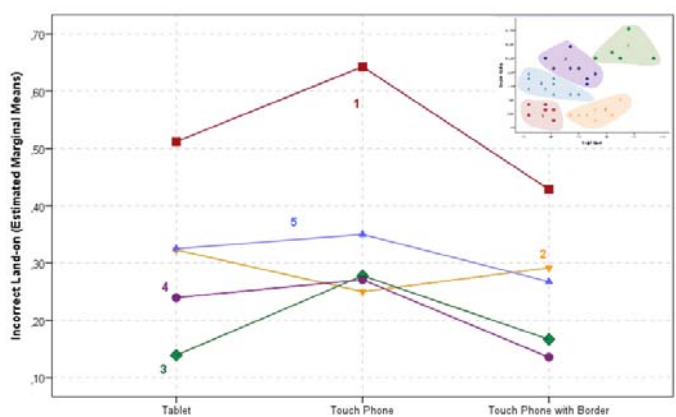

Figure 6.34: Incorrect Land-on Error Rate (Estimated Marginal Means) for Devices for all Cognitive Ability groups (clustered Spatial Ability and Digit Span).

As to Spatial Ability, a minor significant effect was found on Incorrect Land-on Error Rate in the 6-area Tapping setting (F(2,35)=2.747, p=.078). This effect is depicted in Figure 6.35 where besides the significant differences between the three clusters, changes between devices are also observed (albeit not significant). A significant main effect of SA was also found in the 6-area Double Tap setting. People with medium Spatial Ability levels seem to approach the performance of the higher level ones with a smaller device and with the aid of a tactile cue. Higher level SA participants do not have a benefit with the addition of a physical border neither do the lower level ones.

In the 12-area layout, differences between the higher level and medium level SA groups seem to disappear as the higher level ones increase error rate in this more demanding context and approximate the performance of medium SA group, particularly in the borderless settings. In the Touch Phone with Border setting, as the demand decreases, the higher level SA group seems to regain advantage and perform better than the remaining.

Reaction Time presented an isolated significant interaction and main effect of Cognition on the 6-area Double Tapping setting. To understand these results, further investigation

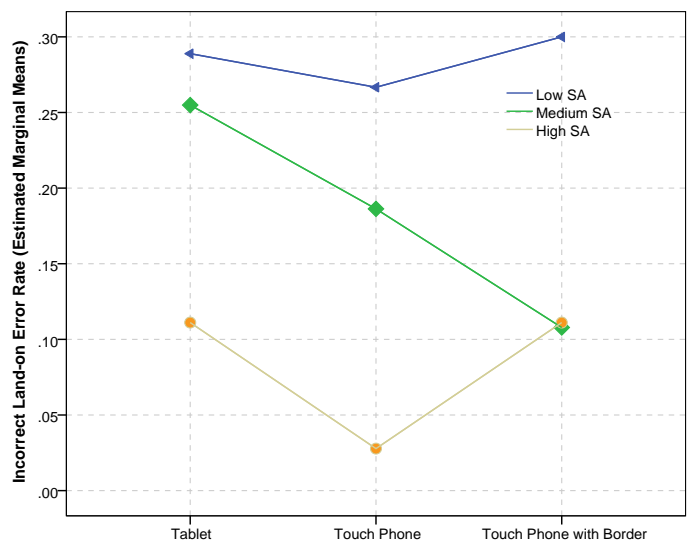

Figure 6.35: Incorrect Land-on Error Rate (Estimated Marginal Means) for all Devices and all Spatial Ability clusters in the 6-area Tap setting.

| | | Grid = 6 | | Grid =12 | |
|---|---|---|---|---|---|
| | Primitive | Device * SA | SA | Device * SA | SA |
| Reaction Time | Tap | p=.592 | p=.141 | F(4,66)=2.123, p=.088 | F(2,33)=2.554, p=.093 |
| | Long | p=.752 | F(2,37)=2.663, p=.083 | F(4,66)=2.238, p=.074 | F(2,33)=2.917, p=.068 |
| | Double | F(4,66)=2.475, p=.053 | F(2,33)=4.859, p<.05 | p=.208 | F(2,33)=2.544, p=.094 |

Table 6.6: Mixed within-between analysis of variance pertaining the effect of Spatial Ability on Reaction Time across the three different touch Device settings

of the separate abilities was required. Counter-intuitively, Digit Span scores, a measure of Attention, did not show a consistent impact on Reaction Times in any layout. Conversely, Spatial ability showed a pervasive tendency: participants from the lower spatial ability group tend to perform slower than the remaining. Table 6.6 presents the statistics on the effect of Spatial Ability on Reaction Time. Spatial Ability shows to have significant or minor significant effects on both the 6-area and 12-area layouts. In sum, one's spatial ability level seems to have influence on the confidence in aiming at a target and performing an action. People with lower spatial abilities are likely to try and circumvent their difficulties with tactile exploration and hence the larger times in Reaction Time. These differences were also larger in the Tablet setting: those who have to explore the device take more time in reaching and exploring the Tablet than in the Touch Phone settings (the phone is held by the participant and the contact with reference points is maintained easing exploration).

Cognitive attributes also revealed to have an impact on the time length of an action. A significant interaction between Device and Long Press Duration was found along with a main effect of the latter in the 12-area layout. Post-hoc Tukey multiple comparisons tests revealed differences to be significant between Cluster 1 and Clusters 2 and 3. Figure 6.36 shows that people from Cluster 1 (low spatial abilities and low digit span scores) tend to perform longer presses. The most cognitively fit perform faster actions but still above the theoretical thresholds currently applied. Once again, larger deviations are visible in the Touch Phone setting. This may be related with two aspects observed in some particular cases: 1) some participants were less confident in this setting and tended to reinforce their actions making a larger distinction between a tap and a long press; 2) in the attempt to verify their position, some participants would maintain the contact and measure their placement. Some of them slightly moved their finger on-screen to improve the certainty of their localization. A main effect of Cognition was also found for the Tap primitive although in this setting the differences are significant between cluster 1 and all the remaining. No interaction was found meaning that the clusters were consistent across all devices.

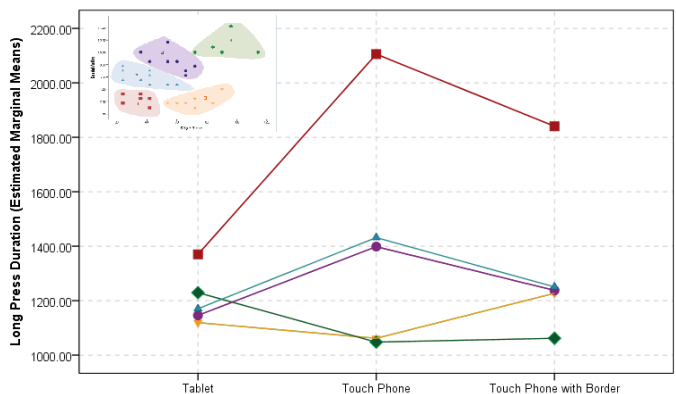

Figure 6.36: Long Press Duration (Estimated marginal means) for All devices and Cognitive Ability groups in the 12-area Tap setting.

In the 6-area layout, a minor effect was found for Tapping while a minor significant interaction was encountered for Long Press. A tendency to perform faster actions was verified in both settings (Tap and Long primitives) in comparison to the more demanding 12-area layout. These goes in line with the aforementioned: accordingly to the demand and uncertainty imposed people vary in the timings applied to perform an action. Furthermore, as demands vary from one device to another, so do the timings employed. Caution is required when designing within a device but also shared models between devices. It is not just the individual but how the individual behaves with the demands imposed.

Looking at the cognitive attributes separately, Digit span scores seem to define most of the encountered differences pertaining Duration of an action. A significant interaction between Device and Digit Span clusters was found in the 12-area Tapping setting (F(6,64)=2.339, p <.05) and in the Long Press Setting (F(6,64)=2.570, p <.05) with comparable trends as the above. As to Spatial Ability, a minor effect was found in the 6-areas Tapping setting and a minor significant interaction was found for the same layout in the Long Press setting. Minor significant effects were also found in the 12-area Tapping (p=.102) and Long Press (p=.116) setting, always with the same tendency: users with higher SA perform faster actions.

Concerning Gestures, no significant interactions nor main effects were found for the combined cognitive component. Inspection of the cognitive attributes separately revealed a significant interaction of Digit Span groups with Gesture Speed ($F_{(6,62)}=2.427$, $p<.05$) along with a significant main effect ($F_{(3,31)}=4.344$, $p<.05$), and a significant interaction with Gesture Size($F_{(6,62)}=3.086$, $p<.05$) along with a minor significant main effect ($F_{(3,31)}=2.363$, $p=0.9$). Although statistically significant, these differences between clusters showed to be practically insignificant between digit span clusters.

A significant main effect of Digit span scores was also found on Gesture Angle Offset ($F_{(3,31)}=3.683$, $p<.05$). In this case, the differences between the lower Digit Span scores group and the remaining is large. This difference is justified by gestures made in the wrong directions made by participants with lower Attention and Memory abilities.

*Cognitive abilities also showed to be paramount to an effective interaction with touch screen devices. Spatial ability plays a relevant role mostly in the acquisition of targets, particularly in the ability to land-on the correct area and on the time people take before issuing an action. This effect is pervasive to both layouts experimented. Attention and Memory, on the other hand, seems to be relevant in the more demanding layouts. Further, primitive requirements like Duration, Gesture Size and Speed, and Gesture Directions (Angle Offset) stress these particular Verbal IQ abilities.*

## 6.4.2. Discussing Abilities and Demands

The preceding sections explored the differences within the blind population and the impact of those differences in the performance a blind person attains with different mobile device settings and their underlining demands. Building on the results reported, we answer the remaining research questions as follows:

**Are individual attributes related with differences in performance?** Relations between individual attributes and demands were found at several levels. More to it, given the assessment of different devices, layouts and primitives and their varying demands, it was possible to observe how people showing different levels of a particular ability stood in surpassing the posed demands. Results showed that the demands posed by a setting may or not be surpassed depending on the user's level of ability. Also, different settings stressed different abilities.

**Which individual attributes play a relevant role in a non-visual mobile touch context?** Impact of individual abilities was verified at all proposed levels: profile, tactile, cognitive and functional. This effect is not constant and in most cases depends on the demand imposed. A person can surpass a certain demand if her set of abilities is higher than the required levels of that demand. When the demand is too high, people from different ability-based groups tend to approximate their performance; when the demands are low, consistency is also achieved but this time at acceptable performance levels. In between, differences are found between groups depending on the ability-demand relationship.

**Which relationships exist between abilities and demands**  In the preceding sections we have looked at how demands imposed by different devices, layouts and primitives are surpassed by groups of people with different characteristics and levels of ability. The set of attributes assessed were already pre-selected taking into account the interviews presented in Chapter 4 and the correlations between them presented in Chapter 5. Table 6.7 presents a summary of the relationships found between individual attributes and metrics. Each marked cell shows that the metric on that row was significantly affected by at least a component of the Ability set represented by the columns.

As to Grid-based layouts, Landing on a target is affected by one's Education Level, Tactile and Cognitive Abilities; Reaction Time is affected by Age-Related attributes, Cognitive Abilities and Mobile User Level; Duration of an action is affected by Education Level, Tactile Abilities and Cognitive Abilities. The Interval between Taps was the only demand that had no significant results for any of the included individual attributes. Still, a minor significant effect of Pressure Sensitivity was found (F(1,34)=3.406, p=.074). User observation showed that several people had difficulties in complying with the requirement of tapping in a short term interval. The ability to do so is likely related with fine motility, particularly, dexterity. This was one limitation of our study: dexterity assessments were not included. Further studies are required to assess the impact of Dexterity in this ability-demand mapping, particularly, in interaction primitives that require repetitive or multi-finger movements.

As to Gesturing, Precision showed to be related with people's Age-Related attributes and Tactile abilities while Angle Offset showed to be impacted by Tactile and Cognitive Abilities. Gesture Speed showed to be related with Age-Related features and Cognitive Abilities while Gesture Size was found impacted by Education Level and Cognitive Abilities.

It is relevant to notice that these relations were made visible by revealing differences of people performance in different devices and settings which by turn are variable in demand. Not only differences were found between people but also between devices for the same users. This means that different people (represented as different sets of abilities) perform differently when confronted with different demands. To reinforce this relationship between abilities and demands, Figure 6.37 shows some of the charts presented before but this time including the baseline, a keypad-based device. As an example, landing-on a target (i.e., pressing a key on the keypad) is less demanding to a blind person than acquiring a target in a tactile clueless touch screen: results confirm this by showing that in the keypad-based setting, differences in individual attributes like tactile or cognitive ones no longer differentiate user performance. On the other hand, time-based restrictions like Reaction Time or Duration of an action (Figure 6.37c) continue to differ.

|  | Age-Related | Education | Tactile | Cognitive | MUL |
|---|---|---|---|---|---|
| *Land-on* |  | ■ | ■ | ■ | ■ |
| *Reaction Time* | ■ |  |  | ■ | ■ |
| *Duration* |  | ■ | ■ | ■ | ■ |
| *Interval* |  |  |  |  |  |
| ▓▓▓▓ | ▓▓▓▓ | ▓▓▓▓ | ▓▓▓▓ | ▓▓▓▓ | ▓▓▓▓ |
| *Precision* | ■ |  | ■ |  |  |
| *Angle Offset* |  |  | ■ | ■ |  |
| *Speed* |  |  |  | ■ |  |
| *Size* |  | ■ | ■ |  |  |

Table 6.7: Resume of mappings between individual attributes and interaction demands

## 6.5. Summary

In this chapter, we present a study performed with 41 blind people that aimed to assess mobile touch demands, both in grid-based layouts target acquisition as in gesture-based primitives, along with identifying which individual abilities were involved for each demand. To stress individual abilities, wide variations in demands were presented under the form of different devices (tablet, touch phone, touch phone with physical border), sizes (different sized grids) and primitives (tap, long press, double tapping, directional gesturing). Further, a baseline key acquisition trial was performed. In the overall, differences were found between device, primitive and size settings, showing that different demands lay around in settings that are currently treated equally by designers. Difficulties in performing the primitives presented (they are the basis for interaction with nowadays touch screen devices) showed that accessibility to mobile touch devices for blind people is still challenged, one that is only surpassed due to the users' own drive along with a pre-required *bag* of abilities. Also, and supporting our user-sensitive stance, dispersion of results in more demanding settings is pervasive throughout our studies. Relationships between demands and abilities were found at several levels. Age-related attributes, education level, tactile and cognitive abilities showed to be relevant to surpass

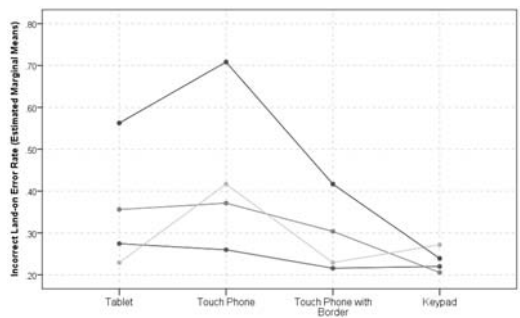

(a) Tactile X 12-Area Tap Land-on Error Rate

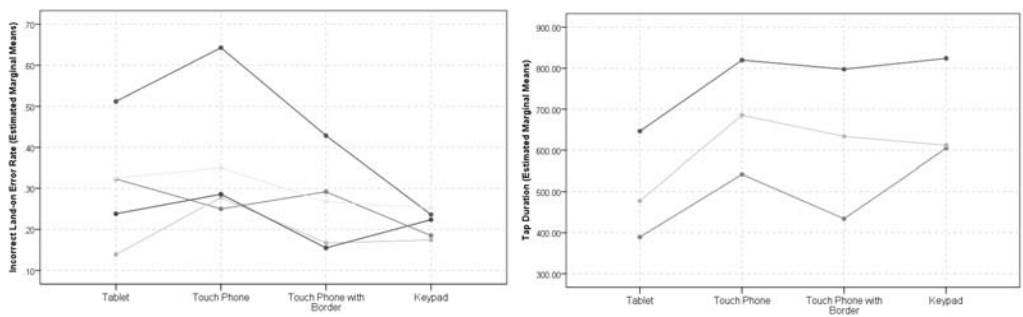

(b) Cognitive X 12-Area Tap Land-on Error Rate    (c) Spatial Ability X 12-Area Tap Duration

Figure 6.37: Land-on Error Rate and Duration for the 12-area Tap setting across all devices (including the baseline keypad)

mobile demands. Particularly, grid-based target acquisition performance showed to be mostly dependent on the user's education, tactile and cognitive abilities while complying with gestures showed to demanding to what concerns Age-related attributes along with tactile and cognitive abilities. Other relationships were found. Further studies are required to create a richer modelling of the relationship between abilities and demands. In the next chapter, we look at how these differences in ability influence the surpassing of mobile touch typing methods, which by turn are compounded by the demands presented here.

# 7

# Revisiting Text-Entry: the High-Level Impact of Individual Differences

Nowadays, it is still common to encounter blind people who are unaware of the possibility to use a touch screen mobile device. Although solutions exist, they still pose several challenges to blind users. Moreover, to foster adoption and enable improvements, these devices should be easy to use from the first contact. Recently, a number of efforts have been made to make these devices more accessible, particularly several text-entry methods have been proposed (Chapter 2). Although each one presents its own advantages and limitations, to our knowledge there are no comprehensive studies that relate them to blind users' individual capabilities. Picking up from the implications between individual abilities and device demands, there is no understanding of the demands imposed by each high-level technique (in this case, pertaining text-entry) and even less about the abilities it supposes. To corroborate our findings (Chapter 6), we performed a study to relate text-entry demands of four different text-entry methods with individual differences among blind people.

| Method | Touch | Gestures | Layout |
|---|---|---|---|
| Apple's VoiceOver | Multi-Touch | Scanning | Fixed |
| Yfantidis 2008 | Single-Touch | Directional | Adaptive |
| Guerreiro et al. 2008b (*NavTouch*) | Single-Touch | Directional | Adaptive |
| Bonner et al. 2010 (*No-Look Notes*) | Single-Touch | Scanning | Fixed |
| Frey et al. 2011 (*BrailleTouch*) | Multi-Touch | N/A | Fixed |
| Oliveira et al. 2011a (*BrailleType*) | Single-Touch | N/A | Fixed |
| Mascetti et al. 2011 (*TypeInBraille*) | Multi-Touch | N/A | Fixed |
| Azenkot et al. 2012 (*Perkinput*) | Multi-Touch | N/A | Adaptive |

Table 7.1: Summary of touch-based input methods for blind users.

## 7.1. Review of Touch Typing Approaches

Overall, there has been an effort to provide to blind and visually impaired users alternative touch-based text-entry methods both within and outside the research community. Table 7.1 illustrates available techniques and their main characteristics (already reviewed in Chapter 2). In fact, different interaction techniques are used, from single to multi-touch primitives, directional and scanning gestures, fixed and adaptive layouts. However, there is no knowledge of which methods are better for each individual user. Most approaches neglect the individual differences among blind people and how they relate to users' performance.

## 7.2. Research Goals

The main purpose of this study is to acknowledge the key role of individual abilities of a blind person in surpassing the demands imposed by different text-entry methods. By doing so and showing that different methods are suited for different people and by revealing which characteristics determine that difference we will be contributing to create a text-entry design space able to cover a wider range of users, thus fostering inclusion. In detail, we aim to understand which are the method's advantages and disadvantages and how individual differences are related with their demands. Further, we intend to identify which individual differences have greater impact in user performance in a first contact and how these differences are revealed in following interactions.

| Method | Layout | Size | Exploration | Selection |
|---|---|---|---|---|
| QWERTY | Fixed | Small | Scan | Split/Double-tap |
| MultiTap | Fixed | Medium | Scan | Split/Double-tap |
| NavTouch | Adaptive | - | Gesture | Split/Double-tap |
| BrailleType | Fixed | Large | Scan | Long Press and Double-tap |

Table 7.2: Text-entry methods characterization.

## 7.3. Text–Entry Methods

In this study, we sought for a set of text-entry methods that could highlight different user capabilities. Looking back to the solutions presented in the related work section (Chapter 2) and our own research (Chapter 3), we selected a set that includes fixed and adaptive layouts, different target sizes and number of on-screen keys, scanning and gesture approaches, and multiple selection mechanisms. This selection was also performed taking in consideration the ability/demand match we intended to corroborate. We then studied blind people using those methods and report their performance, highlighting some individual differences at sensory, cognitive and functional ability.

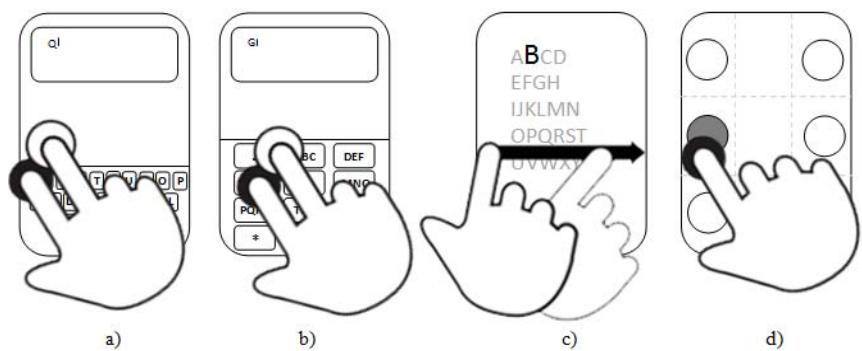

Figure 7.1: Text entry methods included in the evaluation. From left to right: QWERTY, MultiTap, NavTouch, and BrailleType

All text-entry methods, and their characteristics, used in this evaluation are presented in Figure 7.1 and described in Table 7.4 . All methods provide text-to-speech and audio feedback to the users' interactions. The QWERTY text-entry method is identical to Apple's VoiceOver and consists in the traditional computer keyboard layout with a screen reading software. Users can focus the desire key by touching it (painless exploration [1]), and enter the letter by split-tapping or double tapping anywhere. On the strong side, this method enables blind users to input text similarly to a sighted person with a simple screen reading approach. On the other hand, it features a large number of targets of small size, which can be difficult to find, particularly for those who are not proficient with the QWERTY layout. All the other methods were already described earlier in this document.

## 7.4. Procedure

The evaluation was set up with a within-subject design where all participants were evaluated with all four text-entry methods, one method per session, with one week recess between sessions (Figure 7.2). In all sessions, with the help of the experimenter, participants started by learning each method and interacting with it for 15 minutes. They were encouraged to ask questions and allay all doubts. If by the end of 15 minutes the participant was unable to write his name or a simple, common four-letter word, the evaluation was halted.

After the tutorial, participants were instructed to write a set of five sentences as fast and accurately as they could (no accentuation or punctuation). Each sentence comprised 5 words with an average size of 4.48 characters. These sentences were extracted from a written language corpus, and each one had a minimum correlation with language of 0.97. The sentence selection was managed by the application and randomly presented to the user to avoid order effects. The order in which the sessions (methods) were undertaken was also decided randomly to counteract order effects.

All focused and entered characters were registered by the application. The option to delete a character was locked. If a participant made a mistake or was unable to input a certain letter, she/he was told not to worry and simply carry on with the next character. It was made clear to all participants that we were testing the system and not their writing skills. Upon finishing each sentence, the device was handed to the experimenter to load the next random sentence and continue with the evaluation. The session ended with a brief subjective questionnaire on the text-entry method. All these steps were repeated in all sessions (methods).

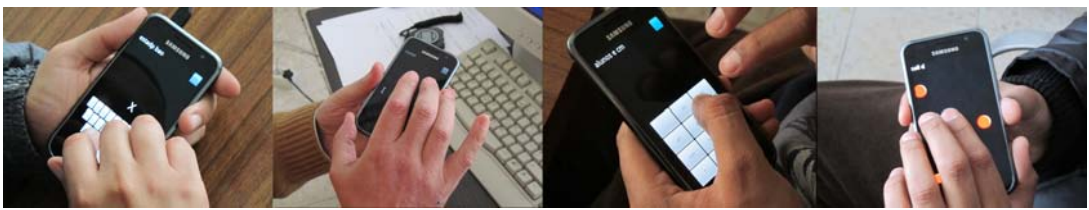

Figure 7.2: Participants testing the four different text-entry method. QWERTY (first), NavTouch (second), MultiTap (third), and BrailleType (fourth).

## 7.5. Apparatus

We used the Samsung Galaxy S touch screen device, which runs Android operating system. This device features a 4 inch capacitive touch screen with multi-touch support. No tactile upper and bottom boundaries were created. All text-entry methods were implemented as Android applications. All audio feedback was given using SVOX Classic TTS, Portuguese language pack. In BrailleType, a timeout of 800ms was used to accept a selection. An application to manage text-entry methods, user sessions and sentences required to type was also implemented. This application informed which sentence to type and logged all the participants' interactions (focus and entry), for later analysis.

## 7.6. Participants

Thirteen blind participants (light perception at most) were recruited from our participant pool (Chapter 5. The participant group was composed of 9 males and 4 females, with ages ranging from 24 to 62 (M=44). Regarding age of onset, the sample presents early-blind participants (e.g., P09) as well as others who have acquired blindness in a late stage in life (e.g., P11). All of the participants knew the Braille alphabet, although one user stated that he did not know how to write with a Perkins Braille typewriter and was not able to read due to poor tactile sensibility and lack of practice. This same user did not use a computer or sent text messages on a mobile phone. With the exception of another user, who was not able to write text on a mobile phone as well, all of the participants, with more or less difficulty, write text messages on their mobile phones and use the computer. Only one of the users had previous experience with mobile touch screen devices. The users' characterization is depicted in Table 7.3.

Exception made for Pressure Sensitivity and Braille Writing Speed, all variables in the table presented a normal distribution (according to the Shapiro-Wilk normality test).

## 7.7. Results

The goal of this study was to assess how different blind people, with different individual attributes, can benefit from a method over others. We start by analysing the different methods from the standpoint of user performance and preference. Then we focus on individual characteristics and how they diverge across methods, and with some case studies, thus giving us a better insight on why certain methods are better suited to a particular person. To conclude, we analyse how three participants behave in an extended version of the study comprehending three more sessions per method in order to draw insights on how the match ability-demand is surpassed with experience.

We report statistical significance with an $\alpha$ value set at 0.05. Nonetheless, when higher level statistical significance is achieved ($p<.01$, $p<.005$ or $p<.001$) we report results at that level, in agreement with the output tables of SPSS statistical analysis application and as it is common in social sciences [Pallant, 2007]. Besides this and given the exploratory scope of this research, we set the $\alpha$ value at .1 and report results at this level as minor significant. The statistics procedures presented in this chapter have their main tables, including p-values, presented in Annex A6.

| U | G | A(O) | PS | SA | VIQ | MP | PC | BR | BW |
|---|---|------|-----|------|-----|------|------|------|------|
| 1 | M | 26(10) | 3.61 | 1.8 | 105 | 15.8 | 45.8 | 49.4 | 26.4 |
| 2 | M | 32(15) | 2.83 | 10.0 | 111 | 11.9 | 44.6 | 21.3 | 13.4 |
| 3 | F | 52(5) | 4.31 | 10.0 | 78 | 4.0 | 11.5 | 8.8 | 14.9 |
| 4 | F | 34(27) | 4.31 | 8.5 | 99 | 12.6 | 41.8 | 2.6 | 8.2 |
| 5 | M | 24(2) | 3.61 | 5.5 | 65 | 14.2 | 45.3 | 63.7 | 27.3 |
| 6 | M | 45(20) | 2.83 | 7.8 | 114 | 6.7 | 21.8 | 9.4 | 11.6 |
| 7 | M | 62(3) | 4.31 | 4.8 | 104 | 7.9 | 23.7 | 64.7 | 25.8 |
| 8 | F | 46(25) | 3.61 | 6.2 | 84 | 7.7 | 20.3 | 26.5 | 17.8 |
| 9 | M | 60(0) | 4.31 | 4.0 | 134 | 9.6 | 24.8 | 80.8 | 13.4 |
| 10 | M | 48(26) | 4.31 | 4.8 | 84 | 10.6 | 33.9 | 19.2 | 22.0 |
| 11 | M | 49(34) | 4.31 | 3.3 | 78 | N/A | N/A | N/A | N/A |
| 12 | F | 49(17) | 4.31 | 5.5 | 78 | 7.1 | 26.7 | 3.8 | 7.9 |
| 13 | M | 46(3) | 4.31 | 7.0 | 84 | N/A | 4.7 | 9.0 | 11.7 |

Table 7.3: Participant characterization.U[User]; G[Gender];A(O)[Age(Onset);PS[Pressure Sensitivity in Newton];SA[Spatial Ability];VIQ[Verbal IQ];MP[Mobile Phone in WPM]; PC[Computer in WPM];BR[Braille Reading in WPM];BW[Braille Writing in WPM]. The lower the PS, the better the tactile sensitivity. The opposite for SA and VIQ.

## 7.7.1. Methods

In this section we focus on the different text-entry methods through the analysis of the user performance in terms of speed and accuracy focusing on the differences revealed between participants. We also examine their preference, opinions and frustrations regarding the presented methods.

**Text-entry Speed.** To assess speed, the words per minute (WPM) text entry measure calculated as (transcribed text - 1) * (60 seconds / time in seconds) / (5 characters per word) was used [Mackenzie et al. 2007]. One participant, after 15 minutes in the practice session was still struggling with the QWERTY and the MultiTap methods, so he did not perform the test with these two methods (P07).

Figure 7.3 shows the users' WPM with the four methods. It is noticeable that the performance of the several users varies widely across the different methods. QWERTY and MultiTap present themselves as faster methods for most users but none of the methods presents a consistent trend, suggesting for individual differences to have a determinant impact (high standard deviations in all methods). Overall, QWERTY was the fastest method (M=2.1, SD=0.7) followed very closely by MultiTap (M=2.0, SD=0.48). BrailleType was the slowest of the methods (M=1.49, SD=0.43) with NavTouch being a little faster (M=1.72, SD=0.55). QWERTY presents a higher value of deviation between the users suggesting that it exposes the differences among the sample group. Given the normality of the data (according to the Shapiro-Wilk normality test) a one-way repeated measures analysis of variance was conducted to see if these differences were significant. There was a statistically significant difference of Method on Text-Entry Speed (F(3,171)=41.00, p<.001). A Bonferroni post-hoc comparison test indicated that QWERTY and MultiTap techniques were significantly faster than NavTouch and BrailleType. QWERTY did not differ significantly from MultiTap, but NavTouch was faster than BrailleType. Even though QWERTY and MultiTap require searching for a specific character or group of characters along the screen, they still proved to be faster as they offer a more direct mapping between input and desired output. Both NavTouch and BrailleType require multiple gestures and inputs to access a specific character, which resulted in slower performances. BrailleType, besides having multiple inputs per character, was hindered by the fact that it uses a timeout system, an aspect that contributed for making the method the worst in terms of speed.

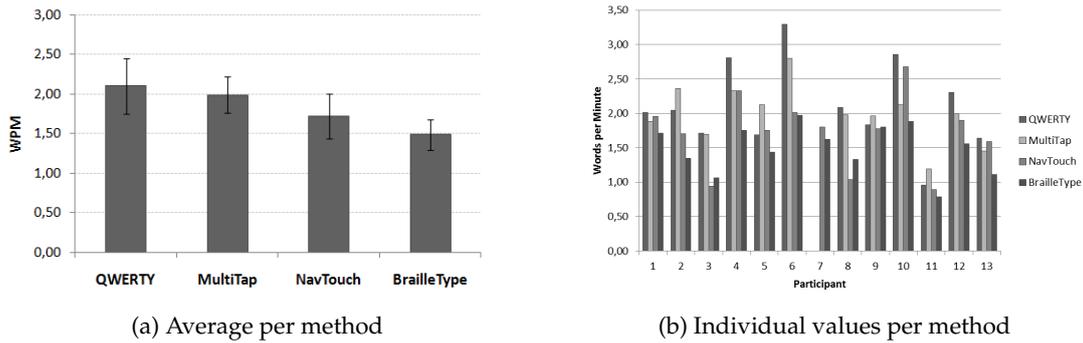

(a) Average per method                    (b) Individual values per method

Figure 7.3: Average text-entry speed (Words per minute) for all methods (left) and text-entry speed per method and individual (right). Error bars denote 95% confidence intervals.

**Text-entry Accuracy.** Accuracy was measured using the the MSD Error Rate, calculated as MSD (presentedText, transcribedText) / Max(—presentedText—, —transcribedText—) * 100. Figure 7.4 presents the MSD Error Rate of individual participants in the different methods. By inspecting the presented chart it is easily verified that some participants faced challenges with particular methods. P03 shows difficulty in dealing with a gesture-based approach (NavTouch) while participants P08 and P11 faced difficulties with a multi-tap one. This also shows that although, in average, there are better methods, in detail, those methods can be totally maladjusted to a particular someone. Since the data did not present a normal distribution, the Friedman test was used verify statistically significant differences among the methods. Results indicated that there was a statistically significant difference in overall Text-Entry Accuracy between the Methods ($\chi^2(3)$=16.265, p <.005). A Wilcoxon Signed Rank Test was used between each pair of methods to understand where these differences resided.

| Method | Easy to comprehend* | Easy to use* | Fast method | Would use |
|---|---|---|---|---|
| QWERTY | 4.0 (2) | 4.0 (2) | 4.0 (3) | 3.0 (3) |
| MultiTap | 4.0 (2) | 4.0 (1) | 3.5 (2) | 4.0 (3) |
| NavTouch | 5.0 (1) | 4.5 (2) | 3.0 (3) | 3.0 (2) |
| BrailleType | 5.0 (1) | 5.0 (1) | 3.0 (1) | 3.5 (1) |

Table 7.4: Questionnaire results for each method (Median, Inter-quartile range). '*' indicates statistical significance.

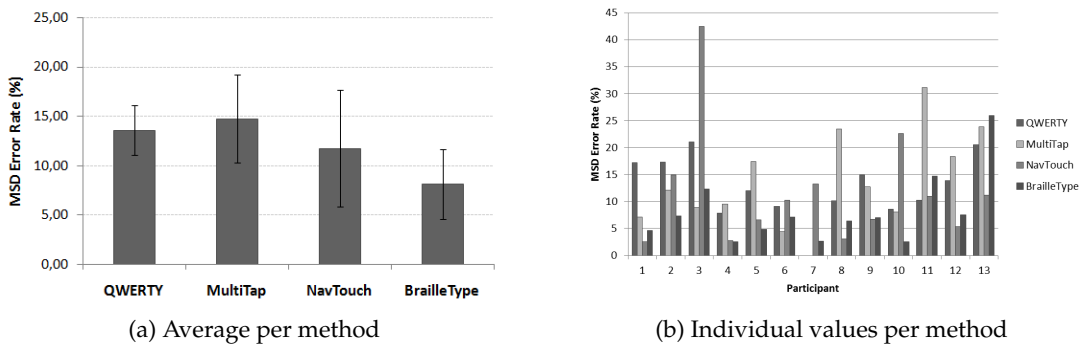

(a) Average per method                    (b) Individual values per method

Figure 7.4: Average text-entry speed (Words per minute) for all methods (left) and text-entry speed per method and individual (right). Error bars denote 95% confidence intervals.

BrailleType and NavTouch were significantly less error prone than both QWERTY and MultiTap. The fastest methods were also the most error prone, while BrailleType, the slowest method, was the one with the best results accuracy-wise.

**Users' Feedback.** User feedback was registered through a brief questionnaire at the end of each session. This questionnaire was composed of four statements to classify using a five-point Likert scale (1=strongly disagree, 5=strongly agree). The participants' ratings to the several methods are shown in Table 7.4. The Wilcoxon Signed Rank Test was used to assess significant differences.

Participants strongly agree that Navtouch is an easier method to understand than Multi-Tap (Z=-2.26, p=.024) and that BrailleType is also easier to understand than both Multi-Tap and QWERTY methods (Z=-2.21, p=.027 and Z=-2.058, p=.040). Users also strongly agree that NavTouch is easier to use than the QWERTY technique (Z=-1.98, p=.047) and that BrailleType is easier than both QWERTY and MultiTap (Z=-2.24, p=.025 and Z=-2.07, p=.039). BrailleType and NavTouch, the methods where users performed fewer mistakes, were also the slowest in terms of WPM, which was reflected in the questionnaire. In terms of preference, MultiTap was the elected followed by BrailleType, probably due to the resemblance to the traditional and familiar multi-tap and Braille methods. However, if we observe the Inter-quartile range values, we can see that there was not a consensus on most methods, in fact, only with BrailleType users seem to collectively agree that they would use the system.

The questionnaire was also composed of an open question about the difficulties faced and general opinion on the text-entry methods. With QWERTY, the main cause of errors and frustration were the proximity and small size of the targets. Most users found them to be a bit too tiny and close to each other, making it hard to select and split-tap the desired one, especially when they had large fingers. Since most users would grab the device with the left hand, and use the other to interact, searching with the index finger and split-tapping with the middle finger, targets near the right edge would also become hard to split-tap. Dexterity problems and some indecision on how to hold and interact with the mobile phone were visible on some users.

With MultiTap most errors occurred due to difficulties in multi-tapping, more specifically in finding the right timing to navigate between characters of a group. This was particularly apparent in the beginning, as some users would tend to not time well their taps, resulting in accepting undesired characters. Even though most are perfectly accustomed to multi-tap on their mobile keypads, some users had difficulty adapting this technique to a sensitive touch device. These adaptation difficulties were also apparent with the Nav-Touch method. Users would frequently touch/rest their fingers on the screen resulting in errors. Some users would also accidentally fail doing the directional gestures, tapping the screen instead of actually doing fling gestures. A concern of some users was the difficulty they found in keeping track of the current text, as they would tend to get confused or even forget the current state of the text as they navigated through the alphabet.

BrailleType, in spite of being the method where fewer errors were committed, they would still happen and their main cause were timeouts. Since focusing each target would read their cell number, but not actually select it until a pre-determined time elapsed, confident users, wanting to write faster, would forget to actually wait for the timeout to select the targets. This resulted in trying to accept incorrect Braille cells. It was evident that most users by the end of the last sentence wanted a shorter timeout, or possibly none whatsoever.

Besides these particular difficulties on each method, common problems such as figuring how to properly hold and interact with the device, as well as involuntary touches were frequent on every method. The general opinion on the methods was in line with what we expected. Users seemed to agree that NavTouch and BrailleType were simpler, easier and safer systems albeit slower (too slow for some participants). On the other hand, the QWERTY and MultiTap methods were perceived as slightly more complex, where errors are more frequent, but that allow writing at a faster pace. It is worth remembering that one participant was unable to use both the QWERTY and MultiTap methods, but had no problems using the other two methods.

## 7.7.2. Individual Differences

Now that we have observed how the different methods fared against each other, we will take a closer look at some individual traits to try to understand if they can explain the differences in the users' performance. In this section, we center our attention in three main groups of characteristics: age related, sensory, cognitive, and a more functional group based on the experience in mobile devices, computer and Braille.

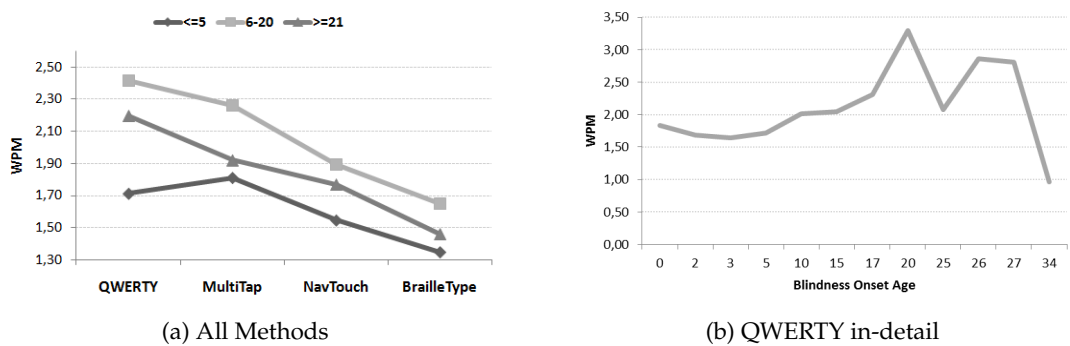

(a) All Methods                    (b) QWERTY in-detail

Figure 7.5: Age of blindness onset impact on text-entry speed (words per minute) for all methods - Left; Individual Age of blindness onset impact on QWERTY text-entry speed (words per minute) - Right.

**Age Related Differences.** In terms of WPM, younger users always performed better than older users, independently of the text-entry method used. This difference was statistically significant for QWERTY (F(1,58)=6.67, p<.05) and MultiTap (F(1,58)=23.12, p<.05) methods. It is interesting to note that although younger users were always faster, the difference between the two age groups is less pronounced on NavTouch and BrailleType methods. In terms of accuracy, younger users also performed better, committing fewer errors whatever the method tested. This difference, however, was only statistically significant for MultiTap method ($\chi^2$(1)=4.75, p<.05). Users, who were blind before the age of 6, had the slowest performance across all methods, as seen in Figure7.5. This difference was statistically significant for QWERTY (F(2,57)=6.096, p<.005) and MultiTap (F(2,57)=5.31, p<.01), with the post-hoc Tukey HSD multiple comparisons test revealing significant differences between the early blind and users who lost their sight between 6 and 20 years of age. Concerning QWERTY, where the differences were most visible, Figure 7.5b presents the relation between each blindness age of onset, in the sample group, and the text-entry speed achieved. One can notice that, until an onset age of 20, the trend is to improve performance. Thus, users with later onset ages are likely to improve performance. This becomes fuzzy after a certain age due to interference with the aforementioned impact of the users' age (older users perform worst). NavTouch and BrailleType methods seem to get smaller differences in performance on different age of onset groups, than the other two methods. The MSD Error Rate of the different groups was significantly different only for QWERTY ($\chi^2(2)$=8.5, p<.05), with users with the oldest age of onset committing fewer errors than the earlier blinds. Just like with the WPM metric, congenitally blind users or that acquired blindness at a very early stage of their lives had the worst performance across all methods.

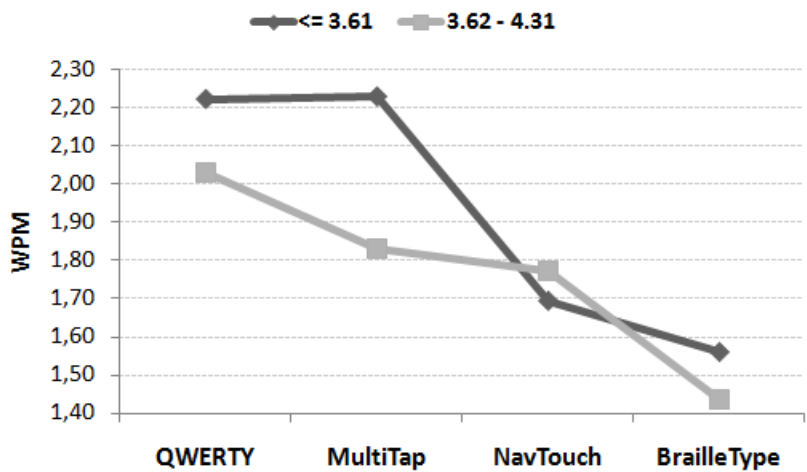

Figure 7.6: Pressure Sensitivity impact on text-entry speed (words per minute).

**Sensory and Cognitive Differences.** Figure 7.6 shows the differences of WPM, for users with different levels of pressure sensitivity. There was a significant statistical difference on the MultiTap method (F(1,58)=11.54, p<.01), as users with better pressure sensitivity performed far better. This was probably due to a combination of the very sensitive nature of the screen and the need for multiple touches of the multi-tap technique. No statistically significant results were found for the MSD Error Rate measure.

For QWERTY and MultiTap, two methods where exploration of the screen is vital, spatial ability was significant (F(2,57)=4.43, p<.05 and F(2,57)=9.95, p<.001, respectively). Participants with the best spatial ability values performed much better than the others, a gap non-existent on NavTouch and BrailleType methods (Figure 7.7). Users with better spatial ability also committed significantly fewer errors on MultiTap ($\chi^2(2)$=12.35, p<.005).

Users with a verbal IQ inferior to 85 were always slower independently of the method. This was significant across all methods (QWERTY: F(2,57)=4.33, p<.05; MultiTap: F(2,57)=7.08, p<.005; NavTouch: F(2,63)=3.66, p<.05; BrailleType: F(2,63)=6.89, p<.005). Users with smaller values of verbal IQ also committed significantly more errors on MultiTap ($\chi^2(2)$=12.56, p<.01) and NavTouch ($\chi^2(2)$=6.81, p<.05) methods. These two methods seem to have had a greater impact of short term memory and attention.

Functional Differences. There was not a statistical significant difference on the QWERTY method, in terms of speed and accuracy, on users with different levels of computer expe-rience. The same is applied to the MultiTap method when comparing users with different levels of mobile device experience. This result suggests that the knowledge acquired from button-based devices do not transfer to their touch counter-parts. However, experience in Braille was significant in terms of Braille reading experience, on the speed of the users with the BrailleType method (F(2,57)=3.60, p<.05). Faster users at reading Braille, and thus knowing extremely well the Braille alphabet, were faster than the others. This suggests that the demand imposed by BrailleType relates with a minimal knowledge of the alphabet.

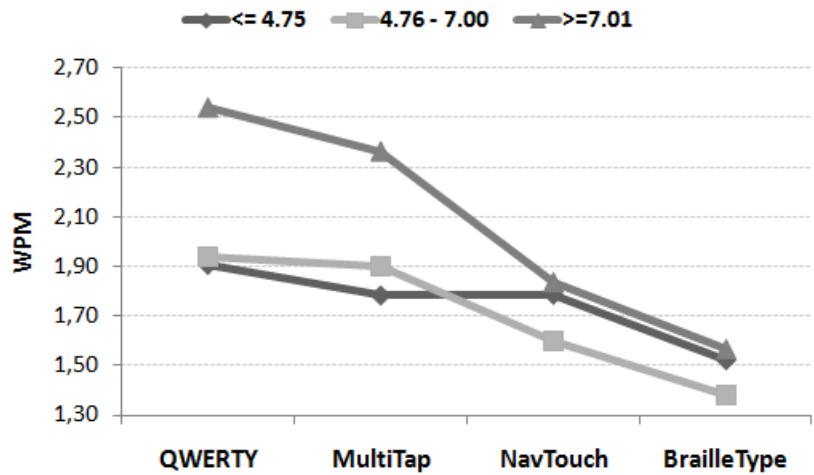

Figure 7.7: Spatial ability impact on text-entry speed (words per minute).

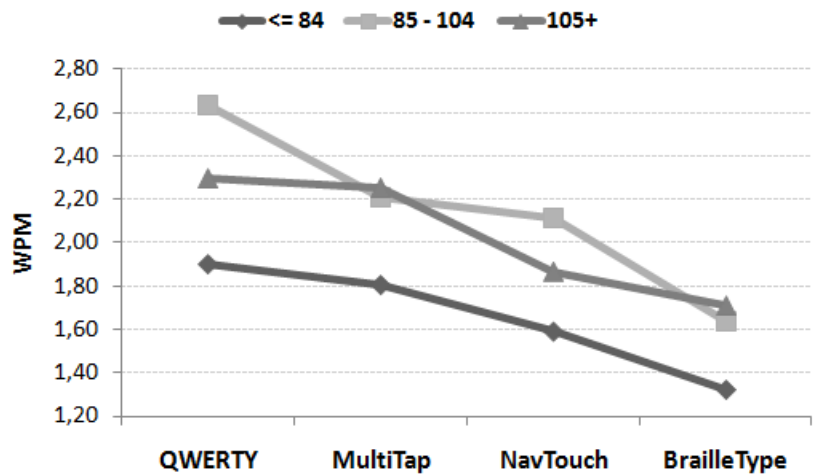

Figure 7.8: Attention and Memory impact on text-entry speed (words per minute).

### 7.7.3. Case Studies

To understand specific behaviors when performing text-entry tasks, in this section we highlight some key observations about specific participants. Starting by looking at the most critical user (Participant 7), the one who was unable to do the test with the QWERTY and MultiTap methods, even after all the practice session time and help from the experimenter. He was an older person, the oldest of the group of participants (62 years old), with an early age of onset (3 years old), bad pressure sensitivity (4.31) and although he had a good verbal IQ (104), he had poor spatial ability (4.75). As we have seen before, these characteristics were significantly related with inferior performances, especially on the two methods the user couldn't cope with, so their combined effect must have contributed for this inability. He was the only user who didn't perform the test in these two methods and, coincidently or not, he was the only user in our study that had this combination of traits. We could argue that maybe he is a Luddite or a technophobe, however the mobile and computer assessments made beforehand would state otherwise (7.9 and 23.7 WPM respectively). The user does have experience with technology, and yet his individual attributes seem to put him in a disadvantage, especially when facing certain methods.

The impact of individual differences can be observed in more cases. Participant P06 and P09 (refer to Figure 6 and 7) have clearly different outcomes both speed-wise and in terms of accuracy. Participant 9 is a congenital blind, with poor pressure sensitivity (4.3) and spatial ability (4.0). These characteristics certainly influenced his performance as he got much better results with NavTouch and BrailleType. In terms of WPM he was constant in all methods, an indication that he had more difficulty with QWERTY and MultiTap, as the other two are clearly slower methods. Although maintaining speed across methods, Participant 1 performed far more errors on the more demanding methods. Participant 6, however, is the opposite: has an older age of onset (20 years old) and much better tactile sensitivity (2.83) and spatial ability (8.0). This is reflected in the results, since he was faster with the more demanding methods, and made as much errors, if not less, with these than with the "safer" methods. The performance on MultiTap, a method highly demanding on spatial ability and pressure sensitivity is a good example of the impact of these individual characteristics, especially if we compare the performance of the two participants. These examples illustrate how important individual attributes are in regards to what methods are most accessible to a certain user.

### 7.7.4. Learning and the Interference of Experience

Given the paucity of experience, with touch screen devices (in our sample group), the aforementioned results are revealing in what concerns the first impact with both the methods and devices. We randomized order between methods but still each user was only evaluated with each method once. To be able to understand how blind users are able to surpass the demands imposed by each method with experience, we designed a small study with three blind users (randomly selected) from the original group. Further, in the light of time restrictions, the set of methods was reduced to two. We selected the method with which the users were able to write faster (QWERTY) and the method where they created sentences with better quality (BrailleType). Three participants wrote consecutive sentences for 45 minutes, for each of the methods, along six sessions, three per method. Each method (randomly chosen) was evaluated in a different week for each participant. BrailleType was significantly slower in the previous studies so, in this evaluation, we chose to reduce the selection time from 800 to 350 milliseconds. As in the previous study, words per minute showed a normal distribution, hence ANOVA results are presented. MSD Error Rate did not show a normal distribution and as such Friedman tests are reported.

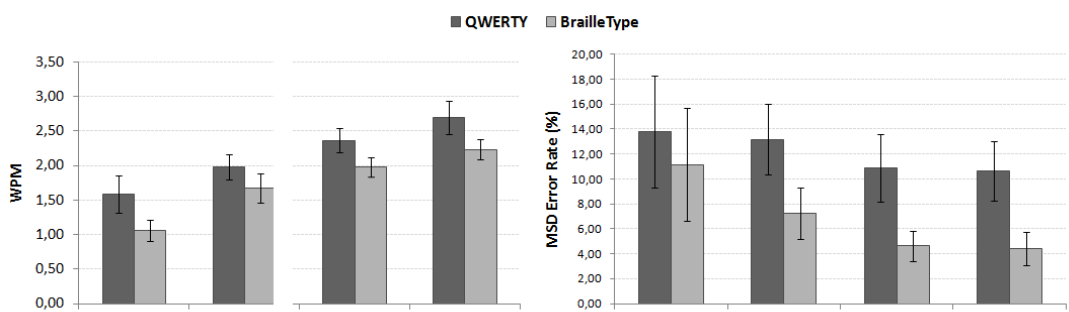

Figure 7.9: Average (all participants) text-entry speed (left) and accuracy (right) for the experimental session and the three following sessions. Error bars denote 95% confidence intervals.

Figure 7.9 presents the average speed and quality values for all three participants, obtained in the main experimental task where they were first confronted with the methods and the three additional session days described in this section. It is possible to observe that speed-wise, overall, participants improved across sessions with both methods. In the end, QWERTY presented a 69.2% increase in relation to the first contact with the method while BrailleType showed improvements of 110.4%. Despite of these improvements, QWERTY was still faster in all three sessions. In average, the difference between methods was reduced to 0.46 words per minute, which, although statistically significant, can be argued to be practically insignificant. Concerning sentence quality, QWERTY showed an overall decrease of 23.1% in MSD Error Rate but looking in detail it seems that it has stabilized from the 2nd to the 3rd day (2.4% decrease). With BrailleType, although we have decreased selection timeout and as such increased the risk of decreasing sentence quality, this method continued to present improvements in the sentence quality. BrailleType was less erroneous than QWERTY, with an overall decrease of 58.4% in MSD Error Rate.

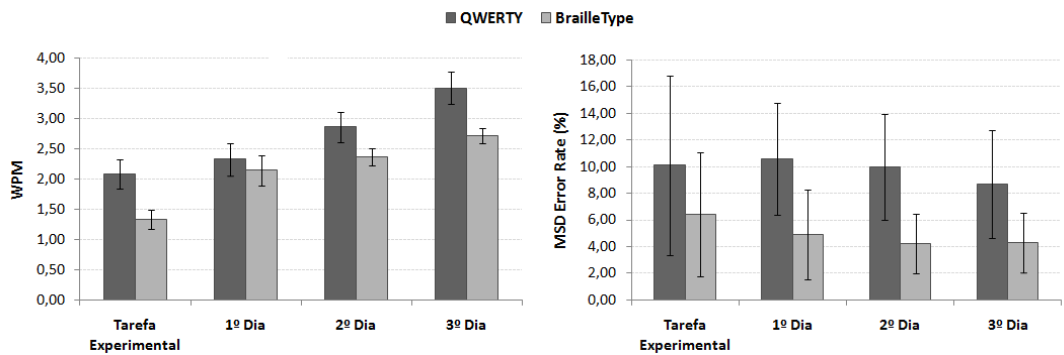

Figure 7.10: Average participant P08 text-entry speed (left) and accuracy (right) for the experimental session and the three following sessions. Error bars denote 95% confidence intervals.

Figure 7.10 depicts the results obtained by participant P08. We can observe that the selection time reduction in BrailleType was beneficial, with this participant getting very close in speed to QWERTY in the first session after the change. However, in the following sessions, the distance between methods started to increase again, with an advantage for QWERTY showing a trend for further improvements. Concerning the quality of transcribed sentences, the values for both methods showed to be relatively constant across sessions. Concerning the timeout at BrailleType, this means that it was indeed a barrier to what concerns speed but not to be seen as a demand as the reduction of time still enabled users to maintain their accuracy. This participant presents average values for almost all measured attributes (refer to Table 7.3). We consider this participant to show the natural average trend for both methods: QWERTY is a faster method and users that are able to cope with its demands will improve performance with experience; BrailleType is a slower method and although users can improve with experience, the trend is to also improve slower. This method eases first impact but gives little space for improvement as it forces the user to wait for the following actions, also reducing his control.

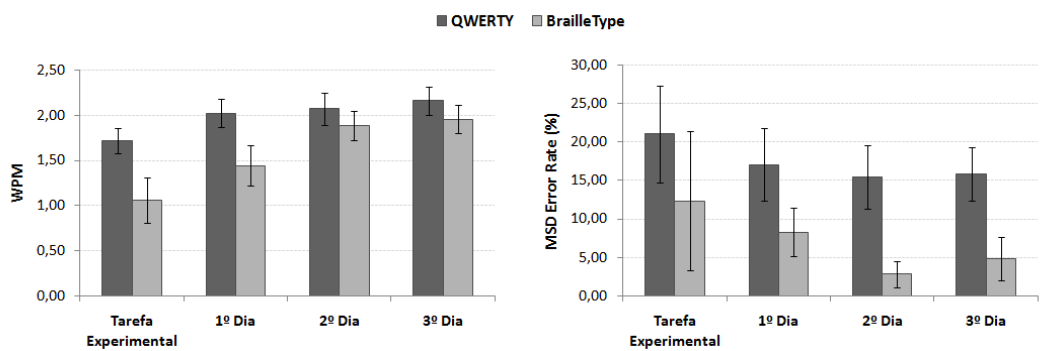

Figure 7.11: Average participant P03 text-entry speed (left) and accuracy (right) for the experimental session and the three following sessions. Error bars denote 95% confidence intervals.

As to participant P03 (Figure 7.11), which showed to be faster with QWERTY in the experimental session, the difference between methods vanished with the timeout reduction and experience. For this user, both these aspects revealed BrailleType as a suitable or even superior alternative. This participant showed few or no improvements across session with QWERTY, but improved his performance with BrailleType from session to session. Sentence quality remained stable with QWERTY but also presented significant improvements with BrailleType. This participant revealed low functional and tactile abilities but stood out for presenting high spatial ability. This may explain the results obtained with QWERTY (the quantity of targets stresses spatial ability) in the experimental session. However, her age along with low verbal IQ values may also explain a difficulty to build over this positive initial step and improve. This difficulty is reflected in other functional abilities like mobile, desktop and Braille text input. With a low value for BrailleType in the experimental session this participant has still space for improvement and given that the method presents lighter demands, we consider that BrailleType is likely to be a better option for her.

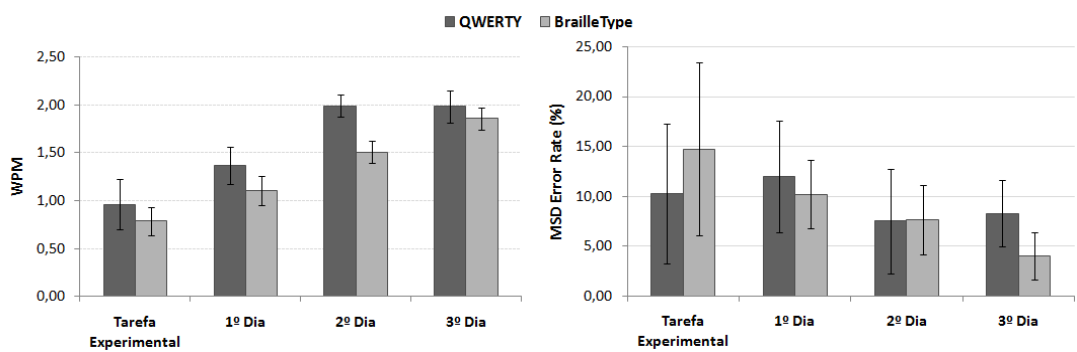

Figure 7.12: Average participant P11 text-entry speed (left) and accuracy (right) for the experimental session and the three following sessions. Error bars denote 95% confidence intervals.

Lastly, participant P11 also showed performance improvements across sessions. These improvements can be observed in Figure 7.12. They occur for both methods exception made for the 3rd day where improvements are only visible for BrailleType pairing its performance with QWERTY. MSD Error Rate decreases consistently for BrailleType across sessions while it presents and unpredictable behavior in QWERTY. P11 presents low cognitive abilities (Verbal IQ and Spatial Ability) which was (exception made for Day 2) translated in a similar performance with both methods. QWERTY is more demanding to what cognitive functions is related and when a user fails to surpass the demands imposed by the method, his performance is likely to be diminished. In this sense, it became similar to a method theoretically slower.

# 7.8. Discussion

The average values obtained for each method suggests that QWERTY (similar to Apple's VoiceOver) and MultiTap (the touch screen counterpart of the original keypad text-entry method) are faster input methods and that NavTouch (a directional approach) and BrailleType (a coding approach), less direct methods, provide a slower but less erroneous experience. In what follows, we organize our discussion beyond this straightforward conclusion:

**Individual differences play a preponderant role in mobile text-entry performance** and should be accounted for when designing interfaces for blind people. Results showed that particular differences were stressed by the methods. These differences were apparent in the overall (e.g., older users perform worst across all methods) but were particularly revealed in some methods suggesting that performance relies in a relation between the individual ability and the mobile demand. The large standard deviations presented in each method, particularly, QWERTY, strengthens this idea as users tend to diverge in performance when their abilities are stressed. Although users improve with experience, results suggest that the evolution is also impacted by the relation between the individual ability and the mobile demand which maintain valid after short-term sessions.

**Text-entry methods are demanding to spatial ability, pressure sensitivity and verbal IQ.** Each method is characterized by a set of demands. This depends mostly on the number of targets, size of those targets, and type of primitives. These demands can be overcome if a set of abilities is available. Conversely, particularly in the first approach with the method, if the ability set required to surpass the demands is not present, the user will fail to perform, leading to exclusion. Besides the differences showed with the users that were somehow able to finalize the evaluations, two of our participants were not able to use the methods even with tutoring. This explains the slow rhythm of adoption presented by a large part of the population. In detail, spatial ability, pressure sensitivity and verbal IQ play an important role in the blind user's ability to use and perform accurately with a touch screen and particularly with touch-based text-entry methods. Also, age and age of blindness onset seem to have an impact in users' overall abilities. Previous experience with mobile and other input devices seem to have a reduced impact, or none, in the users' skill to use a new text-entry method.

**Methods demands and individual abilities are related.** Results showed that text-entry interfaces with a large number of onscreen elements, like QWERTY and MultiTap, are more demanding to what concerns spatial ability. Users with low spatial skills are likely to perform poorly or even be unable to use those methods. On the other hand, NavTouch and MultiTap, are more demanding to what concerns to memory and attention, as the user has to keep track of the evolution within a selection. Also, results suggest that users with low pressure sensitivity have problems with repeated multi-touch inter-actions (e.g.,multi split-tapping). By understanding the users' abilities and limitations along with their consequence performance-wise, we are able to deploy methods able to cope with them, providing alternatives and fostering inclusion.

## 7.9. Summary

In this chapter, we present results of a comparative text-entry method (with QWERTY - similar to VoiceOver, MultiTap, NavTouch and BrailleType) evaluation showing that different methods pose different demands, even at a higher level task. How these demands are overcome largely depends on specific individual attributes. This clearly indicates that different designs suit different blind people. Further, these differences showed to sustain beyond the first impact with the device. This is relevant as if shows that demands are hard to surpass as different abilities need to be put together to achieve success. The difficulties felt by the target population in a preliminary experimentation phase may be prejudicial for adoption which in turn may reduce the person's opportunities. Accessible interfaces should leverage on a deeper understanding of interdependencies between abilities and demands to provide informed design diversity that accounts for individual differences.

# 8

## Conclusions

Mobile devices have become indispensable tools in our daily lives and are now used by everyone in several different situations. They are tools of productivity, leisure, culture and communication going way beyond their initial purpose. Their size, communication capacities and constant availability turns them into trustworthy tools that can support our needs everywhere, anywhere. More to it, such a pervasive tool is now an almost obligatory item. Everyone needs one, everyone wants one.

However, the small size of the devices along with the high number of applications and possibilities is also still a synonym of less usable interfaces than the ones found in a desktop setting where interfaces can be adapted to suit the user's needs and abilities. Furthermore, mobile devices have been in constant evolution concerning hardware and underlining capabilities. This has been a barrier to the increase in maturity of the mobile user interface, which can be considered to be still in its infancy. Only recently, a solid effort has started to be made to improve the interfaces available in mobile devices. What is true is that the end-user is still responsible to adapt himself to the device and its suppositions about the user's abilities and not the other way around, as it should. In general, mobile interfaces are still deployed to fit a common user model, one of a *average user*.

However, every human is different. This diversity has not been given enough attention in mobile user interface design decreasing the user's effectiveness or even hindering the ability to interact. Particularly, disabled target groups, characterized by specific individual differences, can greatly benefit from an effective mobile user interface. However, alternative user interfaces are likely to be misaligned with the users and their identities. These adapted mobile user interfaces are stereotypical disregarding that the relation between the user and the device depends on particular characteristics and not on a general idea. In general, mobile interaction has not evolved to meet the users' requirements.

We focus our attention on blind people. The absence of such an important and integrating sense as vision, in the presence of so demanding interfaces as are the ones in current mobile devices, justifies it. Individual differences among the blind can have significant impacts on how proficient they can become when operating different mobile devices. Although most accessible computing approaches acknowledge the role of user abilities in the inclusion of assistive solutions, the general tendency is to lower the demands imposed to serve the greatest amount of possible users. In sum, the demands imposed are decreased in the overall failing to respect the individuality of each user. On the other hand, current approaches associated with mainstream technologies assume a set of abilities which in turn are not guaranteed to be available in a great part of the population.

Current mobile devices, particularly touch screen devices, present novel opportunities; particularly, they can easily display different interfaces in the same surface (e.g. 12-key keypad, QWERTY keyboard) or adapt to users' preferences and capabilities. However, the knowledge on how to deploy such adaptations is still scarce as there is little understanding of which individual abilities are stressed in the interaction of blind people with mobile devices. This insufficiency translates in a similar lack of knowledge pertaining the impact of a difference in a device, primitive or layout characteristic. Only by assessing abilities and demands together, we can understand how they are related and what dimensions should be considered.

Also, taking into consideration the low costs involved in deploying a diverse set of solutions for mobile touch devices, we argue that the solution to cope with differences in people is by providing differences in opportunities. We consider that both users' abilities as well as interaction challenges should be equally explored to foster inclusive design. To do so, two steps are required: 1) understand which abilities are stressed and how they relate with demands; and 2) provide and map solutions to cover the wide existent set of abilities. Our approach to mobile touch accessibility presented in this dissertation resorts to diversity and ability-demand match as means for inclusion.

To accomplish our goals, we first performed preliminary studies with excluded blind people to assess if by lowering their interfaces demands, we could enable them to use tasks they were unable to use until this intervention. Also, we analysed how they behaved on the long run, particularly by observing and assessing the level of social inclusion they attained. Indeed, these people were only excluded due to their lack of ability and given the adequate methods they were eager to interact.

Upon the recognition that some people had levels of ability in commonly ignored dimen-sions and that by acknowledging those differences we were able to deploy user-sensitive inclusive solutions, we engaged in understanding these individual traits in higher detail. A study performed with a large set of users showed that diversity lingers within the population pertaining profile, tactile, cognitive and functional dimensions.

In what followed, we explored how these differences were translated in terms of performance with different device settings and their demands. The variations on both the user and interface sides enabled us to understand the impact that individual abilities have. Also, it enabled us to understand that slight changes in a setting may have a large influence in the population as a whole and for particular people (within particular levels of ability). Different individual abilities are stressed by different demands. In the overall, this knowledge creates the awareness for the need to create user-sensitive inclusive interfaces and that both individual and interface components present dimensions worth exploring to improve proficiency and accessibility.

The knowledge gathered in this dissertation places us in the position to provide implications for designing more inclusive user-sensitive mobile user interfaces, which are presented above. In what follows, we summarize the benefits and limitations of our approach and outline avenues for future work.

## 8.1. Implications for Designing User–Sensitive Mobile Interfaces

The approaches to provide accessibility to computerized devices share a common goal: to include people with different levels of ability. Methodologies vary from aiming at one universal solution to providing assistive technologies able to cover stereotypical gamuts within the population. Times have changed and a finer design space is possible towards inclusion. The cost of deploying a new application and make it available to the target users is now drastically reduced and, particularly, mobile applications can be made available online and installed without the need for intermediaries. Traditional one-size-fits-all rigid technologies give place to the paradigm envisioned by Alan Newell [Newell and Gregor, 2000]. It is time for *User-Sensitive Inclusive Design*. Further, such a versatile and flexible paradigm presents opportunities for all people as every once in a while we all are situationally impaired [Sears et al., 2003] and demand an alternative or slightly adapted interface.

The research methods applied in this thesis scrutinize individual abilities and mobile settings demands in order to provide a clear view of the dimensions that should be considered when designing for inclusion of blind people in a mobile context. This knowledge is paramount to maximize each individual chances to succeed. At this point, we envision six implications of our research results for the design of more inclusive mobile interfaces for blind people and, in the overall, a more inclusive design space.

## 8.1.1. In–depth Research Methodologies

A **first implication** comes with the research methodologies applied when striving to improve accessibility to technology. Blind people (and disabled people, in general) are themselves outliers in comparison to the remaining population. A blind person standing in the tail of an ability spectrum is an outlier to the blind population itself. By applying simple average comparison statistical procedures we are failing to include the ones that fall out of the central tendency. Those are likely to be the ones that require further attention and alternatives/adaptations to cope with the designed interfaces. This ability demand relationship and the inability of some to surpass interface demands has been long hidden in the *error bars of research*.

Herein, we applied research methodologies that go beyond that first impression of inclusion in different facets: in Chapter 3 we presented a long-term study with a reduced set of blind people. The criteria for participating in this study was just one: exclusion with the traditional approaches used by the remaining population. This long-term study gave us the starting point to start looking at individual abilities. A close and long-term accompaniment of the target users is paramount to outline differences in ability to surpass interface demands. In Chapter 6, we looked at mobile demands in the overall but without letting dispersion unexplained. Several differences were hidden behind this dispersion and a rich set of relationships between abilities and demands was unveiled. This knowledge gives space to possible adaptations and design alternatives that may include otherwise excluded individuals. This hypothesis was verified in Chapter 7 with different text-entry methods being adequate for different ranges of ability.

Our recommendation is that researchers may look at central tendencies to foster evolution and discover new research directions but, to foster inclusion, a second take to broaden the spectrum of people covered is required or at least to have a notion of who is being excluded. This recommendation comes in parallel with a trend verified in the Human-Computer Interaction research field: participant samples are regularly biased [Henze, 2012]. Efforts should be made to balance study samples in order to reveal the dispersion within the population and be able to assess the ability-demand match.

## 8.1.2. Manufacturers and Inclusive hardware design

A **second implication** comes with hardware design and the awareness of manufacturers to the impact they may have in the inclusion of disabled people. Before detailing this implication, two considerations need to be established: 1) keypad mobile devices, particularly 12-key models, which as shown by the results presented in Chapter 6 are generally less demanding to a blind person, tend to disappear and be replaced by touch enriched mobile devices; 2) blind people, as happens with other disabled groups, desire to use devices without a *disabled* connotation. A study by [Shinohara and Wobbrock, 2011] showed that, when possible, *accessibility should be built in mainstream devices*, and researchers should strive do design for social acceptability. Both these considerations indicate that touch screen devices are likely to be the devices used by blind people in the following years (they already are by a part of the population).

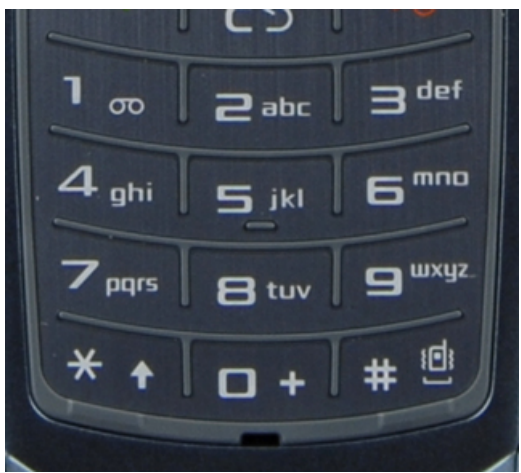
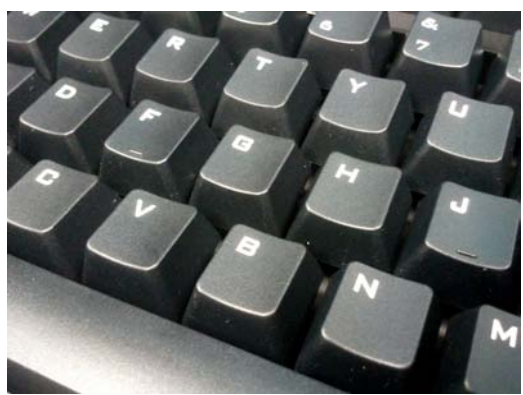

(a) Mobile keypad '5' label                                     (b) Desktop 'F' and 'J' labels

Figure 8.1: Keypad and Desktop tactile cues (physical labels)

Taking this in consideration, an effort to make these devices more accessible is paramount. In our mobile touch settings study (Chapter 6), we have shown that a simple screen border can make a difference in the performance levels blind people attain in the acquisition of on-screen targets. Further, we have also observed and reported that several participants would use all possible tactile cues to improve their spatial awareness. Looking back to desktop keyboards or 12-key mobile keypads, extra tactile cues are added to improve the search and acquisition of keys therein (Figure 8.1). This happens in devices that are already richer than touch screen devices to what tactile cues are concerned. It is important to retain that these labels are socially accepted as they pervade the majority of the devices. In fact, due to their ubiquitousness and subtleness, they are likely to not even be noticed. These socially accepted labels have not yet been applied in the mobile touch screen design space where they are likely to be more relevant. Further studies should focus on the impact of subtle tactile cues (labels, texture, among others) to inform on screen limits, device position or even icon/grid localization.

Subtleness can even be ignored as long as social acceptance is maintained. There is space for the design of assistive but still accepted and aesthetic aids. Figure 8.2 shows the border used in our studies along with a silicone commercially available case. The latter is not meant as an assistive aid, but, in fact, it is one (for those that are able to take advantage

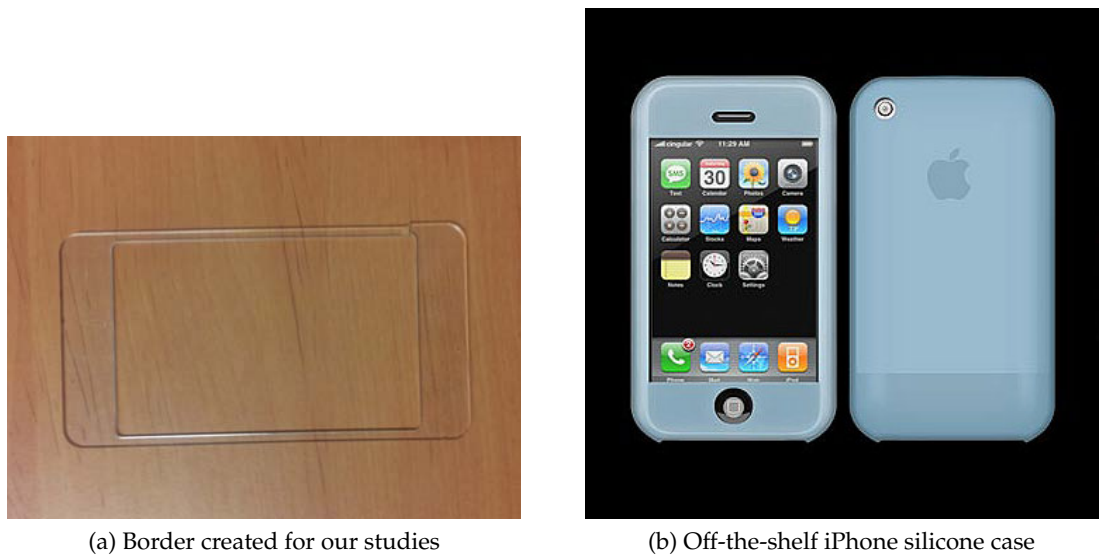

(a) Border created for our studies                    (b) Off-the-shelf iPhone silicone case

Figure 8.2: Mobile touch screen border and commercial silicon case

of tactile cues, as shown in Chapter 6.

Summing up, inclusion can be pursued simply by learning from past experience, particularly, benefits taken from accessibility norms of the previous generations of mobile devices, or even by taking advantage of mainstream but still assistive aids. This can all be achieved while maintaining social acceptance.

### 8.1.3. New sets of abilities and demands

A **third implication** relates to the abilities and demands considered in the design of mobile user interfaces. Related work on individual abilities has long been paying attention to users' age, sex and literacy levels. Respecting demands, a great deal of mobile touch studies have focused on target sizes. There are exceptions. Currently used assistive technologies, particularly screen readers, ignore the diversity and stick to the visual audio replacement ignoring that vision and audio have different characteristics and information is lost in this substitution.

In Chapter 4 and 5 we focused on identifying which individual attributes and abilities were most relevant for a blind person in a mobile interaction setting: other characteristics were revealed. Pressure Sensitivity, Tactile Acuity, Spatial Ability, Attention and Memory, Education level, Mobile experience, all showed to play a part in the overall ability to issue mobile primitives. This impact was revealed differently in different settings, which enabled us to outline device demands. These should be considered when designing inclusive mobile user interfaces. Landing-on a target, making a precise gesture, complying with tap, double tap and press primitive requirements, complying with gesture angle offset, speed and size, direction, location, are all demands that should be considered. Once again, it seems not to be just about Age and it not just about Target Size. In the absence of an integrating sense like vision, these abilities and their relationships with device demands gain higher relevance and should be explored thoroughly. A first take at this was achieved in this thesis, particularly in Chapter 6 and Chapter 7 but further studies should be fostered towards a better comprehension of the ability-demand match within a comprehensive design space.

In sum, individual attributes and abilities have been poorly explored in mobile user interface design. This automatically leads to a lack of knowledge pertaining the demands imposed by current devices, primitives and settings. By assessing individual traits and finding relationships with interaction demands, an iterative path to inclusion of otherwise excluded users is traced.

## 8.1.4. Software Adaptability and the Potential for Inclusion

Our **fourth implication** focuses on software adaptability. Both target acquisition and tough gesturing obey to a set of operating system defined restrictions. Results reported in Chapter 6 showed that people vary in the ability to comply with these requirements. Further, an analysis of the range of two target acquisition restrictions showed that not only some people are unable to cope with the demands but also others would easily comply with more demanding timings. The same is likely to be applicable to offsets, precision, sizes, locations and directions.

Current operating systems and applications are still attached to rigid parametrizations that force the user to adapt to the device instead of the opposite. The barriers exposed in this thesis along with an ability-demand incompatibility are strong vehicles of exclusion. Making a fast and precise gesture within a pre-defined angle offset may be impossible for older blind people with low spatial abilities. This does not mean that the same users are not consistent in their interactions: when consistency happens, an opportunity for adaptation lingers.

The same applies in respect to layouts and the interaction primitives to interact with the device. For users with high cognitive and tactile abilities, dealing with a 12-area icon grid-based layout may be feasible while for other with low cognitive and tactile abilities it may be easier, faster and more comfortable to have two 6-area screens where a flick gesture can be performed to navigate between screens. Chapter 6 revealed these differences to exist. In Chapter 7 we explored how people with different abilities cope with methods that present different demands. Once again, the ability-demand relationship was verified at several levels and we were even able to observe cases where incompatibility between demands and abilities was encountered. Alternatives and parametrization is required.

During the timespan of this research other researchers leaned over personalization and adaptation of touch screen surfaces [Findlater et al., 2011, Findlater and Wobbrock, 2012]. These projects have not focused on disabled populations but the concepts are comparable: devices should learn from their users and adapt accordingly when a consistent behaviour is revealed. This may mean that calibration steps are required which is a low price to pay if performance with mobile user interfaces increases and inclusion is also increased.

## 8.1.5. Different devices, different needs

Related with the previous implication, a **fifth one** relies on the difference in demands between devices and how users cope with those differences. In Chapter 6, we have looked at two different sized devices and differences were found in the overall and for particular ability groups. Error rates differed as well as particular timings like duration of Long Press, Gesture Speed and Size. Looking at interfaces and applications in current tablet and touch phone surfaces, the form factor changes, but the interaction and its restrictions remain similar. Results showed that different behaviours apply for different devices depending on the user's attributes.

Moreover, with the larger space comes the opportunity to enrich it with more options and icons. This is obviously an advantage for some who desire to take full benefits from their screen size and resolution. For others, it may become impossible to deal with. Looking at Apple's iPad and iPhone, there is an effort to maintain operating system versions alike. We argue that to foster inclusion they can be equally deployed but should give space for customization and adaptation.

One promising avenue of research on adaptive interactions with touch screen devices resorts to shared user models [Montague et al., 2011]. The creation of a user model and the ability to share it among applications goes in line with the fourth implication presented here. However, it must be cautiously deployed. Given the differences between devices reported in chapter 6 it is not assured that those models will apply to different settings (different form factors or different tactile cues).

## 8.1.6. Inclusiveness through Informed Diversity

Our **sixth implication** comes with the requirement for a change in the research and market paradigm. This implication aggregates all the remaining. Results obtained in Chapter 7 confirmed that different methods, with underlining divergent demands, suit differently to people with dissimilar levels of ability. All the methods evaluated showed to have advantages and disadvantages. Both the wins and flaws were revealed when confronted with different users (and their abilities). What was shown is that there is no such thing as an ultimate text-entry method. Our take at accessible computing argues that we should not trust average values to select the best method for a sample group. Further, although each method may be adapted, we also believe that the space for improvement within is smaller than the one achieved by complementing a design space with several methods and their advantages. To do so, a better understanding of the methods' demands as well as their wins is required. Understanding a method's demands can only be achieved by exposing it to users with different abilities.

Taking into consideration the low costs involved in deploying a diverse set of solutions for mobile touch devices, we argue that the solution to cope with differences in people is by providing differences in opportunities. We consider that both users' capabilities as well as interaction challenges should be equally explored to foster inclusive design. To do so, two steps are required: 1) understand which abilities are stressed and how they relate with demands; and 2) provide and map solutions to cover the wide existent set of abilities. Our approach to mobile touch accessibility resorts to diversity as a means for inclusion.

We must say that part of this is already taking place: online application markets (e.g., App Store, Android Market) are already recipients of applications, some of them with an accessible focus, that stress different paradigms, interaction methods and primitives, to the same end. Text-entry methods are an example but variety is pervasive to a broad set of applications. What is missing is structure and an awareness of what is available and who it fits.

Our recommendation at a higher applicational level is that researchers lean first at better understanding the design space, and only then provide acceptable and meaningful alternatives. These alternatives should fill the gaps in the ability-demand compatibility design space.

## 8.2. Benefits

Existing mobile assistive technologies for blind people have been following a straightforward approach seeking to overcome the absence of visual feedback with Braille or Audio feedback layers. These approaches have been ignoring the diversity within the population and, as such, have been failing to meet with the needs of a large portion of possible users. This leads to the exclusion of people. Our approach concentrates on the relationship between interface demands and individual abilities.

Our first take at improving the mobile accessibility of blind people has been to lower the demands imposed by current methods but with the requirement of maintaining interaction via mainstream devices, fostering the acceptance of such methods and, hence, social acceptance. These methods have proved to make a difference in the lives of people we were able to interfere with. While a reduced impact was made in terms of the cardinality of users, opportunities for improving the inclusion for the population as a whole through a user-sensitive approach was encountered.

By putting individual abilities to play, we were able to assess variations in device demands. Only by taking these two concepts in consideration will we be able to design interfaces that are in line with what the population requires. This awareness is paramount as it demystifies the notion of the blind person. Calling the attention of the research community to the difference found within the blind population and the impact that those differences translates into, pertaining current devices and primitives, is one of the benefits of this research. Varying the demands has also shown that devices and interfaces can and should be slightly modified to improve accessibility in general. Examples are of subtle tactile cues in the touch devices' hardware design (imitating what is seen in keypad-based devices) and a greater liberty in personalization and adaptation of the user interface, that is still restrictive and follows a one-size fits all paradigm.

Ultimately, by presenting relationships between levels of ability and both low-level and high-level primitives and methods we also laid the groundwork for future adaptations that fill the mobile interface design space and contribute to the inclusion of otherwise excluded slices of the population. Aptitudes like Spatial ability, Peripheral Sensitivity and Memory are now to be considered when designing interfaces for inclusion. This may not mean to develop interfaces specifically for disabled people but yet give space for adaptation with the goal of serving people that vary within the spectrum of implied abilities.

## 8.3. Limitations

While we were able to present relationships between individual attributes and demands, this work still presents limitations worth mentioning.

The biggest limitation of this approach is that it considers a reduced set of demands. The mobile design space is composed of several dimensions whether considering devices, primitives or layouts and several levels and possibilities within. To prove our hypothesis and accomplish our research goals, we focused on a reduced set within those dimensions. Further studies need to be performed to have a better assessment of these levels.

Also, in order to understand the role of individual abilities, we selected a set that was likely (based on our previous work and on the interviews performed) to have a bigger impact on user performance. Still, different demands stress different abilities. One example is Dexterity: we chose not to evaluate it as our settings were not likely to expose differences in that component. Looking retrospectively, dexterity may have had impact in the performance of gestures or even in composed target acquisition primitives like Double Tapping. Others were also excluded from our studies that might have an impact depending on the demands imposed.

One other limitation relates to the scope of the intervention presented here. Our main goal was to identify relationships and create awareness for how the differences within the population are being disrespected by current interfaces and methodologies. Although we believe to have succeeded to create that awareness, there is still a long way to go until a continuous match between interfaces and demands is attained. Such knowledge would require a larger population and as aforementioned a wider set of devices and parametrizations.

## 8.4. Future Work

The results presented in this dissertation pave the way for more inclusive mobile user interfaces for blind people. In what follows, we present future work that, complementary to our work, we did not address

**Fill the design space gaps** Individual attributes and abilities vary within the population. We have showed that device demands also diverge either considering form factor or the presence/absence of tactile cues. The market presents us a wide set of devices with differences in design and layout. Further studies are required to be able to have a finer grained relationship between abilities and demands. To this end, assessments with a larger population and a wider set of devices and demands is required.

**Functional ability assessment** Our work employed low-level assessments to be able to find relationships between levels of ability and device demands. While we consider that such an approach would be feasible for personalization and device selection (each person knowing their ability levels and selecting interfaces accordingly) an opportunity for automatic adaptation lingers. Some adaptations can be performed iteratively and constantly during the regular usage of the device (e.g., long press timings) while others are likely to require a calibration. Games have proven to be suitable platforms for calibrating adaptation. By deploying games that are able to stress Spatial Ability, Memory and Attention, Pressure Sensitivity, among others, and assuring that these enjoyable assessments replicate the real levels of ability, adaptation could be made based on their results.

**Predictive Modelling** Understanding which individual abilities and mobile demands are related gives us the opportunity to match them. In this dissertation, we shed light to what abilities and demands are in place in the current panorama of mobile touch technologies. This showed evidence that these two factors should be assessed carefully when designing for inclusion. An opportunity for future work is to gather data from a wider set of people with different interfaces and parametrizations. With such knowledge, designers and manufacturers will be able to predict beforehand the inclusiveness of a device or interface. Finding the *best* device or interface for a particular person would also be feasible.

**Adaptation**  One important benefit of the studies presented in this dissertation pertains the design of interfaces that respect a population with a broad spectrum of abilities. The goal is to inform designers of the variations within the population so interfaces can be designed to cope with that recognized diversity. This gives space for selection and personalization to one's preferences and abilities. On the other hand, the benefits and particularities of an interface and its suitability for a particular individual may also be automatically identified by the device. The knowledge of the match between interfaces and users along with the ability to functionally assess them (previous item) enables the automatic adaptation of interfaces and parameters to a particular user. This can be done as a personalization step (calibration as aforementioned) or in real-time adapting the interface according to experience.

**Hardware Design**  Added tactile cues (i.e., physical border) showed to have an effect (positive or negative depending on the primitive and ability levels) on user performance. Other smaller physical cues were used by some users to improve their recognition of the devices and positions/directions therein. As with keypad devices and the labelled key '5', touch devices would possibly benefit from the inclusion of subtle tactile cues for non-visual exploration. The study of the impact of such aids along with the selection of type, number and location of physical cues is something to be explored. The results of such studies could be then applied as recommendations for design or even as norms.

**Feedback**  An informative part of the studies performed in this research (particularly Chapter 6) was performed at the lower input level, aiming to restrict the analysis to low-level primitives. However, interaction with mobile devices is compounded of input and output actions performed by the user and the device. The feedback offered by the device influences how the user interacts. If on one hand, we could observe that the conclusions taken from lower-level studies were also visible at the higher-level (i.e., text-entry), a concrete study and parametrization of the feedback offered by the device was not performed. The type feedback given is likely to stress other individual abilities where levels of audition along with cognitive components like reasoning and processing abilities may play an important role. Future work on the feedback offered to the blind user is required (e.g, audio, spatialized audio, vibration). More to it, as performed with text-entry, further studies should focus on high-level interactions with the device, compounded by both input and output primitives (e.g., menu navigation).

**Other Individual Attributes**  In our studies, we included the majority of the individual attributes retrieved from the interview study performed in Chapter 4. Particularly, we have included most that were mentioned to be stressed in a low-level usage of mobile devices. With the study of higher-level tasks, other features like Reasoning, Processing Abilities and Dexterity will gain relevance. A characterization of such attributes and their relationship with mobile high-level demands is something to be explored in the future.

**Other Populations**  We have focused this dissertation on the blind population due to the increased difficulties felt by this group in interacting with extremely visual interfaces. Also, this population reveals differences along several dimensions that justify the attention to the impact that those variations may have. However, all groups of disabled and non-disabled people present differences. Future research should inspect other populations and apply the user-sensitive methodology employed in this dissertation. In parallel to this research, we have looked at motor disabled people and inspected how these users interacted with mobile touch screens.In the future, individual differences within the population should be explored and matched with interface demands. The analysis of other populations will also enable the comparison between groups of people (e.g., visuallly-impaired, motor-impaired and sighted non-motor impaired users).

**Technology Transfer**  Blind people are not the only ones facing difficulties with their devices. In fact, every once in a while, we all are. A situationally-blind person is likely to experience some of the problems felt by a blind one. As problems are comparable, so may be the solutions to overcome them. As such, we envision that the ability-demand match can be extended to different dimensions than the Individual one, particularly, the Environment. Mobile scenarios often force the users for a non-visual access to their devices. In these cases, we can argue that the problems felt are comparable with the ones felt by a blind person. The acknowledgement of such similarities paves the way for new research avenues and market opportunities. One important aspect is that this match places the disabled community paired with the so-called *capable* user and future designs are likely to be beneficial for everyone. We believe that not only we can improve the design of interfaces for disabled people but also that the overall population can benefit from the research and experience gained throughout the years within the accessibility community. In parallel to the work presented in this dissertation, we have engaged in an attempt to transfer technology targeted at blind people with situationally blind people [Lucas et al., 2011]. This first attempt showed that a simple technology transfer approach may not be beneficial. Nonetheless, several accessible design options and more demanding scenarios are left to explore. The success of such technology transfer approaches is also likely to smooth the connotation of assistive technologies and make them more socially acceptable.

**Remove the device**  Mobile devices and their interfaces are designed mostly for a visual interaction. Most of the demands imposed to blind people arise from the need to comply with locations and directions and relations within. One future vector of research relies in removing the device and redesigning the interaction with the user. Speech may not be the sole solution to this problem due to the variety of contexts where mobile interaction takes place but other areas like embodied interaction or mid-air gesture recognition may present opportunities for a more natural and effective interaction for blind people. In particular, we consider that alternative text-entry approaches resorting to Braille can be designed to be used without the need for a mobile touch screen due to the reduced set of targets. Future research may explore how a blind person can input Braille characters by instrumenting the users' fingers, white cane, clothes, or by sensing taps on the users' pockets.

# A1
# Longitudinal Mobile Study

# A1.1. Daily Usage Data

| | User | | | | | | Average | Sum |
|---|---|---|---|---|---|---|---|---|
| | 1 | 2 | 3 | 4 | 5 | | | |
| #Placed Calls | 170 | 1241 | 307 | 154 | 239 | | 196.5 | 2111 |
| #Placed Calls and Talked | 119 | 335 | 134 | 79 | 105 | | 92 | 772 |
| #Received Calls | 70 | 715 | 153 | 23 | 90 | | 56.5 | 1051 |
| #Received Calls and Talked | 53 | 479 | 83 | 21 | 36 | | 28.5 | 672 |
| **Total Calls** | **359** | **2291** | **594** | **256** | **434** | | 345 | 3934 |
| **Total Calls and Talked** | **172** | **814** | **217** | **100** | **141** | | 120.5 | 1444 |
| #Sent Messages | 331 | 625 | 145 | 51 | 14 | | 32.5 | 1166 |
| #Received Messages | 369 | 1101 | 262 | 20 | 37 | | 28.5 | 1789 |
| **Total Messages** | **700** | **1726** | **407** | **71** | **51** | | 61 | 2955 |
| #Added contacts | 25 | 39 | 40 | 15 | 14 | | 14.5 | 133 |

Table A1.1: Communication actions performed during 13 weeks by the 5 participants of the longitudinal study

| TASKS | Date | Hour | Battery | Calculator | Alarm | Add Contact | Delete Contact | Search Contact | Make Call | Receive Call | Receive SMS | Send SMS | Total |
|---|---|---|---|---|---|---|---|---|---|---|---|---|---|
| 1 | 7 | 172 | 115 | 67 | 18 | 25 | 1 | 796 | 119 | 53 | 369 | 331 | 2073 |
| 2 | 9 | 107 | 46 | 26 | 4 | 15 | 0 | 333 | 79 | 21 | 20 | 51 | 711 |
| 3 | 6 | 282 | 159 | 75 | 31 | 39 | 2 | 695 | 335 | 479 | 1101 | 625 | 3829 |
| 4 | 24 | 151 | 142 | 57 | 73 | 40 | 20 | 1057 | 134 | 83 | 262 | 145 | 2188 |
| 5 | 13 | 19 | 89 | 9 | 3 | 14 | 3 | 767 | 105 | 36 | 37 | 14 | 1109 |
| Sum | 59 | 731 | 551 | 234 | 129 | 133 | 26 | 3648 | 772 | 672 | 1789 | 1166 | 7449 |
| % | 0.79 | 9.81 | 7.40 | 3.14 | 1.73 | 1.79 | 0.35 | 48.97 | 10.36 | 9.02 | 24.02 | 15.65 | |

Table A1.2: Overall actions performed during 13 weeks by the 5 participants of the longitudinal study

# A1.2. Text–Entry Sessions Data

| | Best Theoretical KSPC | P01 | P02 | P03 | P04 | P05 |
|---|---|---|---|---|---|---|
| Session 1 | 10.86 | 3.30 | | 6.39 | 4.57 | 4.09 |
| Session 2 | 12.04 | 3.69 | 6.96 | 6.87 | 4.78 | 4.33 |
| Session 3 | 10.33 | 3.47 | 6.10 | 5.27 | 4.82 | |
| Session 4 | 10.67 | 3.31 | 5.69 | 4.98 | 4.54 | |
| Session 5 | 10.67 | 2.88 | 5.45 | 3.79 | 4.13 | 4.94 |
| Session 6 | 11.67 | 3.01 | 5.56 | 3.95 | | 5.28 |
| Session 7 | 10.67 | 2.94 | 4.78 | 3.49 | 3.85 | 5.54 |
| Session 8 | 10.67 | 3.15 | 4.45 | 4.57 | | 6.25 |
| Session 9 | 12.00 | 3.37 | 5.20 | 3.71 | 3.80 | 4.67 |
| Session 10 | 10.67 | 3.27 | 4.71 | 4.44 | 4.09 | 7.00 |
| Session 11 | 9.67 | 2.88 | 4.31 | 3.65 | | 4.85 |
| Session 12 | 10.22 | | 4.76 | 3.63 | | 4.88 |
| Session 13 | 10.67 | 2.93 | 5.36 | 3.67 | 3.48 | 4.72 |

Table A1.3: Keystrokes per character average values per session per participant for the 13 laboratorial sessions performed at the end of each week

| WPM | P01 | P02 | P03 | P04 | P05 |
|---|---|---|---|---|---|
| Session 1 | 2.69 | | 0.96 | 1.26 | 0.72 |
| Session 2 | 2.68 | 0.81 | 0.94 | 1.73 | 0.99 |
| Session 3 | 6.12 | 1.12 | 1.65 | 1.97 | |
| Session 4 | 7.16 | 1.45 | 1.52 | 2.24 | |
| Session 5 | 7.67 | 1.75 | 2.44 | 2.49 | 1.64 |
| Session 6 | 7.83 | 1.75 | 3.20 | | 1.13 |
| Session 7 | 8.47 | 2.57 | 3.05 | 2.36 | 1.48 |
| Session 8 | 7.56 | 2.33 | 2.69 | | 1.05 |
| Session 9 | 6.21 | 2.21 | 3.80 | 2.88 | 1.42 |
| Session 10 | 7.88 | 2.82 | 3.16 | 3.15 | 1.07 |
| Session 11 | 7.40 | 2.88 | 3.56 | | 1.50 |
| Session 12 | | 3.07 | 4.10 | | 1.81 |
| Session 13 | 8.07 | 2.65 | 2.95 | 3.56 | 1.55 |

Table A1.4: Words per minute average values per session per participant for the 13 laboratorial sessions performed at the end of each week

| Error Rate | P01 | P02 | P03 | P04 | P05 |
|---|---|---|---|---|---|
| Session 1 | 0.00 | | 0.04 | 0.00 | 0.05 |
| Session 2 | 0.06 | 0.02 | 0.00 | 0.00 | 0.02 |
| Session 3 | 0.00 | 0.00 | 0.00 | 0.03 | |
| Session 4 | 0.00 | 0.00 | 0.17 | 0.00 | |
| Session 5 | 0.00 | 0.00 | 0.03 | 0.04 | 0.00 |
| Session 6 | 0.00 | 0.18 | 0.00 | | 0.04 |
| Session 7 | 0.00 | 0.00 | 0.00 | 0.06 | 0.00 |
| Session 8 | 0.00 | 0.00 | 0.00 | | 0.00 |
| Session 9 | 0.03 | 0.02 | 0.00 | 0.00 | 0.00 |
| Session 10 | 0.00 | 0.00 | 0.06 | 0.00 | 0.00 |
| Session 11 | 0.00 | 0.00 | 0.00 | | 0.00 |
| Session 12 | | 0.00 | 0.00 | | 0.23 |
| Session 13 | 0.00 | 0.00 | 0.00 | 0.00 | 0.06 |

Table A1.5: Error rate (deletions) average values per session per participant for the 13 laboratorial sessions performed at the end of each week

| MSD ER | P01 | P02 | P03 | P04 | P05 |
|---|---|---|---|---|---|
| Session 1 | 0.00 | | 0.10 | 0.00 | 0.21 |
| Session 2 | 0.00 | 0.02 | 0.00 | 0.00 | 0.11 |
| Session 3 | 0.00 | 0.03 | 0.00 | 0.00 | |
| Session 4 | 0.00 | 0.10 | 0.09 | 0.00 | |
| Session 5 | 0.00 | 0.00 | 0.02 | 0.00 | 0.07 |
| Session 6 | 0.00 | 0.05 | 0.03 | | 0.03 |
| Session 7 | 0.00 | 0.00 | 0.05 | 0.00 | 0.06 |
| Session 8 | 0.00 | 0.09 | 0.00 | | 0.08 |
| Session 9 | 0.00 | 0.00 | 0.00 | 0.00 | 0.00 |
| Session 10 | 0.00 | 0.00 | 0.00 | 0.00 | 0.18 |
| Session 11 | 0.00 | 0.03 | 0.03 | | 0.00 |
| Session 12 | | 0.00 | 0.06 | | 0.14 |
| Session 13 | 0.00 | 0.03 | 0.00 | 0.00 | 0.03 |

Table A1.6: Minimum String Distance Error Rate average values per session per participant for the 13 laboratorial sessions performed at the end of each week

# A2

# Proficient Text-Entry
# Study

# A2.1. Study Script

This annex describes our test plan to evaluate and compare two different mobile text input techniques, i.e. Navtap and Multitap, with blind expert users.

## A2.1.1. Motivation and Goals

The main goal behind this evaluate is to access and compare our text entry method, Navtap, against the traditional method, Multitap. Our past results with novice users showed that Navtap is easy to learn and allows the users to entry text with fewer errors when compared to Multitap. However, we did not perform any evaluation tests with expert users. Particularly, even though NavTap users are less experienced than MultiTap ones, we intend to assess how well do they rate against their peers. This is relevant as we want to evaluate if a more inclusive simpler method can be used with similar or close to similar performance thus fostering the inclusion of its users.

While Multitap has a theoretical advantage over Navtap, we demonstrated that this method is not adequate to some blind users, is error prone and hard to learn.

In this evaluation we want to see how the users' experience influences their performance in the text entry task, with both methods, and if the Multitap's theoretical advantage indeed occurs for expert users.

## A2.1.2. Research Questions

1. How easily and successful can users input text?

2. What are the main difficulties for each method?

3. Which method is faster?

4. Which method is more error prone?

5. How hard is to learn the Multitap method for blind users?

6. How experience influences the users' performance in each input method?

7. How does NavTap stand in relation to MultiTap?

## A2.1.3. User Profile

We will select participants who have some experience in mobile text entry for each method. In this evaluation we will consider a participant that sends at least 5 messages per week for the last 3 months an expert. We will recruit a minimum of 10 participants, 5 for Navtap and 5 for Multitap method.

# A2.1.4. Evaluation Methodology

In this evaluation we will use a between subjects design, i.e each text-entry method will be tested by a unique set of users. We will use 20 minutes of each session to explain it to the participant, conduct a pre questionnaire in order to access the user background and experience and then conduct a post-test debriefing interview.

During the evaluation, each participant will write 10 sentences in his mobile device. All the sentences will be read by the evaluation monitor. The sequence of the sentences will be randomized between participants, so the results will not be biased due to learning effects. In order to guarantee that the participant will not suffer from fatigue effects during the evaluation, he will write half of the sentences before and after lunch.

## Experiment Preparation

- Set up the camera and audio recorder. Check if there is enough battery and storage space available for the session.

## Pre-evaluation settings

- Read the introduction section.
- Read the consent form to the participant and make sure there is a witness when he agree/disagree.
- Fill out the pre questionnaire.
- Ask each participant to write the sentence, "ola tudo bem" before he begins the test, so we can set up the camera position in order to capture the users hands and mobile device.
- Take a picture of the participant to our users' database.

## Tasks

- andar
- aberto
- vamos dois
- ate amanha
- chego atrasado
- podemos correr
- eu vou antes de ti
- o bolo estava bom
- achas que vale a pena ir
- fui com o Antonio correr

## Post–evaluation debriefing

- Interview the participant and follow up on any particular problems that came up.

- Thank the user for his participation.

# A2.1.5. Evaluation Environment and Material

We will use a controlled setting to conduct the sessions. The evaluation will take place at Raquel and Martin Sain Foundation. Besides the participant and evaluation monitor there will be one or two observers. Participants will use their own mobile devices and all interactions will be recorded with a video camera.

# A2.1.6. Monitor Role

The monitor will sit right next to the user while conducting the session. He will introduce the session, conduct a pre questionnaire, make sure the user knows about the consent form and then introduce the tasks. During the tasks execution the monitor should not interfere with the participant, unless he does not remember the current task. Finally, the monitor will take detailed notes about the participants' behaviour and comments.

The monitor will debrief with observers at the end of each session. He will ask observers to contribute their observations about surprises and issues, so that can have an active part on the evaluation.

# A2.1.7. Evaluation Data

To answer the research questions we will collect both performance and preference data during the evaluation sessions.

**Performance:**

- Number of tasks completed without assistance

- Time to complete the tasks

- Number of Errors

- Keystrokes per character

- Path to each character (Navtap)

**Preference:**

- Perceived difficulties
- Participants' behaviours and comments

# A2.1.8. Reporting the Results

The report must address the following points:

- Briefly summarizes the background of the study, including the goals, methodology, logistics, and participant characteristics
- Present findings for the research questions
- Gives quantitative results and discusses specifics as appropriate to the question and the data
- Discuss the implications of the results

# A2.1.9. Monitor Test Plan

This section details the monitor role and all his tasks, before, during and after the evaluation.

**Experiment preparation:**

- Create the record set up
- Turn the laptop on

**Pre-evaluation arrangements:**

**Introduction** Read to the participant:

> Antes de mais, muito obrigado por colaborar connosco. Vou começar por contextualizar este estudo, qual o seu papel e qual o nosso objectivo. Somos alunos de doutoramento do Instituto Superior Técnico e temos vindo a trabalhar juntamente ao com a Fundação Raquel e Martin Sain há alguns anos, com o objectivo de desenvolver um telemóvel adequado a utilizadores com deficiências visuais. Nesta fase estamos a estudar as soluções existentes no mercado, como o TALKS, particularmente no que diz respeito à escrita de texto. O que lhe iremos pedir para fazer é escrever um conjunto de frases no seu telemóvel o mais rápido e preciso que conseguir. Para que este processo não seja demasiado cansativo irá escrever cinco frases na parte da manhã e outras cinco depois de almoço.

**Consent form** Read to the participant:

Antes de continuar gostaria de o informar que iremos usar material de gravação, nomeadamente uma câmara e máquina fotográfica, que irão ser usadas para gravar a sessão e assim permitir uma posterior análise pela equipa de investigação. Há que frisar que todos os resultados serão anónimos e a sua identidade será protegida. Assim: Aceita que a gravação da sessão seja usada para posterior análise pela equipa de investigação? Aceita que os dados da sessão sejam usados pela equipa de investigação em futuras publicações científicas?

**Pre-questionnaire**  (fill the form)

Antes de iniciar a escrita de texto gostaria de lhe fazer algumas perguntas. (ver ficheiro Excel)

**Set up the camera**  Ask each participant to write the sentence, "ola tudo bem" before he begins the test, so you can set up the camera position in order to capture the users hands and mobile device

**Take a picture of the participant**

**Evaluation:**

- Make sure the participant understands that the sentences are to be written without punctuation and accentuation

- Make sure the participant repeats the sentence before he starts the test

- Read the sentences randomly (see excel file)

- Annotate errors, questions, behaviours and comments

- Do not interfere during the task execution

**Post-evaluation debriefing:**

- Interview the participant (see excel file)

- Thank the participant for his availability

# A2.2. Study Report

NavTap has a shorter learning curve than MultiTap [Guerreiro et al., 2008a]. The latter has shown to be hard to learn because it requires higher memorization capabilities and it is therefore error prone. Nevertheless, there are some blind users that are able to reach a high proficiency level with this method, although it can take several months or years of practice. On the other hand, NavTap seems to enable a smooth first approach and to foster adoption as it is simpler. To assess how a simpler inclusive approach stands in relation to the mainstream approach we have performed a comparative study with blind people proficient with NavTap (the group from the previous study after a 4-month experience period) and MultiTap.

## A2.2.1. Research Goals

In this evaluation, we first want to analyse the performance level of a daily text-entry MultiTap user. We then want to assess NavTap's efficiency and effectiveness, using the MultiTap performance results as a baseline. Therefore, we will be able to analyse how users' experience, with both methods, influences their performance. This is relevant as it is vital to understand how an assistive method, theoretically less efficient, behaves after being used in real life scenarios.

Even though MultiTap has a theoretical advantage over NavTap, regarding the average number of required Keystrokes Per Character (KSPC) [MacKenzie and Tanaka-Ishii, 2007], there were no previously published results supporting this in practice. In this evaluation, we observed if this advantage indeed occured for expert users. Additionally, according to [Silfverberg et al., 2000], MultiTap can support rates up to 25 or 27 Words Per Minute (WPM) for expert users. Nevertheless, to our knowledge, no input rates for expert blind users are documented.

## A2.2.2. Participants

This evaluation was performed in the same formation centre for blind people and all participants were trainees there. We found two types of users: those who had learned MultiTap (12 users) and those who could not (5 users). While NavTap has shown to be an alternative solution to those who could not learn MultiTap [Guerreiro et al., 2008a], we have no knowledge about how both methods behave for experts.

In this evaluation we assessed performance after long-term daily usage. Therefore, we selected participants that had sent at least 5 text messages per week for the past 4 months, and were able to manage (add, remove and search) their contact list. The NavTap group was composed of 5 participants, 3 females and 2 males, with ages between 44 and 61 years old. These were the ones from the previous study. NavTap was the first mobile text-entry method that those participants were able to learn and use on a daily basis. On the other hand, the MultiTap group was composed of 12 participants, 9 males and 3 females, with ages between 20 and 58 years old. This group had an average mobile experience of 6.75 years.

## A2.2.3. Procedure

We used a between-subjects evaluation and each participant wrote ten different sentences in the Portuguese language, retrieved from a publicly available corpus. Moreover, sentences had 5 difficulty levels based on their lengths (5, 10, 14, 18 and 24 characters). These lengths have no special justification other than enabling us to consistently vary the phrases difficulty levels.

Sentences were read aloud by the evaluation monitor and repeated by participants before entering them, thus assuring that participants understood them. They were asked to input text as quickly and correctly as possible. The sequence of sentences was randomized between participants, so the results would not be biased due to learning effects. In order to guarantee that participants would not suffer from fatigue effects, they could rest between trials.

Timing began when participants performed the first character of the sentence and ended when they performed the last keystroke. With this procedure the resulting sentences could contain errors. Thus, another of our goals was to observe the quality of the entered phrases with both methods. Also, participants could correct errors by deleting characters. The sentence completion time incorporates the time to correct those errors. Finally, in order to offer a more natural, familiar and realistic scenario, participants used their own mobile devices to enter the proposed sentences.

## A2.2.4. Results

In order to achieve a better understanding of the obtained results, we must emphasize the differences in participants' profiles between text-entry methods. Firstly, NavTap participants were not able to learn the MultiTap method. Secondly, although being considered experienced (4 months usage), they have a much lower level of experience than MultiTap participants (an average usage of 6.75 years). Thus, our goal was to observe what performance level this population can reach when compared to MultiTap's expert participants.

**Keystrokes per Character**

One of our goals with this evaluation was to analyse if MultiTap's theoretical advantage indeed occurs with expert users. The average Keystrokes Per Character rate was 2.31 for MultiTap, and 4.14 for NavTap (Figure A2.1). Indeed, MultiTap requires significant less KSPC than NavTap (Mann-Whitney U test, Z=-3.162, p <.005, r=.08). Also, MultiTap is closer to theoretical values (1.97 KSPC) than NavTap best case scenario (3.11 KSPC). The performance level of NavTap's participants is between the 2-way (using two directions) and 4-way (best case) navigational approaches.

**Text-Entry Speed**

As a consequence of a lower KSPC, MultiTap also outperformed NavTap regarding the words per minute input rate. The average text-entry speeds were 6.64 WPM for MultiTap and 3.82 WPM for NavTap. A Mann-Whitney U test (Z = -1.897, p = .058, r=.05) showed that the MultiTap group was significantly faster than NavTap's.

**Minimum String Distance Error Rate**

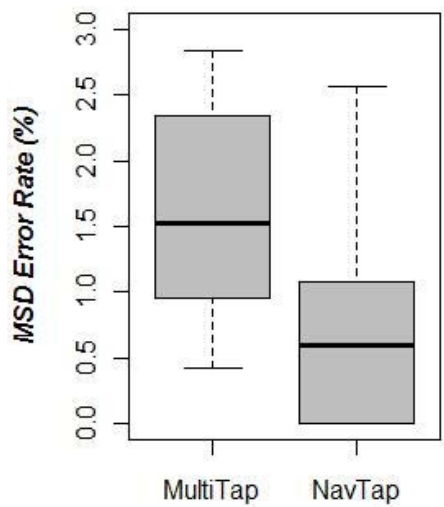

Figure A2.1: Keystrokes per Character by Text-Entry Method (NavTap and MultiTap).

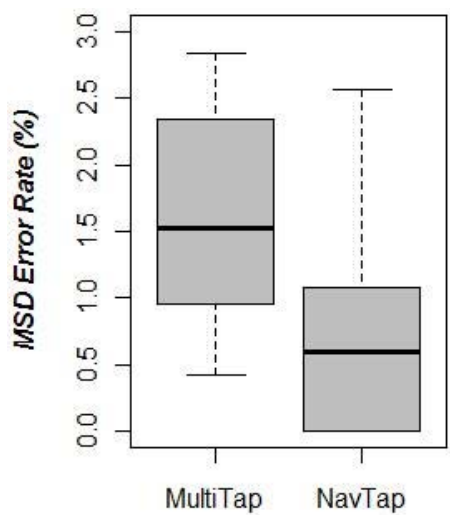

Figure A2.2: Minimum String Distance Error Rate by Text-Entry Method (NavTap and MultiTap).

Recall that our procedure did not force participants to enter phrases correctly. As such, the text-entry speed was influenced by the quality of the transcribed phrases as participants may have entered more or less characters. Nevertheless, it is still helpful to observe the Minimum String Distance (MSD) error rate [MacKenzie and Tanaka-Ishii, 2007] for both methods.

There was a minor significant main effect (Mann-Whitney U test, Z=-1.582, p = 0.114, r=.04) on MSD error rate; NavTap outperformed MultiTap with 0.85% against 2.08%, respectively (Figure A2.2). Sentences transcribed by the NavTap group had a higher quality than MultiTap's.

**Error Rate**

As we said before, transcribed phrases could have some errors. However, participants were free to correct them during the evaluation. For this experiment we considered an error as a character deletion. NavTap showed to be significantly less error prone than MultiTap (Mann-Whitney U test, Z=-2.747, p <.01). The average error rates were 10.03% for MultiTap and 1.32% for NavTap

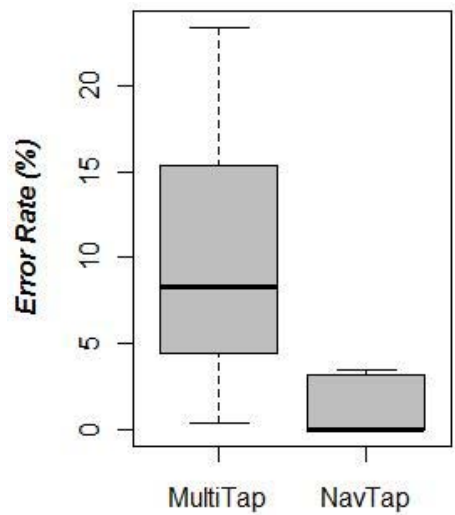

Figure A2.3: Error Rate by Text-Entry Method (NavTap and MultiTap).

(Figure A2.3).

Furthermore, relating these results with MSD error rate, we can see that NavTap's sentences had a higher quality and a lower number of character deletions. Conversely, MultiTap's sentences were of poorer quality and had a higher error rate.

## A2.2.5. Discussion

The main goals of this evaluation were to assess the limitations and main advantages of both MultiTap and NavTap with expert users.

Indeed MultiTap's theoretical advantage occurs for both Keystrokes per Character and, consequently, Words per Minute metrics, meaning higher efficiency. These results were somewhat expected as expert users can take full advantage of a given technique. Being MultiTap a more complex method than NavTap, it enables higher text-entry rates. However, the obtained results are still far from the ones reported by [Silfverberg et al., 2000] for expert users. This may indicate that a simple substitution of visual feedback by audio feedback (i.e. a screen reader) may hinder the learning process. Although sighted experienced users are likely to input text at astonishing rates, during their long-term experience (months or years) they were able to receive proper feedback on their evolution. We believe this learning process to be what differences blind and sighted expert users the most.

On the other hand, NavTap outperforms MultiTap regarding output quality and produces significantly fewer errors. The small number of errors indicates that NavTap is indeed more effective and easier to use. This fact gains higher relevance because NavTap's participants had a lower experience level (less time using the method). Indeed, NavTap was the only mobile text-entry method they were capable to learn and use in a daily basis, much due to the reduced number of produced errors. Results indicate that both methods have different target audiences. On one hand, MultiTap has shown to be a more efficient method with higher text-entry rates. On the other hand, NavTap is an easy to learn and effective technique that has show to be a suitable alternative for those users who were not able to learn MultiTap.

There is a need to identify and comprehend the blind population individual characteristics and their implications for mobile text-entry methods future designs. From our experience, some users are unable to learn current text-entry techniques, revealing an inadequacy of mobile interfaces to their abilities. This is particularly visible on older and adventitious blind users, which require simpler text-entry methods, less demanding in terms of tactile and cognitive abilities.

## A2.3. Study Data

| MultiTap | | | | |
|---|---|---|---|---|
| | WPM | KSPC | MSD Error Rate | Error Rate |
| 1 | 5.76 | 2.05 | 0.02 | 0.02 |
| 2 | 8.61 | 2.32 | 0.02 | 0.10 |
| 3 | 8.65 | 2.18 | 0.01 | 0.05 |
| 4 | 5.30 | 2.69 | 0.08 | 0.18 |
| 5 | 8.03 | 2.12 | 0.01 | 0.06 |
| 6 | 5.79 | 2.46 | 0.01 | 0.16 |
| 7 | 6.20 | 2.34 | 0.02 | 0.14 |
| 8 | 4.17 | 2.34 | 0.03 | 0.14 |
| 9 | 2.43 | 2.95 | 0.01 | 0.23 |
| 10 | 9.37 | 1.98 | 0.01 | 0.00 |
| 11 | 7.84 | 2.11 | 0.00 | 0.04 |
| 12 | 7.51 | 2.21 | 0.02 | 0.06 |

| NavTap | | | | |
|---|---|---|---|---|
| | WPM | KSPC | MSD Error Rate | Error Rate |
| 1 | 2.31 | 3.34 | 0.01 | 0.00 |
| 2 | 8.28 | 3.23 | 0.00 | 0.00 |
| 3 | 2.80 | 4.48 | 0.00 | 0.03 |
| 4 | 2.51 | 5.49 | 0.03 | 0.03 |
| 5 | 3.20 | 4.16 | 0.01 | 0.00 |

Table A2.1: Text-Entry Metrics data for both MultiTap and NavTap methods

# A2.4.  Study Analysis

**Tests of Normality**

| | Method | Kolmogorov-Smirnov[a] | | | Shapiro-Wilk | | |
|---|---|---|---|---|---|---|---|
| | | Statistic | df | Sig. | Statistic | df | Sig. |
| WPM | MultiTap | ,164 | 12 | ,200[*] | ,946 | 12 | ,582 |
| | NavTap | ,397 | 5 | ,010 | ,675 | 5 | ,005 |
| KSPC | MultiTap | ,211 | 12 | ,147 | ,901 | 12 | ,162 |
| | NavTap | ,207 | 5 | ,200[*] | ,927 | 5 | ,577 |
| Error Rate | MultiTap | ,333 | 12 | ,001 | ,672 | 12 | ,000 |
| | NavTap | ,300 | 5 | ,161 | ,833 | 5 | ,146 |
| MSD Error Rate | MultiTap | ,205 | 12 | ,176 | ,950 | 12 | ,637 |
| | NavTap | ,367 | 5 | ,026 | ,684 | 5 | ,006 |

\*. This is a lower bound of the true significance.

a. Lilliefors Significance Correction

Table A2.2: Normality tests for WPM, KSPC, Error Rate and MSD Error Rate metrics

**Test Statistics[a]**

| | WPM | KSPC | MSD Error Rate | Error Rate |
|---|---|---|---|---|
| Mann-Whitney U | 12.000 | .000 | 15.000 | 4.000 |
| Wilcoxon W | 27.000 | 78.000 | 30.000 | 19.000 |
| Z | -1.897 | -3.162 | -1.582 | -2.747 |
| Asymp. Sig. (2-tailed) | .058 | .002 | .114 | .006 |
| Exact Sig. [2*(1-tailed Sig.)] | .064[b] | .000[b] | .130[b] | .004[b] |

a. Grouping Variable: Method

b. Not corrected for ties.

Table A2.3: Null-hypothesis Statistical Testing (non-parametric ) for WPM, KSPC, Error Rate and MSD Error Rate metrics

# A3

# Interview Study

This annex refers to the study performed with ten (10) professionals that have a large experience working closely with blind people. Its main goal was to identify the most relevant individual differences amongst blind people and infer if those differences should be treated with more attention than among the sighted population. The annex is composed of the interview script (in Portuguese), interview form and the interview data pertaining references to relevant individual attributes (induced and spontaneous). Interview transcriptions along with Microsoft Excel 97-2003 data files are available at http://web.ist.utl.pt/tiago.guerreiro/phd.html.

# A3.1. Interview Script

## Guião de Entrevista

➔ *Apresentação*

Bom dia. O meu nome é Tiago Guerreiro, sou aluno de doutoramento do Instituto Superior Técnico e investigador do Instituto de Engenharia e Sistemas Computacionais, Investigação e Desenvolvimento, em Lisboa. Esta entrevista insere-se no contexto dos meus estudos de doutoramento em que procuro identificar e compreender as diferenças individuais da população cega. Para tal, é essencial conseguir obter conhecimento e basear os meus estudos nos anos de experiência que as pessoas envolvidas no dia a dia da população cega têm.

Antes de começar, gostaria de a/o colocar totalmente à vontade. O único objectivo desta entrevista é partilhar experiência e toda a informação será apreciada. Por último, e para me permitir uma análise cuidada posterior à nossa conversa, gostaria de lhe pedir autorização para gravar a entrevista. Esta será completamente privada e a sua identificação anonimizada em qualquer tipo de disseminação efectuada.

➔ *Iniciar Gravação*

Os meus estudos de doutoramento focam na diversidade da população cega. Assim, o meu objectivo máximo consiste na identificação das características que levam a diferenças funcionais, a nível tecnológico, entre pessoas. O foco na população cega é justificado pelo facto de algumas características ganharem maior relevo devido à ausência de retorno visual. Por exemplo, uma pessoa com neuropatia periférica, devido a diabetes, terá problemas em ler Braille como o terá a usar dispositivos com botões pequenos.

Assim, gostaria de lhe fazer algumas perguntas e que partilhasse comigo a sua experiência e opinião. Primeiro, pedia-lhe que me desse alguns dados gerais.

➔ *Preencher Ficha de Entrevista (Dados Gerais)*

➔ *Início das questões*

  1) Há quanto tempo trabalha com pessoas cegas?
  2) Ao longo desses anos, que funções teve/tem junto da população?
     a. Onde trabalhou?
     b. Durante quanto tempo em cada sítio?
     c. Aproximadamente, com quantas pessoas cegas já trabalhou?
     d. (mais do que um local) As populações com que trabalhou eram muito diferentes (lar vs escola de referência)?
  3) Ao longo do seu trabalho com pessoas cegas, tem trabalhado com pessoas muito diferentes?
     a. Pode exemplificar?
     b. O que diferencia mais essas pessoas?
     c.  Em que tarefas se notam essas diferenças?

4) As capacidades funcionais (trabalhar com dispositivos) das pessoas cegas variam muito?
   a. Mais do que entre as pessoas normo-visuais?
   b. Pode enumerar casos?
   c. Com algum dispositivo em particular?

5) Pensando em características possíveis, onde nota maior impacto no sucesso ou insucesso no uso de tecnologia? Exemplos de características podem ser relativas à idade, doenças colaterais, tudo o que determine alteração na execução de tarefas.
   a. Nota algum padrão de casos de sucesso?
   b. E insucesso?
   c. Particularizando, no uso de computadores, quais as pessoas cegas que dominam e aprendem melhor?
   d. Quais serão as características que realçaria nessas pessoas?
   e. E no caso de insucesso, quais as pessoas que têm mais dificuldades?
   f. Quais as características que realça nesses casos?
   g. A nível geral, que características têm mais impacto na diferença entre pessoas cegas a operar computadores?
   h. E no caso dos telemóveis, quais as pessoas cegas que dominam e aprendem melhor?
   i. Alguma característica extra que realça no contexto móvel?
   j. E dificuldade, alguma característica extra a realçar?
   k. Para as características que mencionou, tem alguma forma, algum mecanismo, para avaliar a incapacidade/dificuldade?
   l. E que mecanismos tem para ajudar a equilibrar a capacidade? Alguma ténica em específico?
   m. E as pessoas, como ultrapassam essa dificuldade? Desenvolvem algum mecanismo, compensam ou usam ajuda externa?

6) Em relação a tecnologias assistivas para uso de computadores, que características das pessoas exploram?
   a. Realça alguma tecnologia assistiva em particular?
   b. Que características das pessoas explora essa tecnologia?
   c. Esses pressupostos estão sempre garantidos?
   d. É desajustado em alguns casos? Quais
   e. Os que conseguem, conseguem atingir níveis de sucesso semelhantes às pessoas normo-visuais?
   f. Para além da visão, existem impedimentos extra?
   g. E capacidades extra?

7) Em relação a tecnologias assistivas para uso de telemóveis e assistentes pessoais, que características das pessoas exploram?
   a. Realça alguma tecnologia assistiva em particular?
   b. Que características das pessoas explora essa tecnologia?
   c. Esses pressupostos estão sempre garantidos?
   d. É desajustado em alguns casos? Quais

e.  Os que conseguem, conseguem atingir níveis de sucesso semelhantes às pessoas normo-visuais?

f.  Para além da visão, existem impedimentos extra?

g.  E capacidades extra?

8) No que diz respeito a telemóveis, que tarefas conseguem as pessoas em geral realizar?

a.  Quais é que não conseguem?

b.  Porquê?

c.  Demoram muito a aprender? Porquê?

9) Na sua opinião, quais as características individuais que uma pessoa cega deve ter para se dar bem com tecnologia?

a.  Para ter um bom primeiro impacto?

b.  Para aprender rapidamente?

10) Quais as características dos casos mais dramáticos de insucesso?

a.  Como tentam ultrapassar estes casos?

11) Em relação a situações do dia a dia, que factores acha essenciais para a interacção eficaz com telemóveis por parte de uma pessoa cega?

12) Que situações ou problemas é que dificultam essa interacção? E que impossibilitam?

a.  Porquê?

b.  Quais as exigências extra que transformam a capacidade?

c.  O que fazem as pessoas cegas nessas situações?

➜ *Completar questões (usar grelha de entrevista)*

13) No contacto com computadores ou telemóveis acha que a __________ tem influência no capacidade de controlo tida como normal?

a.  Consegue identificar casos em que a diferença relevante era esta e se traduzia em capacidades muito diferentes?

b.  Esta diferença é drástica, ou seja, um utilizador com pouca _________ poderá nem conseguir operar o dispositivo? Ou, na maioria dos casos, leva apenas a um desempenho mais fraco?

c.  Nestes casos, executa/conhece alguma técnica para ajudar a pessoa a colmatar a incapacidade?

d.  E para a detectar?

e.  E as pessoas cegas, como compensam essa incapacidade?

f.  Acha que ter dispositivos que se adaptem à _________________, por exemplo, fazendo_________, fará com que esta diferença seja esbatida ou mesmo eliminada, ou que é difícil/demasiado complicado eliminar estar limitação?

➜ *Finalizar entrevista*

# A3.2. Interview Form

## Ficha de Entrevista

1) Nome: __________________________________________________________

2) Sexo: _____

3) Idade: _____

4) Habilitações Literárias: _____________________

5) Área de especialização: ____________________

6) Local de Trabalho:_________________________

7) Há quantos anos:___

8) Denominação cargo: _________________________________

9) Área de trabalho: ___________________________________

10) Outros locais e funções: _____________________________________________________

11) Características (a preencher durante entrevista):

| Característica | Discutida | Comentários |
|---|---|---|
| Idade | | |
| Tempo de cegueira | | |
| Onset da Cegueira | | |
| Grau Literacia | | |
| Experiências com tecnologia | | |
| Experiência com telemóveis | | |
| Memória de trabalho | | |
| Memória de curto prazo | | |
| Memória de longo prazo | | |
| Saúde Mental | | |
| Destreza | | |
| Precisão motora | | |
| Força motora | | |
| Sensibilidade periférica | | |
| Coordenação motora | | |
| Audição | | |

# A3.3. Interview Data

| | | I1 | | I2 | | I3 | | I4 | | I5 | | I6 | | I7 | | I8 | | I9 | | I10 | | Total Mentions | | User Mentions | |
|---|---|---|---|---|---|---|---|---|---|---|---|---|---|---|---|---|---|---|---|---|---|---|---|---|---|
| | | Rel | Irrel | Rel | Irrel | Rel | Irrel | Rel | Irrel | Rel | Irrel | Rel | Irrel | Rel | Irrel | Rel | Irrel | Rel | Irrel | Rel | Irrel | relevant | irrelevant | relevant | irrelevant |
| Age | Spontaneous | 1 | | | | 3 | | | | 4 | | 1 | | | | 4 | | | | 1 | | 13 | | 5 | 0 |
| | Induced | | | | | | | 1 | | | | | | 1 | | | | 1 | | | | 4 | | 4 | 0 |
| Time with impairment | Spontaneous | 1 | | | | | | 1 | | 2 | | | | 1 | | | | 1 | | | | 5 | | 3 | 1 |
| | Induced | 1 | | | | 2 | 1 | 1 | | | | 1 | | 2 | | 1 | 1 | 1 | | | | 6 | 1 | 5 | 1 |
| Blindness age of onset | Spontaneous | 3 | | 1 | 2 | 2 | | 2 | | 2 | | 2 | | 4 | | 1 | 1 | 1 | | 2 | | 17 | 2 | 8 | 1 |
| | Induced | | | | | | | 1 | | 1 | 1 | | | | | 1 | | | 1 | | | 3 | 1 | 1 | 1 |
| Literacy Degree | Spontaneous | 2 | | | | 3 | | 1 | | 1 | 1 | | | | | 3 | 1 | | | | | 10 | 2 | 5 | 2 |
| | Induced | | | 1 | | | | | | 1 | 1 | 1 | | 1 | 1 | 1 | | 1 | | 1 | | 2 | 3 | 2 | 3 |
| Experience with technology | Spontaneous | 1 | | | | | | 1 | | 1 | | | | 2 | 1 | | | 1 | | | | 6 | | 5 | 0 |
| | Induced | 1 | | | | 1 | | | | | | 1 | | 1 | 1 | | | | | | | 3 | 1 | 3 | 1 |
| Experience with Mobile devices | Spontaneous | 1 | | | | | | | | | | 1 | | 2 | 2 | | | 1 | | | | 4 | | 2 | 0 |
| | Induced | | | | | | | | | | | | | | | | | | | | | 3 | | 3 | 0 |
| Memory | Spontaneous | 2 | | 4 | | 1 | 2 | | | 1 | | 1 | | 1 | | | | | | | 1 | 11 | | 6 | 0 |
| | Induced | | | | | | | | | 1 | | 1 | | | | | | | | | | 4 | | 4 | 0 |
| Problem Solving Abilities | Spontaneous | 2 | | | | | | | | 2 | | | | | | | | | | | 1 | 5 | | 3 | 0 |
| | Induced | | | 2 | 1 | | | | | 2 | | | | | | | | | | | 2 | 7 | | 4 | 0 |
| Abstract Reasoning | Spontaneous | | | | 1 | | | 1 | | | | | | | | | | | | | 2 | 3 | | 2 | 0 |
| | Induced | | | 3 | 1 | | | | | | | 3 | | | | | | | | | 1 | 8 | | 4 | 0 |
| Verbal Reasoning | Spontaneous | 1 | | | | | | | | | | | | | | | | | | | | 1 | | 1 | 0 |
| | Induced | 1 | | 4 | | | | 1 | | 1 | | | | | | | | 1 | 1 | | | 7 | | 4 | 0 |
| Fine Motor Control | Spontaneous | | | | | | | | | | | | | | | | | | | | | 1 | | 0 | 0 |
| | Induced | 2 | | 1 | | 1 | 4 | | | 1 | | 3 | 1 | 3 | 1 | | | 1 | 1 | 3 | | 19 | | 9 | 0 |
| Peripheral Sensitivity | Spontaneous | | | | | | | 1 | 1 | 1 | | | | | | | | | | 1 | | 2 | 1 | 2 | 1 |
| Motor Precision | Induced | | | | | | | 1 | | 1 | | | | | | | 1 | | | | | 1 | | 1 | 0 |
| | Spontaneous | | | | | | | | | | | | | | | | | | | | | 1 | | 1 | 0 |
| Motor Coordination | Induced | 1 | | | | | | 2 | | | | | | | | 1 | | | | | | 1 | | 2 | 0 |
| Attention | Spontaneous | 1 | | | | | | 2 | | | | | | 1 | | | | | | | | 3 | | 2 | 1 |
| Concentration | Spontaneous | 6 | | 2 | 2 | | | | | 3 | | | 1 | | 3 | | | | | 1 | | 3 | 1 | 6 | 0 |
| | Induced | | | | | | | | | | | | | 5 | | | | 1 | | | | 1 | | 1 | 0 |
| Spatial Ability | Spontaneous | 3 | | | | 1 | | 1 | | 1 | 1 | 1 | 1 | 5 | 2 | 1 | | 1 | | 2 | | 16 | 1 | 8 | 1 |
| Motivation | Spontaneous | | | | | 3 | 2 | | | 1 | 1 | 1 | | 2 | 1 | | | | | | | 4 | | 3 | 0 |
| Hearing | Induced | | | | | 1 | | | | | | | | | | | | | | | | | | 0 | 0 |
| Cause of Blindness | Spontaneous | 1 | | 1 | 1 | 1 | | 1 | | 1 | | | | 1 | | | | | | 3 | | 5 | | 5 | 0 |
| Intelligence | Spontaneous | 1 | | 1 | 1 | 2 | | | | 1 | 1 | | | | | | | 1 | | 5 | | 15 | | 6 | 0 |
| | **Total** | 30 | 1 | 16 | | 24 | 18 | 1 | | 26 | 2 | 15 | 4 | 28 | 19 | 2 | 15 | 1 | 21 | | | 212 | 11 | 125 | 11 |

# A4

# Individual Characterization Study

# A4.1. Normality Assessments

**Tests of Normality**

| | Kolmogorov-Smirnov[a] | | | Shapiro-Wilk | | |
|---|---|---|---|---|---|---|
| | Statistic | df | Sig. | Statistic | df | Sig. |
| Age | .092 | 51 | .200* | .971 | 51 | .248 |
| Onset | .144 | 51 | .010 | .925 | 51 | .003 |
| Years_Impaired | .112 | 51 | .155 | .947 | 51 | .023 |
| Age_Learned_Braille | .135 | 46 | .036 | .924 | 46 | .005 |
| Years_reading_Braille | .128 | 46 | .059 | .949 | 46 | .042 |
| Years_using_Mobile | .125 | 50 | .050 | .972 | 50 | .290 |
| SMS_per_day | .408 | 50 | .000 | .269 | 50 | .000 |
| Used_Acuity | .447 | 51 | .000 | .584 | 51 | .000 |
| Used_Sensitivity | .243 | 51 | .000 | .816 | 51 | .000 |
| Spatial_ability | .142 | 48 | .016 | .941 | 48 | .017 |
| Digit_span | .155 | 48 | .006 | .942 | 48 | .019 |
| Verbal_IQ | .136 | 48 | .026 | .955 | 48 | .061 |
| PC_WPM | .128 | 37 | .129 | .942 | 37 | .052 |
| PC_MSD_ER | .320 | 34 | .000 | .507 | 34 | .000 |
| MOBILE_WPM | .095 | 43 | .200* | .950 | 43 | .060 |
| MOBILE_MSD_ER | .299 | 35 | .000 | .589 | 35 | .000 |
| BRAILLE_READ_WPM | .194 | 45 | .000 | .847 | 45 | .000 |
| BRAILLE_WRITE_WPM | .142 | 45 | .024 | .891 | 45 | .001 |
| BRAILLE_WRITE_MSD | .342 | 39 | .000 | .587 | 39 | .000 |

*. This is a lower bound of the true significance.

a. Lilliefors Significance Correction

Table A4.1: Normality Assessments for continuous variables

# A4.2.  Individual (Profile, Tactile and Cognitive) Correlations

| | | Age | Education Group | Onset | Years Impaired | Braille Reading Rating | Braille Writing Rating | Age Learned Braille | Years reading Braille | Years using Mobile | SMS per day | Used Acuity | Used Sensitivity | Spatial ability | Digit span |
|---|---|---|---|---|---|---|---|---|---|---|---|---|---|---|---|
| Age | Correlation | 1 | -.527** | -.065 | .666** | -.038 | -.110 | .060 | .668** | -.029 | -.342* | .227 | .494** | -.232 | -.099 |
| | Sig. (2-tailed) | | .000 | .649 | .000 | .792 | .443 | .691 | .000 | .840 | .015 | .109 | .000 | .112 | .503 |
| | N | 51 | 51 | 51 | 51 | 51 | 51 | 46 | 46 | 50 | 50 | 51 | 51 | 48 | 48 |
| Education Group | Correlation | -.527** | 1 | -.071 | -0.27259 | .127 | .233 | -.209 | -0.2198 | 0.0844418 | .105354 | -.180 | -.446** | .508** | .513** |
| | Sig. (2-tailed) | 7E-05 | | .622 | .053 | .374 | .100 | .164 | .142 | .560 | .467 | .205 | .001 | .000 | .000 |
| | N | 51 | 51 | 51 | 51 | 51 | 51 | 46 | 46 | 50 | 50 | 51 | 51 | 48 | 48 |
| Onset | Correlation | -.065 | -0.070661 | 1.000 | -.788** | -.517** | -.464** | .713** | -.603** | .277 | -0.07392 | .048 | 0.1676577 | -.052 | -.280 |
| | Sig. (2-tailed) | .649 | .622 | | .000 | .000 | .001 | .000 | .000 | .051 | .610 | .736 | .240 | .724 | .054 |
| | N | 51 | 51 | 51 | 51 | 51 | 51 | 46 | 46 | 50 | 50 | 51 | 51 | 48 | 48 |
| Years Impaired | Correlation | .666** | -0.272593 | -.788** | 1 | .363** | .279* | -.474** | .852** | -.225 | -0.1594 | .104 | 0.1794007 | -.183 | .138 |
| | Sig. (2-tailed) | 1E-07 | .053 | .000 | | .009 | .047 | .001 | .000 | .115 | .269 | .467 | .208 | .212 | .350 |
| | N | 51 | 51 | 51 | 51 | 51 | 51 | 46 | 46 | 50 | 50 | 51 | 51 | 48 | 48 |
| Braille Reading Classification | Correlation | 0 | 0.1270737 | -.517** | .363** | 1.000 | .821** | -.492** | .502** | -.255 | 0.042556 | -.053 | 0.0254972 | -.049 | .366* |
| | Sig. (2-tailed) | .792 | .374 | .000 | .009 | | .000 | .001 | .000 | .073 | .769 | .710 | .859 | .740 | .011 |
| | N | 51 | 51 | 51 | 51 | 51 | 51 | 46 | 46 | 50 | 50 | 51 | 51 | 48 | 48 |
| Braille Writing Classification | Correlation | 0 | 0.2327865 | -.464** | .279* | .821** | 1.000 | -.310* | .299* | -.212 | -.117 | -.019 | .040 | -.018 | .287* |
| | Sig. (2-tailed) | .443 | .100 | .001 | .047 | .000 | | .036 | .043 | .139 | .417 | .893 | .781 | .905 | .048 |
| | N | 51 | 51 | 51 | 51 | 51 | 51 | 46 | 46 | 50 | 50 | 51 | 51 | 48 | 48 |
| Age Learned Braille | Correlation | 0 | -0.208699 | .713** | -.474** | -.492** | -.310* | 1.000 | -.702** | .034 | -0.14122 | .029 | .397** | -.038 | -.464** |
| | Sig. (2-tailed) | .691 | .164 | .000 | .001 | .001 | .036 | | .000 | .825 | .349 | .849 | .006 | .805 | .002 |
| | N | 46 | 46 | 46 | 46 | 46 | 46 | 46 | 46 | 46 | 46 | 46 | 46 | 44 | 44 |
| Years reading Braille | Correlation | .668** | -0.219757 | -.603** | .852** | .502** | .299* | -.702** | 1 | -.212 | -0.11729 | -.019 | 0.0939609 | -0.1388 | 0.27945 |
| | Sig. (2-tailed) | 4E-07 | .142 | .000 | .000 | .000 | .043 | .000 | | .707 | .367 | .414 | .535 | .369 | .066 |
| | N | 46 | 46 | 46 | 46 | 46 | 46 | 46 | 46 | 46 | 46 | 46 | 46 | 44 | 44 |
| Years using Mobile | Correlation | 0 | 0.0844418 | .277 | -.225 | -.255 | -.212 | .034 | -.212 | 1.000 | .09561 | -.064 | -0.193858 | .211 | -.043 |
| | Sig. (2-tailed) | .84 | .560 | .051 | .115 | .073 | .139 | .825 | .707 | | .509 | .661 | .177 | .155 | .776 |
| | N | 50 | 50 | 50 | 50 | 50 | 50 | 46 | 46 | 50 | 50 | 50 | 50 | 47 | 47 |
| SMS per day | Correlation | -.342* | 0.1053538 | -.074 | -0.1594 | .043 | -.117 | -.141 | -0.1361 | .096 | 1 | -.102 | -0.210309 | .095 | .247 |
| | Sig. (2-tailed) | .015 | .467 | .610 | .269 | .769 | .417 | .349 | .367 | .509 | | .483 | .143 | .525 | .094 |
| | N | 50 | 50 | 50 | 50 | 50 | 50 | 46 | 46 | 50 | 50 | 50 | 50 | 47 | 47 |
| Used Acuity | Correlation | 0 | -0.180343 | .048 | 0.104123 | -.053 | -.019 | .029 | 0.12338 | -.064 | -0.10156 | 1 | 0.0762057 | -.371** | -.138 |
| | Sig. (2-tailed) | .109 | .205 | .736 | .467 | .710 | .893 | .849 | .414 | .661 | .483 | | .595 | .009 | .349 |
| | N | 51 | 51 | 51 | 51 | 51 | 51 | 46 | 46 | 50 | 50 | 51 | 51 | 48 | 48 |
| Used Sensitivity | Correlation | .494** | -.446** | .168 | 0.179401 | .025 | .040 | .397** | 0.09396 | -.194 | -0.21031 | .076 | 1 | -.164 | -.369** |
| | Sig. (2-tailed) | 2E-04 | .001 | .240 | .208 | .859 | .781 | .006 | .535 | .177 | .143 | .595 | | .266 | .010 |
| | N | 51 | 51 | 51 | 51 | 51 | 51 | 46 | 46 | 50 | 50 | 51 | 51 | 48 | 48 |
| Spatial ability | Correlation | 0 | .508** | -.052 | -0.18342 | -.049 | -.018 | -.038 | -0.1388 | .211 | .09514 | -.371** | -0.163914 | 1 | .323* |
| | Sig. (2-tailed) | .112 | .000 | .724 | .212 | .740 | .905 | .805 | .369 | .155 | .525 | .009 | .266 | | .025 |
| | N | 48 | 48 | 48 | 48 | 48 | 48 | 44 | 44 | 47 | 47 | 48 | 48 | 48 | 48 |
| Digit span | Correlation | 0 | .513** | -.280 | 0.137857 | .366* | .287* | -.464** | 0.27945 | -.043 | .247 | -.138 | -.369** | .323* | 1.000 |
| | Sig. (2-tailed) | .503 | .000 | .054 | .350 | .011 | .048 | .002 | .066 | .776 | .094 | .349 | .010 | .025 | |
| | N | 48 | 48 | 48 | 48 | 48 | 48 | 44 | 44 | 47 | 47 | 48 | 48 | 48 | 48 |

Table A4.2: Parametric (Pearson) correlations between individual variables (it is advised to look at the normality table to verify if this value should be used)

| | | Age | Education Group | Onset | Years Impaired | Braille Reading Rating | Braille Writing Rating | Age Learned Braille | Years reading Braille | Years using Mobile | SMS per day | Used Acuity | Used Sensitivity | Spatial ability | Digit span |
|---|---|---|---|---|---|---|---|---|---|---|---|---|---|---|---|
| Age | Correlation | 1.000 | -.545** | -0.070 | .613** | 0.002 | -0.051 | 0.114 | .590** | -0.023 | -0.251 | 0.220 | .511** | -0.188 | -0.093 |
| | Sig. (2-tailed) | | 0.000 | 0.627 | 0.000 | 0.986 | 0.725 | 0.452 | 0.000 | 0.873 | 0.079 | 0.121 | 0.000 | 0.201 | 0.531 |
| | N | 51.000 | 51.000 | 51.000 | 51.000 | 51.000 | 51.000 | 46.000 | 46.000 | 50.000 | 50.000 | 51.000 | 51.000 | 48.000 | 48.000 |
| Education Group | Correlation | -.545** | 1.000 | -0.085 | -0.258 | 0.088 | 0.227 | -0.203 | -0.179 | 0.128 | -0.027 | -0.196 | -.428** | .457** | .518** |
| | Sig. (2-tailed) | 0.000 | | 0.551 | 0.067 | 0.537 | 0.110 | 0.176 | 0.234 | 0.376 | 0.853 | 0.167 | 0.002 | 0.001 | 0.000 |
| | N | 51.000 | 51.000 | 51.000 | 51.000 | 51.000 | 51.000 | 46.000 | 46.000 | 50.000 | 50.000 | 51.000 | 51.000 | 48.000 | 48.000 |
| Onset | Correlation | -0.070 | -0.085 | 1.000 | -.797** | -.556** | -.460** | .662** | -.604** | .282* | 0.120 | 0.176 | 0.150 | 0.095 | -0.279 |
| | Sig. (2-tailed) | 0.627 | 0.551 | | 0.000 | 0.000 | 0.001 | 0.000 | 0.000 | 0.047 | 0.408 | 0.217 | 0.295 | 0.521 | 0.055 |
| | N | 51.000 | 51.000 | 51.000 | 51.000 | 51.000 | 51.000 | 46.000 | 46.000 | 50.000 | 50.000 | 51.000 | 51.000 | 48.000 | 48.000 |
| Years Impaired | Correlation | .613** | -0.258 | -.797** | 1.000 | .402** | .337* | -.429** | .812** | -0.203 | -0.244 | -0.003 | 0.151 | -0.125 | 0.130 |
| | Sig. (2-tailed) | 0.000 | 0.067 | 0.000 | | 0.003 | 0.016 | 0.003 | 0.000 | 0.157 | 0.088 | 0.982 | 0.291 | 0.397 | 0.377 |
| | N | 51.000 | 51.000 | 51.000 | 51.000 | 51.000 | 51.000 | 46.000 | 46.000 | 50.000 | 50.000 | 51.000 | 51.000 | 48.000 | 48.000 |
| Braille Reading Classification | Correlation | 0.002 | 0.088 | -.556** | .402** | 1.000 | .749** | -.545** | .512** | -0.271 | -0.012 | -0.104 | -0.039 | -0.083 | .416** |
| | Sig. (2-tailed) | 0.986 | 0.537 | 0.000 | 0.003 | | 0.000 | 0.000 | 0.000 | 0.057 | 0.937 | 0.466 | 0.784 | 0.574 | 0.003 |
| | N | 51.000 | 51.000 | 51.000 | 51.000 | 51.000 | 51.000 | 46.000 | 46.000 | 50.000 | 50.000 | 51.000 | 51.000 | 48.000 | 48.000 |
| Braille Writing Classification | Correlation | -0.051 | 0.227 | -.460** | .337* | .749** | 1.000 | -.359* | .354* | -0.218 | -0.003 | -0.087 | 0.003 | -0.014 | .362* |
| | Sig. (2-tailed) | 0.725 | 0.110 | 0.001 | 0.016 | 0.000 | | 0.014 | 0.016 | 0.128 | 0.985 | 0.543 | 0.982 | 0.923 | 0.011 |
| | N | 51.000 | 51.000 | 51.000 | 51.000 | 51.000 | 51.000 | 46.000 | 46.000 | 50.000 | 50.000 | 51.000 | 51.000 | 48.000 | 48.000 |
| Age Learned Braille | Correlation | 0.114 | -0.203 | .662** | -.429** | -.545** | -.359* | 1.000 | -.714** | 0.129 | 0.057 | 0.096 | .417** | 0.028 | -.457** |
| | Sig. (2-tailed) | 0.452 | 0.176 | 0.000 | 0.003 | 0.000 | 0.014 | | 0.000 | 0.391 | 0.708 | 0.525 | 0.004 | 0.857 | 0.002 |
| | N | 46.000 | 46.000 | 46.000 | 46.000 | 46.000 | 46.000 | 46.000 | 46.000 | 46.000 | 46.000 | 46.000 | 46.000 | 44.000 | 44.000 |
| Years reading Braille | Correlation | .590** | -0.179 | -.604** | .812** | .512** | .354* | -.714** | 1.000 | -0.108 | -0.228 | 0.051 | 0.054 | -0.107 | 0.275 |
| | Sig. (2-tailed) | 0.000 | 0.234 | 0.000 | 0.000 | 0.000 | 0.016 | 0.000 | | 0.475 | 0.128 | 0.736 | 0.723 | 0.491 | 0.070 |
| | N | 46.000 | 46.000 | 46.000 | 46.000 | 46.000 | 46.000 | 46.000 | 46.000 | 46.000 | 46.000 | 46.000 | 46.000 | 44.000 | 44.000 |
| Years using Mobile | Correlation | -0.023 | 0.128 | .282* | -0.203 | -0.271 | -0.218 | 0.129 | -0.108 | 1.000 | 0.073 | -0.053 | -0.220 | 0.178 | -0.002 |
| | Sig. (2-tailed) | 0.873 | 0.376 | 0.047 | 0.157 | 0.057 | 0.128 | 0.391 | 0.475 | | 0.612 | 0.716 | 0.124 | 0.231 | 0.989 |
| | N | 50.000 | 50.000 | 50.000 | 50.000 | 50.000 | 50.000 | 46.000 | 46.000 | 50.000 | 50.000 | 50.000 | 50.000 | 47.000 | 47.000 |
| SMS per day | Correlation | -0.251 | -0.027 | 0.120 | -0.244 | -0.012 | -0.003 | 0.057 | -0.228 | 0.073 | 1.000 | 0.005 | -0.035 | -0.021 | -0.117 |
| | Sig. (2-tailed) | 0.079 | 0.853 | 0.408 | 0.088 | 0.937 | 0.985 | 0.708 | 0.128 | 0.612 | | 0.972 | 0.809 | 0.888 | 0.434 |
| | N | 50.000 | 50.000 | 50.000 | 50.000 | 50.000 | 50.000 | 46.000 | 46.000 | 50.000 | 50.000 | 50.000 | 50.000 | 47.000 | 47.000 |
| Used Acuity | Correlation | 0.220 | -0.196 | 0.176 | -0.003 | -0.104 | -0.087 | 0.096 | 0.051 | -0.053 | 0.005 | 1.000 | 0.081 | -.395** | -0.161 |
| | Sig. (2-tailed) | 0.121 | 0.167 | 0.217 | 0.982 | 0.466 | 0.543 | 0.525 | 0.736 | 0.716 | 0.972 | | 0.572 | 0.005 | 0.275 |
| | N | 51.000 | 51.000 | 51.000 | 51.000 | 51.000 | 51.000 | 46.000 | 46.000 | 50.000 | 50.000 | 51.000 | 51.000 | 48.000 | 48.000 |
| Used Sensitivity | Correlation | .511** | -.428** | 0.150 | 0.151 | -0.039 | 0.003 | .417** | 0.054 | -0.220 | -0.035 | 0.081 | 1.000 | -0.122 | -.354* |
| | Sig. (2-tailed) | 0.000 | 0.002 | 0.295 | 0.291 | 0.784 | 0.982 | 0.004 | 0.723 | 0.124 | 0.809 | 0.572 | | 0.408 | 0.014 |
| | N | 51.000 | 51.000 | 51.000 | 51.000 | 51.000 | 51.000 | 46.000 | 46.000 | 50.000 | 50.000 | 51.000 | 51.000 | 48.000 | 48.000 |
| Spatial ability | Correlation | -0.188 | .457** | -0.125 | -0.083 | -0.083 | -0.014 | 0.028 | -0.107 | 0.178 | -0.021 | -.395** | -0.122 | 1.000 | 0.202 |
| | Sig. (2-tailed) | 0.201 | 0.001 | 0.521 | 0.397 | 0.574 | 0.923 | 0.857 | 0.491 | 0.231 | 0.888 | 0.005 | 0.408 | | 0.168 |
| | N | 48.000 | 48.000 | 48.000 | 48.000 | 48.000 | 48.000 | 44.000 | 44.000 | 47.000 | 47.000 | 48.000 | 48.000 | 48.000 | 48.000 |
| Digit span | Correlation | -0.093 | .518** | -0.279 | 0.130 | .416** | .362* | -.457** | 0.275 | -0.002 | -0.117 | -0.161 | -.354* | 0.202 | 1.000 |
| | Sig. (2-tailed) | 0.531 | 0.000 | 0.055 | 0.377 | 0.003 | 0.011 | 0.002 | 0.070 | 0.989 | 0.434 | 0.275 | 0.014 | 0.168 | |
| | N | 48.000 | 48.000 | 48.000 | 48.000 | 48.000 | 48.000 | 44.000 | 44.000 | 47.000 | 47.000 | 48.000 | 48.000 | 48.000 | 48.000 |

Table A4.3: Non-Parametric (Spearman) correlations between individual variables (it is advised to look at the normality table to verify if this value should be used)

# A4.3.  Individual–Functional Correlations

| | | Age | Education Group | Onset | Years Impaired | Desktop WPM | Desktop MSD Error Rate | Used Acuity | Used Sensitivity | Spatial ability | Digit span |
|---|---|---|---|---|---|---|---|---|---|---|---|
| Age | Correlation | 1 | -.527** | -.065 | .666** | -.585** | .112 | .227 | .494** | -.232 | -.099 |
| | Sig. (2-tailed) | | .000 | .649 | .000 | .000 | .528 | .109 | .000 | .112 | .503 |
| | N | 51 | 51 | 51 | 51 | 37 | 34 | 51 | 51 | 48 | 48 |
| Education Group | Correlation | -.527** | 1 | -.071 | -.273 | .537** | -.073 | -.180 | -.446** | .508** | .513** |
| | Sig. (2-tailed) | .000 | | .622 | .053 | .001 | .682 | .205 | .001 | .000 | .000 |
| | N | 51 | 51 | 51 | 51 | 37 | 34 | 51 | 51 | 48 | 48 |
| Onset | Correlation | -.065 | -.071 | 1 | -.788** | .030 | .023 | .048 | .168 | .052 | -.280 |
| | Sig. (2-tailed) | .649 | .622 | | .000 | .862 | .897 | .736 | .240 | .724 | .054 |
| | N | 51 | 51 | 51 | 51 | 37 | 34 | 51 | 51 | 48 | 48 |
| Years Impaired | Correlation | .666** | -.273 | -.788** | 1 | -.406* | .058 | .104 | .179 | -.183 | .138 |
| | Sig. (2-tailed) | .000 | .053 | .000 | | .013 | .743 | .467 | .212 | .212 | .350 |
| | N | 51 | 51 | 51 | 51 | 37 | 34 | 51 | 51 | 48 | 48 |
| **Desktop WPM** | Correlation | -.585** | .537** | .030 | -.406* | 1 | -.406* | -.334* | -.282 | .468** | .391* |
| | Sig. (2-tailed) | .000 | .001 | .862 | .013 | | .017 | .044 | .090 | .003 | .017 |
| | N | 37 | 37 | 37 | 37 | 37 | 34 | 37 | 37 | 37 | 37 |
| **Desktop MSD Error Rate** | Correlation | .112 | -.073 | .023 | .058 | -.406* | 1 | .380* | -.067 | -.221 | -.214 |
| | Sig. (2-tailed) | .528 | .682 | .897 | .743 | .017 | | .027 | .707 | .209 | .224 |
| | N | 34 | 34 | 34 | 34 | 34 | 34 | 34 | 34 | 34 | 34 |
| Used Acuity | Correlation | .227 | -.180 | .048 | .104 | -.334* | .380* | 1 | .076 | -.371** | -.138 |
| | Sig. (2-tailed) | .109 | .205 | .736 | .467 | .044 | .027 | | .595 | .009 | .349 |
| | N | 51 | 51 | 51 | 51 | 37 | 34 | 51 | 51 | 48 | 48 |
| Used Sensitivity | Correlation | .494** | -.446** | .168 | .179 | -.282 | -.067 | .076 | 1 | -.164 | -.369** |
| | Sig. (2-tailed) | .000 | .001 | .240 | .208 | .090 | .707 | .595 | | .266 | .010 |
| | N | 51 | 51 | 51 | 51 | 37 | 34 | 51 | 51 | 48 | 48 |
| Spatial ability | Correlation | -.232 | .508** | .052 | -.183 | .468** | -.221 | -.371** | -.164 | 1 | .323* |
| | Sig. (2-tailed) | .112 | .000 | .724 | .212 | .003 | .209 | .009 | .266 | | .025 |
| | N | 48 | 48 | 48 | 48 | 37 | 34 | 48 | 48 | 48 | 48 |
| Digit span | Correlation | -.099 | .513** | -.280 | .138 | .391* | -.214 | -.138 | -.369** | .323* | 1 |
| | Sig. (2-tailed) | .503 | .000 | .054 | .350 | .017 | .224 | .349 | .010 | .025 | |
| | N | 48 | 48 | 48 | 48 | 37 | 34 | 48 | 48 | 48 | 48 |

Table A4.4: Parametric (Pearson) correlations between desktop functional and individual variables (it is advised to look at the normality table to verify if this value should be used)

| | | Age | Education Group | Onset | Years Impaired | Desktop WPM | Desktop MSD Error Rate | Used Acuity | Used Sensitivity | Spatial ability | Digit span |
|---|---|---|---|---|---|---|---|---|---|---|---|
| Age | Correlation | 1.000 | -.545** | -.070 | .613** | -.533** | .081 | .220 | .511** | -.188 | -.093 |
| | Sig. (2-tailed) | | .000 | .627 | .000 | .001 | .650 | .121 | .000 | .201 | .531 |
| | N | 51 | 51 | 51 | 51 | 37 | 34 | 51 | 51 | 48 | 48 |
| Education Group | Correlation | -.545** | 1.000 | -.085 | -.258 | .463** | .048 | -.196 | -.428** | .457** | .518** |
| | Sig. (2-tailed) | .000 | | .551 | .067 | .004 | .787 | .167 | .002 | .001 | .000 |
| | N | 51 | 51 | 51 | 51 | 37 | 34 | 51 | 51 | 48 | 48 |
| Onset | Correlation | -.070 | -.085 | 1.000 | -.797** | .044 | .023 | .176 | .150 | .095 | -.279 |
| | Sig. (2-tailed) | .627 | .551 | | .000 | .795 | .897 | .217 | .295 | .521 | .055 |
| | N | 51 | 51 | 51 | 51 | 37 | 34 | 51 | 51 | 48 | 48 |
| Years Impaired | Correlation | .613** | -.258 | -.797** | 1.000 | -.413* | .093 | -.003 | .151 | -.125 | .130 |
| | Sig. (2-tailed) | .000 | .067 | .000 | | .011 | .601 | .982 | .291 | .397 | .377 |
| | N | 51 | 51 | 51 | 51 | 37 | 34 | 51 | 51 | 48 | 48 |
| Desktop WPM | Correlation | -.533** | .463** | .044 | -.413* | 1.000 | -.533** | -.307 | -.257 | .399* | .303 |
| | Sig. (2-tailed) | .001 | .004 | .795 | .011 | | .001 | .064 | .124 | .014 | .068 |
| | N | 37 | 37 | 37 | 37 | 37 | 37 | 37 | 37 | 37 | 37 |
| Desktop MSD Error Rate | Correlation | .081 | .048 | .023 | .093 | -.533** | 1.000 | .194 | .074 | .081 | -.267 |
| | Sig. (2-tailed) | .650 | .787 | .897 | .601 | .001 | | .271 | .677 | .650 | .126 |
| | N | 34 | 34 | 34 | 34 | 34 | 34 | 34 | 34 | 34 | 34 |
| Used Acuity | Correlation | .220 | -.196 | .176 | -.003 | -.307 | .194 | 1.000 | .081 | -.395** | -.161 |
| | Sig. (2-tailed) | .121 | .167 | .217 | .982 | .064 | .271 | | .572 | .005 | .275 |
| | N | 51 | 51 | 51 | 51 | 37 | 34 | 51 | 51 | 48 | 48 |
| Used Sensitivity | Correlation | .511** | -.428** | .150 | .151 | -.257 | .074 | .081 | 1.000 | -.122 | -.354* |
| | Sig. (2-tailed) | .000 | .002 | .295 | .291 | .124 | .677 | .572 | | .408 | .014 |
| | N | 51 | 51 | 51 | 51 | 37 | 34 | 51 | 51 | 48 | 48 |
| Spatial ability | Correlation | -.188 | .457** | .095 | -.125 | .399* | .081 | -.395** | -.122 | 1.000 | .202 |
| | Sig. (2-tailed) | .201 | .001 | .521 | .397 | .014 | .650 | .005 | .408 | | .168 |
| | N | 48 | 48 | 48 | 48 | 37 | 34 | 48 | 48 | 48 | 48 |
| Digit span | Correlation | -.093 | .518** | -.279 | .130 | .303 | -.267 | -.161 | -.354* | .202 | 1.000 |
| | Sig. (2-tailed) | .531 | .000 | .055 | .377 | .068 | .126 | .275 | .014 | .168 | |
| | N | 48 | 48 | 48 | 48 | 37 | 34 | 48 | 48 | 48 | 48 |

Table A4.5: Non-Parametric (Spearman) correlations between desktop functional and individual variables (it is advised to look at the normality table to verify if this value should be used)

| | | Age | Education Group | Onset | Years Impaired | **Mobile WPM** | **Mobile MSD** | Years using Mobile | SMS per day | Used Acuity | Used Sensitivity | Spatial ability | Digit span |
|---|---|---|---|---|---|---|---|---|---|---|---|---|---|
| Age | Correlation | 1 | -.527** | -.065 | .666** | -.618** | .311 | -.029 | -.342* | .227 | .494** | -.232 | -.099 |
| | Sig. (2-tailed) | | .000 | .649 | .000 | .000 | .069 | .840 | .015 | .109 | .000 | .112 | .503 |
| | N | 51 | 51 | 51 | 51 | 43 | 35 | 50 | 50 | 51 | 51 | 48 | 48 |
| Education Group | Correlation | -.527** | 1 | -.071 | -.273 | .373* | -.263 | .084 | .105 | -.180 | -.446** | .508** | .513** |
| | Sig. (2-tailed) | .000 | | .622 | .053 | .014 | .127 | .560 | .467 | .205 | .001 | .000 | .000 |
| | N | 51 | 51 | 51 | 51 | 43 | 35 | 50 | 50 | 51 | 51 | 48 | 48 |
| Onset | Correlation | -.065 | -.071 | 1 | -.788** | -.023 | .013 | .277 | -.074 | .048 | .168 | .052 | -.280 |
| | Sig. (2-tailed) | .649 | .622 | | .000 | .885 | .941 | .051 | .610 | .736 | .240 | .724 | .054 |
| | N | 51 | 51 | 51 | 51 | 43 | 35 | 50 | 50 | 51 | 51 | 48 | 48 |
| Years Impaired | Correlation | .666** | -.273 | -.788** | 1 | -.372* | .196 | -.225 | -.159 | .104 | .179 | -.183 | .138 |
| | Sig. (2-tailed) | .000 | .053 | .000 | | .014 | .258 | .115 | .269 | .467 | .208 | .212 | .350 |
| | N | 51 | 51 | 51 | 51 | 43 | 35 | 50 | 50 | 51 | 51 | 48 | 48 |
| **Mobile WPM** | Correlation | -.618** | .373* | -.023 | -.372* | 1 | -.431** | .225 | .489** | -.232 | -.384* | .301 | .379* |
| | Sig. (2-tailed) | .000 | .014 | .885 | .014 | | .010 | .151 | .001 | .135 | .011 | .050 | .012 |
| | N | 43 | 43 | 43 | 43 | 43 | 35 | 42 | 42 | 43 | 43 | 43 | 43 |
| **Mobile MSD Error Rate** | Correlation | .311 | -.263 | .013 | .196 | -.431** | 1 | .266 | -.139 | .405* | -.177 | -.207 | -.280 |
| | Sig. (2-tailed) | .069 | .127 | .941 | .258 | .010 | | .128 | .433 | .016 | .310 | .234 | .104 |
| | N | 35 | 35 | 35 | 35 | 35 | 35 | 34 | 34 | 35 | 35 | 35 | 35 |
| Years using Mobile | Correlation | -.029 | .084 | .277 | -.225 | .225 | .266 | 1 | .096 | -.064 | -.194 | .211 | -.043 |
| | Sig. (2-tailed) | .840 | .560 | .051 | .115 | .151 | .128 | | .509 | .661 | .177 | .155 | .776 |
| | N | 50 | 50 | 50 | 50 | 42 | 34 | 50 | 50 | 50 | 50 | 47 | 47 |
| SMS per day | Correlation | -.342* | .105 | -.074 | -.159 | .489** | -.139 | .096 | 1 | -.102 | -.210 | .095 | .247 |
| | Sig. (2-tailed) | .015 | .467 | .610 | .269 | .001 | .433 | .509 | | .483 | .143 | .525 | .094 |
| | N | 50 | 50 | 50 | 50 | 42 | 34 | 50 | 50 | 50 | 50 | 47 | 47 |
| Used Acuity | Correlation | .227 | -.180 | .048 | .104 | -.232 | .405* | -.064 | -.102 | 1 | .076 | -.211 | -.138 |
| | Sig. (2-tailed) | .109 | .205 | .736 | .467 | .135 | .016 | .661 | .483 | | .595 | .009 | .349 |
| | N | 51 | 51 | 51 | 51 | 43 | 35 | 50 | 50 | 51 | 51 | 48 | 48 |
| Used Sensitivity | Correlation | .494** | -.446** | .168 | .179 | -.384* | -.177 | -.194 | -.210 | .076 | 1 | -.164 | -.369** |
| | Sig. (2-tailed) | .000 | .001 | .240 | .208 | .011 | .310 | .177 | .143 | .595 | | .266 | .010 |
| | N | 51 | 51 | 51 | 51 | 43 | 35 | 50 | 50 | 51 | 51 | 48 | 48 |
| Spatial ability | Correlation | -.232 | .508** | .052 | -.183 | .301 | -.207 | .211 | .095 | -.371** | -.164 | 1 | .323* |
| | Sig. (2-tailed) | .112 | .000 | .724 | .212 | .050 | .234 | .155 | .525 | .009 | .266 | | .025 |
| | N | 48 | 48 | 48 | 48 | 43 | 35 | 47 | 47 | 48 | 48 | 48 | 48 |
| Digit span | Correlation | -.099 | .513** | -.280 | .138 | .379* | -.280 | -.043 | .247 | -.138 | -.369** | .323* | 1 |
| | Sig. (2-tailed) | .503 | .000 | .054 | .350 | .012 | .104 | .776 | .094 | .349 | .010 | .025 | |
| | N | 48 | 48 | 48 | 48 | 43 | 35 | 47 | 47 | 48 | 48 | 48 | 48 |

Table A4.6: Parametric (Pearson) correlations between mobile functional and individual variables (it is advised to look at the normality table to verify if this value should be used)

| | | Age | Education Group | Onset | Years Impaired | **Mobile WPM** | **Mobile MSD** | Years using Mobile | SMS per day | Used Acuity | Used Sensitivity | Spatial ability | Digit span |
|---|---|---|---|---|---|---|---|---|---|---|---|---|---|
| Age | Correlation | 1.000 | -.545** | -.070 | .613** | -.598** | .328 | -.023 | -.251 | .220 | .511** | -.188 | -.093 |
| | Sig. (2-tailed) | | .000 | .627 | .000 | .000 | .054 | .873 | .079 | .121 | .000 | .201 | .531 |
| | N | 51 | 51 | 51 | 51 | 43 | 35 | 50 | 50 | 51 | 51 | 48 | 48 |
| Education Group | Correlation | -.545** | 1 | -.085 | -0.2582 | .375* | -.342* | .128 | -.027 | -.196 | -.428** | .457** | .518** |
| | Sig. (2-tailed) | 4E-05 | | .551 | .067 | .013 | .044 | .376 | .853 | .167 | .002 | .001 | .000 |
| | N | 51 | 51 | 51 | 51 | 43 | 35 | 50 | 50 | 51 | 51 | 48 | 48 |
| Onset | Correlation | -0.070 | -0.085445 | 1.000 | -.797** | 0.03672 | .232 | .282* | .120 | .176 | 0.1496229 | .095 | -.279 |
| | Sig. (2-tailed) | 0.627 | .551 | | .000 | .815 | .179 | .047 | .408 | .217 | .295 | .521 | .055 |
| | N | 51 | 51 | 51 | 51 | 43 | 35 | 50 | 50 | 51 | 51 | 48 | 48 |
| Years Impaired | Correlation | .613** | -0.258201 | -.797** | 1 | -.357* | .077 | -.203 | -.244 | .176 | 0.1506563 | -.125 | .130 |
| | Sig. (2-tailed) | 2E-06 | .067 | .000 | | .019 | .659 | .157 | .088 | .982 | .291 | .397 | .377 |
| | N | 51 | 51 | 51 | 51 | 43 | 35 | 50 | 50 | 51 | 51 | 48 | 48 |
| **Mobile WPM** | Correlation | -.598** | .375* | .037 | -.357* | 1 | -.599** | .178 | .626** | -.194 | -.355* | .285 | .298 |
| | Sig. (2-tailed) | 2E-05 | .013 | .815 | .019 | | .000 | .259 | .000 | .213 | .019 | .064 | .052 |
| | N | 43 | 43 | 43 | 43 | 43 | 35 | 42 | 43 | 43 | 43 | 43 | 43 |
| **Mobile MSD Error Rate** | Correlation | 0.328 | -.342* | .232 | 0.077382 | -.599** | 1.000 | .148 | -.331 | .322 | -0.061957 | -.085 | -.376* |
| | Sig. (2-tailed) | .054 | .044 | .179 | .659 | .000 | | .405 | .056 | .059 | .724 | .625 | .026 |
| | N | 35 | 35 | 35 | 35 | 35 | 35 | 34 | 34 | 35 | 35 | 35 | 35 |
| Years using Mobile | Correlation | -0.023 | 0.1278754 | .282* | -0.20318 | 0.17822 | .148 | 1.000 | .073 | -.053 | -0.220409 | .178 | -.002 |
| | Sig. (2-tailed) | .873 | .376 | .047 | .157 | .259 | .405 | | .612 | .716 | .124 | .231 | .989 |
| | N | 50 | 50 | 50 | 50 | 42 | 34 | 50 | 50 | 50 | 50 | 47 | 47 |
| SMS per day | Correlation | -0.251 | -0.02684 | .120 | -0.2438 | .626** | -.331 | .073 | 1.000 | .005 | -0.035011 | -.021 | -.117 |
| | Sig. (2-tailed) | .079 | .853 | .408 | .088 | .000 | .056 | .612 | | .972 | .809 | .888 | .434 |
| | N | 50 | 50 | 50 | 50 | 42 | 34 | 50 | 50 | 50 | 50 | 47 | 47 |
| Used Acuity | Correlation | 0.220 | -0.196273 | .176 | -0.00332 | -0.19391 | .322 | -.053 | .005 | 1.000 | 0.0809077 | -.395** | -.161 |
| | Sig. (2-tailed) | 0.121 | .167 | .217 | .982 | .213 | .059 | .716 | .972 | | .572 | .005 | .275 |
| | N | 51 | 51 | 51 | 51 | 43 | 35 | 50 | 50 | 51 | 51 | 48 | 48 |
| Used Sensitivity | Correlation | .511** | -.428** | .150 | 0.150656 | -.355* | -.062 | -.220 | -.035 | .081 | 1 | -.122 | -.354* |
| | Sig. (2-tailed) | 1E-04 | .002 | .295 | .291 | .019 | .724 | .124 | .809 | .572 | | .408 | .014 |
| | N | 51 | 51 | 51 | 51 | 43 | 35 | 50 | 50 | 51 | 51 | 48 | 48 |
| Spatial ability | Correlation | -0.188 | .457** | .095 | -0.125 | 0.28523 | -.085 | .178 | -.021 | -.395** | -0.122248 | 1.000 | .202 |
| | Sig. (2-tailed) | 0.201 | .001 | .521 | .397 | .064 | .625 | .231 | .888 | .005 | .408 | | .168 |
| | N | 48 | 48 | 48 | 48 | 43 | 35 | 47 | 47 | 48 | 48 | 48 | 48 |
| Digit span | Correlation | -0.093 | .518** | -.279 | 0.130499 | 0.2983 | -.376* | -.002 | -.117 | -.161 | -.354* | .202 | 1.000 |
| | Sig. (2-tailed) | 0.531 | .000 | .055 | .377 | .052 | .026 | .989 | .434 | .275 | .014 | .168 | |
| | N | 48 | 48 | 48 | 48 | 43 | 35 | 47 | 47 | 48 | 48 | 48 | 48 |

Table A4.7: Non-Parametric (Spearman) correlations between mobile functional and individual variables (it is advised to look at the normality table to verify if this value should be used)

| | | Age | Education Group | Onset | Years Impaired | Braille Reading Rating | Braille Writing Rating | Age Learned Braille | Years reading Braille | **Braille Read WPM** | **Braille Write WPM** | **Braille Write MSD Error** | Used Acuity | Used Sensitivity | Spatial ability | Digit span |
|---|---|---|---|---|---|---|---|---|---|---|---|---|---|---|---|---|
| Age | Correlation | 1 | -.527** | -.065 | .666** | -.038 | -.110 | .060 | .668** | -.038 | -.207 | .089 | .227 | .494** | -.232 | -.099 |
| | Sig. (2-tailed) | | .000 | .649 | .000 | .792 | .443 | .691 | .000 | .805 | .173 | .588 | .109 | .000 | .112 | .503 |
| | N | 51 | 51 | 51 | 51 | 51 | 51 | 46 | 46 | 45 | 45 | 39 | 51 | 51 | 48 | 48 |
| Education Group | Correlation | -.527** | 1 | -.071 | -.273 | .127 | .233 | -.209 | -.220 | .232 | .413** | -.205 | -.180 | -.446** | .508** | .513** |
| | Sig. (2-tailed) | .000 | | .622 | .053 | .374 | .100 | .164 | .142 | .125 | .005 | .210 | .205 | .000 | .000 | .000 |
| | N | 51 | 51 | 51 | 51 | 51 | 51 | 46 | 46 | 45 | 45 | 39 | 51 | 51 | 48 | 48 |
| Onset | Correlation | -.065 | -.071 | 1 | -.788** | -.517** | -.464** | .713** | -.603** | -.679** | -.560** | .305 | .048 | .168 | .052 | -.280 |
| | Sig. (2-tailed) | .649 | .622 | | .000 | .000 | .001 | .000 | .000 | .000 | .000 | .059 | .736 | .240 | .724 | .054 |
| | N | 51 | 51 | 51 | 51 | 51 | 51 | 46 | 46 | 45 | 45 | 39 | 51 | 51 | 48 | 48 |
| Years Impaired | Correlation | .666** | -.273 | -.788** | 1 | .363** | .279* | -.474** | .852** | .461** | .273 | -.148 | .104 | .179 | -.183 | .138 |
| | Sig. (2-tailed) | .000 | .053 | .000 | | .009 | .047 | .001 | .000 | .001 | .070 | .369 | .467 | .208 | .212 | .350 |
| | N | 51 | 51 | 51 | 51 | 51 | 51 | 46 | 46 | 45 | 45 | 39 | 51 | 51 | 48 | 48 |
| Braille Reading Classification | Correlation | -.038 | .127 | -.517** | .363** | 1 | .821** | -.492** | .502** | .728** | .627** | -.686** | -.053 | .025 | -.049 | .366* |
| | Sig. (2-tailed) | .792 | .374 | .000 | .009 | | .000 | .001 | .000 | .000 | .000 | .000 | .710 | .859 | .740 | .011 |
| | N | 51 | 51 | 51 | 51 | 51 | 51 | 46 | 46 | 45 | 45 | 39 | 51 | 51 | 48 | 48 |
| Braille Writing Classification | Correlation | -.110 | .233 | -.464** | .279* | .821** | 1 | -.310* | .299* | .587** | .664** | -.644** | -.019 | .040 | -.018 | .287* |
| | Sig. (2-tailed) | .443 | .100 | .001 | .047 | .000 | | .036 | .043 | .000 | .000 | .000 | .893 | .781 | .905 | .048 |
| | N | 51 | 51 | 51 | 51 | 51 | 51 | 46 | 46 | 45 | 45 | 39 | 51 | 51 | 48 | 48 |
| Age Learned Braille | Correlation | .060 | -.209 | .713** | -.474** | -.492** | -.310* | 1 | -.702** | -.539** | .300 | .554** | .029 | .397** | -.038 | -.464** |
| | Sig. (2-tailed) | .691 | .164 | .000 | .001 | .001 | .036 | | .000 | .000 | .000 | .000 | .849 | .006 | .805 | .002 |
| | N | 46 | 46 | 46 | 46 | 46 | 46 | 46 | 46 | 41 | 41 | 37 | 46 | 46 | 44 | 44 |
| Years reading Braille | Correlation | .668** | -.220 | -.603** | .852** | .502** | .299* | -.702** | 1 | .546** | .300 | -.450** | .123 | .094 | -.139 | .279 |
| | Sig. (2-tailed) | .000 | .142 | .000 | .000 | .000 | .043 | .000 | | .000 | .057 | .005 | .414 | .535 | .369 | .066 |
| | N | 46 | 46 | 46 | 46 | 46 | 46 | 46 | 46 | 41 | 41 | 37 | 46 | 46 | 44 | 44 |
| **Braille Read WPM** | Correlation | -.038 | .232 | -.679** | .461** | .728** | .587** | -.702** | .546** | 1 | .825** | -.455** | -.164 | -.163 | -.067 | .479** |
| | Sig. (2-tailed) | .805 | .125 | .000 | .001 | .000 | .000 | .000 | .000 | | .000 | .004 | .280 | .286 | .664 | .001 |
| | N | 45 | 45 | 45 | 45 | 45 | 45 | 41 | 41 | 45 | 45 | 39 | 45 | 45 | 45 | 45 |
| **Braille Write WPM** | Correlation | -.207 | .413** | -.560** | .273 | .627** | .664** | -.539** | .300 | .825** | 1 | -.448** | -.321* | -.232 | .131 | .425** |
| | Sig. (2-tailed) | .173 | .005 | .000 | .070 | .000 | .000 | .000 | .057 | .000 | | .004 | .031 | .126 | .390 | .004 |
| | N | 45 | 45 | 45 | 45 | 45 | 45 | 41 | 41 | 45 | 45 | 39 | 45 | 45 | 45 | 45 |
| **Braille Write MSD Error** | Correlation | .089 | -.205 | .305 | -.148 | -.686** | -.644** | .554** | -.450** | -.455** | -.448** | 1 | -.106 | -.055 | -.014 | -.168 |
| | Sig. (2-tailed) | .588 | .210 | .059 | .369 | .000 | .000 | .004 | .005 | .004 | .004 | | .522 | .742 | .934 | .308 |
| | N | 39 | 39 | 39 | 39 | 39 | 39 | 37 | 37 | 39 | 39 | 39 | 39 | 39 | 39 | 39 |
| Used Acuity | Correlation | .227 | -.180 | .048 | .104 | -.053 | -.019 | .029 | .123 | -.164 | -.321* | -.106 | 1 | .076 | -.371** | -.138 |
| | Sig. (2-tailed) | .109 | .205 | .736 | .467 | .710 | .893 | .849 | .414 | .280 | .031 | .522 | | .595 | .009 | .349 |
| | N | 51 | 51 | 51 | 51 | 51 | 51 | 46 | 46 | 45 | 45 | 39 | 51 | 51 | 48 | 48 |
| Used Sensitivity | Correlation | .494** | -.446** | .168 | .179 | .025 | .040 | .397** | .094 | -.163 | -.232 | -.055 | .076 | 1 | -.164 | -.369** |
| | Sig. (2-tailed) | .000 | .001 | .240 | .208 | .859 | .781 | .006 | .535 | .286 | .126 | .742 | .595 | | .266 | .010 |
| | N | 51 | 51 | 51 | 51 | 51 | 51 | 46 | 46 | 45 | 45 | 39 | 51 | 51 | 48 | 48 |
| Spatial ability | Correlation | -.232 | .508** | .052 | -.183 | -.049 | -.018 | -.038 | -.139 | -.067 | .131 | -.014 | -.371** | -.164 | 1 | .323* |
| | Sig. (2-tailed) | .112 | .000 | .724 | .212 | .740 | .905 | .805 | .369 | .664 | .390 | .934 | .009 | .266 | | .025 |
| | N | 48 | 48 | 48 | 48 | 48 | 48 | 44 | 44 | 45 | 45 | 39 | 48 | 48 | 48 | 48 |
| Digit span | Correlation | -.099 | .513** | -.280 | .138 | .366* | .287* | -.464** | .279 | .479** | .425** | -.168 | -.138 | -.369** | .323* | 1 |
| | Sig. (2-tailed) | .503 | .000 | .054 | .350 | .011 | .048 | .002 | .066 | .001 | .004 | .308 | .349 | .010 | .025 | |
| | N | 48 | 48 | 48 | 48 | 48 | 48 | 44 | 44 | 45 | 45 | 39 | 48 | 48 | 48 | 48 |

Table A4.8: Parametric (Pearson) correlations between Braille functional and individual variables (it is advised to look at the normality table to verify if this value should be used)

| | | Age | Education Group | Onset | Years Impaired | Braille Reading Rating | Braille Writing Rating | Age Learned Braille | Years reading Braille | **Braille Read WPM** | **Braille Write WPM** | **Braille Write MSD Error** | Used Acuity | Used Sensitivity | Spatial ability | Digit span |
|---|---|---|---|---|---|---|---|---|---|---|---|---|---|---|---|---|
| Age | Correlation | 1.000 | -.545** | -.070 | .613** | .002 | -.051 | .114 | .590** | .004 | -.127 | -.088 | .220 | .511** | -.188 | -.093 |
| | Sig (2-tailed) | | .000 | .627 | .000 | .986 | .725 | .452 | .000 | .979 | .404 | .596 | .121 | .000 | .201 | .531 |
| | N | 51 | 51 | 51 | 51 | 51 | 51 | 46 | 45 | 45 | 45 | 39 | 51 | 51 | 48 | 48 |
| Education Group | Correlation | -.545** | 1.000 | -.085 | -.258 | .088 | .227 | -.203 | -.179 | .217 | .360** | -.171 | -.196 | -.428** | .457** | .518** |
| | Sig (2-tailed) | .000 | | .551 | .067 | .537 | .110 | .176 | .234 | .152 | .015 | .297 | .167 | .002 | .001 | .000 |
| | N | 51 | 51 | 51 | 51 | 51 | 51 | 46 | 46 | 45 | 45 | 39 | 51 | 51 | 48 | 48 |
| Onset | Correlation | -0.07 | -0.085 | 1.000 | -.797** | -.556** | -.460** | .662** | -.604** | -.740** | -.633** | .562** | .176 | 0.1496229 | 0.0949 | -0.279 |
| | Sig (2-tailed) | .627 | 0.5510644 | | .000 | .000 | .001 | .000 | .000 | .000 | .000 | .000 | .217 | .295 | .521 | .055 |
| | N | 51 | 51 | 51 | 51 | 51 | 51 | 46 | 46 | 45 | 45 | 39 | 51 | 51 | 48 | 48 |
| Years Impaired | Correlation | .613** | -.258 | -.797** | 1.000 | .402** | .337** | -.429** | .812** | .514** | .382** | -.665** | -.003 | 0.1506563 | -0.125 | 0.13 |
| | Sig (2-tailed) | .000 | 0.0673448 | .000 | | .003 | .016 | .003 | .000 | .000 | .010 | .013 | .982 | .291 | .397 | .377 |
| | N | 51 | 51 | 51 | 51 | 51 | 51 | 46 | 46 | 45 | 45 | 39 | 51 | 51 | 48 | 48 |
| Braille Reading Classification | Correlation | 0.002 | 0.088 | -.556** | .402** | 1.000 | .749** | -.545** | .512** | .813** | .711** | -.665** | -.104 | -0.039349 | -0.083 | .416** |
| | Sig (2-tailed) | .986 | 0.5372599 | .000 | .003 | | .000 | .000 | .000 | .000 | .000 | .000 | .466 | .784 | .574 | .003 |
| | N | 51 | 51 | 51 | 51 | 51 | 51 | 46 | 46 | 45 | 45 | 39 | 51 | 51 | 48 | 48 |
| Braille Writing Classification | Correlation | -0.051 | 0.227 | -.460** | .337** | .749** | 1.000 | -.359** | .354** | .653** | .709** | -.474** | -.087 | 0.0032984 | -0.014 | .362** |
| | Sig (2-tailed) | .725 | 0.1097026 | .001 | .016 | .000 | | .014 | .016 | .000 | .000 | .002 | .543 | .982 | .923 | .011 |
| | N | 51 | 51 | 51 | 51 | 51 | 51 | 46 | 46 | 45 | 45 | 39 | 51 | 51 | 48 | 48 |
| Age Learned Braille | Correlation | 0.114 | -0.203 | .662** | -.429** | -.545** | -.359** | 1.000 | -.714** | -.725** | -.569** | .659** | .096 | .417** | 0.0279 | -.457** |
| | Sig (2-tailed) | .452 | 0.1757877 | .000 | .003 | .000 | .014 | | .000 | .000 | .000 | .000 | .525 | .004 | .857 | .002 |
| | N | 46 | 46 | 46 | 46 | 46 | 46 | 46 | 46 | 41 | 41 | 37 | 46 | 46 | 44 | 44 |
| Years reading Braille | Correlation | .590** | -0.179 | -.604** | .812** | .512** | .354** | -.714** | 1.000 | .612** | .442** | -.547** | .051 | 0.0536456 | -0.107 | 0.275 |
| | Sig (2-tailed) | .000 | 0.2341941 | .000 | .000 | .000 | .016 | .000 | | .000 | .004 | .000 | .736 | .723 | .491 | .070 |
| | N | 46 | 46 | 46 | 46 | 46 | 46 | 46 | 46 | 41 | 41 | 37 | 46 | 46 | 44 | 44 |
| **Braille Read WPM** | Correlation | 0.004 | 0.217 | -.740** | .514** | .813** | .653** | -.725** | .612** | 1.000 | .839** | -.644** | -.298* | -0.136177 | 0.059 | .498** |
| | Sig (2-tailed) | .979 | 0.1518399 | .000 | .000 | .000 | .000 | .000 | .000 | | .000 | .000 | .047 | .372 | .700 | .000 |
| | N | 45 | 45 | 45 | 45 | 45 | 45 | 41 | 41 | 45 | 45 | 39 | 45 | 45 | 45 | 45 |
| **Braille Write WPM** | Correlation | -0.127 | .360* | -.633** | .382** | .711** | .709** | -.569** | .442** | .839** | 1 | -.487** | -.439** | -0.133377 | 0.1892 | .395** |
| | Sig (2-tailed) | .404 | 0.0152518 | .000 | .010 | .000 | .000 | .000 | .004 | .000 | | .002 | .003 | .382 | .213 | .007 |
| | N | 45 | 45 | 45 | 45 | 45 | 45 | 41 | 41 | 45 | 45 | 39 | 45 | 45 | 45 | 45 |
| **Braille Write MSD Error** | Correlation | -0.088 | -0.171 | .562** | -.665** | -.665** | -.474** | .659** | -.547** | -.644** | -.487** | 1.000 | -.050 | 0.0550918 | 0.0474 | -.346* |
| | Sig (2-tailed) | .596 | 0.2973693 | .000 | .013 | .000 | .002 | .000 | .000 | .000 | .002 | | .762 | .739 | .774 | .031 |
| | N | 39 | 39 | 39 | 39 | 39 | 39 | 37 | 37 | 39 | 39 | 39 | 39 | 39 | 39 | 39 |
| Used Acuity | Correlation | 0.22 | -0.196 | .176 | -.003 | -.104 | -.087 | .096 | .051 | -.298* | -.439** | -.050 | 1.000 | 0.0809077 | -.395** | -0.161 |
| | Sig (2-tailed) | .121 | 0.1674666 | .217 | .982 | .466 | .543 | .525 | .736 | .047 | .003 | .762 | | .572 | .005 | .275 |
| | N | 51 | 51 | 51 | 51 | 51 | 51 | 46 | 46 | 45 | 45 | 39 | 51 | 51 | 48 | 48 |
| Used Sensitivity | Correlation | .511** | -.428** | .150 | .151 | -.039 | .003 | .417** | .054 | -.136 | -0.1334 | .055 | .081 | 1 | -0.122 | -.354* |
| | Sig (2-tailed) | .000 | 0.0017407 | .295 | .291 | .784 | .982 | .004 | .723 | .372 | .382 | .739 | .572 | | .408 | .014 |
| | N | 51 | 51 | 51 | 51 | 51 | 51 | 46 | 46 | 45 | 45 | 39 | 51 | 51 | 48 | 48 |
| Spatial ability | Correlation | -0.188 | .457** | .095 | -.125 | -.083 | -.014 | .028 | -.107 | .059 | 0.18916 | .047 | -.395** | -0.122248 | 1 | 0.202 |
| | Sig (2-tailed) | .201 | 0.0010891 | .521 | .397 | .574 | .923 | .857 | .491 | .700 | .213 | .774 | .005 | .408 | | .168 |
| | N | 48 | 48 | 48 | 48 | 48 | 48 | 44 | 44 | 45 | 45 | 39 | 48 | 48 | 48 | 48 |
| Digit span | Correlation | -.093 | .518** | -.279 | .130 | .416** | .362** | -.457** | .275 | .498** | .395** | -.346* | -.161 | -.354* | .202 | 1.000 |
| | Sig (2-tailed) | .531 | .000 | .055 | .377 | .003 | .011 | .002 | .070 | .000 | .007 | .031 | .275 | .014 | .168 | |
| | N | 48 | 48 | 48 | 48 | 48 | 48 | 44 | 44 | 45 | 45 | 39 | 48 | 48 | 48 | 48 |

Table A4.9: Non-Parametric (Spearman) correlations between Braille functional and individual variables (it is advised to look at the normality table to verify if this value should be used)

# A5

## Touch Primitives Study

# A5.1. Landing on a target statistics resume

| Source | | df | F | Sig. |
|---|---|---|---|---|
| Device | Sphericity Assumed | 1 | .000 | .995 |
| Error(Device) | Sphericity Assumed | 32 | | |
| Grid | Sphericity Assumed | 1 | 12.468 | .001 |
| Error(Grid) | Sphericity Assumed | 32 | | |
| Primitive | Sphericity Assumed | 2 | 1.326 | .273 |
| Error(Primitive) | Sphericity Assumed | 64 | | |
| Device * Grid | Sphericity Assumed | 1 | .486 | .491 |
| Error(Device*Grid) | Sphericity Assumed | 32 | | |
| Device * Primitive | Sphericity Assumed | 2 | .602 | .551 |
| Error(Device*Primitive) | Sphericity Assumed | 64 | | |
| Grid * Primitive | Sphericity Assumed | 2 | .948 | .393 |
| Error(Grid*Primitive) | Sphericity Assumed | 64 | | |
| Device * Grid * Primitive | Sphericity Assumed | 2 | .195 | .823 |
| Error (Device*Grid*Primitive) | Sphericity Assumed | 64 | | |

Table A5.1: Three-way RM ANOVA within subjects effects table (Device X Grid X Primitive on Incorrect Land Error Rate)

| Source | | df | F | Sig. |
|---|---|---|---|---|
| Device | Sphericity Assumed | 1 | .002 | .966 |
| | Greenhouse-Geisser | 1.000 | .002 | .966 |
| | Huynh-Feldt | 1.000 | .002 | .966 |
| | Lower-bound | 1.000 | .002 | .966 |
| Error(Device) | Sphericity Assumed | 35 | | |
| | Greenhouse-Geisser | 35.000 | | |
| | Huynh-Feldt | 35.000 | | |
| | Lower-bound | 35.000 | | |
| Grid | Sphericity Assumed | 1 | 11.336 | .002 |
| | Greenhouse-Geisser | 1.000 | 11.336 | .002 |
| | Huynh-Feldt | 1.000 | 11.336 | .002 |
| | Lower-bound | 1.000 | 11.336 | .002 |
| Error(Grid) | Sphericity Assumed | 35 | | |
| | Greenhouse-Geisser | 35.000 | | |
| | Huynh-Feldt | 35.000 | | |
| | Lower-bound | 35.000 | | |
| Device * Grid | Sphericity Assumed | 1 | 2.300 | .138 |
| | Greenhouse-Geisser | 1.000 | 2.300 | .138 |
| | Huynh-Feldt | 1.000 | 2.300 | .138 |
| | Lower-bound | 1.000 | 2.300 | .138 |
| Error(Device*Grid) | Sphericity Assumed | 35 | | |
| | Greenhouse-Geisser | 35.000 | | |
| | Huynh-Feldt | 35.000 | | |
| | Lower-bound | 35.000 | | |

Table A5.2: Two-way RM ANOVA within subjects effects table (Device X Grid on Incorrect Land Error Rate) between Tablet and Touch Phone for Tapping

| Source | | df | F | Sig. |
|---|---|---|---|---|
| Device | Sphericity Assumed | 1 | 5.767 | .021 |
| | Greenhouse-Geisser | 1.000 | 5.767 | .021 |
| | Huynh-Feldt | 1.000 | 5.767 | .021 |
| | Lower-bound | 1.000 | 5.767 | .021 |
| Error(Device) | Sphericity Assumed | 39 | | |
| | Greenhouse-Geisser | 39.000 | | |
| | Huynh-Feldt | 39.000 | | |
| | Lower-bound | 39.000 | | |
| Grid | Sphericity Assumed | 1 | 9.371 | .004 |
| | Greenhouse-Geisser | 1.000 | 9.371 | .004 |
| | Huynh-Feldt | 1.000 | 9.371 | .004 |
| | Lower-bound | 1.000 | 9.371 | .004 |
| Error(Grid) | Sphericity Assumed | 39 | | |
| | Greenhouse-Geisser | 39.000 | | |
| | Huynh-Feldt | 39.000 | | |
| | Lower-bound | 39.000 | | |
| Device * Grid | Sphericity Assumed | 1 | 2.919 | .096 |
| | Greenhouse-Geisser | 1.000 | 2.919 | .096 |
| | Huynh-Feldt | 1.000 | 2.919 | .096 |
| | Lower-bound | 1.000 | 2.919 | .096 |
| Error(Device*Grid) | Sphericity Assumed | 39 | | |
| | Greenhouse-Geisser | 39.000 | | |
| | Huynh-Feldt | 39.000 | | |
| | Lower-bound | 39.000 | | |

Table A5.3: Two-way RM ANOVA within subjects effects table (Device X Grid on Incorrect Land Error Rate) between Touch Phone and Touch Phone with border for Tapping

| Source | | df | F | Sig. |
|---|---|---|---|---|
| Device | Sphericity Assumed | 2 | 4.620 | .013 |
| | Greenhouse-Geisser | 1.999 | 4.620 | .013 |
| | Huynh-Feldt | 2.000 | 4.620 | .013 |
| | Lower-bound | 1.000 | 4.620 | .039 |
| Error(Device) | Sphericity Assumed | 70 | | |
| | Greenhouse-Geisser | 69.959 | | |
| | Huynh-Feldt | 70.000 | | |
| | Lower-bound | 35.000 | | |
| Areas | Sphericity Assumed | 1 | 43.199 | .000 |
| | Greenhouse-Geisser | 1.000 | 43.199 | .000 |
| | Huynh-Feldt | 1.000 | 43.199 | .000 |
| | Lower-bound | 1.000 | 43.199 | .000 |
| Error(Areas) | Sphericity Assumed | 35 | | |
| | Greenhouse-Geisser | 35.000 | | |
| | Huynh-Feldt | 35.000 | | |
| | Lower-bound | 35.000 | | |
| Device * Areas | Sphericity Assumed | 2 | .126 | .882 |
| | Greenhouse-Geisser | 1.789 | .126 | .861 |
| | Huynh-Feldt | 1.879 | .126 | .870 |
| | Lower-bound | 1.000 | .126 | .725 |
| Error(Device*Areas) | Sphericity Assumed | 70 | | |
| | Greenhouse-Geisser | 62.626 | | |
| | Huynh-Feldt | 65.778 | | |
| | Lower-bound | 35.000 | | |

Table A5.4: Two-way RM ANOVA within subjects effects table (Device X Area (middle vs edges) on Incorrect Land Error Rate) between devices for Tapping

| Source | | df | F | Sig. |
|---|---|---|---|---|
| Device | Sphericity Assumed | 2 | 4.008 | .022 |
| | Greenhouse-Geisser | 1.975 | 4.008 | .023 |
| | Huynh-Feldt | 2.000 | 4.008 | .022 |
| | Lower-bound | 1.000 | 4.008 | .053 |
| Error(Device) | Sphericity Assumed | 70 | | |
| | Greenhouse-Geisser | 69.142 | | |
| | Huynh-Feldt | 70.000 | | |
| | Lower-bound | 35.000 | | |
| Areas | Sphericity Assumed | 3 | 26.662 | .000 |
| | Greenhouse-Geisser | 2.461 | 26.662 | .000 |
| | Huynh-Feldt | 2.661 | 26.662 | .000 |
| | Lower-bound | 1.000 | 26.662 | .000 |
| Error(Areas) | Sphericity Assumed | 105 | | |
| | Greenhouse-Geisser | 86.120 | | |
| | Huynh-Feldt | 93.127 | | |
| | Lower-bound | 35.000 | | |
| Device * Areas | Sphericity Assumed | 6 | 2.086 | .056 |
| | Greenhouse-Geisser | 4.535 | 2.086 | .077 |
| | Huynh-Feldt | 5.293 | 2.086 | .065 |
| | Lower-bound | 1.000 | 2.086 | .158 |
| Error(Device*Areas) | Sphericity Assumed | 210 | | |
| | Greenhouse-Geisser | 158.721 | | |
| | Huynh-Feldt | 185.260 | | |
| | Lower-bound | 35.000 | | |

Table A5.5: Two-way RM ANOVA within subjects effects table (Device X Area (rows) on Incorrect Land Error Rate) between devices for Tapping

# A5.2. Lifting of a target statistics resume

| Source | | df | F | Sig. |
|---|---|---|---|---|
| Device | Sphericity Assumed | 1 | .493 | .487 |
| | Greenhouse-Geisser | 1.000 | .493 | .487 |
| | Huynh-Feldt | 1.000 | .493 | .487 |
| | Lower-bound | 1.000 | .493 | .487 |
| Error(Device) | Sphericity Assumed | 35 | | |
| | Greenhouse-Geisser | 35.000 | | |
| | Huynh-Feldt | 35.000 | | |
| | Lower-bound | 35.000 | | |
| Grid | Sphericity Assumed | 1 | 1.346 | .254 |
| | Greenhouse-Geisser | 1.000 | 1.346 | .254 |
| | Huynh-Feldt | 1.000 | 1.346 | .254 |
| | Lower-bound | 1.000 | 1.346 | .254 |
| Error(Grid) | Sphericity Assumed | 35 | | |
| | Greenhouse-Geisser | 35.000 | | |
| | Huynh-Feldt | 35.000 | | |
| | Lower-bound | 35.000 | | |
| Device * Grid | Sphericity Assumed | 1 | .493 | .487 |
| | Greenhouse-Geisser | 1.000 | .493 | .487 |
| | Huynh-Feldt | 1.000 | .493 | .487 |
| | Lower-bound | 1.000 | .493 | .487 |
| Error(Device*Grid) | Sphericity Assumed | 35 | | |
| | Greenhouse-Geisser | 35.000 | | |
| | Huynh-Feldt | 35.000 | | |
| | Lower-bound | 35.000 | | |

Table A5.6: Two-way RM ANOVA within subjects effects table (Device X Grid on Incorrect Lift Error Rate) between Tablet and Touch Phone for Tapping

| Source | | df | F | Sig. |
|---|---|---|---|---|
| Device | Sphericity Assumed | 1 | 1.296 | .263 |
| | Greenhouse-Geisser | 1.000 | 1.296 | .263 |
| | Huynh-Feldt | 1.000 | 1.296 | .263 |
| | Lower-bound | 1.000 | 1.296 | .263 |
| Error(Device) | Sphericity Assumed | 35 | | |
| | Greenhouse-Geisser | 35.000 | | |
| | Huynh-Feldt | 35.000 | | |
| | Lower-bound | 35.000 | | |
| Grid | Sphericity Assumed | 1 | .000 | 1.000 |
| | Greenhouse-Geisser | 1.000 | .000 | 1.000 |
| | Huynh-Feldt | 1.000 | .000 | 1.000 |
| | Lower-bound | 1.000 | .000 | 1.000 |
| Error(Grid) | Sphericity Assumed | 35 | | |
| | Greenhouse-Geisser | 35.000 | | |
| | Huynh-Feldt | 35.000 | | |
| | Lower-bound | 35.000 | | |
| Device * Grid | Sphericity Assumed | 1 | 3.855 | .058 |
| | Greenhouse-Geisser | 1.000 | 3.855 | .058 |
| | Huynh-Feldt | 1.000 | 3.855 | .058 |
| | Lower-bound | 1.000 | 3.855 | .058 |
| Error(Device*Grid) | Sphericity Assumed | 35 | | |
| | Greenhouse-Geisser | 35.000 | | |
| | Huynh-Feldt | 35.000 | | |
| | Lower-bound | 35.000 | | |

Table A5.7: Two-way RM ANOVA within subjects effects table (Device X Grid on Incorrect Lift Error Rate) between Tablet and Touch Phone for Long Pressing

| Source | | df | F | Sig. |
|---|---|---|---|---|
| Device | Sphericity Assumed | 1 | 3.804 | .059 |
| | Greenhouse-Geisser | 1.000 | 3.804 | .059 |
| | Huynh-Feldt | 1.000 | 3.804 | .059 |
| | Lower-bound | 1.000 | 3.804 | .059 |
| Error(Device) | Sphericity Assumed | 34 | | |
| | Greenhouse-Geisser | 34.000 | | |
| | Huynh-Feldt | 34.000 | | |
| | Lower-bound | 34.000 | | |
| Grid | Sphericity Assumed | 1 | 1.090 | .304 |
| | Greenhouse-Geisser | 1.000 | 1.090 | .304 |
| | Huynh-Feldt | 1.000 | 1.090 | .304 |
| | Lower-bound | 1.000 | 1.090 | .304 |
| Error(Grid) | Sphericity Assumed | 34 | | |
| | Greenhouse-Geisser | 34.000 | | |
| | Huynh-Feldt | 34.000 | | |
| | Lower-bound | 34.000 | | |
| Device * Grid | Sphericity Assumed | 1 | .042 | .838 |
| | Greenhouse-Geisser | 1.000 | .042 | .838 |
| | Huynh-Feldt | 1.000 | .042 | .838 |
| | Lower-bound | 1.000 | .042 | .838 |
| Error(Device*Grid) | Sphericity Assumed | 34 | | |
| | Greenhouse-Geisser | 34.000 | | |
| | Huynh-Feldt | 34.000 | | |
| | Lower-bound | 34.000 | | |

Table A5.8: Two-way RM ANOVA within subjects effects table (Device X Grid on Incorrect Lift Error Rate) between Tablet and Touch Phone for Double Tapping

| Source | | df | F | Sig. |
|---|---|---|---|---|
| Device | Sphericity Assumed | 1 | 1.000 | .323 |
| | Greenhouse-Geisser | 1.000 | 1.000 | .323 |
| | Huynh-Feldt | 1.000 | 1.000 | .323 |
| | Lower-bound | 1.000 | 1.000 | .323 |
| Error(Device) | Sphericity Assumed | 39 | | |
| | Greenhouse-Geisser | 39.000 | | |
| | Huynh-Feldt | 39.000 | | |
| | Lower-bound | 39.000 | | |
| Grid | Sphericity Assumed | 1 | 2.786 | .103 |
| | Greenhouse-Geisser | 1.000 | 2.786 | .103 |
| | Huynh-Feldt | 1.000 | 2.786 | .103 |
| | Lower-bound | 1.000 | 2.786 | .103 |
| Error(Grid) | Sphericity Assumed | 39 | | |
| | Greenhouse-Geisser | 39.000 | | |
| | Huynh-Feldt | 39.000 | | |
| | Lower-bound | 39.000 | | |
| Device * Grid | Sphericity Assumed | 1 | 1.000 | .323 |
| | Greenhouse-Geisser | 1.000 | 1.000 | .323 |
| | Huynh-Feldt | 1.000 | 1.000 | .323 |
| | Lower-bound | 1.000 | 1.000 | .323 |
| Error(Device*Grid) | Sphericity Assumed | 39 | | |
| | Greenhouse-Geisser | 39.000 | | |
| | Huynh-Feldt | 39.000 | | |
| | Lower-bound | 39.000 | | |

Table A5.9: Two-way RM ANOVA within subjects effects table (Device X Grid on Incorrect Lift Error Rate) between Touch Phone and Touch Phone with border for Double Tapping

# A5.3. Reaction Times for grid layouts statistics re-sume

| Source | | df | F | Sig. |
|---|---|---|---|---|
| Device | Sphericity Assumed | 1 | 9.631 | .004 |
| | Greenhouse-Geisser | 1.000 | 9.631 | .004 |
| | Huynh-Feldt | 1.000 | 9.631 | .004 |
| | Lower-bound | 1.000 | 9.631 | .004 |
| Error(Device) | Sphericity Assumed | 35 | | |
| | Greenhouse-Geisser | 35.000 | | |
| | Huynh-Feldt | 35.000 | | |
| | Lower-bound | 35.000 | | |
| Grid | Sphericity Assumed | 1 | .002 | .965 |
| | Greenhouse-Geisser | 1.000 | .002 | .965 |
| | Huynh-Feldt | 1.000 | .002 | .965 |
| | Lower-bound | 1.000 | .002 | .965 |
| Error(Grid) | Sphericity Assumed | 35 | | |
| | Greenhouse-Geisser | 35.000 | | |
| | Huynh-Feldt | 35.000 | | |
| | Lower-bound | 35.000 | | |
| Device * Grid | Sphericity Assumed | 1 | 1.022 | .319 |
| | Greenhouse-Geisser | 1.000 | 1.022 | .319 |
| | Huynh-Feldt | 1.000 | 1.022 | .319 |
| | Lower-bound | 1.000 | 1.022 | .319 |
| Error(Device*Grid) | Sphericity Assumed | 35 | | |
| | Greenhouse-Geisser | 35.000 | | |
| | Huynh-Feldt | 35.000 | | |
| | Lower-bound | 35.000 | | |

Table A5.10: Two-way RM ANOVA within subjects effects table (Device X Grid on Reaction Time) between Tablet and Touch Phone for Tapping

| Source | | df | F | Sig. |
|---|---|---|---|---|
| Device | Sphericity Assumed | 1 | 6.607 | .015 |
| | Greenhouse-Geisser | 1.000 | 6.607 | .015 |
| | Huynh-Feldt | 1.000 | 6.607 | .015 |
| | Lower-bound | 1.000 | 6.607 | .015 |
| Error(Device) | Sphericity Assumed | 35 | | |
| | Greenhouse-Geisser | 35.000 | | |
| | Huynh-Feldt | 35.000 | | |
| | Lower-bound | 35.000 | | |
| Grid | Sphericity Assumed | 1 | .699 | .409 |
| | Greenhouse-Geisser | 1.000 | .699 | .409 |
| | Huynh-Feldt | 1.000 | .699 | .409 |
| | Lower-bound | 1.000 | .699 | .409 |
| Error(Grid) | Sphericity Assumed | 35 | | |
| | Greenhouse-Geisser | 35.000 | | |
| | Huynh-Feldt | 35.000 | | |
| | Lower-bound | 35.000 | | |
| Device * Grid | Sphericity Assumed | 1 | .364 | .550 |
| | Greenhouse-Geisser | 1.000 | .364 | .550 |
| | Huynh-Feldt | 1.000 | .364 | .550 |
| | Lower-bound | 1.000 | .364 | .550 |
| Error(Device*Grid) | Sphericity Assumed | 35 | | |
| | Greenhouse-Geisser | 35.000 | | |
| | Huynh-Feldt | 35.000 | | |
| | Lower-bound | 35.000 | | |

Table A5.11: Two-way RM ANOVA within subjects effects table (Device X Grid on Reaction Time) between Tablet and Touch Phone for Long Pressing

| Source | | df | F | Sig. |
|---|---|---|---|---|
| Device | Sphericity Assumed | 1 | 12.125 | .001 |
| | Greenhouse-Geisser | 1.000 | 12.125 | .001 |
| | Huynh-Feldt | 1.000 | 12.125 | .001 |
| | Lower-bound | 1.000 | 12.125 | .001 |
| Error(Device) | Sphericity Assumed | 34 | | |
| | Greenhouse-Geisser | 34.000 | | |
| | Huynh-Feldt | 34.000 | | |
| | Lower-bound | 34.000 | | |
| Grid | Sphericity Assumed | 1 | .157 | .694 |
| | Greenhouse-Geisser | 1.000 | .157 | .694 |
| | Huynh-Feldt | 1.000 | .157 | .694 |
| | Lower-bound | 1.000 | .157 | .694 |
| Error(Grid) | Sphericity Assumed | 34 | | |
| | Greenhouse-Geisser | 34.000 | | |
| | Huynh-Feldt | 34.000 | | |
| | Lower-bound | 34.000 | | |
| Device * Grid | Sphericity Assumed | 1 | .777 | .384 |
| | Greenhouse-Geisser | 1.000 | .777 | .384 |
| | Huynh-Feldt | 1.000 | .777 | .384 |
| | Lower-bound | 1.000 | .777 | .384 |
| Error(Device*Grid) | Sphericity Assumed | 34 | | |
| | Greenhouse-Geisser | 34.000 | | |
| | Huynh-Feldt | 34.000 | | |
| | Lower-bound | 34.000 | | |

Table A5.12: Two-way RM ANOVA within subjects effects table (Device X Grid on Reaction Time) between Tablet and Touch Phone for Double Tapping

| Source | | df | F | Sig. |
|---|---|---|---|---|
| Device | Sphericity Assumed | 1 | .007 | .933 |
| | Greenhouse-Geisser | 1.000 | .007 | .933 |
| | Huynh-Feldt | 1.000 | .007 | .933 |
| | Lower-bound | 1.000 | .007 | .933 |
| Error(Device) | Sphericity Assumed | 39 | | |
| | Greenhouse-Geisser | 39.000 | | |
| | Huynh-Feldt | 39.000 | | |
| | Lower-bound | 39.000 | | |
| Grid | Sphericity Assumed | 1 | .432 | .515 |
| | Greenhouse-Geisser | 1.000 | .432 | .515 |
| | Huynh-Feldt | 1.000 | .432 | .515 |
| | Lower-bound | 1.000 | .432 | .515 |
| Error(Grid) | Sphericity Assumed | 39 | | |
| | Greenhouse-Geisser | 39.000 | | |
| | Huynh-Feldt | 39.000 | | |
| | Lower-bound | 39.000 | | |
| Device * Grid | Sphericity Assumed | 1 | 3.184 | .082 |
| | Greenhouse-Geisser | 1.000 | 3.184 | .082 |
| | Huynh-Feldt | 1.000 | 3.184 | .082 |
| | Lower-bound | 1.000 | 3.184 | .082 |
| Error(Device*Grid) | Sphericity Assumed | 39 | | |
| | Greenhouse-Geisser | 39.000 | | |
| | Huynh-Feldt | 39.000 | | |
| | Lower-bound | 39.000 | | |

Table A5.13: Two-way RM ANOVA within subjects effects table (Device X Grid on Reaction Time) between Touch Phone and Touch Phone with border for Tapping

| Source | | df | F | Sig. |
|---|---|---|---|---|
| Device | Sphericity Assumed | 1 | 4.414 | .042 |
| | Greenhouse-Geisser | 1.000 | 4.414 | .042 |
| | Huynh-Feldt | 1.000 | 4.414 | .042 |
| | Lower-bound | 1.000 | 4.414 | .042 |
| Error(Device) | Sphericity Assumed | 38 | | |
| | Greenhouse-Geisser | 38.000 | | |
| | Huynh-Feldt | 38.000 | | |
| | Lower-bound | 38.000 | | |
| Grid | Sphericity Assumed | 1 | .148 | .703 |
| | Greenhouse-Geisser | 1.000 | .148 | .703 |
| | Huynh-Feldt | 1.000 | .148 | .703 |
| | Lower-bound | 1.000 | .148 | .703 |
| Error(Grid) | Sphericity Assumed | 38 | | |
| | Greenhouse-Geisser | 38.000 | | |
| | Huynh-Feldt | 38.000 | | |
| | Lower-bound | 38.000 | | |
| Device * Grid | Sphericity Assumed | 1 | .114 | .737 |
| | Greenhouse-Geisser | 1.000 | .114 | .737 |
| | Huynh-Feldt | 1.000 | .114 | .737 |
| | Lower-bound | 1.000 | .114 | .737 |
| Error(Device*Grid) | Sphericity Assumed | 38 | | |
| | Greenhouse-Geisser | 38.000 | | |
| | Huynh-Feldt | 38.000 | | |
| | Lower-bound | 38.000 | | |

Table A5.14: Two-way RM ANOVA within subjects effects table (Device X Grid on Reaction Time) between Touch Phone and Touch Phone with border for Long Pressing

| Source | | df | F | Sig. |
|---|---|---|---|---|
| Device | Sphericity Assumed | 1 | .792 | .379 |
| | Greenhouse-Geisser | 1.000 | .792 | .379 |
| | Huynh-Feldt | 1.000 | .792 | .379 |
| | Lower-bound | 1.000 | .792 | .379 |
| Error(Device) | Sphericity Assumed | 38 | | |
| | Greenhouse-Geisser | 38.000 | | |
| | Huynh-Feldt | 38.000 | | |
| | Lower-bound | 38.000 | | |
| Grid | Sphericity Assumed | 1 | 4.952 | .032 |
| | Greenhouse-Geisser | 1.000 | 4.952 | .032 |
| | Huynh-Feldt | 1.000 | 4.952 | .032 |
| | Lower-bound | 1.000 | 4.952 | .032 |
| Error(Grid) | Sphericity Assumed | 38 | | |
| | Greenhouse-Geisser | 38.000 | | |
| | Huynh-Feldt | 38.000 | | |
| | Lower-bound | 38.000 | | |
| Device * Grid | Sphericity Assumed | 1 | .152 | .699 |
| | Greenhouse-Geisser | 1.000 | .152 | .699 |
| | Huynh-Feldt | 1.000 | .152 | .699 |
| | Lower-bound | 1.000 | .152 | .699 |
| Error(Device*Grid) | Sphericity Assumed | 38 | | |
| | Greenhouse-Geisser | 38.000 | | |
| | Huynh-Feldt | 38.000 | | |
| | Lower-bound | 38.000 | | |

Table A5.15: Two-way RM ANOVA within subjects effects table (Device X Grid on Reaction Time) between Touch Phone and Touch Phone with border for Double Tapping

# A5.4. Primitive requirements for grid layouts statistics resume

**Tests of Within-Subjects Effects**

Measure: MEASURE_1

| Source | | df | F | Sig. |
|---|---|---|---|---|
| Device | Sphericity Assumed | 1 | .542 | .467 |
| | Greenhouse-Geisser | 1.000 | .542 | .467 |
| | Huynh-Feldt | 1.000 | .542 | .467 |
| | Lower-bound | 1.000 | .542 | .467 |
| Error(Device) | Sphericity Assumed | 35 | | |
| | Greenhouse-Geisser | 35.000 | | |
| | Huynh-Feldt | 35.000 | | |
| | Lower-bound | 35.000 | | |
| Grid | Sphericity Assumed | 1 | .407 | .528 |
| | Greenhouse-Geisser | 1.000 | .407 | .528 |
| | Huynh-Feldt | 1.000 | .407 | .528 |
| | Lower-bound | 1.000 | .407 | .528 |
| Error(Grid) | Sphericity Assumed | 35 | | |
| | Greenhouse-Geisser | 35.000 | | |
| | Huynh-Feldt | 35.000 | | |
| | Lower-bound | 35.000 | | |
| Device * Grid | Sphericity Assumed | 1 | .000 | 1.000 |
| | Greenhouse-Geisser | 1.000 | .000 | 1.000 |
| | Huynh-Feldt | 1.000 | .000 | 1.000 |
| | Lower-bound | 1.000 | .000 | 1.000 |
| Error(Device*Grid) | Sphericity Assumed | 35 | | |
| | Greenhouse-Geisser | 35.000 | | |
| | Huynh-Feldt | 35.000 | | |
| | Lower-bound | 35.000 | | |

Table A5.16: Two-way RM ANOVA within subjects effects table (Device X Grid on Automatic Detection) between Tablet and Touch Phone for Tapping

**Tests of Within-Subjects Effects**

Measure: MEASURE_1

| Source | | df | F | Sig. |
|---|---|---|---|---|
| Device | Sphericity Assumed | 1 | 1.918 | .175 |
| | Greenhouse-Geisser | 1.000 | 1.918 | .175 |
| | Huynh-Feldt | 1.000 | 1.918 | .175 |
| | Lower-bound | 1.000 | 1.918 | .175 |
| Error(Device) | Sphericity Assumed | 35 | | |
| | Greenhouse-Geisser | 35.000 | | |
| | Huynh-Feldt | 35.000 | | |
| | Lower-bound | 35.000 | | |
| Grid | Sphericity Assumed | 1 | 1.229 | .275 |
| | Greenhouse-Geisser | 1.000 | 1.229 | .275 |
| | Huynh-Feldt | 1.000 | 1.229 | .275 |
| | Lower-bound | 1.000 | 1.229 | .275 |
| Error(Grid) | Sphericity Assumed | 35 | | |
| | Greenhouse-Geisser | 35.000 | | |
| | Huynh-Feldt | 35.000 | | |
| | Lower-bound | 35.000 | | |
| Device * Grid | Sphericity Assumed | 1 | .440 | .511 |
| | Greenhouse-Geisser | 1.000 | .440 | .511 |
| | Huynh-Feldt | 1.000 | .440 | .511 |
| | Lower-bound | 1.000 | .440 | .511 |
| Error(Device*Grid) | Sphericity Assumed | 35 | | |
| | Greenhouse-Geisser | 35.000 | | |
| | Huynh-Feldt | 35.000 | | |
| | Lower-bound | 35.000 | | |

Table A5.17: Two-way RM ANOVA within subjects effects table (Device X Grid on Automatic Detection) between Tablet and Touch Phone for Long Pressing

**Tests of Within-Subjects Effects**

Measure: MEASURE_1

| Source | | df | F | Sig. |
|---|---|---|---|---|
| Device | Sphericity Assumed | 1 | 4.475 | .042 |
| | Greenhouse-Geisser | 1.000 | 4.475 | .042 |
| | Huynh-Feldt | 1.000 | 4.475 | .042 |
| | Lower-bound | 1.000 | 4.475 | .042 |
| Error(Device) | Sphericity Assumed | 34 | | |
| | Greenhouse-Geisser | 34.000 | | |
| | Huynh-Feldt | 34.000 | | |
| | Lower-bound | 34.000 | | |
| Grid | Sphericity Assumed | 1 | .036 | .850 |
| | Greenhouse-Geisser | 1.000 | .036 | .850 |
| | Huynh-Feldt | 1.000 | .036 | .850 |
| | Lower-bound | 1.000 | .036 | .850 |
| Error(Grid) | Sphericity Assumed | 34 | | |
| | Greenhouse-Geisser | 34.000 | | |
| | Huynh-Feldt | 34.000 | | |
| | Lower-bound | 34.000 | | |
| Device * Grid | Sphericity Assumed | 1 | 2.574 | .118 |
| | Greenhouse-Geisser | 1.000 | 2.574 | .118 |
| | Huynh-Feldt | 1.000 | 2.574 | .118 |
| | Lower-bound | 1.000 | 2.574 | .118 |
| Error(Device*Grid) | Sphericity Assumed | 34 | | |
| | Greenhouse-Geisser | 34.000 | | |
| | Huynh-Feldt | 34.000 | | |
| | Lower-bound | 34.000 | | |

Table A5.18: Two-way RM ANOVA within subjects effects table (Device X Grid on Automatic Detection) between Tablet and Touch Phone for Double Tapping

**Tests of Within-Subjects Effects**

Measure: MEASURE_1

| Source | | df | F | Sig. |
|---|---|---|---|---|
| Device | Sphericity Assumed | 1 | 12.614 | .001 |
| | Greenhouse-Geisser | 1.000 | 12.614 | .001 |
| | Huynh-Feldt | 1.000 | 12.614 | .001 |
| | Lower-bound | 1.000 | 12.614 | .001 |
| Error(Device) | Sphericity Assumed | 35 | | |
| | Greenhouse-Geisser | 35.000 | | |
| | Huynh-Feldt | 35.000 | | |
| | Lower-bound | 35.000 | | |
| Grid | Sphericity Assumed | 1 | 7.346 | .010 |
| | Greenhouse-Geisser | 1.000 | 7.346 | .010 |
| | Huynh-Feldt | 1.000 | 7.346 | .010 |
| | Lower-bound | 1.000 | 7.346 | .010 |
| Error(Grid) | Sphericity Assumed | 35 | | |
| | Greenhouse-Geisser | 35.000 | | |
| | Huynh-Feldt | 35.000 | | |
| | Lower-bound | 35.000 | | |
| Device * Grid | Sphericity Assumed | 1 | 2.463 | .126 |
| | Greenhouse-Geisser | 1.000 | 2.463 | .126 |
| | Huynh-Feldt | 1.000 | 2.463 | .126 |
| | Lower-bound | 1.000 | 2.463 | .126 |
| Error(Device*Grid) | Sphericity Assumed | 35 | | |
| | Greenhouse-Geisser | 35.000 | | |
| | Huynh-Feldt | 35.000 | | |
| | Lower-bound | 35.000 | | |

Table A5.19: Two-way RM ANOVA within subjects effects table (Device X Grid on Duration) between Tablet and Touch Phone for Tapping

**Tests of Within-Subjects Effects**

Measure: MEASURE_1

| Source | | df | F | Sig. |
|---|---|---|---|---|
| Device | Sphericity Assumed | 1 | 9.174 | .005 |
| | Greenhouse-Geisser | 1.000 | 9.174 | .005 |
| | Huynh-Feldt | 1.000 | 9.174 | .005 |
| | Lower-bound | 1.000 | 9.174 | .005 |
| Error(Device) | Sphericity Assumed | 35 | | |
| | Greenhouse-Geisser | 35.000 | | |
| | Huynh-Feldt | 35.000 | | |
| | Lower-bound | 35.000 | | |
| Grid | Sphericity Assumed | 1 | 3.014 | .091 |
| | Greenhouse-Geisser | 1.000 | 3.014 | .091 |
| | Huynh-Feldt | 1.000 | 3.014 | .091 |
| | Lower-bound | 1.000 | 3.014 | .091 |
| Error(Grid) | Sphericity Assumed | 35 | | |
| | Greenhouse-Geisser | 35.000 | | |
| | Huynh-Feldt | 35.000 | | |
| | Lower-bound | 35.000 | | |
| Device * Grid | Sphericity Assumed | 1 | .147 | .704 |
| | Greenhouse-Geisser | 1.000 | .147 | .704 |
| | Huynh-Feldt | 1.000 | .147 | .704 |
| | Lower-bound | 1.000 | .147 | .704 |
| Error(Device*Grid) | Sphericity Assumed | 35 | | |
| | Greenhouse-Geisser | 35.000 | | |
| | Huynh-Feldt | 35.000 | | |
| | Lower-bound | 35.000 | | |

Table A5.20: Two-way RM ANOVA within subjects effects table (Device X Grid on Duration) between Tablet and Touch Phone for Long Pressing

**Tests of Within-Subjects Effects**

Measure: MEASURE_1

| Source | | df | F | Sig. |
|---|---|---|---|---|
| Device | Sphericity Assumed | 1 | 2.040 | .162 |
| | Greenhouse-Geisser | 1.000 | 2.040 | .162 |
| | Huynh-Feldt | 1.000 | 2.040 | .162 |
| | Lower-bound | 1.000 | 2.040 | .162 |
| Error(Device) | Sphericity Assumed | 34 | | |
| | Greenhouse-Geisser | 34.000 | | |
| | Huynh-Feldt | 34.000 | | |
| | Lower-bound | 34.000 | | |
| Grid | Sphericity Assumed | 1 | .048 | .828 |
| | Greenhouse-Geisser | 1.000 | .048 | .828 |
| | Huynh-Feldt | 1.000 | .048 | .828 |
| | Lower-bound | 1.000 | .048 | .828 |
| Error(Grid) | Sphericity Assumed | 34 | | |
| | Greenhouse-Geisser | 34.000 | | |
| | Huynh-Feldt | 34.000 | | |
| | Lower-bound | 34.000 | | |
| Device * Grid | Sphericity Assumed | 1 | .644 | .428 |
| | Greenhouse-Geisser | 1.000 | .644 | .428 |
| | Huynh-Feldt | 1.000 | .644 | .428 |
| | Lower-bound | 1.000 | .644 | .428 |
| Error(Device*Grid) | Sphericity Assumed | 34 | | |
| | Greenhouse-Geisser | 34.000 | | |
| | Huynh-Feldt | 34.000 | | |
| | Lower-bound | 34.000 | | |

Table A5.21: Two-way RM ANOVA within subjects effects table (Device X Grid on Interval) between Tablet and Touch Phone for Double Tapping

**Tests of Within-Subjects Effects**

Measure: MEASURE_1

| Source | | df | F | Sig. |
|---|---|---|---|---|
| Device | Sphericity Assumed | 1 | 3.783 | .059 |
| | Greenhouse-Geisser | 1.000 | 3.783 | .059 |
| | Huynh-Feldt | 1.000 | 3.783 | .059 |
| | Lower-bound | 1.000 | 3.783 | .059 |
| Error(Device) | Sphericity Assumed | 39 | | |
| | Greenhouse-Geisser | 39.000 | | |
| | Huynh-Feldt | 39.000 | | |
| | Lower-bound | 39.000 | | |
| Grid | Sphericity Assumed | 1 | .529 | .472 |
| | Greenhouse-Geisser | 1.000 | .529 | .472 |
| | Huynh-Feldt | 1.000 | .529 | .472 |
| | Lower-bound | 1.000 | .529 | .472 |
| Error(Grid) | Sphericity Assumed | 39 | | |
| | Greenhouse-Geisser | 39.000 | | |
| | Huynh-Feldt | 39.000 | | |
| | Lower-bound | 39.000 | | |
| Device * Grid | Sphericity Assumed | 1 | 1.402 | .244 |
| | Greenhouse-Geisser | 1.000 | 1.402 | .244 |
| | Huynh-Feldt | 1.000 | 1.402 | .244 |
| | Lower-bound | 1.000 | 1.402 | .244 |
| Error(Device*Grid) | Sphericity Assumed | 39 | | |
| | Greenhouse-Geisser | 39.000 | | |
| | Huynh-Feldt | 39.000 | | |
| | Lower-bound | 39.000 | | |

Table A5.22: Two-way RM ANOVA within subjects effects table (Device X Grid on Automatic Detection) between Touch Phone and Touch Phone with Border for Tapping

**Tests of Within-Subjects Effects**

Measure: MEASURE_1

| Source | | df | F | Sig. |
|---|---|---|---|---|
| Device | Sphericity Assumed | 1 | .006 | .938 |
| | Greenhouse-Geisser | 1.000 | .006 | .938 |
| | Huynh-Feldt | 1.000 | .006 | .938 |
| | Lower-bound | 1.000 | .006 | .938 |
| Error(Device) | Sphericity Assumed | 38 | | |
| | Greenhouse-Geisser | 38.000 | | |
| | Huynh-Feldt | 38.000 | | |
| | Lower-bound | 38.000 | | |
| Grid | Sphericity Assumed | 1 | .031 | .860 |
| | Greenhouse-Geisser | 1.000 | .031 | .860 |
| | Huynh-Feldt | 1.000 | .031 | .860 |
| | Lower-bound | 1.000 | .031 | .860 |
| Error(Grid) | Sphericity Assumed | 38 | | |
| | Greenhouse-Geisser | 38.000 | | |
| | Huynh-Feldt | 38.000 | | |
| | Lower-bound | 38.000 | | |
| Device * Grid | Sphericity Assumed | 1 | 1.957 | .170 |
| | Greenhouse-Geisser | 1.000 | 1.957 | .170 |
| | Huynh-Feldt | 1.000 | 1.957 | .170 |
| | Lower-bound | 1.000 | 1.957 | .170 |
| Error(Device*Grid) | Sphericity Assumed | 38 | | |
| | Greenhouse-Geisser | 38.000 | | |
| | Huynh-Feldt | 38.000 | | |
| | Lower-bound | 38.000 | | |

Table A5.23: Two-way RM ANOVA within subjects effects table (Device X Grid on Automatic Detection) between Touch Phone and Touch Phone with Border for Long Pressing

**Tests of Within-Subjects Effects**

Measure: MEASURE_1

| Source | | df | F | Sig. |
|---|---|---|---|---|
| Device | Sphericity Assumed | 1 | .370 | .547 |
| | Greenhouse-Geisser | 1.000 | .370 | .547 |
| | Huynh-Feldt | 1.000 | .370 | .547 |
| | Lower-bound | 1.000 | .370 | .547 |
| Error(Device) | Sphericity Assumed | 38 | | |
| | Greenhouse-Geisser | 38.000 | | |
| | Huynh-Feldt | 38.000 | | |
| | Lower-bound | 38.000 | | |
| Grid | Sphericity Assumed | 1 | .148 | .703 |
| | Greenhouse-Geisser | 1.000 | .148 | .703 |
| | Huynh-Feldt | 1.000 | .148 | .703 |
| | Lower-bound | 1.000 | .148 | .703 |
| Error(Grid) | Sphericity Assumed | 38 | | |
| | Greenhouse-Geisser | 38.000 | | |
| | Huynh-Feldt | 38.000 | | |
| | Lower-bound | 38.000 | | |
| Device * Grid | Sphericity Assumed | 1 | 2.152 | .151 |
| | Greenhouse-Geisser | 1.000 | 2.152 | .151 |
| | Huynh-Feldt | 1.000 | 2.152 | .151 |
| | Lower-bound | 1.000 | 2.152 | .151 |
| Error(Device*Grid) | Sphericity Assumed | 38 | | |
| | Greenhouse-Geisser | 38.000 | | |
| | Huynh-Feldt | 38.000 | | |
| | Lower-bound | 38.000 | | |

Table A5.24: Two-way RM ANOVA within subjects effects table (Device X Grid on Automatic Detection) between Touch Phone and Touch Phone with Border for Double Tapping

| Source | | df | F | Sig. |
|---|---|---|---|---|
| Device | Sphericity Assumed | 1 | 1.666 | .204 |
| | Greenhouse-Geisser | 1.000 | 1.666 | .204 |
| | Huynh-Feldt | 1.000 | 1.666 | .204 |
| | Lower-bound | 1.000 | 1.666 | .204 |
| Error(Device) | Sphericity Assumed | 39 | | |
| | Greenhouse-Geisser | 39.000 | | |
| | Huynh-Feldt | 39.000 | | |
| | Lower-bound | 39.000 | | |
| Grid | Sphericity Assumed | 1 | 7.891 | .008 |
| | Greenhouse-Geisser | 1.000 | 7.891 | .008 |
| | Huynh-Feldt | 1.000 | 7.891 | .008 |
| | Lower-bound | 1.000 | 7.891 | .008 |
| Error(Grid) | Sphericity Assumed | 39 | | |
| | Greenhouse-Geisser | 39.000 | | |
| | Huynh-Feldt | 39.000 | | |
| | Lower-bound | 39.000 | | |
| Device * Grid | Sphericity Assumed | 1 | 1.655 | .206 |
| | Greenhouse-Geisser | 1.000 | 1.655 | .206 |
| | Huynh-Feldt | 1.000 | 1.655 | .206 |
| | Lower-bound | 1.000 | 1.655 | .206 |
| Error(Device*Grid) | Sphericity Assumed | 39 | | |
| | Greenhouse-Geisser | 39.000 | | |
| | Huynh-Feldt | 39.000 | | |
| | Lower-bound | 39.000 | | |

Table A5.25: Two-way RM ANOVA within subjects effects table (Device X Grid on Duration) between Touch Phone and Touch Phone with Border for Tapping

| Source | | df | F | Sig. |
|---|---|---|---|---|
| Device | Sphericity Assumed | 1 | .093 | .762 |
| | Greenhouse-Geisser | 1.000 | .093 | .762 |
| | Huynh-Feldt | 1.000 | .093 | .762 |
| | Lower-bound | 1.000 | .093 | .762 |
| Error(Device) | Sphericity Assumed | 38 | | |
| | Greenhouse-Geisser | 38.000 | | |
| | Huynh-Feldt | 38.000 | | |
| | Lower-bound | 38.000 | | |
| Grid | Sphericity Assumed | 1 | .358 | .553 |
| | Greenhouse-Geisser | 1.000 | .358 | .553 |
| | Huynh-Feldt | 1.000 | .358 | .553 |
| | Lower-bound | 1.000 | .358 | .553 |
| Error(Grid) | Sphericity Assumed | 38 | | |
| | Greenhouse-Geisser | 38.000 | | |
| | Huynh-Feldt | 38.000 | | |
| | Lower-bound | 38.000 | | |
| Device * Grid | Sphericity Assumed | 1 | .525 | .473 |
| | Greenhouse-Geisser | 1.000 | .525 | .473 |
| | Huynh-Feldt | 1.000 | .525 | .473 |
| | Lower-bound | 1.000 | .525 | .473 |
| Error(Device*Grid) | Sphericity Assumed | 38 | | |
| | Greenhouse-Geisser | 38.000 | | |
| | Huynh-Feldt | 38.000 | | |
| | Lower-bound | 38.000 | | |

Table A5.26: Two-way RM ANOVA within subjects effects table (Device X Grid on Duration) between Touch Phone and Touch Phone with Border for Long Pressing

| Source | | df | F | Sig. |
|---|---|---|---|---|
| Device | Sphericity Assumed | 1 | 2.159 | .150 |
| | Greenhouse-Geisser | 1.000 | 2.159 | .150 |
| | Huynh-Feldt | 1.000 | 2.159 | .150 |
| | Lower-bound | 1.000 | 2.159 | .150 |
| Error(Device) | Sphericity Assumed | 38 | | |
| | Greenhouse-Geisser | 38.000 | | |
| | Huynh-Feldt | 38.000 | | |
| | Lower-bound | 38.000 | | |
| Grid | Sphericity Assumed | 1 | .092 | .763 |
| | Greenhouse-Geisser | 1.000 | .092 | .763 |
| | Huynh-Feldt | 1.000 | .092 | .763 |
| | Lower-bound | 1.000 | .092 | .763 |
| Error(Grid) | Sphericity Assumed | 38 | | |
| | Greenhouse-Geisser | 38.000 | | |
| | Huynh-Feldt | 38.000 | | |
| | Lower-bound | 38.000 | | |
| Device * Grid | Sphericity Assumed | 1 | .006 | .939 |
| | Greenhouse-Geisser | 1.000 | .006 | .939 |
| | Huynh-Feldt | 1.000 | .006 | .939 |
| | Lower-bound | 1.000 | .006 | .939 |
| Error(Device*Grid) | Sphericity Assumed | 38 | | |
| | Greenhouse-Geisser | 38.000 | | |
| | Huynh-Feldt | 38.000 | | |
| | Lower-bound | 38.000 | | |

Table A5.27: Two-way RM ANOVA within subjects effects table (Device X Grid on Duration) between Touch Phone and Touch Phone with Border for Double Tapping

| Source | | df | F | Sig. |
|--------|--------|-----|------|------|
| Device | Sphericity Assumed | 1 | .013 | .909 |
| | Greenhouse-Geisser | 1.000 | .013 | .909 |
| | Huynh-Feldt | 1.000 | .013 | .909 |
| | Lower-bound | 1.000 | .013 | .909 |
| Error(Device) | Sphericity Assumed | 38 | | |
| | Greenhouse-Geisser | 38.000 | | |
| | Huynh-Feldt | 38.000 | | |
| | Lower-bound | 38.000 | | |
| Grid | Sphericity Assumed | 1 | .211 | .649 |
| | Greenhouse-Geisser | 1.000 | .211 | .649 |
| | Huynh-Feldt | 1.000 | .211 | .649 |
| | Lower-bound | 1.000 | .211 | .649 |
| Error(Grid) | Sphericity Assumed | 38 | | |
| | Greenhouse-Geisser | 38.000 | | |
| | Huynh-Feldt | 38.000 | | |
| | Lower-bound | 38.000 | | |
| Device * Grid | Sphericity Assumed | 1 | .582 | .450 |
| | Greenhouse-Geisser | 1.000 | .582 | .450 |
| | Huynh-Feldt | 1.000 | .582 | .450 |
| | Lower-bound | 1.000 | .582 | .450 |
| Error(Device*Grid) | Sphericity Assumed | 38 | | |
| | Greenhouse-Geisser | 38.000 | | |
| | Huynh-Feldt | 38.000 | | |
| | Lower-bound | 38.000 | | |

Table A5.28: Two-way RM ANOVA within subjects effects table (Device X Grid on Interval) between Touch Phone and Touch Phone with Border for Double Tapping

## A5.5. Overall 12–Tapping statistics resume

| Source | | df | F | Sig. |
|--------|--------|-----|-------|------|
| Device | Sphericity Assumed | 3 | 6.427 | .000 |
| | Greenhouse-Geisser | 2.639 | 6.427 | .001 |
| | Huynh-Feldt | 2.881 | 6.427 | .001 |
| | Lower-bound | 1.000 | 6.427 | .016 |
| Error(Device) | Sphericity Assumed | 102 | | |
| | Greenhouse-Geisser | 89.721 | | |
| | Huynh-Feldt | 97.963 | | |
| | Lower-bound | 34.000 | | |

Table A5.29: One-way RM ANOVA within subjects effects table (Device on Incorrect Land Error Rate) between all devices (includes 12-key physical keypad) for Tapping

# A5.6.  Gesturing Automatic Detection statistics resume

| Source | | df | F | Sig. |
|---|---|---|---|---|
| Device | Sphericity Assumed | 1 | 50.895 | .000 |
| | Greenhouse-Geisser | 1.000 | 50.895 | .000 |
| | Huynh-Feldt | 1.000 | 50.895 | .000 |
| | Lower-bound | 1.000 | 50.895 | .000 |
| Error(Device) | Sphericity Assumed | 34 | | |
| | Greenhouse-Geisser | 34.000 | | |
| | Huynh-Feldt | 34.000 | | |
| | Lower-bound | 34.000 | | |

Table A5.30: One-way RM ANOVA within subjects effects table (Device on Automatic Area Recognition) between Tablet and Touch Phone for Gesturing

| Source | | df | F | Sig. |
|---|---|---|---|---|
| Device | Sphericity Assumed | 1 | .590 | .448 |
| | Greenhouse-Geisser | 1.000 | .590 | .448 |
| | Huynh-Feldt | 1.000 | .590 | .448 |
| | Lower-bound | 1.000 | .590 | .448 |
| Error(Device) | Sphericity Assumed | 34 | | |
| | Greenhouse-Geisser | 34.000 | | |
| | Huynh-Feldt | 34.000 | | |
| | Lower-bound | 34.000 | | |

Table A5.31: One-way RM ANOVA within subjects effects table (Device on Automatic Direction Recognition) between Tablet and Touch Phone for Gesturing

| Source | | df | F | Sig. |
|---|---|---|---|---|
| Device | Sphericity Assumed | 1 | 15.968 | .000 |
| | Greenhouse-Geisser | 1.000 | 15.968 | .000 |
| | Huynh-Feldt | 1.000 | 15.968 | .000 |
| | Lower-bound | 1.000 | 15.968 | .000 |
| Error(Device) | Sphericity Assumed | 39 | | |
| | Greenhouse-Geisser | 39.000 | | |
| | Huynh-Feldt | 39.000 | | |
| | Lower-bound | 39.000 | | |

Table A5.32: One-way RM ANOVA within subjects effects table (Device on Automatic Area Recognition) between Touch Phone and Touch Phone with Border for Gesturing

| Source | | df | F | Sig. |
|---|---|---|---|---|
| Device | Sphericity Assumed | 1 | 16.357 | .000 |
| | Greenhouse-Geisser | 1.000 | 16.357 | .000 |
| | Huynh-Feldt | 1.000 | 16.357 | .000 |
| | Lower-bound | 1.000 | 16.357 | .000 |
| Error(Device) | Sphericity Assumed | 39 | | |
| | Greenhouse-Geisser | 39.000 | | |
| | Huynh-Feldt | 39.000 | | |
| | Lower-bound | 39.000 | | |

Table A5.33: One-way RM ANOVA within subjects effects table (Device on Direction Recognition) between Touch Phone and Touch Phone with Border for Gesturing

| Source | | df | F | Sig. |
|---|---|---|---|---|
| Device | Sphericity Assumed | 2 | 11.717 | .000 |
| | Greenhouse-Geisser | 1.774 | 11.717 | .000 |
| | Huynh-Feldt | 1.865 | 11.717 | .000 |
| | Lower-bound | 1.000 | 11.717 | .002 |
| Error(Device) | Sphericity Assumed | 68 | | |
| | Greenhouse-Geisser | 60.329 | | |
| | Huynh-Feldt | 63.412 | | |
| | Lower-bound | 34.000 | | |
| Area | Sphericity Assumed | 1 | 12.245 | .001 |
| | Greenhouse-Geisser | 1.000 | 12.245 | .001 |
| | Huynh-Feldt | 1.000 | 12.245 | .001 |
| | Lower-bound | 1.000 | 12.245 | .001 |
| Error(Area) | Sphericity Assumed | 34 | | |
| | Greenhouse-Geisser | 34.000 | | |
| | Huynh-Feldt | 34.000 | | |
| | Lower-bound | 34.000 | | |
| Device * Area | Sphericity Assumed | 2 | .107 | .899 |
| | Greenhouse-Geisser | 1.791 | .107 | .878 |
| | Huynh-Feldt | 1.884 | .107 | .888 |
| | Lower-bound | 1.000 | .107 | .746 |
| Error(Device*Area) | Sphericity Assumed | 68 | | |
| | Greenhouse-Geisser | 60.893 | | |
| | Huynh-Feldt | 64.058 | | |
| | Lower-bound | 34.000 | | |

Table A5.34: Two-way RM ANOVA within subjects effects table (Device X Area (middle vs edges) on Automatic Direction Recognition) between devices for Gesturing

| Source | | df | F | Sig. |
|---|---|---|---|---|
| Device | Sphericity Assumed | 2 | 11.956 | .000 |
| | Greenhouse-Geisser | 1.809 | 11.956 | .000 |
| | Huynh-Feldt | 1.904 | 11.956 | .000 |
| | Lower-bound | 1.000 | 11.956 | .001 |
| Error(Device) | Sphericity Assumed | 68 | | |
| | Greenhouse-Geisser | 61.492 | | |
| | Huynh-Feldt | 64.744 | | |
| | Lower-bound | 34.000 | | |
| Area | Sphericity Assumed | 1 | 8.992 | .005 |
| | Greenhouse-Geisser | 1.000 | 8.992 | .005 |
| | Huynh-Feldt | 1.000 | 8.992 | .005 |
| | Lower-bound | 1.000 | 8.992 | .005 |
| Error(Area) | Sphericity Assumed | 34 | | |
| | Greenhouse-Geisser | 34.000 | | |
| | Huynh-Feldt | 34.000 | | |
| | Lower-bound | 34.000 | | |
| Device * Area | Sphericity Assumed | 2 | 1.427 | .247 |
| | Greenhouse-Geisser | 1.785 | 1.427 | .248 |
| | Huynh-Feldt | 1.877 | 1.427 | .247 |
| | Lower-bound | 1.000 | 1.427 | .240 |
| Error(Device*Area) | Sphericity Assumed | 68 | | |
| | Greenhouse-Geisser | 60.673 | | |
| | Huynh-Feldt | 63.806 | | |
| | Lower-bound | 34.000 | | |

Table A5.35: Two-way RM ANOVA within subjects effects table (Device X Area (vertical vs horizontal) on Automatic Direction Recognition) between devices for Gesturing

| Source | | df | F | Sig. |
|---|---|---|---|---|
| Device | Sphericity Assumed | 2 | 69.912 | .000 |
| | Greenhouse-Geisser | 1.740 | 69.912 | .000 |
| | Huynh-Feldt | 1.825 | 69.912 | .000 |
| | Lower-bound | 1.000 | 69.912 | .000 |
| Error(Device) | Sphericity Assumed | 68 | | |
| | Greenhouse-Geisser | 59.145 | | |
| | Huynh-Feldt | 62.059 | | |
| | Lower-bound | 34.000 | | |
| Area | Sphericity Assumed | 1 | .479 | .494 |
| | Greenhouse-Geisser | 1.000 | .479 | .494 |
| | Huynh-Feldt | 1.000 | .479 | .494 |
| | Lower-bound | 1.000 | .479 | .494 |
| Error(Area) | Sphericity Assumed | 34 | | |
| | Greenhouse-Geisser | 34.000 | | |
| | Huynh-Feldt | 34.000 | | |
| | Lower-bound | 34.000 | | |
| Device * Area | Sphericity Assumed | 2 | 12.427 | .000 |
| | Greenhouse-Geisser | 1.767 | 12.427 | .000 |
| | Huynh-Feldt | 1.857 | 12.427 | .000 |
| | Lower-bound | 1.000 | 12.427 | .001 |
| Error(Device*Area) | Sphericity Assumed | 68 | | |
| | Greenhouse-Geisser | 60.094 | | |
| | Huynh-Feldt | 63.144 | | |
| | Lower-bound | 34.000 | | |

Table A5.36: Two-way RM ANOVA within subjects effects table (Device X Area (vertical vs horizontal) on Automatic Area Recognition) between devices for Gesturing

# A5.7.  Gesturing Reaction Times statistics resume

| Source | | df | F | Sig. |
|---|---|---|---|---|
| Device | Sphericity Assumed | 1 | ,321 | ,575 |
| | Greenhouse-Geisser | 1,000 | ,321 | ,575 |
| | Huynh-Feldt | 1,000 | ,321 | ,575 |
| | Lower-bound | 1,000 | ,321 | ,575 |
| Error(Device) | Sphericity Assumed | 34 | | |
| | Greenhouse-Geisser | 34,000 | | |
| | Huynh-Feldt | 34,000 | | |
| | Lower-bound | 34,000 | | |

Table A5.37:  One-way RM ANOVA within subjects effects table (Device on Reaction Time) between Tablet and Touch Phone for Gesturing

| Source | | df | F | Sig. |
|---|---|---|---|---|
| Device | Sphericity Assumed | 1 | 8,213 | ,007 |
| | Greenhouse-Geisser | 1,000 | 8,213 | ,007 |
| | Huynh-Feldt | 1,000 | 8,213 | ,007 |
| | Lower-bound | 1,000 | 8,213 | ,007 |
| Error(Device) | Sphericity Assumed | 39 | | |
| | Greenhouse-Geisser | 39,000 | | |
| | Huynh-Feldt | 39,000 | | |
| | Lower-bound | 39,000 | | |

Table A5.38:  One-way RM ANOVA within subjects effects table (Device on Reaction Time) between Touch Phone and Touch Phone with Border for Gesturing

| Source | | df | F | Sig. |
|---|---|---|---|---|
| Device | Sphericity Assumed | 2 | 2.441 | .095 |
| | Greenhouse-Geisser | 1.633 | 2.441 | .106 |
| | Huynh-Feldt | 1.704 | 2.441 | .104 |
| | Lower-bound | 1.000 | 2.441 | .127 |
| Error(Device) | Sphericity Assumed | 68 | | |
| | Greenhouse-Geisser | 55.525 | | |
| | Huynh-Feldt | 57.941 | | |
| | Lower-bound | 34.000 | | |
| Area | Sphericity Assumed | 1 | 81.063 | .000 |
| | Greenhouse-Geisser | 1.000 | 81.063 | .000 |
| | Huynh-Feldt | 1.000 | 81.063 | .000 |
| | Lower-bound | 1.000 | 81.063 | .000 |
| Error(Area) | Sphericity Assumed | 34 | | |
| | Greenhouse-Geisser | 34.000 | | |
| | Huynh-Feldt | 34.000 | | |
| | Lower-bound | 34.000 | | |
| Device * Area | Sphericity Assumed | 2 | 3.617 | .032 |
| | Greenhouse-Geisser | 1.967 | 3.617 | .033 |
| | Huynh-Feldt | 2.000 | 3.617 | .032 |
| | Lower-bound | 1.000 | 3.617 | .066 |
| Error(Device*Area) | Sphericity Assumed | 68 | | |
| | Greenhouse-Geisser | 66.869 | | |
| | Huynh-Feldt | 68.000 | | |
| | Lower-bound | 34.000 | | |

Table A5.39: Two-way RM ANOVA within subjects effects table (Device X Area (vertical vs horizontal) on Reaction Time) between devices for Gesturing

| Source | | df | F | Sig. |
|---|---|---|---|---|
| Device | Sphericity Assumed | 2 | 3.495 | .036 |
| | Greenhouse-Geisser | 1.625 | 3.495 | .046 |
| | Huynh-Feldt | 1.695 | 3.495 | .044 |
| | Lower-bound | 1.000 | 3.495 | .070 |
| Error(Device) | Sphericity Assumed | 68 | | |
| | Greenhouse-Geisser | 55.247 | | |
| | Huynh-Feldt | 57.626 | | |
| | Lower-bound | 34.000 | | |
| Area | Sphericity Assumed | 1 | 2.627 | .114 |
| | Greenhouse-Geisser | 1.000 | 2.627 | .114 |
| | Huynh-Feldt | 1.000 | 2.627 | .114 |
| | Lower-bound | 1.000 | 2.627 | .114 |
| Error(Area) | Sphericity Assumed | 34 | | |
| | Greenhouse-Geisser | 34.000 | | |
| | Huynh-Feldt | 34.000 | | |
| | Lower-bound | 34.000 | | |
| Device * Area | Sphericity Assumed | 2 | .387 | .680 |
| | Greenhouse-Geisser | 1.645 | .387 | .640 |
| | Huynh-Feldt | 1.718 | .387 | .649 |
| | Lower-bound | 1.000 | .387 | .538 |
| Error(Device*Area) | Sphericity Assumed | 68 | | |
| | Greenhouse-Geisser | 55.940 | | |
| | Huynh-Feldt | 58.412 | | |
| | Lower-bound | 34.000 | | |

Table A5.40: Two-way RM ANOVA within subjects effects table (Device X Area (middle vs edges) on Reaction Time) between devices for Gesturing

# A5.8.  Gesturing Primitive Requirements statistics resume

| Source | | df | F | Sig. |
|---|---|---|---|---|
| Device | Sphericity Assumed | 1 | 15,251 | ,000 |
| | Greenhouse-Geisser | 1,000 | 15,251 | ,000 |
| | Huynh-Feldt | 1,000 | 15,251 | ,000 |
| | Lower-bound | 1,000 | 15,251 | ,000 |
| Error(Device) | Sphericity Assumed | 34 | | |
| | Greenhouse-Geisser | 34,000 | | |
| | Huynh-Feldt | 34,000 | | |
| | Lower-bound | 34,000 | | |

Table A5.41: One-way RM ANOVA within subjects effects table (Device on Precision) between Tablet and Touch Phone for Gesturing

| Source | | df | F | Sig. |
|---|---|---|---|---|
| Device | Sphericity Assumed | 2 | 81.803 | .000 |
| | Greenhouse-Geisser | 1.269 | 81.803 | .000 |
| | Huynh-Feldt | 1.296 | 81.803 | .000 |
| | Lower-bound | 1.000 | 81.803 | .000 |
| Error(Device) | Sphericity Assumed | 68 | | |
| | Greenhouse-Geisser | 43.148 | | |
| | Huynh-Feldt | 44.061 | | |
| | Lower-bound | 34.000 | | |

Table A5.42: One-way RM ANOVA within subjects effects table (Device on Distance) between Tablet and Touch Phone for Gesturing

| Source | | df | F | Sig. |
|---|---|---|---|---|
| Device | Sphericity Assumed | 2 | 1.244 | .295 |
| | Greenhouse-Geisser | 1.982 | 1.244 | .295 |
| | Huynh-Feldt | 2.000 | 1.244 | .295 |
| | Lower-bound | 1.000 | 1.244 | .273 |
| Error(Device) | Sphericity Assumed | 68 | | |
| | Greenhouse-Geisser | 67.395 | | |
| | Huynh-Feldt | 68.000 | | |
| | Lower-bound | 34.000 | | |

Table A5.43: One-way RM ANOVA within subjects effects table (Device on Angle Offset) between Tablet and Touch Phone for Gesturing

| Source | | df | F | Sig. |
|--------|--------|-----|--------|------|
| Device | Sphericity Assumed | 1 | 70,804 | ,000 |
| | Greenhouse-Geisser | 1,000 | 70,804 | ,000 |
| | Huynh-Feldt | 1,000 | 70,804 | ,000 |
| | Lower-bound | 1,000 | 70,804 | ,000 |
| Error(Device) | Sphericity Assumed | 34 | | |
| | Greenhouse-Geisser | 34,000 | | |
| | Huynh-Feldt | 34,000 | | |
| | Lower-bound | 34,000 | | |

Table A5.44: One-way RM ANOVA within subjects effects table (Device on Velocity) between Tablet and Touch Phone for Gesturing

| Source | | df | F | Sig. |
|--------|--------|-----|-------|------|
| Device | Sphericity Assumed | 1 | 2,343 | ,134 |
| | Greenhouse-Geisser | 1,000 | 2,343 | ,134 |
| | Huynh-Feldt | 1,000 | 2,343 | ,134 |
| | Lower-bound | 1,000 | 2,343 | ,134 |
| Error(Device) | Sphericity Assumed | 39 | | |
| | Greenhouse-Geisser | 39,000 | | |
| | Huynh-Feldt | 39,000 | | |
| | Lower-bound | 39,000 | | |

Table A5.45: One-way RM ANOVA within subjects effects table (Device on Precision) between Touch Phone and Touch Phone with Border for Gesturing

| Source | | df | F | Sig. |
|--------|--------|-----|------|------|
| Device | Sphericity Assumed | 1 | ,036 | ,851 |
| | Greenhouse-Geisser | 1,000 | ,036 | ,851 |
| | Huynh-Feldt | 1,000 | ,036 | ,851 |
| | Lower-bound | 1,000 | ,036 | ,851 |
| Error(Device) | Sphericity Assumed | 39 | | |
| | Greenhouse-Geisser | 39,000 | | |
| | Huynh-Feldt | 39,000 | | |
| | Lower-bound | 39,000 | | |

Table A5.46: One-way RM ANOVA within subjects effects table (Device on Distance) between Touch Phone and Touch Phone with Border for Gesturing

| Source | | df | F | Sig. |
|---|---|---|---|---|
| Device | Sphericity Assumed | 1 | ,079 | ,780 |
| | Greenhouse-Geisser | 1,000 | ,079 | ,780 |
| | Huynh-Feldt | 1,000 | ,079 | ,780 |
| | Lower-bound | 1,000 | ,079 | ,780 |
| Error(Device) | Sphericity Assumed | 39 | | |
| | Greenhouse-Geisser | 39,000 | | |
| | Huynh-Feldt | 39,000 | | |
| | Lower-bound | 39,000 | | |

Table A5.47: One-way RM ANOVA within subjects effects table (Device on Angle Offset) between Touch Phone and Touch Phone with Border for Gesturing

| Source | | df | F | Sig. |
|---|---|---|---|---|
| Device | Sphericity Assumed | 1 | 4.924 | .032 |
| | Greenhouse-Geisser | 1.000 | 4.924 | .032 |
| | Huynh-Feldt | 1.000 | 4.924 | .032 |
| | Lower-bound | 1.000 | 4.924 | .032 |
| Error(Device) | Sphericity Assumed | 39 | | |
| | Greenhouse-Geisser | 39.000 | | |
| | Huynh-Feldt | 39.000 | | |
| | Lower-bound | 39.000 | | |

Table A5.48: One-way RM ANOVA within subjects effects table (Device on Velocity) between Touch Phone and Touch Phone with Border for Gesturing

| Source | | df | F | Sig. |
|---|---|---|---|---|
| Device | Sphericity Assumed | 2 | 8.906 | .000 |
| | Greenhouse-Geisser | 1.894 | 8.906 | .000 |
| | Huynh-Feldt | 2.000 | 8.906 | .000 |
| | Lower-bound | 1.000 | 8.906 | .005 |
| Error(Device) | Sphericity Assumed | 68 | | |
| | Greenhouse-Geisser | 64.407 | | |
| | Huynh-Feldt | 68.000 | | |
| | Lower-bound | 34.000 | | |
| Area | Sphericity Assumed | 1 | 2.175 | .149 |
| | Greenhouse-Geisser | 1.000 | 2.175 | .149 |
| | Huynh-Feldt | 1.000 | 2.175 | .149 |
| | Lower-bound | 1.000 | 2.175 | .149 |
| Error(Area) | Sphericity Assumed | 34 | | |
| | Greenhouse-Geisser | 34.000 | | |
| | Huynh-Feldt | 34.000 | | |
| | Lower-bound | 34.000 | | |
| Device * Area | Sphericity Assumed | 2 | .400 | .672 |
| | Greenhouse-Geisser | 1.923 | .400 | .664 |
| | Huynh-Feldt | 2.000 | .400 | .672 |
| | Lower-bound | 1.000 | .400 | .531 |
| Error(Device*Area) | Sphericity Assumed | 68 | | |
| | Greenhouse-Geisser | 65.376 | | |
| | Huynh-Feldt | 68.000 | | |
| | Lower-bound | 34.000 | | |

Table A5.49: Two-way RM ANOVA within subjects effects table (Device X Area (middle vs edges) on Precision) between devices for Gesturing

| Source | | df | F | Sig. |
|---|---|---|---|---|
| Device | Sphericity Assumed | 2 | 10.795 | .000 |
| | Greenhouse-Geisser | 1.785 | 10.795 | .000 |
| | Huynh-Feldt | 1.877 | 10.795 | .000 |
| | Lower-bound | 1.000 | 10.795 | .002 |
| Error(Device) | Sphericity Assumed | 68 | | |
| | Greenhouse-Geisser | 60.686 | | |
| | Huynh-Feldt | 63.822 | | |
| | Lower-bound | 34.000 | | |
| Area | Sphericity Assumed | 1 | 1.251 | .271 |
| | Greenhouse-Geisser | 1.000 | 1.251 | .271 |
| | Huynh-Feldt | 1.000 | 1.251 | .271 |
| | Lower-bound | 1.000 | 1.251 | .271 |
| Error(Area) | Sphericity Assumed | 34 | | |
| | Greenhouse-Geisser | 34.000 | | |
| | Huynh-Feldt | 34.000 | | |
| | Lower-bound | 34.000 | | |
| Device * Area | Sphericity Assumed | 2 | .624 | .539 |
| | Greenhouse-Geisser | 1.975 | .624 | .537 |
| | Huynh-Feldt | 2.000 | .624 | .539 |
| | Lower-bound | 1.000 | .624 | .435 |
| Error(Device*Area) | Sphericity Assumed | 68 | | |
| | Greenhouse-Geisser | 67.151 | | |
| | Huynh-Feldt | 68.000 | | |
| | Lower-bound | 34.000 | | |

Table A5.50: Two-way RM ANOVA within subjects effects table (Device X Area (vertical vs horizontal) on Precision) between devices for Gesturing

| Source | | df | F | Sig. |
|---|---|---|---|---|
| Device | Sphericity Assumed | 2 | 1.636 | .202 |
| | Greenhouse-Geisser | 1.964 | 1.636 | .203 |
| | Huynh-Feldt | 2.000 | 1.636 | .202 |
| | Lower-bound | 1.000 | 1.636 | .209 |
| Error(Device) | Sphericity Assumed | 68 | | |
| | Greenhouse-Geisser | 66.782 | | |
| | Huynh-Feldt | 68.000 | | |
| | Lower-bound | 34.000 | | |
| Area | Sphericity Assumed | 1 | 2.123 | .154 |
| | Greenhouse-Geisser | 1.000 | 2.123 | .154 |
| | Huynh-Feldt | 1.000 | 2.123 | .154 |
| | Lower-bound | 1.000 | 2.123 | .154 |
| Error(Area) | Sphericity Assumed | 34 | | |
| | Greenhouse-Geisser | 34.000 | | |
| | Huynh-Feldt | 34.000 | | |
| | Lower-bound | 34.000 | | |
| Device * Area | Sphericity Assumed | 2 | .078 | .925 |
| | Greenhouse-Geisser | 1.448 | .078 | .867 |
| | Huynh-Feldt | 1.496 | .078 | .874 |
| | Lower-bound | 1.000 | .078 | .782 |
| Error(Device*Area) | Sphericity Assumed | 68 | | |
| | Greenhouse-Geisser | 49.237 | | |
| | Huynh-Feldt | 50.851 | | |
| | Lower-bound | 34.000 | | |

Table A5.51: Two-way RM ANOVA within subjects effects table (Device X Area (middle vs edges) on Angle Offset) between devices for Gesturing

| Source | | df | F | Sig. |
|---|---|---|---|---|
| Device | Sphericity Assumed | 2 | 1.223 | .301 |
| | Greenhouse-Geisser | 1.982 | 1.223 | .301 |
| | Huynh-Feldt | 2.000 | 1.223 | .301 |
| | Lower-bound | 1.000 | 1.223 | .277 |
| Error(Device) | Sphericity Assumed | 68 | | |
| | Greenhouse-Geisser | 67.373 | | |
| | Huynh-Feldt | 68.000 | | |
| | Lower-bound | 34.000 | | |
| Area | Sphericity Assumed | 1 | 16.036 | .000 |
| | Greenhouse-Geisser | 1.000 | 16.036 | .000 |
| | Huynh-Feldt | 1.000 | 16.036 | .000 |
| | Lower-bound | 1.000 | 16.036 | .000 |
| Error(Area) | Sphericity Assumed | 34 | | |
| | Greenhouse-Geisser | 34.000 | | |
| | Huynh-Feldt | 34.000 | | |
| | Lower-bound | 34.000 | | |
| Device * Area | Sphericity Assumed | 2 | .001 | .999 |
| | Greenhouse-Geisser | 1.865 | .001 | .999 |
| | Huynh-Feldt | 1.969 | .001 | .999 |
| | Lower-bound | 1.000 | .001 | .976 |
| Error(Device*Area) | Sphericity Assumed | 68 | | |
| | Greenhouse-Geisser | 63.408 | | |
| | Huynh-Feldt | 66.946 | | |
| | Lower-bound | 34.000 | | |

Table A5.52: Two-way RM ANOVA within subjects effects table (Device X Area (vertical vs horizontal) on Angle Offset) between devices for Gesturing

| Source | | df | F | Sig. |
|---|---|---|---|---|
| Device | Sphericity Assumed | 2 | 81.730 | .000 |
| | Greenhouse-Geisser | 1.273 | 81.730 | .000 |
| | Huynh-Feldt | 1.300 | 81.730 | .000 |
| | Lower-bound | 1.000 | 81.730 | .000 |
| Error(Device) | Sphericity Assumed | 68 | | |
| | Greenhouse-Geisser | 43.272 | | |
| | Huynh-Feldt | 44.199 | | |
| | Lower-bound | 34.000 | | |
| Area | Sphericity Assumed | 1 | 48.347 | .000 |
| | Greenhouse-Geisser | 1.000 | 48.347 | .000 |
| | Huynh-Feldt | 1.000 | 48.347 | .000 |
| | Lower-bound | 1.000 | 48.347 | .000 |
| Error(Area) | Sphericity Assumed | 34 | | |
| | Greenhouse-Geisser | 34.000 | | |
| | Huynh-Feldt | 34.000 | | |
| | Lower-bound | 34.000 | | |
| Device * Area | Sphericity Assumed | 2 | 1.562 | .217 |
| | Greenhouse-Geisser | 1.609 | 1.562 | .221 |
| | Huynh-Feldt | 1.676 | 1.562 | .220 |
| | Lower-bound | 1.000 | 1.562 | .220 |
| Error(Device*Area) | Sphericity Assumed | 68 | | |
| | Greenhouse-Geisser | 54.695 | | |
| | Huynh-Feldt | 57.001 | | |
| | Lower-bound | 34.000 | | |

Table A5.53: Two-way RM ANOVA within subjects effects table (Device X Area (vertical vs horizontal) on Gesture Size) between devices for Gesturing

| Source | | df | F | Sig. |
|---|---|---|---|---|
| Device | Sphericity Assumed | 2 | 81.219 | .000 |
| | Greenhouse-Geisser | 1.266 | 81.219 | .000 |
| | Huynh-Feldt | 1.293 | 81.219 | .000 |
| | Lower-bound | 1.000 | 81.219 | .000 |
| Error(Device) | Sphericity Assumed | 68 | | |
| | Greenhouse-Geisser | 43.047 | | |
| | Huynh-Feldt | 43.950 | | |
| | Lower-bound | 34.000 | | |
| Area | Sphericity Assumed | 1 | 4.165 | .049 |
| | Greenhouse-Geisser | 1.000 | 4.165 | .049 |
| | Huynh-Feldt | 1.000 | 4.165 | .049 |
| | Lower-bound | 1.000 | 4.165 | .049 |
| Error(Area) | Sphericity Assumed | 34 | | |
| | Greenhouse-Geisser | 34.000 | | |
| | Huynh-Feldt | 34.000 | | |
| | Lower-bound | 34.000 | | |
| Device * Area | Sphericity Assumed | 2 | 4.116 | .021 |
| | Greenhouse-Geisser | 1.464 | 4.116 | .033 |
| | Huynh-Feldt | 1.513 | 4.116 | .032 |
| | Lower-bound | 1.000 | 4.116 | .050 |
| Error(Device*Area) | Sphericity Assumed | 68 | | |
| | Greenhouse-Geisser | 49.764 | | |
| | Huynh-Feldt | 51.443 | | |
| | Lower-bound | 34.000 | | |

Table A5.54: Two-way RM ANOVA within subjects effects table (Device X Area (middle vs edges) on Gesture Size) between devices for Gesturing

| Source | | df | F | Sig. |
|---|---|---|---|---|
| Device | Sphericity Assumed | 2 | 62.533 | .000 |
| | Greenhouse-Geisser | 1.239 | 62.533 | .000 |
| | Huynh-Feldt | 1.263 | 62.533 | .000 |
| | Lower-bound | 1.000 | 62.533 | .000 |
| Error(Device) | Sphericity Assumed | 68 | | |
| | Greenhouse-Geisser | 42.133 | | |
| | Huynh-Feldt | 42.937 | | |
| | Lower-bound | 34.000 | | |
| Area | Sphericity Assumed | 1 | 20.981 | .000 |
| | Greenhouse-Geisser | 1.000 | 20.981 | .000 |
| | Huynh-Feldt | 1.000 | 20.981 | .000 |
| | Lower-bound | 1.000 | 20.981 | .000 |
| Error(Area) | Sphericity Assumed | 34 | | |
| | Greenhouse-Geisser | 34.000 | | |
| | Huynh-Feldt | 34.000 | | |
| | Lower-bound | 34.000 | | |
| Device * Area | Sphericity Assumed | 2 | 4.965 | .010 |
| | Greenhouse-Geisser | 1.363 | 4.965 | .021 |
| | Huynh-Feldt | 1.400 | 4.965 | .020 |
| | Lower-bound | 1.000 | 4.965 | .033 |
| Error(Device*Area) | Sphericity Assumed | 68 | | |
| | Greenhouse-Geisser | 46.345 | | |
| | Huynh-Feldt | 47.617 | | |
| | Lower-bound | 34.000 | | |

Table A5.55: Two-way RM ANOVA within subjects effects table (Device X Area (middle vs edges) on Gesture Velocity) between devices for Gesturing

| Source | | df | F | Sig. |
|---|---|---|---|---|
| Device | Sphericity Assumed | 2 | 67.768 | .000 |
| | Greenhouse-Geisser | 1.274 | 67.768 | .000 |
| | Huynh-Feldt | 1.302 | 67.768 | .000 |
| | Lower-bound | 1.000 | 67.768 | .000 |
| Error(Device) | Sphericity Assumed | 68 | | |
| | Greenhouse-Geisser | 43.332 | | |
| | Huynh-Feldt | 44.266 | | |
| | Lower-bound | 34.000 | | |
| Area | Sphericity Assumed | 1 | 15.281 | .000 |
| | Greenhouse-Geisser | 1.000 | 15.281 | .000 |
| | Huynh-Feldt | 1.000 | 15.281 | .000 |
| | Lower-bound | 1.000 | 15.281 | .000 |
| Error(Area) | Sphericity Assumed | 34 | | |
| | Greenhouse-Geisser | 34.000 | | |
| | Huynh-Feldt | 34.000 | | |
| | Lower-bound | 34.000 | | |
| Device * Area | Sphericity Assumed | 2 | 14.099 | .000 |
| | Greenhouse-Geisser | 1.670 | 14.099 | .000 |
| | Huynh-Feldt | 1.746 | 14.099 | .000 |
| | Lower-bound | 1.000 | 14.099 | .001 |
| Error(Device*Area) | Sphericity Assumed | 68 | | |
| | Greenhouse-Geisser | 56.778 | | |
| | Huynh-Feldt | 59.364 | | |
| | Lower-bound | 34.000 | | |

Table A5.56: Two-way RM ANOVA within subjects effects table (Device X Area (vertical vs horizontal) on Gesture Velocity) between devices for Gesturing

# A6
## Touch Typing Study

This annex presents the main results to the statistics procedures applied in the touch typing study analysis. Includes normality assessments and null-hypothesis statistical testing (NHST) main SPSS tables.

# A6.1. Touch Methods' Performance Analysis

**Tests of Normality**

| | Method | Kolmogorov-Smirnov[a] | | | Shapiro-Wilk | | |
|---|---|---|---|---|---|---|---|
| | | Statistic | df | Sig. | Statistic | df | Sig. |
| WPM | *VoiceOver | ,095 | 55 | ,200[*] | ,968 | 55 | ,148 |
| | MT VoiceOver | ,098 | 58 | ,200[*] | ,983 | 58 | ,594 |
| | Navtouch | ,096 | 60 | ,200[*] | ,977 | 60 | ,306 |
| | BrailleTouch | ,075 | 60 | ,200[*] | ,986 | 60 | ,700 |
| MSD_Error | *VoiceOver | ,091 | 55 | ,200[*] | ,937 | 55 | ,006 |
| | MT VoiceOver | ,102 | 58 | ,200[*] | ,924 | 58 | ,001 |
| | Navtouch | ,166 | 60 | ,000 | ,830 | 60 | ,000 |
| | BrailleTouch | ,193 | 60 | ,000 | ,789 | 60 | ,000 |

*. This is a lower bound of the true significance.

a. Lilliefors Significance Correction

Table A6.1: Normality tests for WPM and MSD Error Rate metrics

## A6.1.1. Words Per Minute Analysis

**Tests of Within-Subjects Effects**

Measure: MEASURE_1

| Source | | df | Mean Square | F | Sig. |
|---|---|---|---|---|---|
| metodos | Sphericity Assumed | 3 | 4.533 | 41.002 | .000 |
| | Greenhouse-Geisser | 2.658 | 5.117 | 41.002 | .000 |
| | Huynh-Feldt | 2.800 | 4.857 | 41.002 | .000 |
| | Lower-bound | 1.000 | 13.599 | 41.002 | .000 |
| Error(metodos) | Sphericity Assumed | 171 | .111 | | |
| | Greenhouse-Geisser | 151.485 | .125 | | |
| | Huynh-Feldt | 159.583 | .118 | | |
| | Lower-bound | 57.000 | .332 | | |

Table A6.2: Null-hypothesis Statistical Testing (non-parametric ) for WPM

**Pairwise Comparisons**

Measure: MEASURE_1

| (I) Method | (J) Method | Mean Difference (I-J) | Std. Error | Sig.[b] | 95% Confidence Interval for Difference[b] | |
|---|---|---|---|---|---|---|
| | | | | | Lower Bound | Upper Bound |
| 1 | 2 | ,111 | ,065 | ,547 | -,066 | ,288 |
| | 3 | ,374[*] | ,072 | ,000 | ,178 | ,569 |
| | 4 | ,625[*] | ,065 | ,000 | ,448 | ,801 |
| 2 | 1 | -,111 | ,065 | ,547 | -,288 | ,066 |
| | 3 | ,263[*] | ,064 | ,001 | ,086 | ,439 |
| | 4 | ,514[*] | ,051 | ,000 | ,373 | ,654 |
| 3 | 1 | -,374[*] | ,072 | ,000 | -,569 | -,178 |
| | 2 | -,263[*] | ,064 | ,001 | -,439 | -,086 |
| | 4 | ,251[*] | ,051 | ,000 | ,112 | ,390 |
| 4 | 1 | -,625[*] | ,065 | ,000 | -,801 | -,448 |
| | 2 | -,514[*] | ,051 | ,000 | -,654 | -,373 |
| | 3 | -,251[*] | ,051 | ,000 | -,390 | -,112 |

Based on estimated marginal means

*. The mean difference is significant at the ,05 level.

b. Adjustment for multiple comparisons: Bonferroni.

Table A6.3: Post-hoc tests for WPM

## A6.1.2. MSD Error Rate Analysis

**Test Statistics[a]**

| N | 58 |
|---|---|
| Chi-Square | 16,265 |
| df | 3 |
| Asymp. Sig. | ,001 |

a. Friedman Test

Table A6.4: Null-hypothesis Statistical Testing (non-parametric ) for MSD Error Rate

**Test Statistics[a]**

|   | NavTouch_M SD - BrailleTouch_ MSD | mtVoiceOver_ MSD - BrailleTouch_ MSD | VoiceOver_M SD - BrailleTouch_ MSD | VoiceOver_W PM - NavTouch_M SD |
|---|---|---|---|---|
| Z | -1,025[b] | -3,687[b] | -3,616[b] | -5,234[c] |
| Asymp. Sig. (2-tailed) | ,306 | ,000 | ,000 | ,000 |

**Test Statistics[a]**

|   | NavTouch_M SD - mtVoiceOver_ WPM |
|---|---|
| Z | -5,168[b] |
| Asymp. Sig. (2-tailed) | ,000 |

a. Wilcoxon Signed Ranks Test

b. Based on negative ranks.

c. Based on positive ranks.

Table A6.5: Post-hoc tests for MSD Error Rate

# A6.2.  Individual Abilities and Methods Analysis

## A6.2.1.  Age-related attributes

### Age of Blindness Onset

**Onset ~ WPM**

**ANOVA**

| | | Sum of Squares | df | F | Sig. |
|---|---|---|---|---|---|
| VoiceOver_WPM | Between Groups | 5.158 | 2 | 6.096 | .004 |
| | Within Groups | 24.116 | 57 | | |
| | Total | 29.274 | 59 | | |
| mtVoiceOver_WPM | Between Groups | 2.150 | 2 | 5.305 | .008 |
| | Within Groups | 11.146 | 55 | | |
| | Total | 13.295 | 57 | | |
| NavTouch_WPM | Between Groups | 1.394 | 2 | 2.426 | .097 |
| | Within Groups | 17.817 | 62 | | |
| | Total | 19.211 | 64 | | |
| BrailleTouch_WPM | Between Groups | .757 | 2 | 2.117 | .129 |
| | Within Groups | 11.080 | 62 | | |
| | Total | 11.837 | 64 | | |

**Multiple Comparisons**

Tukey HSD

| Dependent Variable | (I) Onset (Binned) | (J) Onset (Binned) | Mean Difference (I-J) | Std. Error | Sig. |
|---|---|---|---|---|---|
| VoiceOver_WPM | <= 5 | 6 - 20 | -.69205* | .20323 | .003 |
| | | 21+ | -.47694 | .20595 | .062 |
| | 6 - 20 | <= 5 | .69205* | .20323 | .003 |
| | | 21+ | .21511 | .20838 | .560 |
| | 21+ | <= 5 | .47694 | .20595 | .062 |
| | | 6 - 20 | -.21511 | .20838 | .560 |
| mtVoiceOver_WPM | <= 5 | 6 - 20 | -.44892* | .14421 | .008 |
| | | 21+ | -.10947 | .14605 | .735 |
| | 6 - 20 | <= 5 | .44892* | .14421 | .008 |
| | | 21+ | .33945 | .14421 | .057 |
| | 21+ | <= 5 | .10947 | .14605 | .735 |
| | | 6 - 20 | -.33945 | .14421 | .057 |
| NavTouch_WPM | <= 5 | 6 - 20 | -.34273 | .15944 | .088 |
| | | 21+ | -.21923 | .16180 | .371 |
| | 6 - 20 | <= 5 | .34273 | .15944 | .088 |
| | | 21+ | .12350 | .17174 | .753 |
| | 21+ | <= 5 | .21923 | .16180 | .371 |
| | | 6 - 20 | -.12350 | .17174 | .753 |
| BrailleTouch_WPM | <= 5 | 6 - 20 | -.25442 | .12573 | .115 |
| | | 21+ | -.06745 | .12759 | .858 |
| | 6 - 20 | <= 5 | .25442 | .12573 | .115 |
| | | 21+ | .18697 | .13543 | .357 |
| | 21+ | <= 5 | .06745 | .12759 | .858 |
| | | 6 - 20 | -.18697 | .13543 | .357 |

*. The mean difference is significant at the 0.05 level.

### Onset ~ MSD Error Rate

**Test Statistics[a,b]**

|  | VoiceOver_ MSD | mtVoiceOver _MSD | NavTouch_M SD | BrailleTouch _MSD |
|---|---|---|---|---|
| Chi-Square | 8.524 | 2.562 | 1.456 | 4.022 |
| df | 2 | 2 | 2 | 2 |
| Asymp. Sig. | .014 | .278 | .483 | .134 |

a. Kruskal Wallis Test

b. Grouping Variable: Onset (Binned)

## Post-hoc (<= 5 vs 6-20)

**Ranks**

| | Onset (Binned) | N | Mean Rank | Sum of Ranks |
|---|---|---|---|---|
| VoiceOver_MSD | <= 5 | 26 | 25.37 | 659.50 |
| | 6 - 20 | 20 | 21.08 | 421.50 |
| | Total | 46 | | |

**Test Statistics[a]**

| | VoiceOver_ MSD |
|---|---|
| Mann-Whitney U | 211.500 |
| Wilcoxon W | 421.500 |
| Z | -1.075 |
| Asymp. Sig. (2-tailed) | .282 |

a. Grouping Variable: Onset (Binned)

## Post-hoc (<= 5 vs 21+)

**Ranks**

| | Onset (Binned) | N | Mean Rank | Sum of Ranks |
|---|---|---|---|---|
| VoiceOver_MSD | <= 5 | 26 | 27.52 | 715.50 |
| | 21+ | 19 | 16.82 | 319.50 |
| | Total | 45 | | |

**Test Statistics[a]**

| | VoiceOver_ MSD |
|---|---|
| Mann-Whitney U | 129.500 |
| Wilcoxon W | 319.500 |
| Z | -2.702 |
| Asymp. Sig. (2-tailed) | .007 |

a. Grouping Variable: Onset (Binned)

# A6.2.2. Tactile Abilities

## Pressure Sensitivity

### Pressure Sensitivity ~ WPM

**ANOVA**

|  |  | df | Mean Square | F | Sig. |
|---|---|---|---|---|---|
| VoiceOver_WPM | Between Groups | 1 | .623 | 1.260 | .266 |
|  | Within Groups | 58 | .494 |  |  |
|  | Total | 59 |  |  |  |
| mtVoiceOver_WPM | Between Groups | 1 | 2.271 | 11.538 | .001 |
|  | Within Groups | 56 | .197 |  |  |
|  | Total | 57 |  |  |  |
| NavTouch_WPM | Between Groups | 1 | .029 | .095 | .759 |
|  | Within Groups | 63 | .304 |  |  |
|  | Total | 64 |  |  |  |
| BrailleTouch_WPM | Between Groups | 1 | .189 | 1.020 | .316 |
|  | Within Groups | 63 | .185 |  |  |
|  | Total | 64 |  |  |  |

### Pressure Sensitivity ~ MSD Error Rate

**Test Statistics[a]**

|  | VoiceOver_MSD | mtVoiceOver_MSD | NavTouch_MSD | BrailleTouch_MSD |
|---|---|---|---|---|
| Mann-Whitney U | 427.500 | 434.500 | 324.000 | 363.500 |
| Wilcoxon W | 752.500 | 759.500 | 649.000 | 688.500 |
| Z | -.978 | -.570 | -1.718 | -1.115 |
| Asymp. Sig. (2-tailed) | .328 | .569 | .086 | .265 |

a. Grouping Variable: Pressure (Binned)

# A6.2.3. Cognitive Abilities

## Spatial Ability

### Spatial Ability ~ WPM

**ANOVA**

| | | df | Mean Square | F | Sig. |
|---|---|---|---|---|---|
| VoiceOver_WPM | Between Groups | 2 | 1.969 | 4.429 | .016 |
| | Within Groups | 57 | .445 | | |
| | Total | 59 | | | |
| mtVoiceOver_WPM | Between Groups | 2 | 1.766 | 9.949 | .000 |
| | Within Groups | 55 | .178 | | |
| | Total | 57 | | | |
| NavTouch_WPM | Between Groups | 2 | .215 | .709 | .496 |
| | Within Groups | 62 | .303 | | |
| | Total | 64 | | | |
| BrailleTouch_WPM | Between Groups | 2 | .172 | .930 | .400 |
| | Within Groups | 62 | .185 | | |
| | Total | 64 | | | |

### Post Hoc Tests

Tukey HSD

| Dependent Variable | (I) SpatialHab (Binned) | (J) SpatialHab (Binned) | Std. Error | Sig. |
|---|---|---|---|---|
| VoiceOver_WPM | <= 4,75 | 4,76 - 7,00 | .21110 | .988 |
| | | 7,01+ | .20831 | .026 |
| | 4,76 - 7,00 | <= 4,75 | .21110 | .988 |
| | | 7,01+ | .21359 | .043 |
| | 7,01+ | <= 4,75 | .20831 | .026 |
| | | 4,76 - 7,00 | .21359 | .043 |
| mtVoiceOver_WPM | <= 4,75 | 4,76 - 7,00 | .13340 | .651 |
| | | 7,01+ | .13533 | .000 |
| | 4,76 - 7,00 | <= 4,75 | .13340 | .651 |
| | | 7,01+ | .13858 | .004 |
| | 7,01+ | <= 4,75 | .13533 | .000 |
| | | 4,76 - 7,00 | .13858 | .004 |
| NavTouch_WPM | <= 4,75 | 4,76 - 7,00 | .16612 | .481 |
| | | 7,01+ | .16370 | .964 |
| | 4,76 - 7,00 | <= 4,75 | .16612 | .481 |
| | | 7,01+ | .17633 | .671 |
| | 7,01+ | <= 4,75 | .16370 | .964 |
| | | 4,76 - 7,00 | .17633 | .671 |
| BrailleTouch_WPM | <= 4,75 | 4,76 - 7,00 | .12994 | .422 |
| | | 7,01+ | .12805 | .997 |
| | 4,76 - 7,00 | <= 4,75 | .12994 | .422 |
| | | 7,01+ | .13792 | .505 |
| | 7,01+ | <= 4,75 | .12805 | .997 |
| | | 4,76 - 7,00 | .13792 | .505 |

**Spatial Ability ~ MSD Error Rate**

**Test Statistics[a,b]**

|  | VoiceOver_MSD | mtVoiceOver_MSD | NavTouch_MSD | BrailleTouch_MSD |
|---|---|---|---|---|
| Chi-Square | .634 | 12.353 | 4.417 | 1.704 |
| df | 2 | 2 | 2 | 2 |
| Asymp. Sig. | .728 | .002 | .110 | .427 |

a. Kruskal Wallis Test

b. Grouping Variable: SpatialHab (Binned)

# Post-hoc (<= 4.75 vs 4.76-7.00)

**Test Statistics[a]**

|  | mtVoiceOver_MSD |
|---|---|
| Mann-Whitney U | 141.500 |
| Wilcoxon W | 492.500 |
| Z | -2.427 |
| Asymp. Sig. (2-tailed) | .015 |

a. Grouping Variable: SpatialHab (Binned)

# Post-hoc (4.76-7.00 vs 7.01+)

**Test Statistics[a]**

|  | mtVoiceOver_MSD |
|---|---|
| Mann-Whitney U | 51.000 |
| Wilcoxon W | 222.000 |
| Z | -3.649 |
| Asymp. Sig. (2-tailed) | .000 |
| Exact Sig. [2*(1-tailed Sig.)] | .000[b] |

a. Grouping Variable: SpatialHab (Binned)

b. Not corrected for ties.

# Verbal IQ

**ANOVA**

|  |  | df | Mean Square | F | Sig. |
|---|---|---|---|---|---|
| VoiceOver_WPM | Between Groups | 2 | 1.930 | 4.329 | .018 |
|  | Within Groups | 57 | .446 |  |  |
|  | Total | 59 |  |  |  |
| mtVoiceOver_WPM | Between Groups | 2 | 1.362 | 7.084 | .002 |
|  | Within Groups | 55 | .192 |  |  |
|  | Total | 57 |  |  |  |
| NavTouch_WPM | Between Groups | 2 | 1.015 | 3.663 | .031 |
|  | Within Groups | 62 | .277 |  |  |
|  | Total | 64 |  |  |  |
| BrailleTouch_WPM | Between Groups | 2 | 1.076 | 6.887 | .002 |
|  | Within Groups | 62 | .156 |  |  |
|  | Total | 64 |  |  |  |

**Multiple Comparisons**

Tukey HSD

| Dependent Variable | (I) VerbalIQ (Binned) | (J) VerbalIQ (Binned) | Std. Error | Sig. |
|---|---|---|---|---|
| VoiceOver_WPM | <= 84 | 85 - 104 | .29567 | .041 |
|  |  | 105+ | .18816 | .096 |
|  | 85 - 104 | <= 84 | .29567 | .041 |
|  |  | 105+ | .31081 | .527 |
|  | 105+ | <= 84 | .18816 | .096 |
|  |  | 85 - 104 | .31081 | .527 |
| mtVoiceOver_WPM | <= 84 | 85 - 104 | .19505 | .105 |
|  |  | 105+ | .12497 | .002 |
|  | 85 - 104 | <= 84 | .19505 | .105 |
|  |  | 105+ | .20408 | .979 |
|  | 105+ | <= 84 | .12497 | .002 |
|  |  | 85 - 104 | .20408 | .979 |
| NavTouch_WPM | <= 84 | 85 - 104 | .18260 | .067 |
|  |  | 105+ | .14835 | .102 |
|  | 85 - 104 | <= 84 | .18260 | .067 |
|  |  | 105+ | .19761 | .850 |
|  | 105+ | <= 84 | .14835 | .102 |
|  |  | 85 - 104 | .19761 | .850 |
| BrailleTouch_WPM | <= 84 | 85 - 104 | .13710 | .069 |
|  |  | 105+ | .11138 | .003 |
|  | 85 - 104 | <= 84 | .13710 | .069 |
|  |  | 105+ | .14836 | .858 |
|  | 105+ | <= 84 | .11138 | .003 |
|  |  | 85 - 104 | .14836 | .858 |

## Verbal IQ ~ MSD Error Rate

**Test Statistics[a,b]**

|  | VoiceOver_MSD | mtVoiceOver_MSD | NavTouch_MSD | BrailleTouch_MSD |
|---|---|---|---|---|
| Chi-Square | 1.230 | 12.561 | 6.811 | 3.073 |
| df | 2 | 2 | 2 | 2 |
| Asymp. Sig. | .541 | .002 | .033 | .215 |

a. Kruskal Wallis Test

b. Grouping Variable: VerbalIQ (Binned)

# A6.2.4. Functional Abilities

## Personal Computer Typing Experience

### PC Experience ~ WPM

**ANOVA**

VoiceOver_WPM

|  | Sum of Squares | df | Mean Square | F | Sig. |
|---|---|---|---|---|---|
| Between Groups | .342 | 2 | .171 | .414 | .663 |
| Within Groups | 21.455 | 52 | .413 |  |  |
| Total | 21.797 | 54 |  |  |  |

## PC Experience ~ MSD Error Rate

**Test Statistics[a,b]**

|  | VoiceOver_ MSD |
|---|---|
| Chi-Square | 1.572 |
| df | 2 |
| Asymp. Sig. | .456 |

a. Kruskal Wallis Test

b. Grouping Variable: pcExp (Binned)

## Braille Experience

### Braille Reading ~ WPM

**ANOVA**

BrailleTouch_WPM

|  | Sum of Squares | df | Mean Square | F | Sig. |
|---|---|---|---|---|---|
| Between Groups | 1.015 | 2 | .507 | 3.602 | .034 |
| Within Groups | 8.029 | 57 | .141 |  |  |
| Total | 9.044 | 59 |  |  |  |

**Post Hoc Tests**

**Multiple Comparisons**

Dependent Variable: BrailleTouch_WPM

Tukey HSD

| (I) BrailleRead (Binned) | (J) BrailleRead (Binned) | Std. Error | Sig. |
|---|---|---|---|
| <= 9,00 | 9,01 - 49,41 | .11363 | .028 |
| | 49,42+ | .12588 | .224 |
| 9,01 - 49,41 | <= 9,00 | .11363 | .028 |
| | 49,42+ | .12113 | .737 |
| 49,42+ | <= 9,00 | .12588 | .224 |
| | 9,01 - 49,41 | .12113 | .737 |

# Braille Reading ~ MSD Error Rate

## Test Statistics[a,b]

| | BrailleTouch_MSD |
|---|---|
| Chi-Square | 3.353 |
| df | 2 |
| Asymp. Sig. | .187 |

a. Kruskal Wallis Test

b. Grouping Variable: BrailleRead (Binned)

## Braille Writing ~ WPM

**ANOVA**

BrailleTouch_WPM

| | Sum of Squares | df | Mean Square | F | Sig. |
|---|---|---|---|---|---|
| Between Groups | .101 | 2 | .050 | .322 | .726 |
| Within Groups | 8.943 | 57 | .157 | | |
| Total | 9.044 | 59 | | | |

# Braille Writing ~ MSD Error Rate

## Test Statistics[a,b]

| | BrailleTouch_MSD |
|---|---|
| Chi-Square | 2.319 |
| df | 2 |
| Asymp. Sig. | .314 |

a. Kruskal Wallis Test

b. Grouping Variable: BrailleWrite (Binned)

# Mobile Keypad Typing Experience

## Mobile Experience ~ WPM

**ANOVA**

mtVoiceOver_WPM

| | Sum of Squares | df | Mean Square | F | Sig. |
|---|---|---|---|---|---|
| Between Groups | .022 | 2 | .011 | .074 | .929 |
| Within Groups | 6.799 | 45 | .151 | | |
| Total | 6.822 | 47 | | | |

## Mobile Experience ~ MSD Error Rate

**Test Statistics[a,b]**

| | mtVoiceOver MSD |
|---|---|
| Chi-Square | 2.535 |
| df | 2 |
| Asymp. Sig. | .281 |

a. Kruskal Wallis Test

b. Grouping Variable: MobileExp (Binned)